\documentclass[12pt]{article}
\usepackage[inner=2.5cm,outer=2.5cm,bottom=2.5cm]{geometry}
\usepackage{amssymb}
\usepackage{amsthm} 
\usepackage{graphicx,psfrag,epsf,epstopdf}
\usepackage{enumerate}
\usepackage{siunitx}
\usepackage{natbib} 
\usepackage{amsmath,amsfonts,amssymb,amsthm,algorithm}
\usepackage{bm}    
\usepackage{threeparttable}   
\usepackage{tcolorbox}    
\usepackage{color,float}    
\usepackage{multirow}   
\usepackage{caption}
\usepackage{subcaption}
\usepackage[toc,page]{appendix}        
\usepackage{hyperref}
\hypersetup{     
	colorlinks=true,
	linkcolor=blue,
	citecolor=blue 
}
\usepackage{setspace}
\date{February 1, 2023}
 
\usepackage[figuresright]{rotating}

\newtheorem{theorem}{Theorem} 
     
\newtheorem{definition}{Definition}

\newtheorem{lemma}{Lemma}

\newcounter{examplecounter}

\def \R {\mathbb{R}}

\def\1{1\!{\rm l}}

\author{
  Yunyun Wang\footnote{
    Department of Econometrics and Business Statistics, Monash University    
    (\href{mailto:yunyun.wang@monash.edu}{yunyun.wang@monash.edu})} 
  \and
  Tatsushi Oka\footnote{ 
    AI Lab, CyberAgent
    (\href{mailto:oka\_tatsushi@cyberagent.co.jp}{oka\_tatsushi@cyberagent.co.jp})
  } 
  \and 
  Dan Zhu\footnote{
    Department of Econometrics and Business Statistics, Monash University    
    (\href{mailto:dan.zhu@monash.edu}{dan.zhu@monash.edu})} 
  \vspace{0.5cm}
}

\title{
  \textbf{\LARGE Distributional Vector Autoregression: \\Eliciting Macro and Financial Dependence}\thanks{
    We are grateful to 
    Catherine Forbes,     
    Jiti Gao,
    Kostas Kardaras,    
    Seojeong Lee, 
    Myung Hwan Seo,
    Qiwei Yao,     
    and the other participants who attended the seminars held at
    the London School of Economics,
    Queen Mary University of London,
    Seoul National University,    
    University of New South Wales,
    the University of Queensland, as well as the Time Series and Forecasting Symposium (TSF2022), for their valuable comments and suggestions.}
}

\begin{document}

\maketitle
\thispagestyle{empty}

\begin{abstract}
  Vector autoregression is an essential tool in empirical macroeconomics and finance
  for understanding the dynamic
  interdependencies among multivariate time series.
  In this study, we expand the scope of vector autoregression
  by 
  incorporating a multivariate distributional regression framework and introducing a distributional impulse response function,
  providing a comprehensive view of dynamic heterogeneity.
  We propose a straightforward yet flexible estimation method 
  and establish its asymptotic properties
  under weak dependence assumptions.
  Our empirical analysis examines the conditional joint distribution of GDP growth and financial conditions in the United States, with
  a focus on 
  the global financial crisis.
  Our results show that 
  tight financial conditions lead to a multimodal conditional joint distribution of GDP growth and financial conditions,
  and 
  easing financial conditions significantly impacts long-term GDP growth,
  while improving the GDP growth during the global financial crisis has limited effect on financial conditions.   
\end{abstract}
 
\vspace{0.5cm}
\noindent%
 \vspace{0.2cm} 

\clearpage
\setcounter{page}{1}
\setstretch{1.5}
\section{Introduction}

Since the seminal work of \cite{sims1980macroeconomics}, vector autoregression (VAR) has emerged as an essential tool in empirical macroeconomics and finance to facilitate the basic quantitative description, forecasting, and structural analysis of multivariate time series  \citep[see][]{litterman1986forecasting,stock2001vector}.
The standard VAR is built upon the mean regression for multivariate systems, often with multivariate Gaussian errors. It enables insightful structural analysis, most notably through impulse response functions (IRFs). However, sharp macroeconomic downturns triggered by the recent financial crisis and the pandemic have led to increasing interest in exploring the distributional features of multivariate time series. This line of research moves beyond the traditional focus on mean estimation and studies the the distributional effect of a shock, such as the growth-at-risk of economic activity.

This study proposes a semiparametric distributional VAR model that serves as a flexible alternative for analyzing the distributional properties of multivariate time series.
We introduce an estimation method that combines 
the distribution factorization and the distributional regression (DR) approach.
Unlike traditional parametric models, our approach does not impose a global parametric assumption on either the marginal or joint distributions of the variables, conditional on their past values.
The framework employs regression models that can incorporate a moderately large number of conditional variables, capturing the influence of past events on the entire response distribution. Additionally, we introduce a distributional counterpart to the commonly used IRFs to examine how the conditional distributions evolve after a perturbation in the distribution of a variable in the system at a specific point in time. This enables us to identify the heterogeneity and nonlinearity in the response dynamics.

Our work builds on and contributes to several strands of literature.
Firstly, the fundamental basis of our estimation method is DR, which is a semiparametric method for marginal conditional distributions. \cite{williams1972analysis} first introduced DR
to analyze ordered categorical outcomes using multiple binary regressions.
It was later extended by \citet{foresi1995conditional} to characterize any conditional distribution, and a local version was proposed by \citet{hall1999methods}.
\citet{chernozhukov2013inference} established the uniform validity of the inference for the entire conditional distribution.
The DR approach has also been explored by \citet{rothe2013misspecification}  and \citet{chernozhukov2020network}, among others.
More recently, the DR method has been extended to the conditional multivariate distributions of 
independent cross-sectional data. 
\cite{meier2020multivariate} considered the direct application of the DR method to estimate the joint conditional distribution,
while \cite{wang2022bivariate} proposed a method based on the DR and distribution factorization, which addressed a possible computational issue arising from
the direct DR application.
Our study further extends the scope of the DR approach
to analyze the multivariate time-series data
and 
their dynamic interdependencies.
We also provide asymptotic properties of the proposed estimator under the $\beta$-mixing condition.   

Our paper also contributes to the literature on
the structural vector autoregression (SVAR) model.
Structural interpretations of VAR models require additional identifying assumptions based on institutional knowledge, economic theory, or other external constraints on the model responses \citep{blanchard1989dynamic,rubio2010structural,kilian2017structural}. A commonly used identification scheme in the VAR literature is to assume a lower triangular form for the contemporaneous variance matrix of the endogenous variables \citep{sims1980macroeconomics,primiceri2005time}.
Our study emlopyes a similar structural identification scheme based on a triangular assumption. 
When estimating the multiperiod conditional forecasting distributions for impulse responses analysis, we adopt the concepts of local projection \citep{jorda2005estimation} and direct multistep forecasting \citep{mccracken2019empirical}.
This approach allows us to estimate a distinct multivariate distribution at each forecast horizon, rather than using an iterative forecast.
Recent research by \cite{plagborg2021local} has shown that the local projection and VAR models are conceptually equivalent in estimating the impulse response functions.

This paper also contributes to the growing literature on extending the quantile regression framework to VAR models, building on previous works by \cite{koenker1978regression}
and \cite{koenker2004unit, koenker2006quantile}.
More specifically, 
\cite{white2015var} developed a multivariate autoregression model of the quantiles to directly study the degree of tail interdependence among multivariate time series. \cite{montes2019multivariate} suggested a reduced form quantile VAR model based on the directional quantiles framework and estimated a quantile impulse response function (QIRF) to explore the dynamic effects for a fixed collection of quantile indices. \cite{chavleishvili2019forecasting} proposed a quantile SVAR model and analyzed QIRFs based on fixed sample paths.
Our proposed approach differs from existing Quantile VAR models in that we target the joint conditional distribution, while Quantile VAR is modeled by a system of
quantile-regression equations,  
similar to mean VAR models.
In some applications, our approach provides a straightforward way to analyze joint distributional features and dynamic propagation of shocks on the entire joint distribution conditional on past events.
However, it should be noted that the conditional quantile and distribution of a continuous random variable are equivalent up to an inverse transformation.
Therefore, both quantile and distributional VAR can extract similar information from multivariate time series and can be used depending on the research goal.

This paper applies the proposed framework to analyze the joint distribution of the real GDP growth rate and the national financial condition index (NFCI) of the United States (U.S.), conditional upon past lagged variables. Insightful work by 
\cite{adrian2021multimodality} previously studied the same dataset, using a nonparametric kernel framework to estimate the joint conditional distribution and a density IRF. The key findings from this work suggest that the joint distribution is unimodal during normal times but exhibits clear multimodality
during the Great Recession. \cite{adrian2019vulnerable} studied the distribution of GDP growth given the past financial and economic conditions using quantile regressions, suggesting that the distribution exhibits much more variation over time in the median and lower tail compared to the upper tail.
Our approach complements theirs by allowing for more lagged variables as conditional regressors, which is not possible in nonparametric approaches due to the curse of dimensionality.
Our empirical results confirm their key findings even after conditioning more lagged variables. 
We also find  multimodality only appears in short-term forecasts and resolves within a few quarters with relatively mild tightness.

We further investigate the DIRFs for the possible policy effect on the moments, quantiles, and entire distributions of all variables during the Great Recession. Our findings indicate that if the policies implemented in 2008:Q3 had been successful in preventing financial tightening in 2008:Q4, the likelihood of adverse real GDP growth and tight NFCI would have been improved in 2009:Q1-Q2 and reduced in 2009:Q3-Q4, which is consistent with the findings of \citet{adrian2021multimodality}. Additional evidence from the mixed-frequency model suggests that an impulse on the NFCI has a long-term effect on the NFCI and real GDP growth. However, our analysis of a distributional impulse on the real GDP growth suggests a different result that limiting the likelihood of negative real GDP growth in 2008:Q4 only increases the likelihood of positive economic activity in the short run but has almost no effect on the NFCI even in the subsequent quarter.

The remainder of this paper is organized as follows. Section \ref{sec: Model and DIRF} introduces the proposed multivariate model and DIRF.
Section \ref{sec: estimation-asymptotic} explains the estimation procedures and
presents the estimators' asymptotic results.
In Section \ref{sec: empirical}, we apply our approach to study the U.S. time series data on macroeconomic and financial conditions. Section \ref{sec: con} concludes the paper.
The proofs of the theoritical results and additional empirical analysis are provided in Appendix.

\section{Multivariate Distributional Regression}\label{sec: Model and DIRF}

We introduce a semiparametric regression approach for conditional multivariate distributions
and explain the DIRF, which describes the dynamic effect of a shock on the entire distribution of  multivariate time series.

In what follows, $\mathbb{R}$ denotes the the set of real numbers, and $\1\{\cdot\}$ denotes the indicator function
taking the value 1 if the condition inside $\{\cdot\}$ is satisfied and
0 otherwise. We use $\ell^{\infty}(D)$ to refer to the collection of all real-valued bounded functions defined on an arbitrary set $D$.
We denote by $\|\cdot\|$ the Euclidean norm for vectors
and use $a^\top$ to represent the transpose of a vector $a$.

\subsection{Model}
\label{subsec: modelspecification}

Suppose that we observe a stationary time series
$\{(Y_{t}, Z_{t})\}_{t=1}^{T}$
with a sample size of $T$,
where 
$Y_{t}=(Y_{1t},\ldots,Y_{Jt})^{\top}$
is a $J$-dimensional outcome variables
and $Z_{t}$ is a $k\times1$ vector
of conditioning variables.
We denote supports
$\mathcal{Y}
:=
\times_{j=1}^{J}
\mathcal{Y}_{j} 
\subset
\mathbb{R}^{J}$
and 
$\mathcal{Z}\subset\mathbb{R}^{k}$
for $Y_{t}$ and $Z_{t}$, respectively,
where $\mathcal{Y}_{j}$ denotes
the support of $Y_{jt}$.
Given the multivariate time series,
the objective is to estimate 
the conditional joint distribution
$F_{Y_{t}|Z_{t}}(y|z)$
of $Y_{t}$ 
given $Z_{t}=z$,  
for $(y,z) \in \mathcal{Y}{\times}\mathcal{Z}$.
This study mainly focuses on 
the VAR setup
wherein 
$Z_{t}$ comprises only lagged dependent variables,
while our framework
can also be applied to other cases, such as VAR models with proxy or instrumental variables \citep{bloom2009impact,jurado2015measuring}.

We propose a semiparametric estimation method for joint conditional distributions.
We first apply distribution factorization,
which expresses 
the joint conditional distribution
using a collection of  marginal conditional distributions
through a hierarchical structure,
and then estimate
conditional marginal distributions.
Specifically, we define
\[
  X_{1t}:=(1,Z_{t}^{\top})^{\top}
  \ \ \mathrm{and} \ \
  X_{jt}:=(1,Z_{t}^{\top},Y_{1t},\dots,Y_{j-1,t})^{\top}
  \ \ \mathrm{for} \ j=2,\dots, J,
\]
with supports
$\mathcal{X}_j\subset \mathbb{R}^{k+j}$
for $j=1,\dots,J$.
Let 
$F_{Y_{jt}|X_{jt}}$ be the marginal conditional distribution of $Y_{jt}$
given
$X_{jt}$. 
Subsequently, 
the distribution factorization yields 
the existence of a transformation
$\rho:
\times_{j=1}^{J} \ell^{\infty}(\mathcal{Y}_{j}{\times}\mathcal{X}_{j})
\to
\ell^{\infty}(\mathcal{Y}{\times}\mathcal{Z})
$,
given by
\begin{equation}
  \label{eq-joint}
  F_{Y_{t}|Z_{t}}=\rho(
  F_{Y_{1t}|X_{1t}},
  \dots, 
  F_{Y_{Jt}|X_{Jt}}).
\end{equation}
This expression is useful because a multivariate joint distribution 
can be obtained by separately modeling these $J$ marginal conditional distributions. 
 There are many different transformations for the distribution factorization,
each of which is mathematically valid and
relevant depending on its empirical applications.
This is because
the transformation $\rho(\cdot)$ in
(\ref{eq-joint})
depends on the ordering of $Y_{t}$ coordinates. 
Selecting a different permutation-based ordering 
yields an alternative transformation. We shall discuss this ordering issue later in Section \ref{subsec: DIRF} for the purpose of structural analysis. 

In the following example, we illustrate the transformation in the bivariate outcome case.

\vspace{0.2cm}
\paragraph{Example 1.}
\label{example1}
  Let 
  $Y_t=(Y_{1t},Y_{2t})^{\top}$
  and the
  predetermined variables
  be up to two lags
  or
  $Z_{t}= (Y_{t-1}^{\top}, Y_{t-2}^{\top})^{\top}$.
  The conditional joint distribution of $Y_{t}$ given $Z_{t}$
  is then characterized via the following distribution factorization:
  \begin{eqnarray*}
    F_{Y_{t}|Z_{t}}
    =
    \int
    F_{Y_{2t}|X_{2t}}
    dF_{Y_{1t}|X_{1t}},
  \end{eqnarray*}
  where
  $
  F_{Y_{1t}|X_{1t}}
  $
  and
  $F_{Y_{2t}|X_{2t}}$
  are conditional marginal distributions
  with
  $X_{1t}=(1, Z_{t}^{\top})^{\top}$
  and
  $X_{2t}=(1, Z_{t}^{\top}, Y_{1t})^{\top}$. \qed
\vspace{0.2cm}

When studying an univariate conditional distribution, a common method is to assume an appropriate parametric distribution based on sample information, often with the conditioning variables affecting only its location or scale parameters. However, when observations exhibit complex statistical features, such as long-tails and extreme skewness, selecting an appropriate parametric model becomes challenging. In this study, we apply the DR method, 
which does not impose restrictive global parametric assumptions.

The DR approach characterizes the entire conditional distribution of an outcome variable, conditional upon a vector of covariates, by fitting a collection of parametric linear-index models over the outcome locations. Specifically,
for the estimation of the $j$-th conditional distribution $F_{Y_{jt}|X_{jt}}$,
we consider, for any $(y_j, x_{j}) \in \mathcal{Y}_{j}{\times}\mathcal{X}_j$,
\begin{equation}\label{model-DR}
  F_{Y_{jt}|X_{jt}}(y_j|x_{j})=\Lambda\big(\phi_{j}({x_{j}})^{\top}\theta_{j}(y_j)\big),
\end{equation}
where $\Lambda: \R \to [0,1]$ is a known link function such as a logistic or probit function\footnote{In practice, for each $Y_{jt}$, one can choose different link functions, while we use the same notation for simplification. For sufficiently rich transformation of the covariates, one can approximate the conditional distribution function arbitrarily well without extra concern about the choice of the link function.}, $\phi_{j}: \mathcal{X}_j\mapsto\mathbb{R}^{d_j}$ is a transformation and $\theta_{j}(y_j)$ is a $d_j\times 1$ vector of unknown parameters specific to the location $y_j$.  The entire conditional distribution of $Y_{jt}$ is
characterised by considering different locations over the support $\mathcal{Y}_j$,
and the set of marginal conditional distributions results in a joint conditional distribution by the transformation in (\ref{eq-joint}).
The proposed model is sufficiently flexible in its ability to incorporate covariates and set regression coefficients for each outcome location.
 
\subsection{Distributional Impulse Response Function}\label{subsec: DIRF}

IRFs are standard structural analysis tools that characterize the dynamic propagation of contemporaneous shocks on multivariate time series in empirical macroeconomics and finance.
\cite{sims1980macroeconomics} originally proposed IRFs
using the moving average representation of VAR,
whereas 
\cite{jorda2005estimation} introduced the local projection approach,
which evaluates the dynamic effects of shocks under the multistep ahead forecast framework.
\cite{plagborg2021local} recently proved that the local projections and VARs estimate the same impulse responses in population.
Unlike the VAR literature, which has traditionally considered mean IFRs, the recent literature
explores the dynamic effect of a shock on the entire distribution using QIRFs \citep{montes2019multivariate,chavleishvili2019forecasting} and density IRF
\citep{adrian2021multimodality}. 

We consider a local projection approach
by integrating the conditional distribution of observable variables with respect to a counterfactual distribution to develop the DIRFs. 
The proposed approach can be viewed as a dynamic extension of 
\citet{chernozhukov2013inference},
who considered the counterfactual unconditional distributions  
for program evaluation with cross-sectional observations.
Given a non-negative integer $h$, 
the baseline joint distribution $F_{Y_{t+h}|Z_{t}}$
of $h$-ahead outcomes $Y_{t+h}$ conditional on $Z_{t}$
is written as 
\begin{eqnarray*}
  F_{Y_{t+h}|Z_{t}}
  =
  \int
  F_{Y_{t+h}|Y_{t}, Z_{t}}
  d
  F_{Y_{t}| Z_{t}},
\end{eqnarray*}
where
$F_{Y_{t}| Z_{t}}$ and $F_{Y_{t+h}|Y_{t},Z_{t}}$
are two different conditional distributions of the observed 
variables that are identified from the data and characterized in a manner similar to the proposed semiparametric approach.
When estimating $F_{Y_{t+h}|Y_{t},Z_{t}}$, the concept of local projection is adopted, that is, we estimate different models for different horizons $h$ 
by regressing $Y_{t+h}$ on $(Y_{t}, Z_{t})$ with the DR approach. 

We consider a scenario where an alternative conditional distribution,
$G_{Y_{t}| Z_{t}}$, is used instead of
the actual distribution $F_{Y_{t}| Z_{t}}$.
Throughout the paper, we assume that
the counterfactual distribution 
$G_{Y_{t}| Z_{t}}$
is supported by a subset
of $\mathcal{Y}_{t}$
for identification purposes.
Under the scenario with the distribution $G_{Y_{t}| Z_{t}}$,
the counterfactual conditional joint distribution is defined as 
\begin{eqnarray*}
  F_{Y_{t+h}|Z_{t}}^{\ast}
  :=
  \int
  F_{Y_{t+h}|Y_{t}, Z_{t}}
  d
  G_{Y_{t}| Z_{t}}. 
\end{eqnarray*}
 
In addition, the baseline and counterfactual marginal distributions of the $j$-th variable $Y_{j,t+h}$ can be defined in the similar way with
\begin{eqnarray}
      \label{eq:vvv-q}
	F_{Y_{j,t+h}|Z_{t}}
	=
	\int
	F_{Y_{j,t+h}|Y_{t}, Z_{t}}
	d
	F_{Y_{t}| Z_{t}}
  \ \ \ \ \mathrm{and} \ \ \ \ 
	F^*_{Y_{j,t+h}|Z_{t}}
	:=
	\int
    F_{Y_{j,t+h}|Y_{t}, Z_{t}}
	d
	G_{Y_{t}| Z_{t}},
\end{eqnarray}
where the conditional distribution $F_{Y_{j,t+h}|Y_{t}, Z_{t}}$ can be modeled using the univariate DR approach by regressing $Y_{j,t+h}$ on $(Y_{t}, Z_{t})$.

We consider a distributional change of only one element of $Y_{t}$,
which benefits the analysis and interpretation in empirical applications,
to set up the counterfactual joint distribution $G_{Y_{t}|Z_{t}}$.
For instance,
we replace
the actual $i$-th marginal distribution
$F_{Y_{it}|X_{it}}$
with a counterfactual marginal distribution $G_{Y_{it}|X_{it}}$;
thus, the counterfactual joint distribution is given by
$G_{Y_{t}|Z_{t}}=\rho(
F_{Y_{1t}|X_{1t}},
\dots,
G_{Y_{it}|X_{it}},
\dots,
F_{Y_{Jt}|X_{Jt}})
$
under (\ref{eq-joint}).  When applying the proposed semiparametric approach to characterize $G_{Y_{t}|Z_{t}}$, the ordering of the observables is required for distribution factorization to identify the shock. Under the SVAR setting, one standard identification scheme is the recursive short-run restriction, which assumes a lower triangular form for the contemporaneous covariance matrix of the endogenous variables \citep{sims1980macroeconomics}. Our triangular assumption is similar to that of the structural identification scheme. Other identification strategies in the SVAR literature include the long-run restriction \citep{blanchard1989dynamic}, sign restriction \citep{antolin2018narrative}, and identification using instrumental variables \citep{jurado2015measuring}.

We formally define the DIRF as follows.
\begin{definition}
	 The distributional impulse response function that describles the effect of the shock on the joint distribution of $Y_t$ after $h$ periods is defined as,
	  \begin{equation}\label{eq:DIRF-joint}
		DIR_{h}:=
		F_{Y_{t+h}|Z_{t}}^{\ast}
		-
		F_{Y_{t+h}|Z_{t}}.
	\end{equation}  	
The distribution impulse response function of the $j$-th variable after $h$ periods is
defined as 
\begin{equation}\label{eq:DIRF-marginal}
	DIR_{j,h}:=
	F_{Y_{j,t+h}|Z_{t}}^{\ast}
	-
	F_{Y_{j,t+h}|Z_{t}}.
\end{equation}  
\end{definition}
	
\vspace{0.5cm} 

The proposed framework is general in several ways.
First, standard impulse response analysis
is often conducted by evaluating the effects of a one-unit change
on $h$-ahead outcomes. This is a special case of the counterfactual scenarios in which
the counterfactual joint distribution 
$G_{Y_{t}|Z_{t}}$ can take a degenerate distribution or a point mass at one value
for a variable of interest.
Next, when estimating the DIRF, the identification restriction is only required for constructing the counterfactual distribution $G_{Y_{t}|Z_{t}}$ in order to identify the shock, but it is unnecessary for $F_{Y_{t+h}|Y_{t},Z_{t}}$. 
Finally, given distributional information, other statistics of interest, such as the mean, standard deviation, quantiles, can be easily obtained. Therefore, the proposed DIRF is 
sufficiently flexible for researchers to investigate other impulse response functions generally considered in the literature.  Specifically,
the mean IRF for the $j$-th variable is given by
\begin{align*}
  MIRF_{j,h}&:=
              \int y_{j,t+h}
              d F_{Y_{j,t+h}|Z_{t}}^{\ast}(y_{j,t+h})
              -
              \int y_{j,t+h}
              d F_{Y_{j,t+h}|Z_{t}}(y_{j,t+h}),
\end{align*}
and the $\tau$-th quantile IRF of the $j$-th element for $\tau \in (0,1)$ is given by
\begin{align*}
  QIRF_{j,h}(\tau)&:=
                    F_{Y_{j,t+h}|Z_{t}}^{\ast -1}(\tau)
                    -
                    F_{Y_{j,t+h}|Z_{t}}^{-1}(\tau),
\end{align*}
where 
$F_{Y_{j,t+h}|Z_{t}}^{\ast -1}(\cdot)$
and 
$F_{Y_{j,t+h}|Z_{t}}^{-1}(\cdot)$
are the quantile functions
as inverse of the $j$-th variable's distribution functions 
$F_{Y_{j,t+h}|Z_{t}}^{\ast}(\cdot)$
and 
$F_{Y_{j,t+h}|Z_{t}}(\cdot)$,
respectively.

\paragraph{Example 1 \textmd{(continued)}.}
We illustrate how the proposed framework works in the bivariate case. 
First, the joint baseline distribution $F_{Y_{t+h}|Z_{t}}$ is written as 
  \begin{eqnarray*}
    F_{Y_{t+h}|Z_{t}}(y_{t+h}|z_{t})
    =\int F_{Y_{t+h}|Y_t,Z_{t}}(y_{t+h}|y_{t},z_{t})
    dF_{Y_t|Z_{t}}(y_t|z_{t}),
  \end{eqnarray*}
  Letting
  $G_{Y_{2t}|X_{2t}}$ 
  be a counterfactual marginal distribution for $Y_{2t}$
  given $X_{2t}$,
  we obtain the counterfactual joint distribution
  at time $t$:
  \begin{eqnarray*}
    G_{Y_{t}|Z_{t}} = \int F_{Y_{1t}|X_{1t}} d G_{Y_{2t}|X_{2t}} .
  \end{eqnarray*}
  Then, the joint counterfactual distribution after $h$ periods
  is given by
  \begin{eqnarray*}
    F^*_{Y_{t+h}|Z_{t}}(y_{t+h}|z_{t})
    =\int
    F_{Y_{t+h}|Y_t, Z_{t}}(y_{t+h}|y_{t},z_{t})
    dG_{Y_t|Z_{t}}(y_t|z_{t}).
  \end{eqnarray*}
  The difference between $F_{Y_{t+h}|Z_{t}^*}$ and $F_{Y_{t+h}|Z_{t}}$
  leads to $DIR_{h}$ in (\ref{eq:DIRF-joint}). 
  As a special case,
  we can set the counterfactual marginal distribution to 
  be a degenerate distribution with $\Pr(Y_{2t} = y_{2t}^*|X_{2t})=1$.
  In this case, the counterfactual distribution
  can be reduced to 
  $
    \int
    F_{Y_{t+h}|Y_{1t},Y_{2t},Z_{t}}(y_{t+h}|y_{1t},y_{2t}^*, z_{t})
    dF_{Y_{1t}|X_{1t}}(y_{1t}|x_{1t}).
  $ 
  \qed

\section{Estimation and Asymptotic Properties}\label{sec: estimation-asymptotic}
We introduce the estimation procedures of the conditional distribution and the DIRF and then provide the asymptotic properties for the estimators of conditional distribution functions and their transformations. 

\subsection{Estimation}\label{subsec: estimation}
For estimating the multivariate joint distributions, the primary step is to estimate
a collection of univariate conditional distributions using the DR approach. In this study, we estimate Model (\ref{model-DR}) using a binary choice model for the binary outcome $\1\{Y_{jt}\leq y_j\}$
with $y_{j} \in \mathcal{Y}_{j}$
under the maximum likelihood framework
for each $j \in \{1, \dots, J\}$.
The estimators of the unknown parameters are defined as the
maximizer of a log-likelihood function as follows:
\begin{equation}\label{obj-estimation}
  \widehat{\theta}_{j}(y_j)
  =\arg\max_{\theta_j \in \Theta_{j}}
  \widehat{\ell}_{y,j}(\theta_j),
\end{equation}
where
$\Theta_{j} \subset \mathbb{R}^{d_j}$ is the parameter space
and the log likelihood is given by 
\[
  \widehat{\ell}_{y,j}(\theta_j)
  :=
  \frac{1}{T}\sum_{t=1}^{T}
  \big [
  \1\{Y_{jt}\leq y_j\}\ln\Lambda\big(
  \phi_{j}(X_{jt})^{\top}\theta_j\big)+\1\{Y_{jt}>y_j\}\ln\big(1-\Lambda\big(
  \phi_{j} (X_{jt})^{\top}\theta_j\big)\big)
  \big ].
\] 
The conditional distribution estimator of $Y_{jt}$ given $X_{jt}=x_{j}$ is
given by 
\begin{equation}\label{eq:estimate-DR}
  \widehat{F}_{Y_{jt}|X_{jt}}(y_j|x_{j}) :=
  \Lambda\big(\phi_{j}(x_{j})^{\top}\widehat{\theta}_{j}(y_j)\big).
\end{equation}
In practice, a sufficient number of discrete points of the support $\mathcal{Y}_j$ can be selected to obtain the estimator $\widehat{F}_{Y_{jt}|X_{jt}}(y_j|x_{j})$.
One important property is that
the map $y_{j} \mapsto F_{Y_{jt}|X_{jt}}(y_j|x_{j})$ is non-decreasing by definition. However, the estimated distribution function
$\widehat{F}_{Y_{jt}|X_{jt}}(\cdot|x_{j})$ does not necessarily satisfy monotonicity in finite samples. We monotonize the conditional distribution estimators at different locations using the rearrangement method proposed by \cite{chernozhukov2009improving}. This procedure can yield finite-sample improvement
\cite[for instance, see][]{chetverikov2018econometrics} and permit a straightforward application of the functional delta method
when transforming the estimated distributions using Hadamard differentiable maps. 

Given the estimators of marginal conditional distributions
and
the transformation in (\ref{eq-joint}),
the conditional joint distribution can then be estimated as 
\begin{equation}\label{eq: estimate-joint}
  \widehat{F}_{Y_{t}|Z_{t}}=\rho(
  \widehat{F}_{Y_{1t}|X_{1t}},
  \widehat{F}_{Y_{2t}|X_{2t}},
  \ldots\\
  \widehat{F}_{Y_{Jt}|X_{Jt}}).
\end{equation}
For different horizons $h$, we estimate the conditional joint distribution $F_{Y_{t+h}|Y_{t}, Z_{t}}$ in the similar way and denote the estimator by $\widehat{F}_{Y_{t+h}|Y_{t}, Z_{t}}$. If we consider a marginal counterfactual distribution $G_{Y_{it}|X_{it}}$ for the $i$-th element $Y_{it}$, 
the joint counterfactual distribution $G_{Y_{t}|Z_{t}}$ can be estimated with
$\widehat{G}_{Y_{t}|Z_{t}}=\rho(
\widehat{F}_{Y_{1t}|X_{1t}},
\dots,
G_{Y_{it}|X_{it}},
\dots,
\widehat{F}_{Y_{Jt}|X_{Jt}}).
$
These estimated distributions enable us to obtain the estimator
of the actual and conterfactual joint distributions:\footnote{
  Alternavily, we can estimate the actual distribution $\widehat{F}_{Y_{t+h}|Z_{t}}$ diretctly using the transformation in equation (\ref{eq: estimate-joint}). To maintain comparability
between the estimators of the actual and counterfactual distributions,
we use the same estimators for both distributions
for our empirical application, as described in the main text. 
}
\begin{eqnarray*}
  \widehat{F}_{Y_{t+h}|Z_{t}}
  =
  \int
  \widehat{F}_{Y_{t+h}|Y_{t}, Z_{t}}
  d
  \widehat{F}_{Y_{t}| Z_{t}},
  \ \ \mathrm{and} \ \ 
  \widehat{F}^*_{Y_{t+h}|Z_{t}}
  =
  \int
  \widehat{F}_{Y_{t+h}|Y_{t}, Z_{t}}
  d
  \widehat{G}_{Y_{t}| Z_{t}}. 
\end{eqnarray*}
We apply the univariate DR approach to estimate the conditional distribution $F_{Y_{j,t+h}|Y_{t}, Z_{t}}$, with the estimator denoted as $\widehat{F}_{Y_{j,t+h}|Y_{t}, Z_{t}}$. The actual and conterfactual marginal distributions of $Y_{j,t+h}$ can be estimated in a similar way by
replacing the conditional distributions
with the estimated counterparts in  (\ref{eq:vvv-q}).
Finally, the $DIR_{h}$ and $DIR_{j,h}$ can be estimated by the difference between the corresponding baseline and counterfactual distributions.

\subsection{Asymptotic Properties}\label{subsec: asym}
In this subsection, we provide the asymptotic properties of
the conditional joint distribution estimators and
the estimator of DIRFs.
The proofs of all theorems in this subsection are provided in Appendix \ref{sec: appendix-A}.

Let $\ell_{y, j}(\cdot)$ be the population log-likelihood; we
define the true parameters $\theta_{j}(y_j)$ as the
solution to the following maximization problem:
\begin{equation}\label{population-likelihood} 
  \theta_{j}(y_j)=\arg\max_{\theta_j\in \Theta_{j}}\ell_{y, j}(\theta_j).
\end{equation}
A vector of the true parameters
related to $J$ conditional marginal distributions
and
a vector of the corresponding estimators 
are respectively  given by
\[
\theta(y):=\big(\theta_{1}(y_1)^{\top},\theta_{2}(y_2)^{\top},\ldots,\theta_{J}(y_J)^{\top}\big)^{\top}
  \ \ \mathrm{and} \ \
  \widehat{\theta}(y):=
  \big(
  \widehat{\theta}_{1}(y_1)^{\top},
  \widehat{\theta}_{2}(y_2)^{\top},\ldots,\widehat{\theta}_{J}(y_J)^{\top}\big)^{\top}.
\]
Also,
let $\Theta := \times_{j=1}^{J}\Theta_{j}$
be the parameter space for $\theta(y)$ and $\widehat{\theta}(y)$.
We denote the second derivative of the population log-likelihood at the
true parameters by $H_{j}(y_{j}):=\nabla^{2} \ell_{y, j}
\big(\theta_{j}(y_j) \big)$.

The following assumptions are imposed to obtain the asymptotic results: 

\vspace{0.5cm}
\paragraph{Assumptions} 
\begin{itemize}
\item[A1.]
  The time series $\{(Y_{t}, Z_{t})\}_{t=1}^{T}$ are strictly stationary $\beta$-mixing or absolutely regular process, 
  with $\beta$-mixing coefficients $\{\beta_{k}\}$ satisfying the condition that $\sum_{k>0} \beta_k<\infty$.
  The supports $\mathcal{Y}$ and $\mathcal{Z}$
  are compact subsets of $\mathbb{R}^{J}$ and $\mathbb{R}^{k}$, respectively. 
  
\item [A2.] The link function $\Lambda(\cdot)$ is twice continuously
  differentiable with its first derivative $\lambda(\cdot)$. The log-likelihood 
  function $\theta_{j}\mapsto\widehat{\ell}_{y,j}(\theta_{j})$
  is uniformly concave for any $y_j\in\mathcal{Y}_{j}$ with  $j = 1,\ldots,J$.
  
\item [A3.] The true parameters $\theta_{j}(\cdot)$ are contained
  in the interior of the compact parameter space $\Theta_{j}$
  for every $j = 1,\ldots, J$.
   
\item[A4.] 
  The conditional density function $f_{Y_{jt}|X_{jt}}(y_j|x_{j})$
  is uniformly bounded in $\mathcal{Y}_j{\times}\mathcal{X}_j$ and continuous in $\mathcal{Y}_j$
  for every $j\in\{1,\ldots,J\}$.
\end{itemize}
\vspace{0.5cm}

Assumption A1 requires that the time series are $\beta$-mixing sequences, which allows for heteroscedasticity and serial dependence. In the theorems presented below, we establish the weak convergence of the empirical processes by utilizing the result in \cite{rio1998processus}. The requirement of this result is that $\beta$-mixing sequences satisfy $\sum_{k>0} \beta_k<\infty$, which is a weaker condition than the one considered in \cite{arcones1994central}, where it is required that $\beta_{k} = O(k^{-c})$ for some $c>1$.
Furthermore, to obtain the limit processes, a compact support is required, which can be satisfied in our empirical application.

Assumption A2 ensures that standard optimization procedures based on derivatives can be used to obtain the maximum likelihood estimators, and that both the logit and probit links satisfy this assumption. Additionaly, this assumption implies that the maximum eigenvalue of the Hessian matrix $H_j(y_j)$ is strictly negative uniformly over its support \citep[see][]{boyd2004convex},
which with Assumption A3 guarantees that the true parameters exist uniquely.
Even when Model (\ref{model-DR}) is misspecified, under assumptions A2 and A3, the true parameters can be considered as pseudo-parameters satisfying the first-order condition, $\nabla\ell_{y,j}(\theta_{j}(y_j))=0$; thus, the parameter estimators can be interpreted under the quasi-likelihood framework for each $y_j\in\mathcal{Y}_j$ \citep[see][]{Huber1967,White1982}. Assumption A4 is necessary to obtain the limit process of the estimators
of the joint conditional distribution
and the DIRFs over the supports for statistical inference.

For the maximum likelihood estimation in (\ref{obj-estimation}), 
we use the first derivative of the objective function
$\nabla \widehat{\ell}_{y, j}(\theta_j)$ for each $j = 1, \dots, J$.
We define,
for
$(\theta,y) \in \Theta {\times} \mathcal{Y}$,
\begin{eqnarray*}
  \widehat{\Psi}_{y}(\theta)
  :=   
  \big [
  \widehat{\Psi}_{y, 1}(\theta_{1})^{\top}
  \dots, 
  \widehat{\Psi}_{y, J}(\theta_{J})^{\top}
  \big ]^{\top}, 
\end{eqnarray*}
where  
$\widehat{\Psi}_{y, j}(\theta_{j} ):=
\sqrt{T}\nabla \widehat{\ell}_{y, j}(\theta_j)$
is written as 
\[
  \widehat{\Psi}_{y, j}(\theta_{j})
  =
  \frac{1}{\sqrt{T}}
  \sum_{t=1}^{T}
  \big[\Lambda\big(\phi_{j}(X_{jt})^{\top}\theta_j\big)-\1\{Y_{jt}\leq y_j\}\big]R\big(
  \phi_{j} (X_{jt})^{\top}\theta_j\big) \phi_{j}(X_{jt}),
\]
with $R(u):=\lambda(u)/\big\{\Lambda(u)[1-\Lambda(u)]\big\}$.
 
In the below theorem, we obtain the joint limit process of the DR estimators.

\vspace{0.5cm}
\begin{theorem}
  \label{theorem:beta}
  Suppose that Assumptions A1-A3 hold. Then, we have
  \[
    \sqrt{T}
    \big(
    \widehat{\theta}(\cdot)-\theta(\cdot)
    \big)
    \rightsquigarrow\mathbb{B}(\cdot)\ \ \ \text{in}\ \ \times_{j=1}^{J}\ell^{\infty}(\mathcal{Y}_{j})^{d_j}
  \]
  where $\mathbb{B}(\cdot)$ is a $\sum_{j=1}^{J}d_{j}$-dimensional tight mean-zero Gaussian process over $\mathcal{Y}$.
  For any $y, y' \in\mathcal{Y}$, the covariance kernel of $\mathbb{B}(\cdot)$ is given by
  $H(y)^{-1}\Sigma(y, y')H(y')^{-1}$,    
  where
  $H(y):=
  \mathrm{diag}
  \big(
  \{ H_{j}(y_{j}) \}_{j=1}^{J}
  \big)$  
  and
  $\Sigma(y,y')
  :=
  \lim_{T \to \infty}
  \mathbb{E}[
  \widehat{\Psi}_{y}
  \big( \theta(y) \big)
  \widehat{\Psi}_{y'}
  \big( \theta(y') \big)^{\top}
  ].$
\end{theorem}
\vspace{0.5cm}

The result in Theorem 1 shows that the covariance kernel exhibits
a sandwich form
owing to possible miss-specification
under the quasi-likelihood framework.
Additionally,
the covariance kernel depends on
the long-run covariance matrix
in the presence of serial dependence.
Since the limit process 
depends on unknown nuisance parameters,
a moving block bootstrap \citep{kunsch1989jackknife,liu1992efficiency},
stationary bootstrap \citep{politis1994stationary}
or sub-sampling \citep{politis1997subsampling} can be used for practical inference.

We introduce a map from the DR parameters to 
a collection of the marginal conditional distributions. 
For each $j = 1, \dots, J$,
we define a map 
$\varphi_{j}:
\mathbb{D}_{\varphi_{j}}
\subset
\mathbb{D}_{j}:=\ell^{\infty}(\mathcal{Y}_{j})^{d_j}
\mapsto
\mathbb{S}_{\varphi_{j}}
\subset 
\ell^{\infty}(\mathcal{X}_{j} {\times} \mathcal{Y}_{j}) 
$, as 
\begin{eqnarray*}
  \varphi_{j}(b_{j})(x_{j},y_{j})
  :=
  \Lambda
  \big(
  \phi_{j}(x_{j})^{\top}
  b_{j}(y_j)
  \big).
\end{eqnarray*}
Let
$
\varphi(b)
:=
\big[
\varphi_{1}(b_{1}),
\ldots,
\varphi_{J}(b_{J})
\big]^{\top}
$
for
$b = (b_{1}^{\top}, \dots, b_{J}^{\top})^{\top}
\in
\mathbb{D}_{\varphi} \subset \mathbb{D}
$,
where 
$\mathbb{D}_{\varphi}
:=
\times_{j=1}^{J} \mathbb{D}_{\varphi_{j}}
$
and
$\mathbb{D}
:=
\times_{j=1}^{J} \mathbb{D}_{j}
$.
Then,
using the map
$\varphi:
\mathbb{D}_{\varphi} \mapsto 
\mathbb{S}_\varphi
:=
\times_{j=1}^{J} \mathbb{S}_{\varphi_{j}}
$,
we can write 
\begin{eqnarray*}
  \varphi(\widehat{\theta})
  =
  (\widehat{F}_{Y_{1t}|X_{1t}},\widehat{F}_{Y_{2t}|X_{2t}},\ldots,\widehat{F}_{Y_{Jt}|X_{Jt}})^{\top}
  \ \mathrm{and} \
  \varphi(\theta)
  =
  (F_{Y_{1t}|X_{1t}},F_{Y_{2t}|X_{2t}},\ldots,F_{Y_{Jt}|X_{Jt}})^{\top}.
\end{eqnarray*}
The map $\varphi$ is Hadamard differentiable at
$\theta \in \mathbb{D}_{\varphi}$
tangentially to $\mathbb{D}$
with its Hadamard derivative, given by,
\begin{eqnarray*}
  \varphi_{\theta(\cdot)}'(b)
  :=
  \big[
  \varphi_{1, \theta_{1}(\cdot)}'(b_{1}),
  \ldots,
  \varphi_{J,  \theta_{J}(\cdot)}'(b_{J})
  \big]^{\top},
\end{eqnarray*}
where 
$
\varphi'_{j, \theta_{j}(\cdot)}(b_j)(x_{j},y_{j})
:=
\lambda\big(\phi_{j}(x_{j})^{\top}\theta_{j}(y_j)\big)\phi_{j} (x_{j})^{\top}b_{j}(y_j)
$
for $j = 1, \dots, J$.

The theorem below 
provides the joint asymptotic distribution of the $J$ univariate distribution function
estimators, applying the functional delta method with the Hadamard derivative in the above
display. Furthermore, we can derive the asymptotic distribution of the estimator of any distributional characteristic that can be obtained through Hadamard differentiable maps.
\begin{theorem}\label{theorem: joint}
  Suppose that Assumptions A1-A3 hold. Then, 
  \begin{enumerate}
  \item[(a)]we have
    \[
      \sqrt{T}\left(
        \begin{array}{c}
          \widehat{F}_{Y_{1t}|X_{1t}}-F_{Y_{1t}|X_{1t}}\\
          \vdots\\
          \widehat{F}_{Y_{Jt}|X_{Jt}}-F_{Y_{Jt}|X_{Jt}}
        \end{array}
      \right)\rightsquigarrow\varphi'_{\theta(\cdot)}(\mathbb{B})\ \ \ \text{in}\ \ \times_{j=1}^J\ell^{\infty}(\mathcal{Y}_j{\times}\mathcal{X}_j),
    \]
    where $\mathbb{B}$ is the tight mean-zero Gaussian process defined in Theorem \ref{theorem:beta}.
     
  \item[(b)]
    additionally, if a map $\nu: \mathbb{S}_\varphi\mapsto\ell^{\infty}(\mathcal{Z}{\times}\mathcal{Y})$ is Hadamard differentiable at $(F_{Y_{1t}|X_{1t}},\ldots,F_{Y_{Jt}|X_{Jt}})$ tangentially to $\varphi'_{\theta(\cdot)}(\mathbb{D})$ with the Hadamard derative $\nu'_{F_{Y_{1t}|X_{1t}},\ldots,F_{Y_{Jt}|X_{Jt}}}$, then
    \[
      \sqrt{T} \big\{\nu(\widehat{F}_{Y_{1t}|X_{1t}},\ldots,\widehat{F}_{Y_{Jt}|X_{Jt}})-\nu(F_{Y_{1t}|X_{1t}},\ldots,F_{Y_{Jt}|X_{Jt}})\big\}\rightsquigarrow\nu'_{F_{Y_{1t}|X_{1t}},\ldots,F_{Y_{Jt}|X_{Jt}}}\circ\varphi'_{\theta(\cdot)}(\mathbb{B}),
    \]
  in $\ell^{\infty}(\mathcal{Z}{\times}\mathcal{Y})$.      
  \end{enumerate}
\end{theorem} 
\vspace{0.5cm}

As the composition of Hadamard differentiable transformations remains Hadamard differentiable \citep[Lemma 3.9.3,][]{van1996weak}, Theorem \ref{theorem: joint} can be applied to all these distributional characteristics of interest.
As explained in Section \ref{subsec: estimation}, the conditional distributions $F_{Y_{t}|Z_{t}}$, $F_{Y_{t+h}|Y_{t}, Z_{t}}$ and $G_{Y_{t}|Z_{t}}$ can be obtained via Hadamard differentiable transformations of several univariate conditional distributions
under Assumption A4. Furthermore, the DIRFs, $DIR_{h}$ and $DIR_{j,h}$, are Hadamard differentiable transformations of these conditional distributions.
Theorem \ref{theorem: joint}(b) can be applied to perform statistical inference on the estimators of the conditional joint distribution and the DIRFs.

\section{Macroeconomic and Financial Dependence}\label{sec: empirical}

We apply the proposed approach to examine the time series data of macroeconomic and financial conditions in the U.S. The real GDP growth and the NFCI are used as indicators to measure the state of the economy and financial sector. A growing body of research has attempted to investigate the macro-financial interactions during recessions via VAR models of the GDP growth and NFCI \citep{carriero2020capturing, clark2021tail, gertler2018happened}. The present empirical study extends this direction to conduct a more comprehensive distributional analysis. We address two main questions in this section. First, does the joint distribution of the GDP growth and NFCI conditional on past lagged information change during financial stress? Secondly, how does their joint distribution respond to a distributional shock in macro and financial conditions? 
\subsection{Data and Modeling Specification}
The real GDP growth is computed using quarterly real GDP data from the Bureau of Economic Analysis\footnote{The data is downloaded from FRED \href{https://fred.stlouisfed.org/series/A191RL1Q225SBEA}{https://fred.stlouisfed.org/series/A191RL1Q225SBEA}}. The NFCI is a weighted average of 105 measures of national financial activity, each expressed relative to their sample averages and scaled by their sample standard deviations, which are released weekly by the Federal Reserve Bank of Chicago\footnote{More details about NFCI are available at \href{https://www.chicagofed.org/publications/nfci/index}{https://www.chicagofed.org/publications/nfci/index}}. 
A positive NFCI suggests that the financial conditions are tighter than average. We use data from these two indices for the period 1973:Q1 to 2019:Q1, for analysis. These time series are released at different frequencies. Following \cite{adrian2021multimodality}, we convert the NFCI data into quarterly observations by averaging each quarter's weekly observations.
We consider a bivariate model for the outcome
$Y_{t} = (Y_{1t}, Y_{2t})^{\top}$,
with $Y_{1t}$ representing the quarterly NFCI and $Y_{2t}$ representing the real GDP growth. 

We apply the proposed multivariate DR approach to estimate horizon-specific multiperiod forecasting distributions for each variable. Specifically, we consider two-lag information $Z_{t}=(Y_{t-1}^{\top},Y_{t-2}^{\top})^{\top}$ to estimate the joint conditional distribution $F_{Y_t|Z_t}$ and three-lag information to develop the $h$-ahead forecasting distribution $F_{Y_{t+h}|Y_t, Z_t}$ for different $h$. DIRFs can then be estimated based on these conditional distributions, which enables the design of different counterfactual scenarios to investigate the possible policy effect on the entire distributions of both the NFCI and real GDP growth over time. 

\subsection{Multiperiod Ahead Conditional Distribution Forecast}\label{subsec: MDF}

  Focusing on the one-quarter and one-year ahead horizons, we present the out-of-sample performance of the multivariate DR approach in estimating the multiperiod ahead forecasting distributions of the NFCI and real GDP growth. 

  \begin{figure}[H]
	\captionsetup[subfigure]{aboveskip=-3pt,belowskip=0pt}
	\centering
	\caption{Empirical CDF of the Out-of-sample PITs\label{fig: PITs-Q}}
	\begin{subfigure}[b]{0.45\textwidth}
		\centering
		\caption{GDP}     		
		\includegraphics[width=0.95\textwidth]{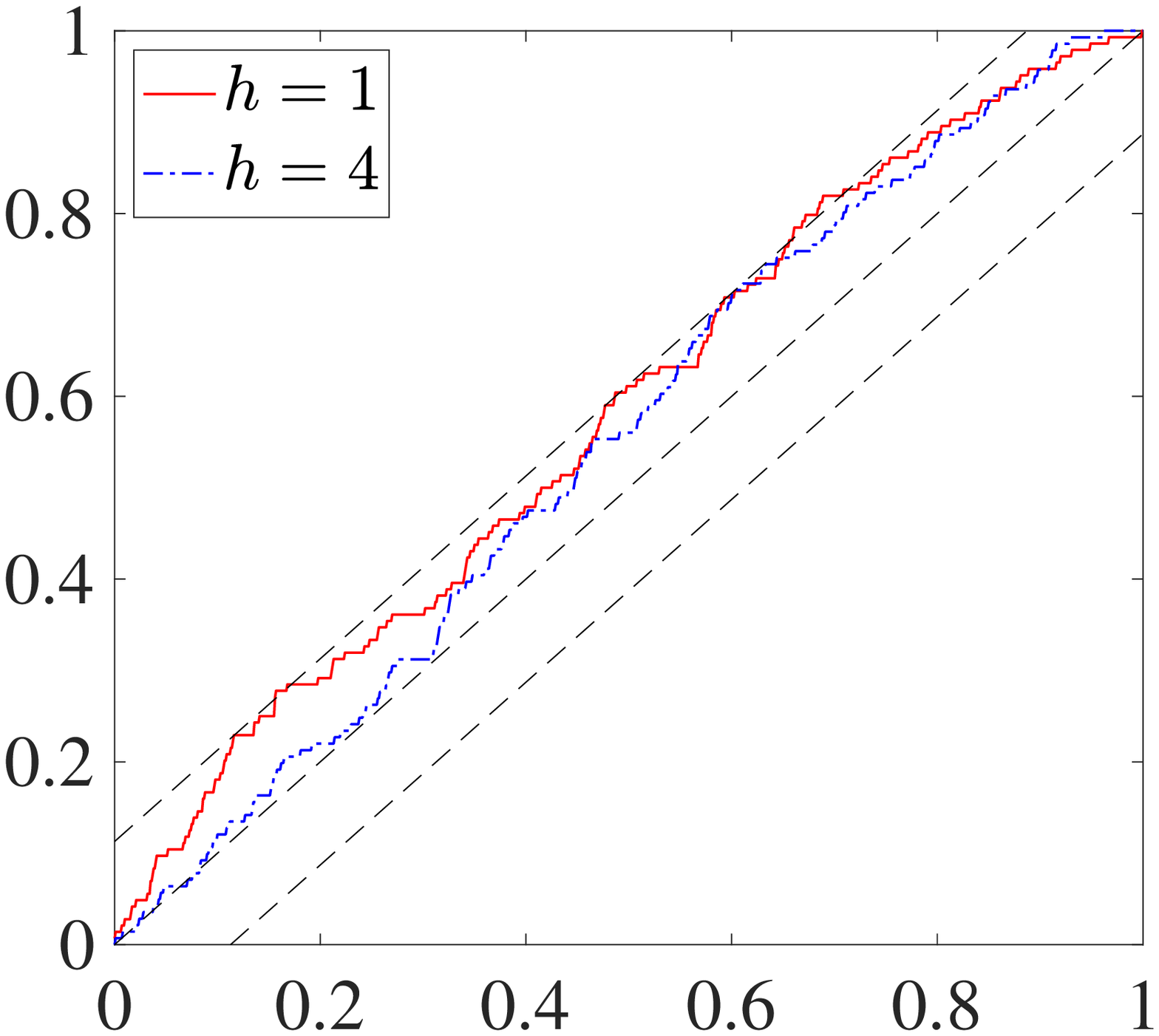}      		
		\label{GDP-h1}
	\end{subfigure}
	\begin{subfigure}[b]{0.45\textwidth}
		\centering
		\caption{NFCI}		
		\includegraphics[width=0.95\textwidth]{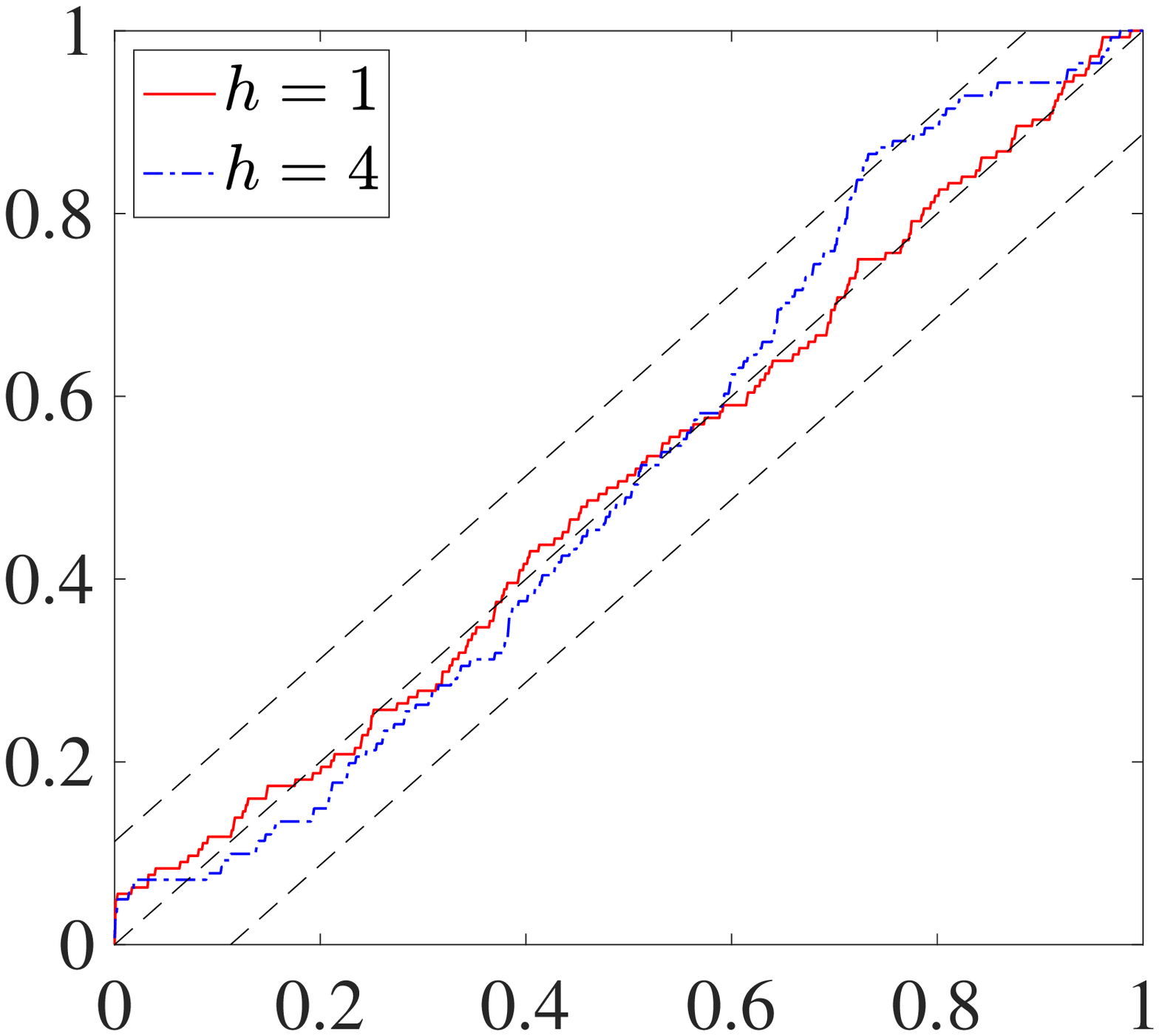}
		\label{GDP-h4}
	\end{subfigure}
	\begin{minipage}{.9\linewidth} 
		\linespread{1}\footnotesize
		\textit{Notes}: This figure reports the empirical CDF of the PITs by the DR approach for one-quarter-ahead (red solid line), and one-year-ahead (blue dotdash line), plus the CDF of the PITs under the null hypothesis of correct calibration (the 45-degree line) and the 95\% confidence bands (dashed line) of the \cite{rossi2019alternative} PITs test.
	\end{minipage}
\end{figure}

  First, using the expanding window beginning with the estimation of the sample ranging from 1973:Q1 to 1982:Q3, we evaluate the out-of-sample performance of the distribution forecasts by analyzing the probability integral transform (PIT), which reflects the percentage of observations below any given quantile. In a perfectly calibrated model, the fraction of realizations below any given quantile of the predictive distribution exactly equals the quantile probability, thus the cumulative distribution of the PITs is a 45-degree line. The closer the empirical cumulative distribution of the PITs is to the 45-degree line, the better the model is calibrated. For different forecast horizons, the empirical distribution of PITs together with 95\% confidence bands of \cite{rossi2019alternative} PITs test\footnote{Under the null of uniformity and independence of the PITs, we use the asymptotic critical value for a 5\% significance level, 1.34, suggested by \cite{rossi2019alternative} to construct the confidence bands.} for the predicted marginals of the real GDP growth and NFCI are shown in Figure \ref{fig: PITs-Q}. This illustrates that the empirical distributions of the PITs by the proposed approach are all well within the confidence intervals. 
 
  \begin{figure}[H]
 	\captionsetup[subfigure]{aboveskip=-3pt,belowskip=0pt}
 	\centering
 	\caption{Out-of-sample Predicted Distributions\label{fig: predicted-dis}}
 	\begin{subfigure}[b]{0.40\textwidth}
 		\centering
 		\caption{GDP: one-quarter-ahead}     		
 		\includegraphics[width=0.95\textwidth]{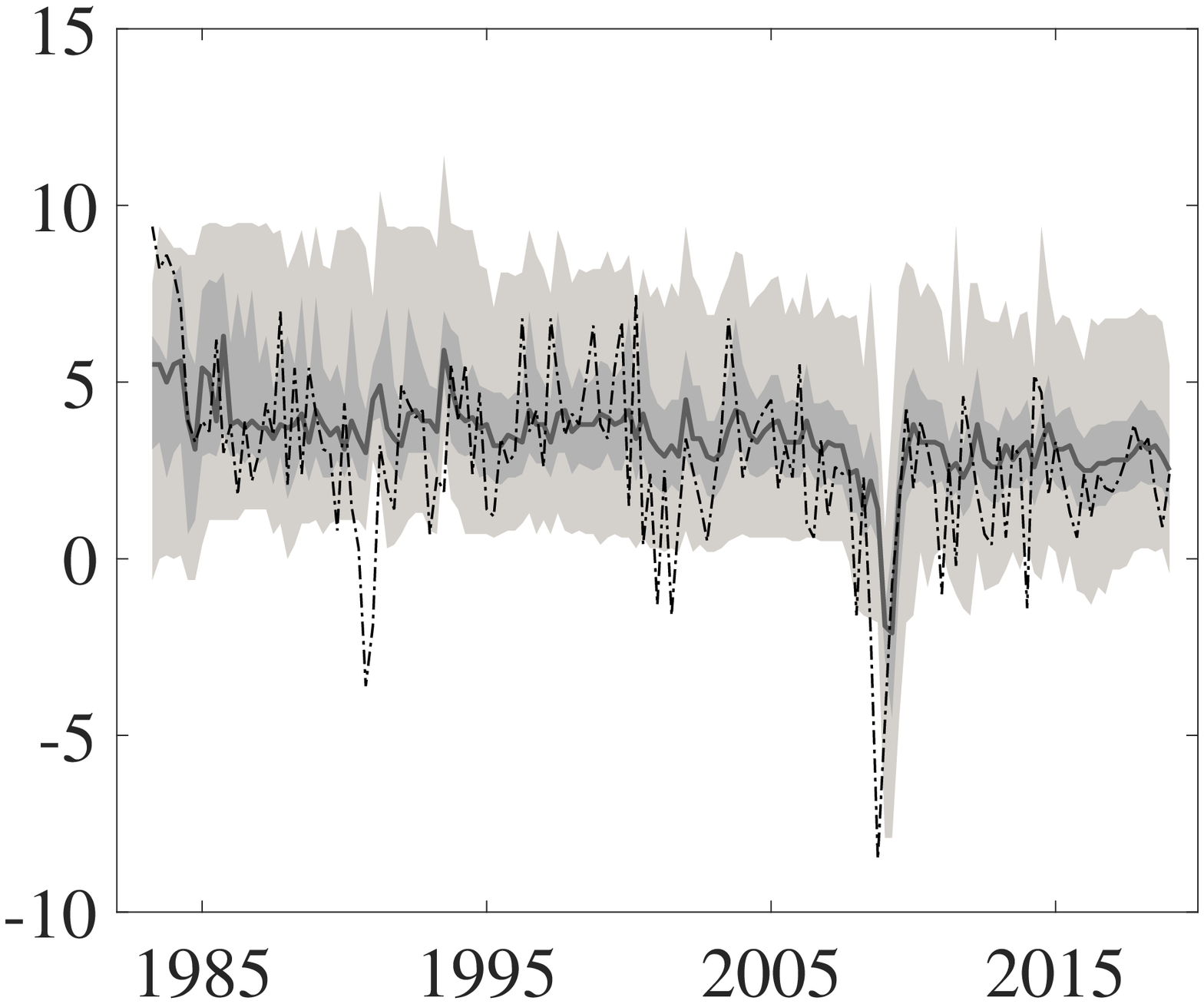}      		
 		\label{GDP-trend-h1}
 	\end{subfigure}
 	\begin{subfigure}[b]{0.40\textwidth}
 		\centering
 		\caption{GDP: one-year-ahead}		
 		\includegraphics[width=0.95\textwidth]{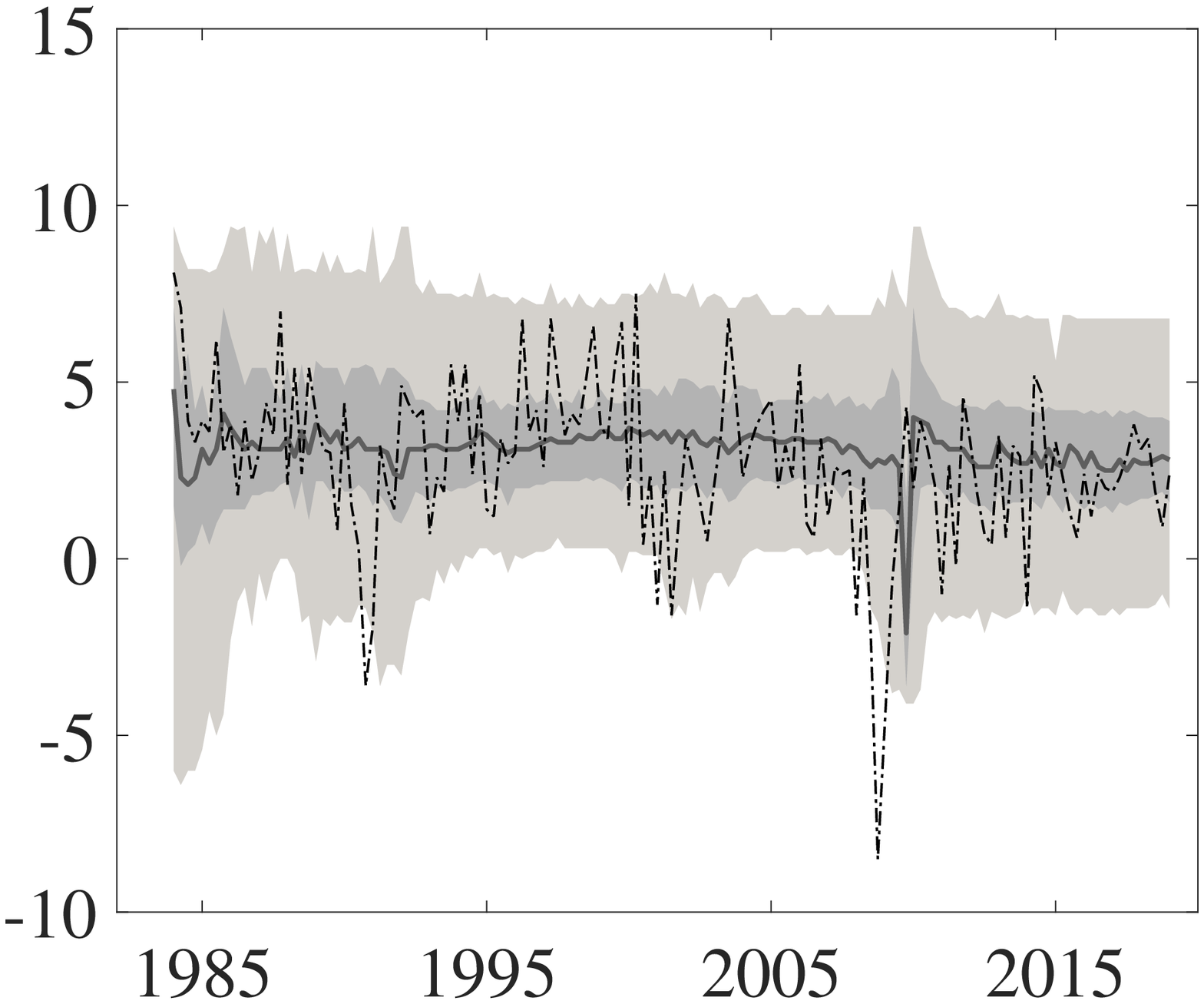}
 		\label{GDP-trend-h4}
 	\end{subfigure}
 	\hfill
 	\begin{subfigure}[b]{0.40\textwidth}
 		\centering
 		\caption{NFCI: one-quarter-ahead}		
 		\includegraphics[width=0.95\textwidth]{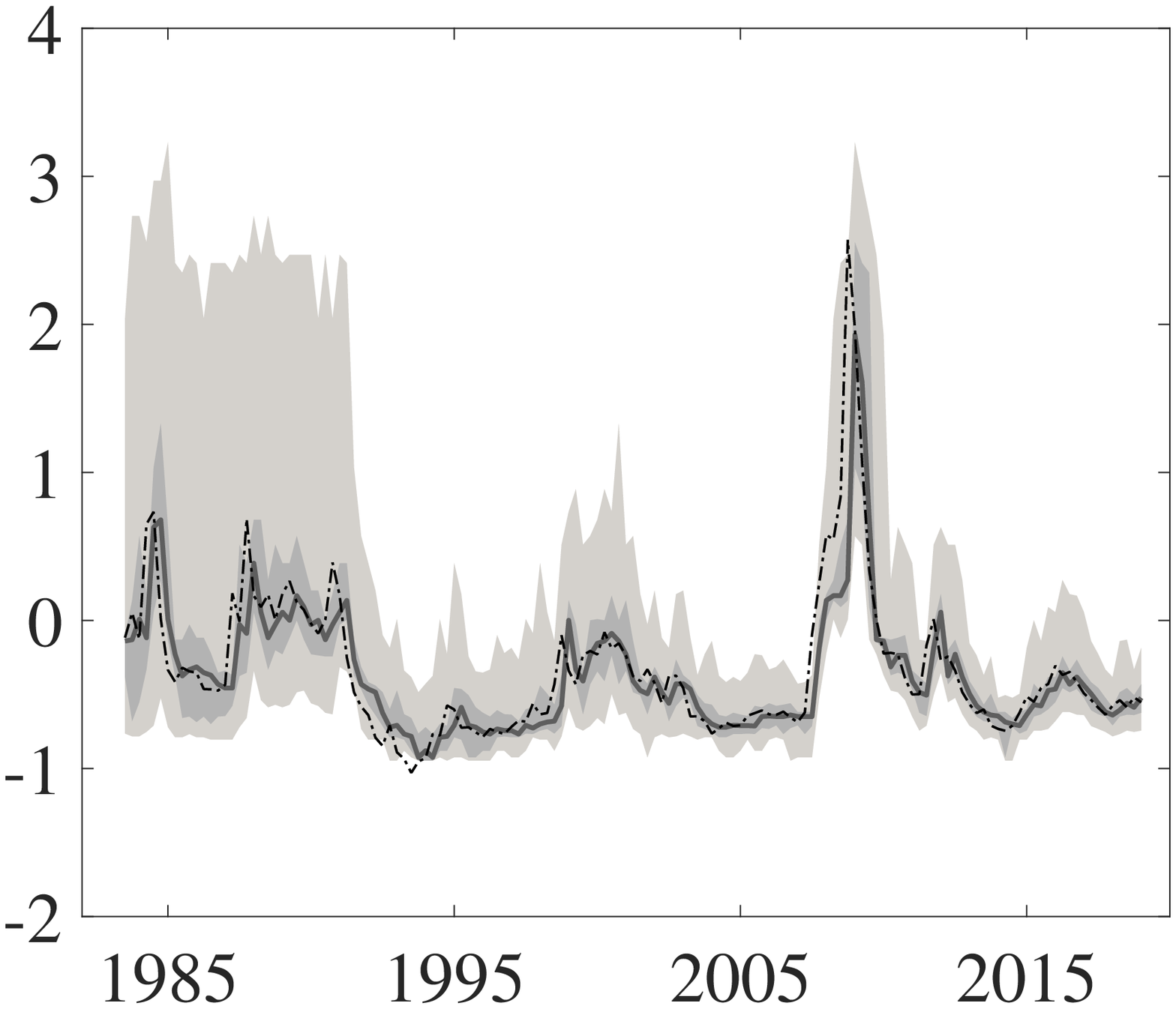}
 		\label{NFCI-trend-h1}
 	\end{subfigure}          
 	\begin{subfigure}[b]{0.40\textwidth}
 		\centering
 		\caption{NFCI: one-year-ahead}
 		\includegraphics[width=0.95\textwidth]{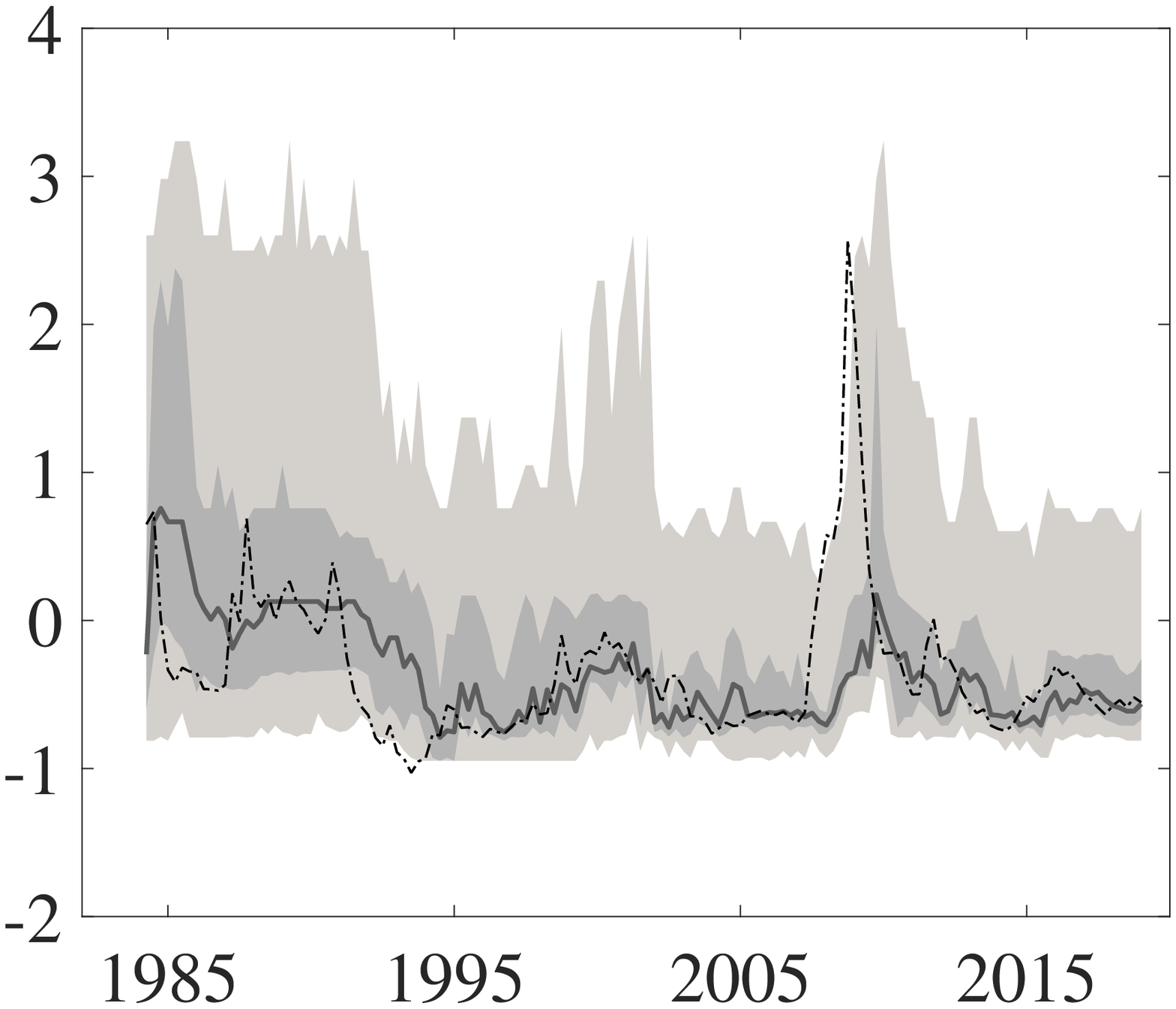}
 		\label{NFCI-trend-h4}
 	\end{subfigure}
    \vspace{0.3cm}
 	\begin{minipage}{.75\linewidth} 
 		\linespread{1}\footnotesize
 		\textit{Notes}: Each panel displays the predicted marginal distributions over time. The light gray area represents the 5th-95th percentile intervals, the dark gray area indicates the 25th-75th percentile intervals, and the solid line represents the median of the distribution. The actual observations are illustrated with dotted lines. 
 	\end{minipage}
 \end{figure}

Additionally, we examine the estimated one-quarter and one-year ahead marginal distributions of the real GDP growth and NFCI using the expanding window (out-of-sample). Based on the predicted distributions, the 5th to 95th and 25th to 75th percentile intervals, the median, along with the data realizations are plotted in Figure \ref{fig: predicted-dis}. The distribution evolution of the real GDP growth shows that the median and lower tail (downside risk) exhibit significant time-series variation compared to the upper tail (upside risk). Comparing the predicted quantiles to the realizations reveals that the possibilities of adverse GDP growth and tight financial conditions can be detected by the predicted distributions in real time.

\begin{figure}[H]
 	\captionsetup[subfigure]{aboveskip=-3pt,belowskip=0pt}
 	\centering
 	\caption{In-sample Conditional Correlation between NFCI and real GDP Growth\label{fig: correlation}}
 	\includegraphics[width=0.8\textwidth]{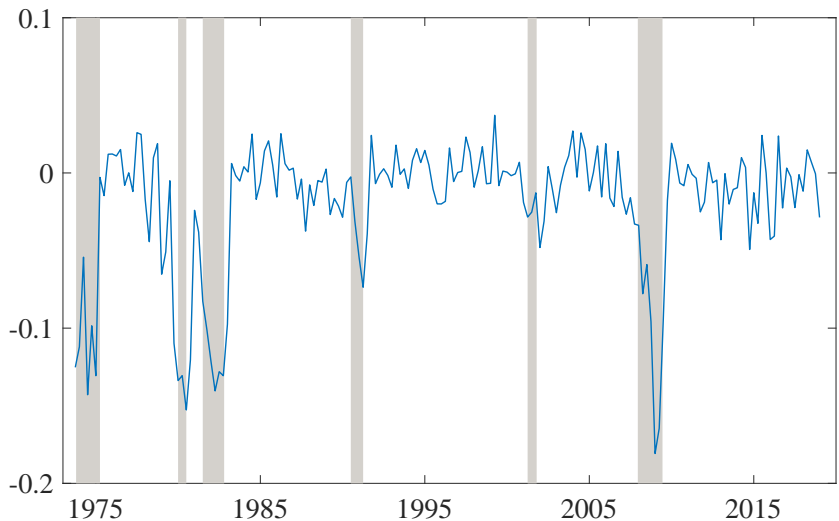}    
 	\begin{minipage}{.7\linewidth} 
 		\linespread{1}\footnotesize
 		\textit{Notes}: The figure plots the correlation coefficients computed based on the one-step-ahead forecasting distributions. Shaded areas indicate U.S. recessions. 
 	\end{minipage} 
 \end{figure}

Based on the realizations of the real GDP growth and the NFCI, we find that when the NFCI is relatively loose, the economy evolves as usual. Simultaneously, extreme tightening of the NFCI coincides with extremely adverse GDP growth. In Figure \ref{fig: correlation}, we plot the in-sample conditional correlation coefficients between the real GDP growth and NFCI. The correlation coefficient fluctuates around 0 during the normal times but becomes significantly negative during recession periods. This suggests a nonlinear relationship between financial conditions and real activity, with their conditional joint distribution behaving very differently during normal times and recessions. 
 
\subsection{Multimodality in Macro-Financial Dynamics During the Great Recession }

\begin{figure}[H]
	\captionsetup[subfigure]{aboveskip=-2pt,belowskip=-10pt, labelformat=empty}
	\centering 
	\caption{Contour Plots of the Joint Distribution during the Great Recession} \label{fig: contour-plots}	
	\begin{minipage}[t]{\textwidth}
		\centering \footnotesize(a) $Y_{t+h}$ given $(Y_t,Z_t)$ for $t$=2008:Q3\\
		\begin{subfigure}[t]{0.24\textwidth} 	
			\caption{\footnotesize $h=1$}
			\includegraphics[width=0.95\textwidth]{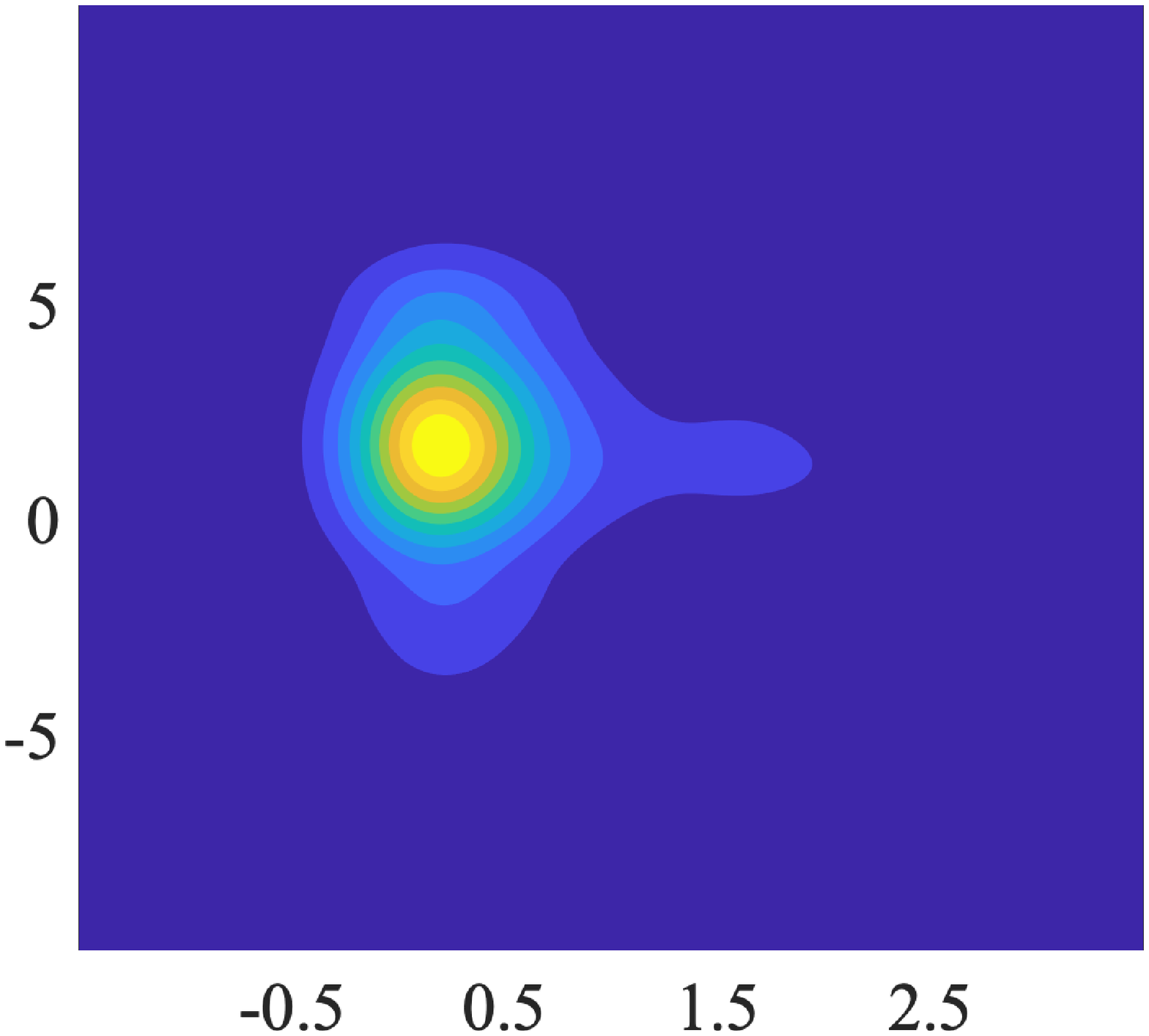}	
		\end{subfigure}
		\begin{subfigure}[t]{0.23\textwidth}
			\caption{\footnotesize $h=2$}	
			\includegraphics[width=0.95\textwidth]{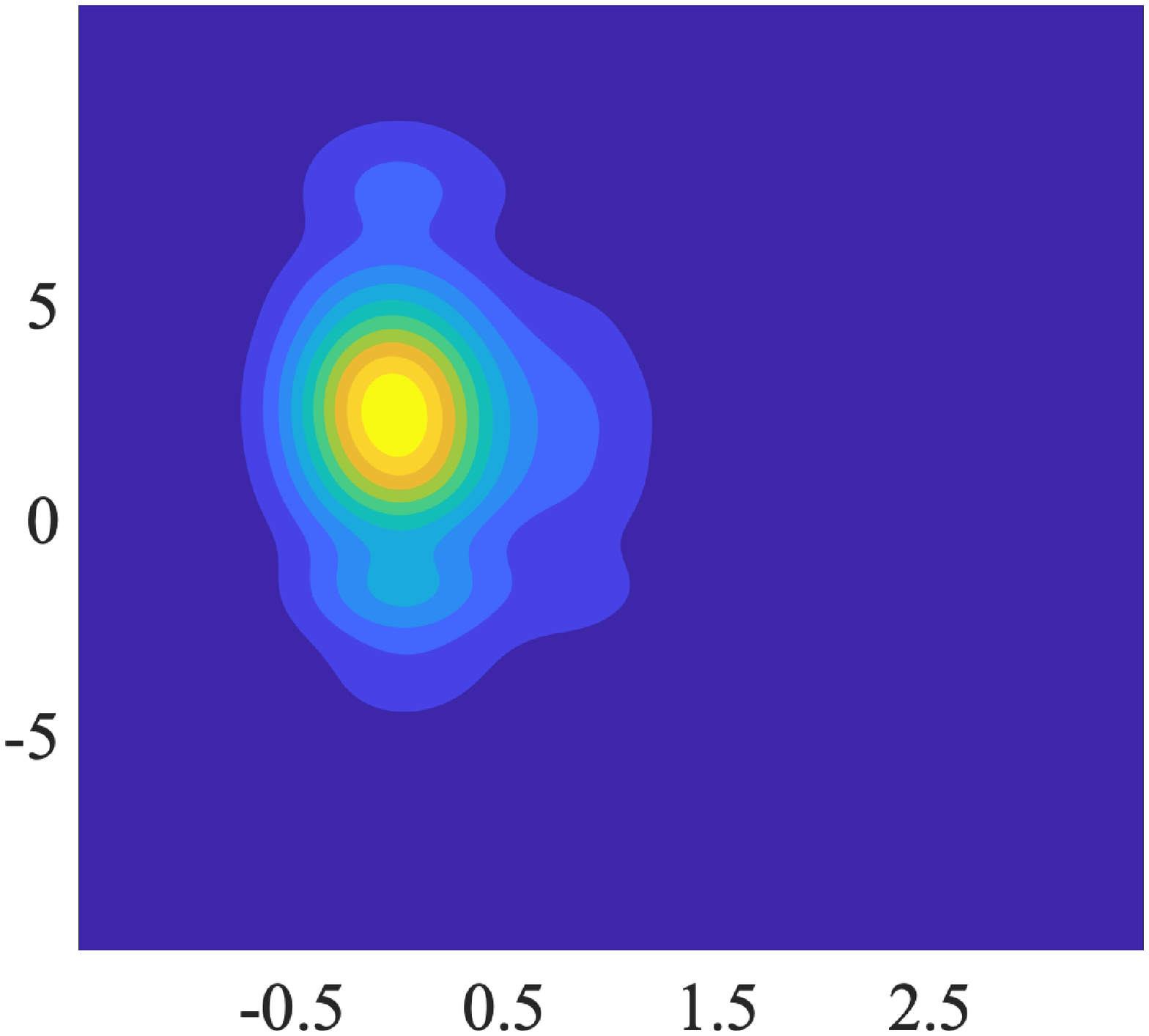}
		\end{subfigure}
		\begin{subfigure}[t]{0.23\textwidth}
			\caption{\footnotesize $h=3$}	
			\includegraphics[width=0.95\textwidth]{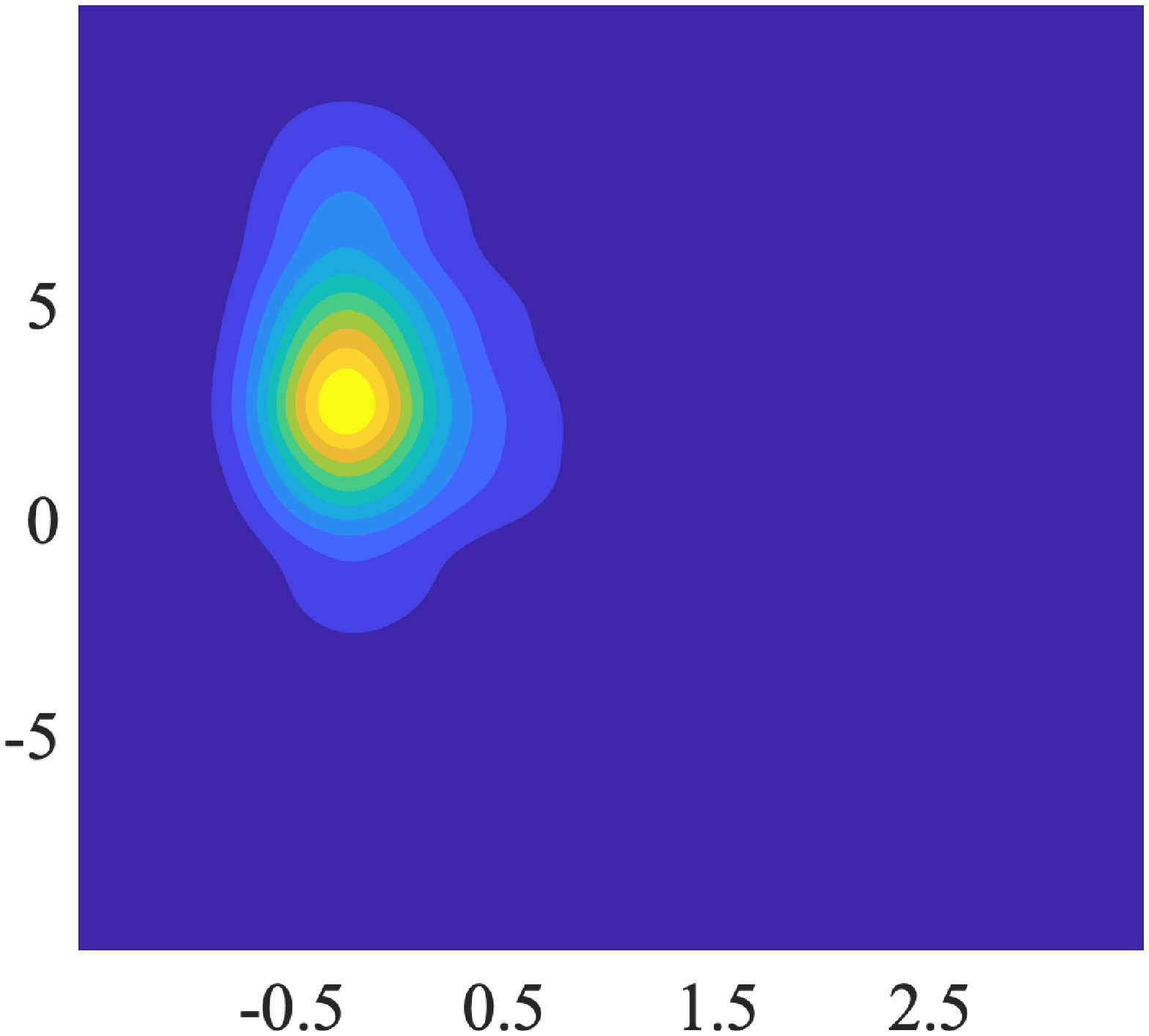}
		\end{subfigure}
		\begin{subfigure}[t]{0.23\textwidth}
			\caption{\footnotesize $h=4$}
			\includegraphics[width=0.95\textwidth]{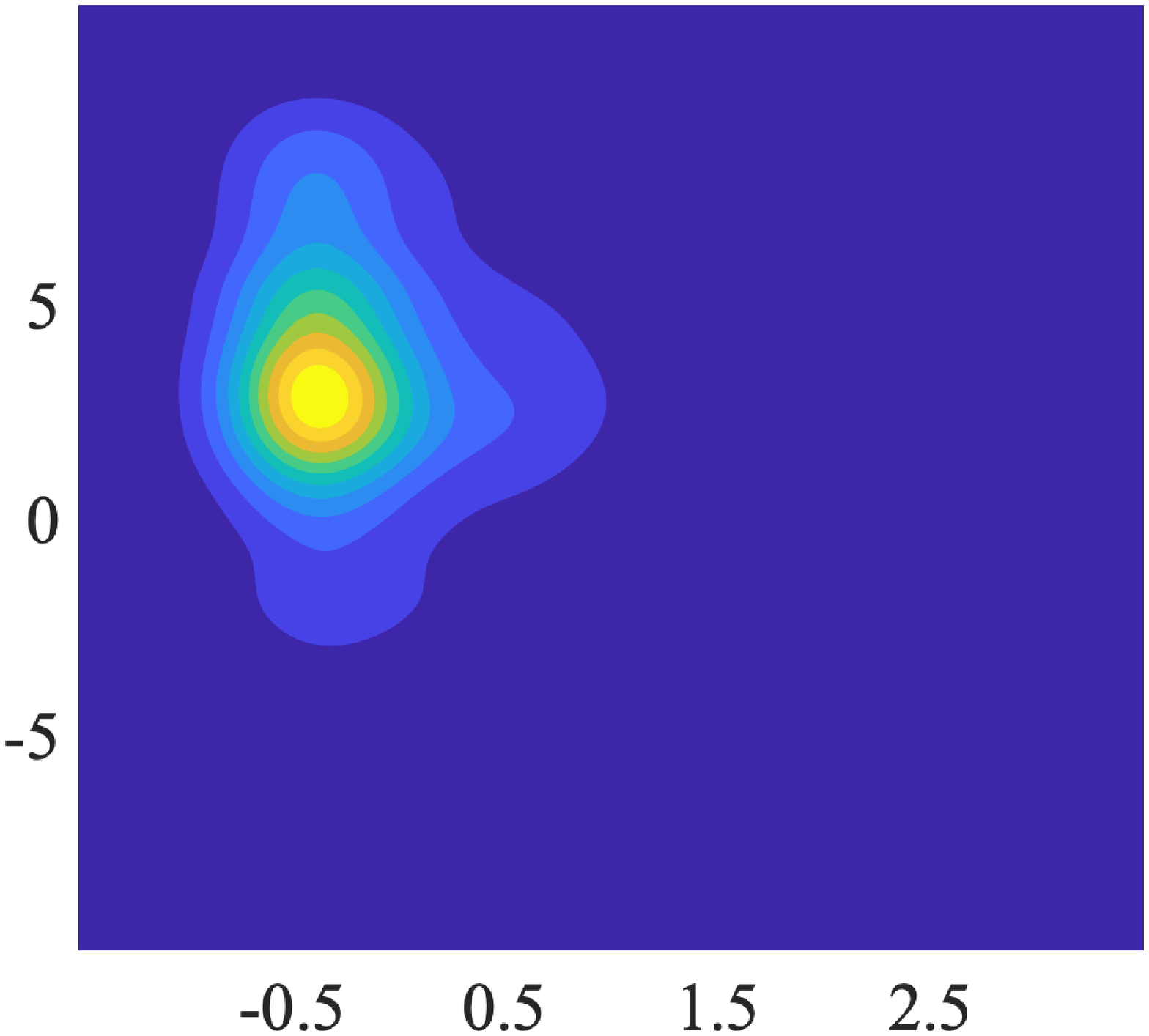}
		\end{subfigure}
		\vspace{2mm}
	\end{minipage}
	\begin{minipage}[t]{\textwidth}
		\centering \footnotesize(b) $Y_{t+h}$ given $(Y_t,Z_t)$ for $t$=2008:Q4 \\
		\begin{subfigure}[t]{0.23\textwidth}
			\caption{\footnotesize $h=1$}	
			\includegraphics[width=0.95\textwidth]{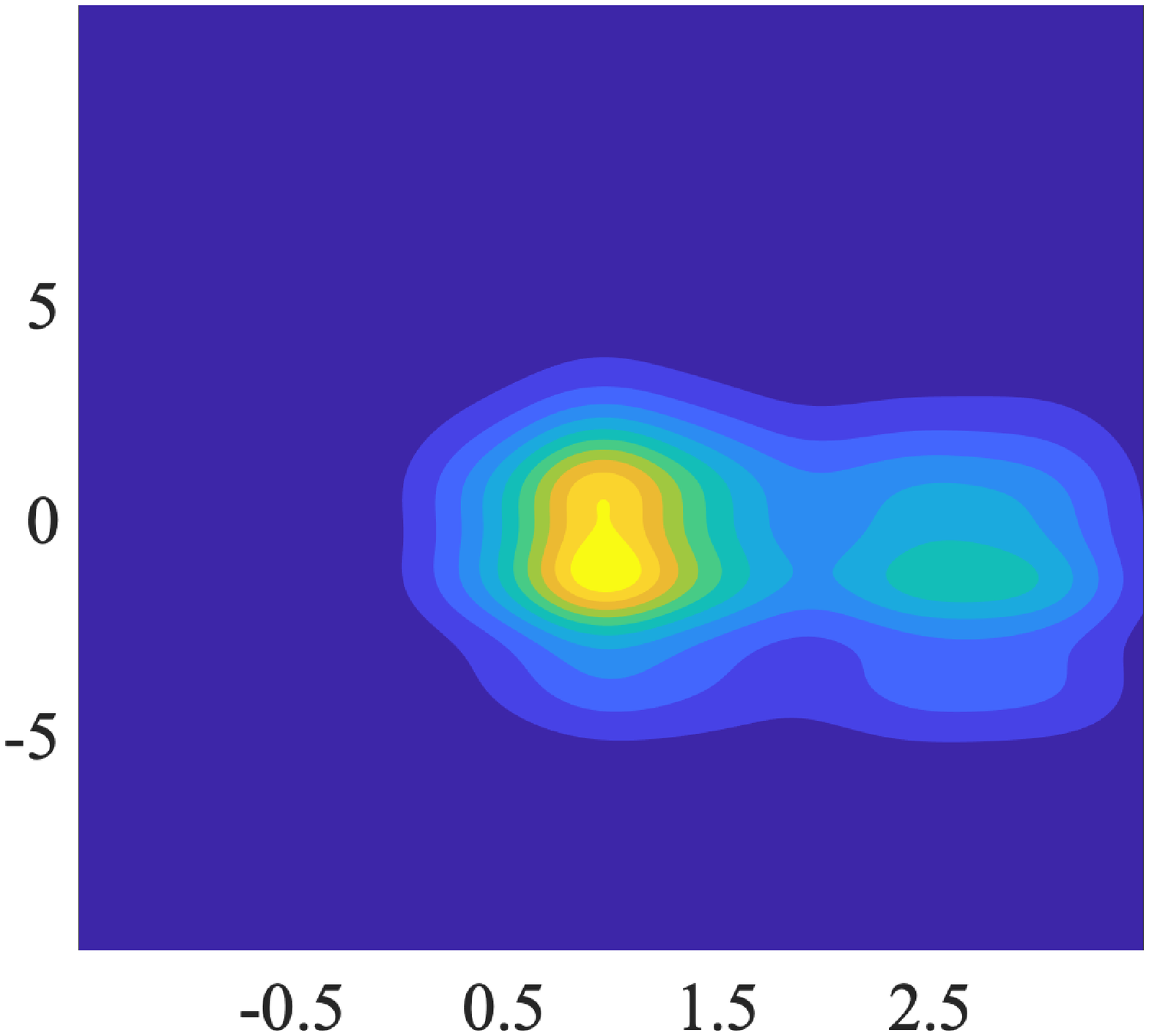}     		
		\end{subfigure}
		\begin{subfigure}[t]{0.23\textwidth}
			\caption{\footnotesize $h=2$}
			\includegraphics[width=0.95\textwidth]{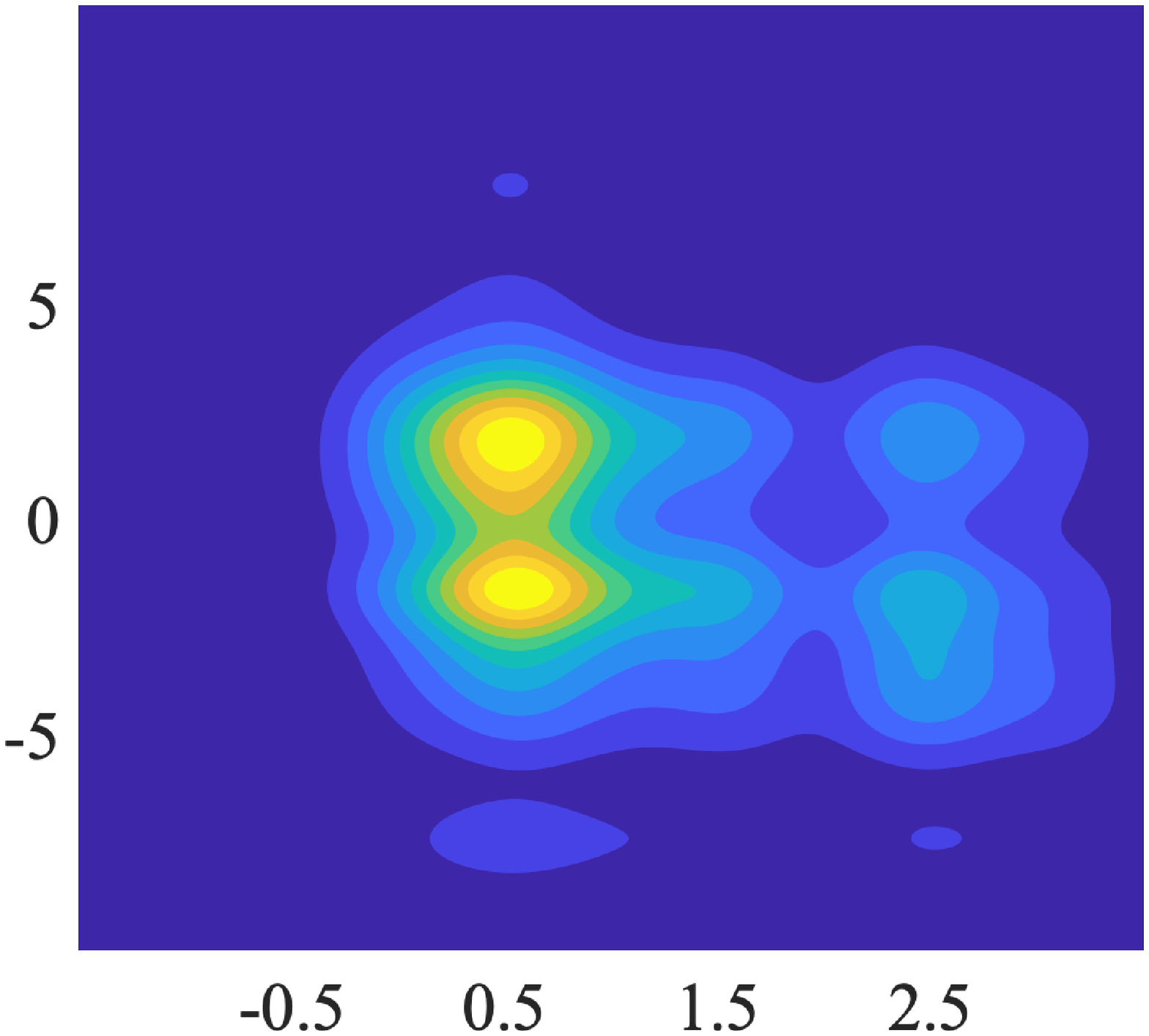}
		\end{subfigure}
		\begin{subfigure}[t]{0.23\textwidth}
			\caption{\footnotesize $h=3$}
			\includegraphics[width=0.95\textwidth]{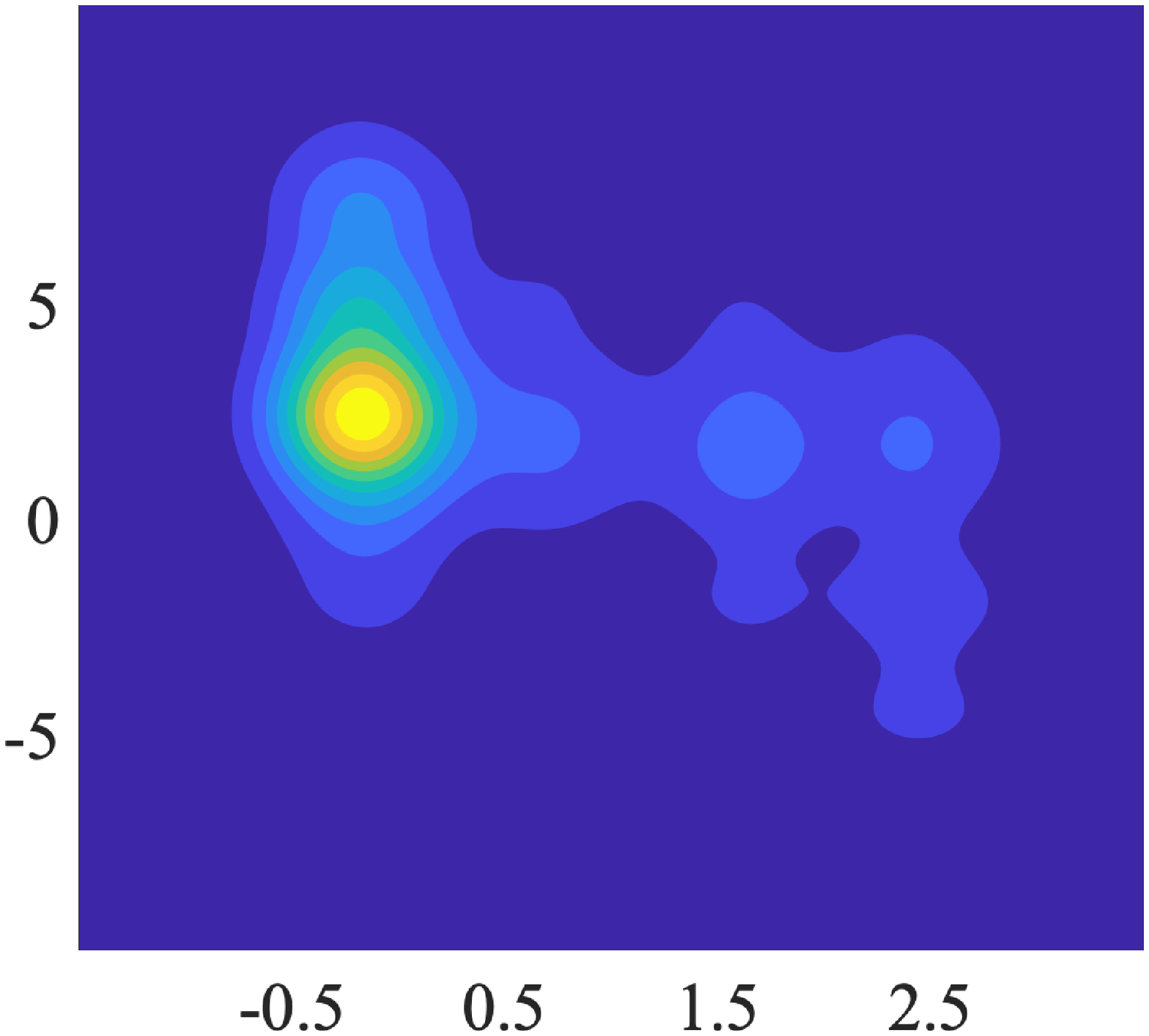}
		\end{subfigure}
		\begin{subfigure}[t]{0.23\textwidth}
			\caption{\footnotesize $h=4$}
			\includegraphics[width=0.95\textwidth]{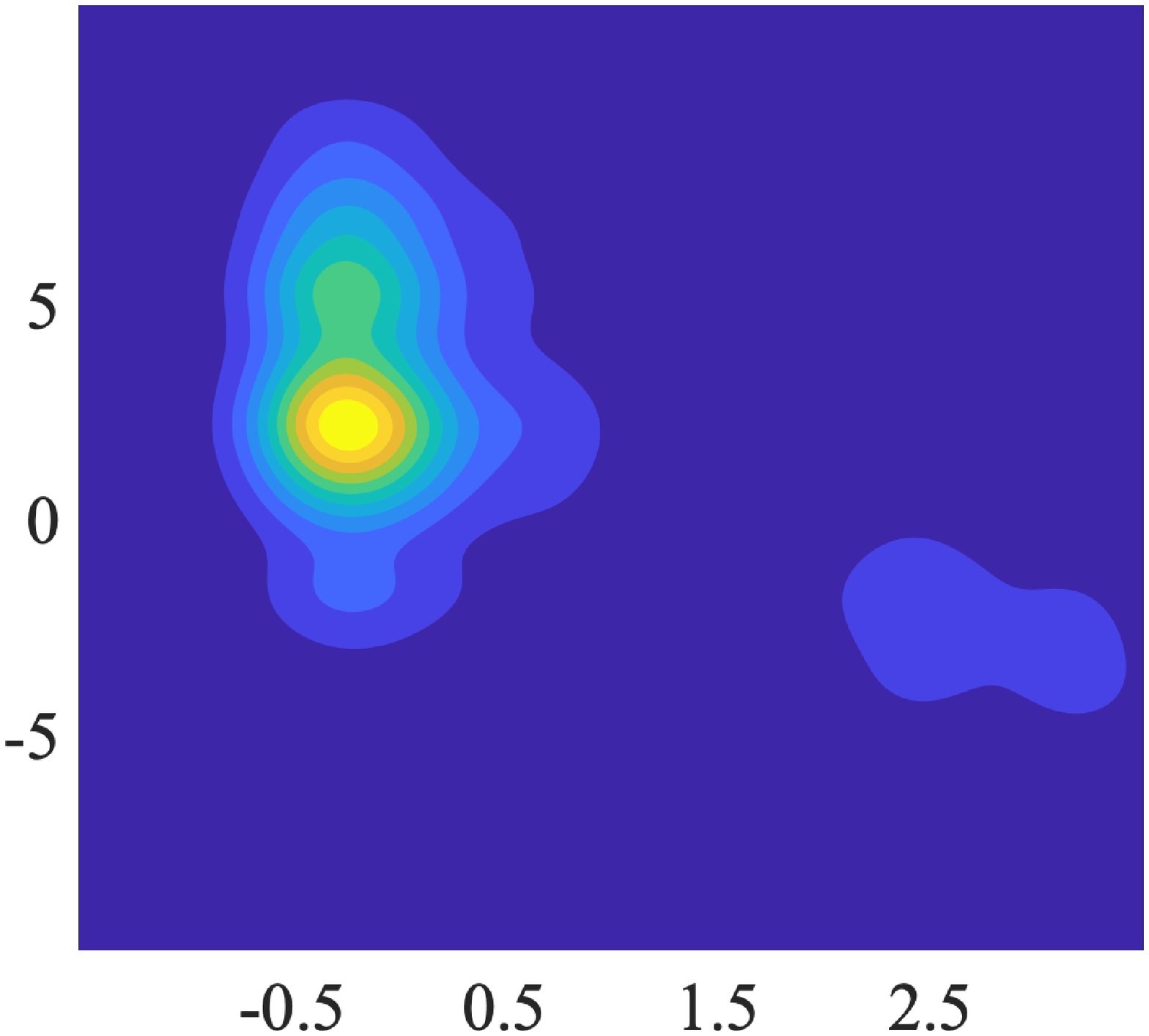}
		\end{subfigure}   
		\vspace{2mm}
	\end{minipage}
	\begin{minipage}[t]{\textwidth}
		\centering \footnotesize(c) $Y_{t+h}$ given $(Y_t,Z_t)$ for $t$=2009:Q1 \\		
		\begin{subfigure}[t]{0.23\textwidth}
			\caption{\footnotesize $h=1$}	
			\includegraphics[width=0.95\textwidth]{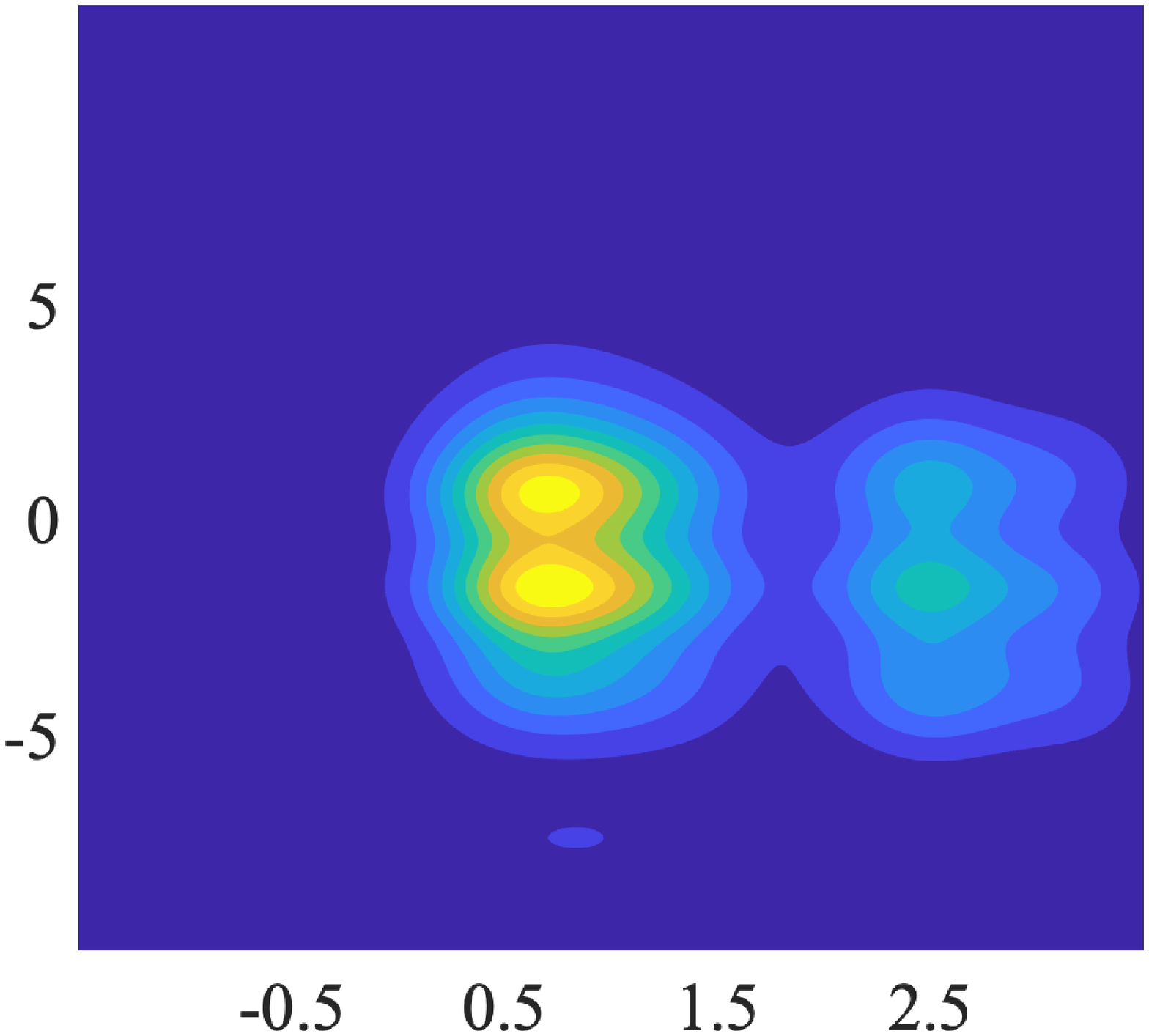}     		
		\end{subfigure}
		\begin{subfigure}[t]{0.23\textwidth}
			\caption{\footnotesize $h=2$}	
			\includegraphics[width=0.95\textwidth]{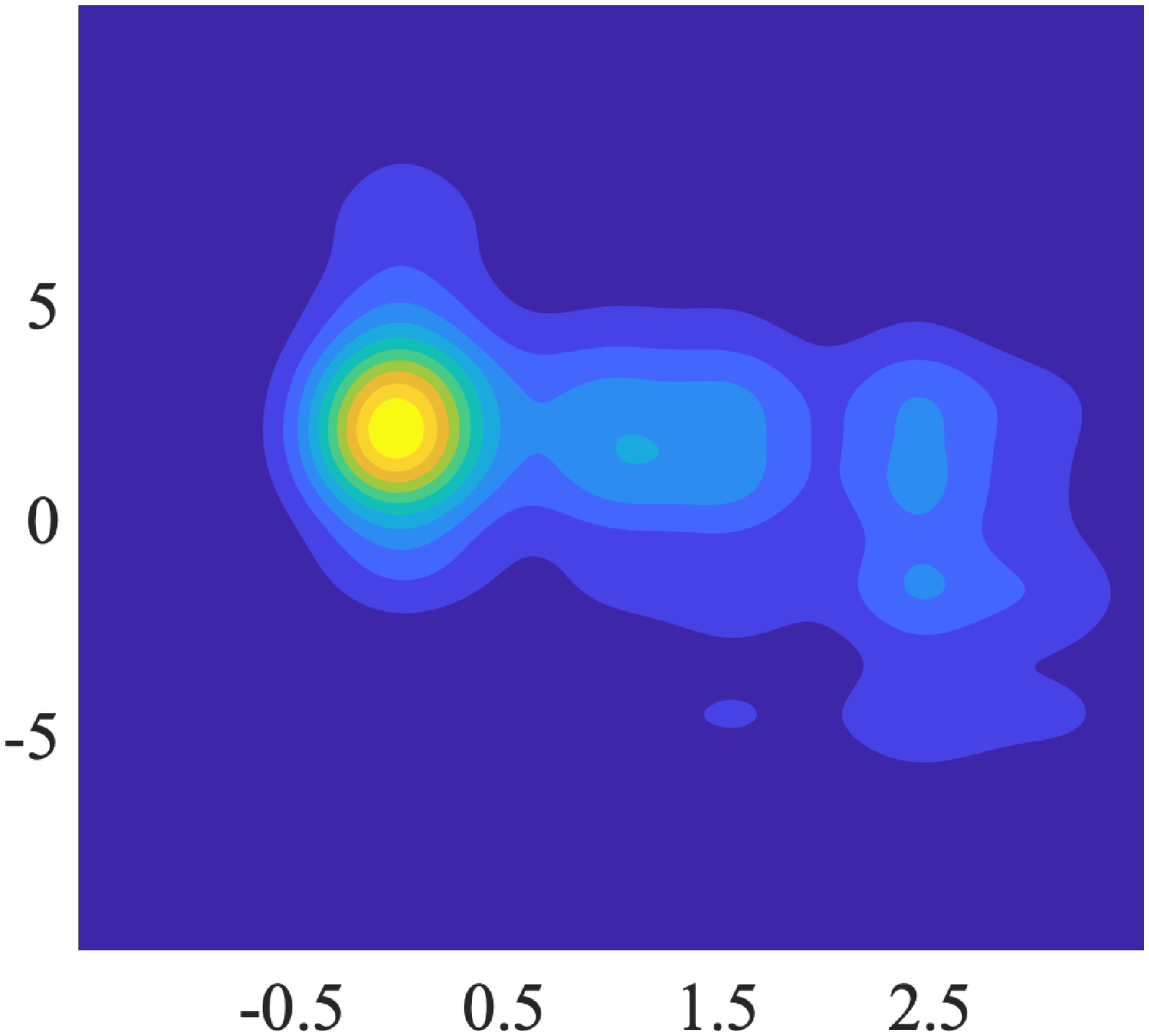}
		\end{subfigure}
		\begin{subfigure}[t]{0.23\textwidth}
			\caption{\footnotesize $h=3$}
			\includegraphics[width=0.95\textwidth]{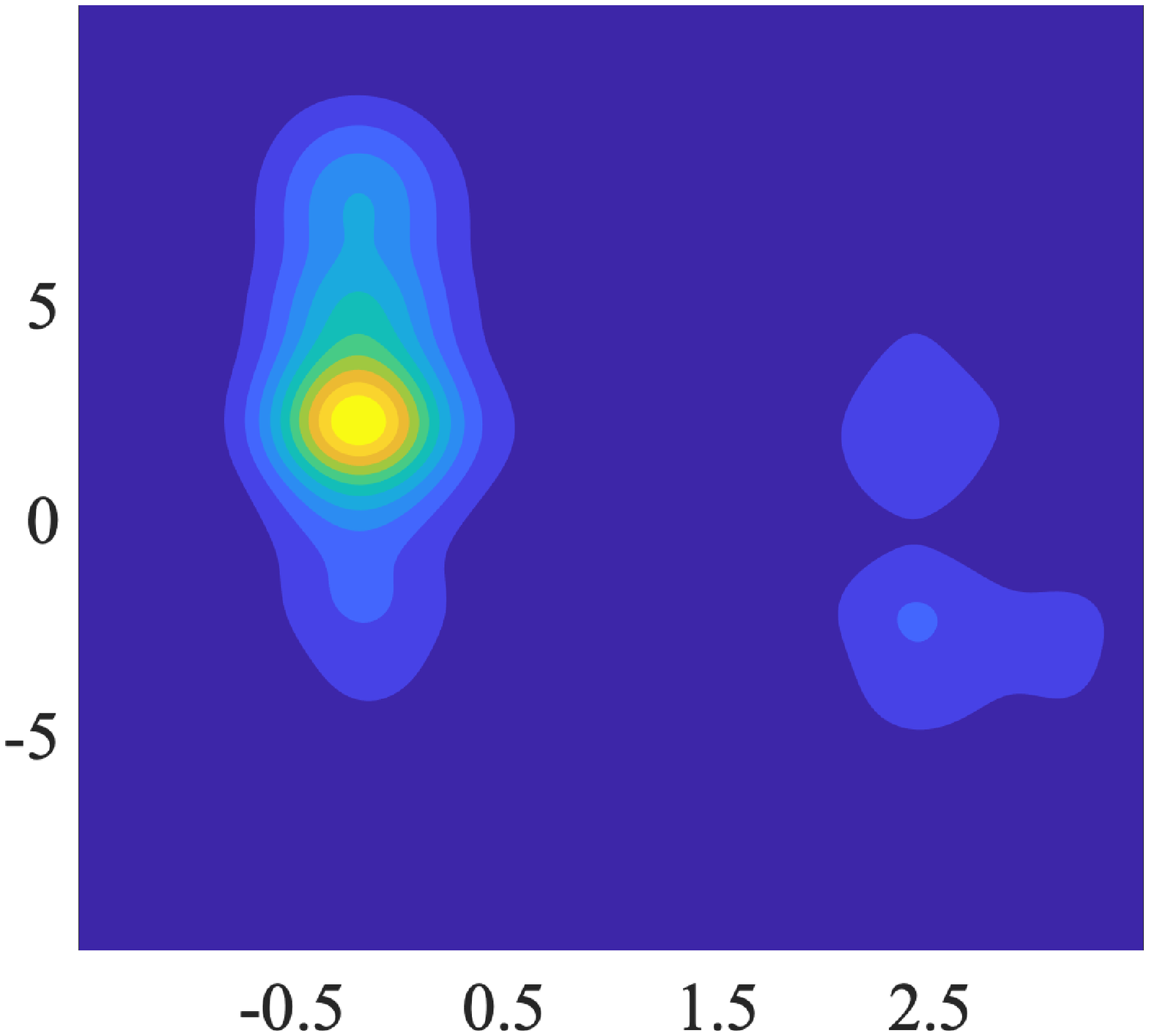}
		\end{subfigure}
		\begin{subfigure}[t]{0.23\textwidth}
			\caption{\footnotesize $h=4$}
			\includegraphics[width=0.95\textwidth]{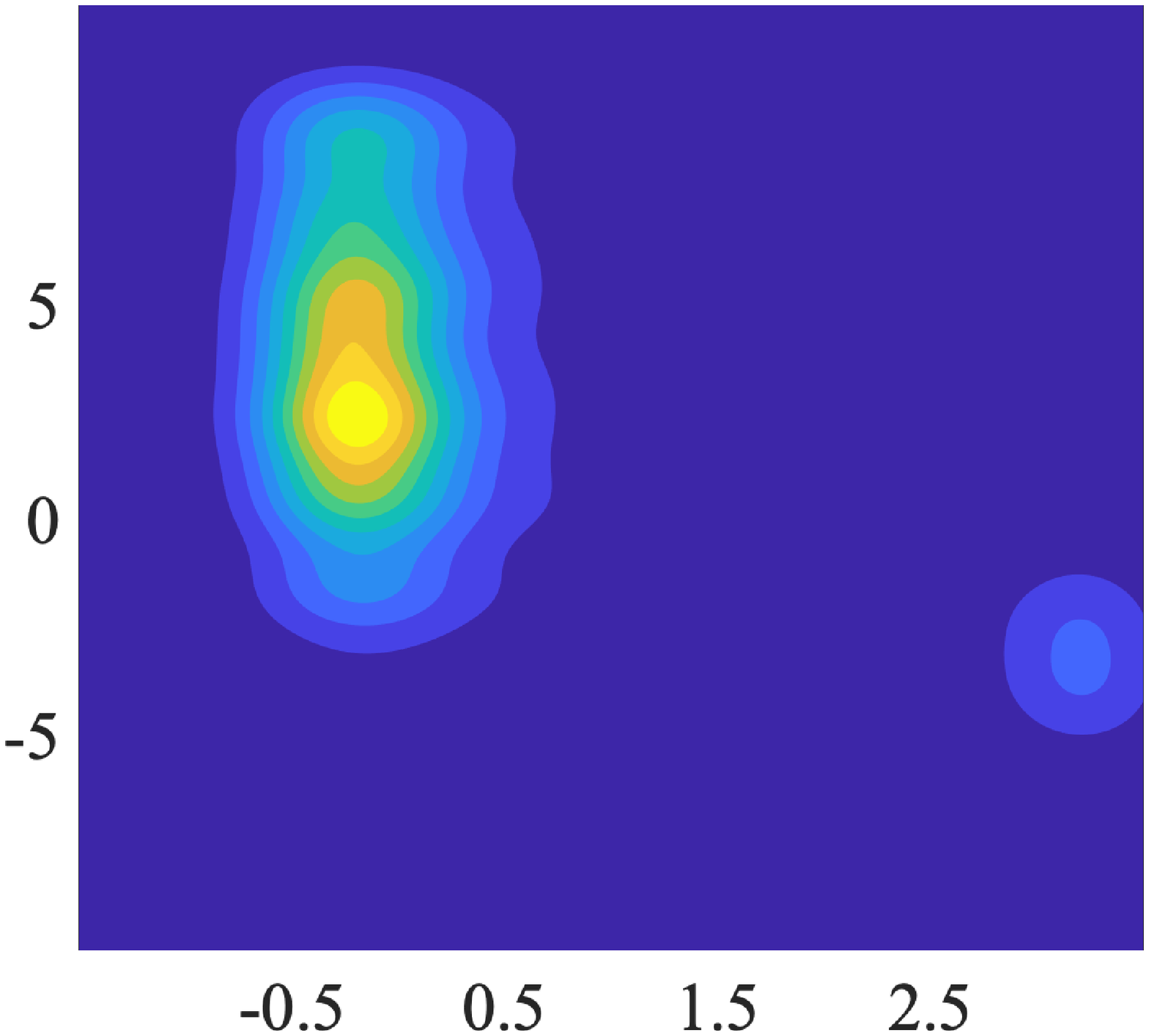}
		\end{subfigure} 
		\vspace{2mm}
	\end{minipage}
	\begin{minipage}[t]{\textwidth}
		\centering \footnotesize(d) $Y_{t+h}$ given $(Y_t,Z_t)$ for $t$=2009:Q2 \\		
		\begin{subfigure}[t]{0.23\textwidth}
			\caption{\footnotesize $h=1$}	
			\includegraphics[width=0.95\textwidth]{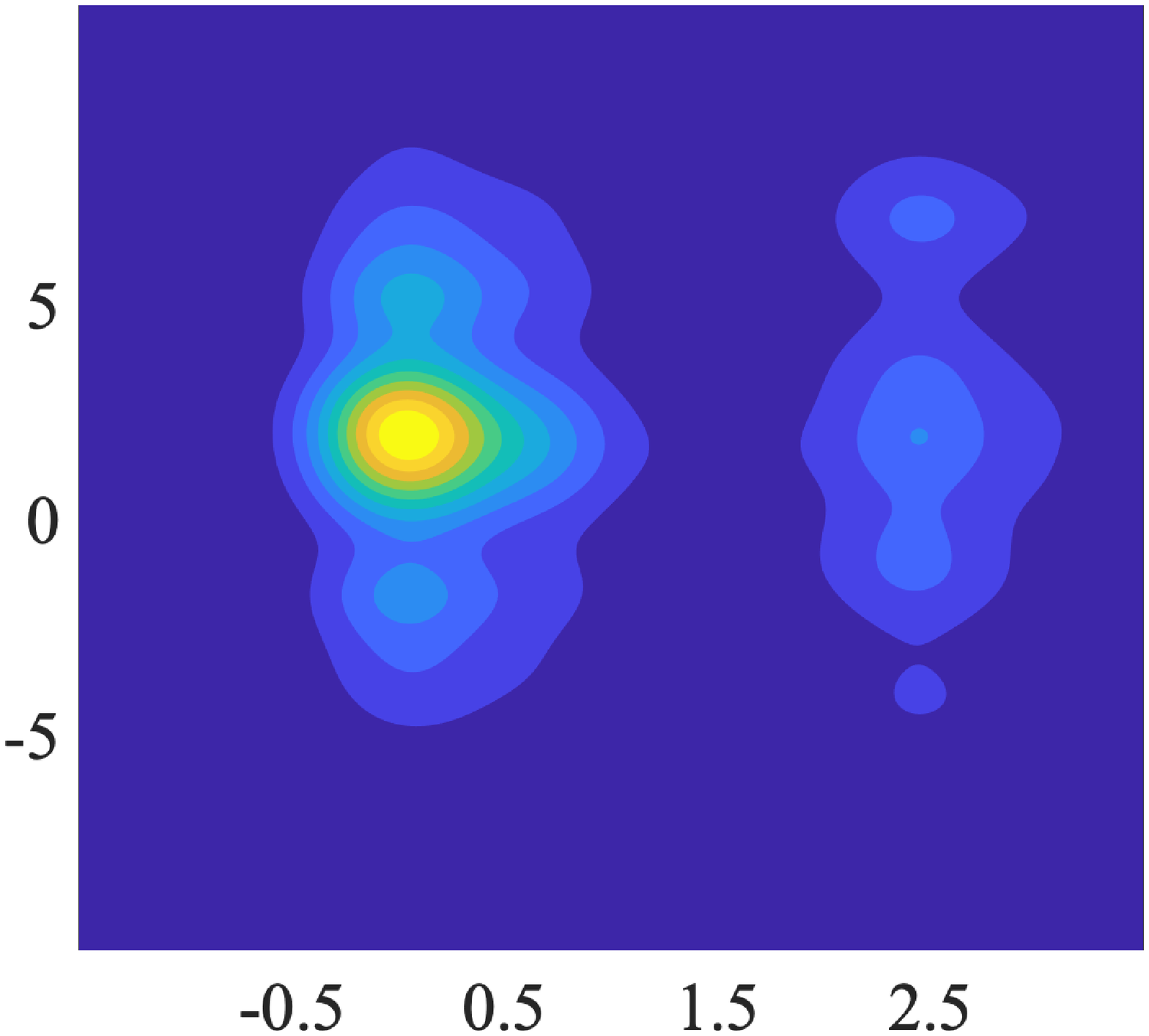}     		
		\end{subfigure}
		\begin{subfigure}[t]{0.23\textwidth}
			\caption{\footnotesize $h=2$}
			\includegraphics[width=0.95\textwidth]{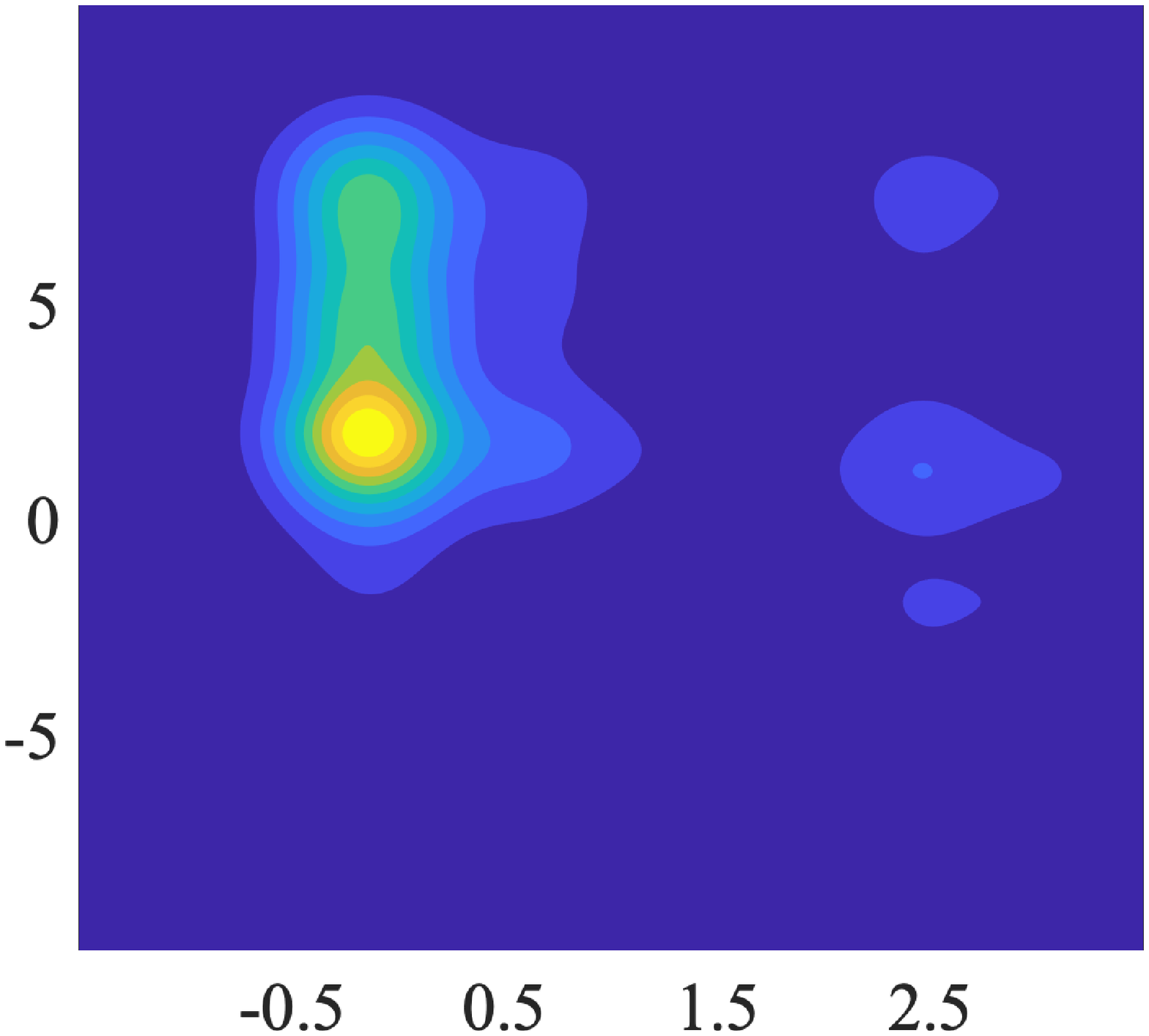}
		\end{subfigure}
		\begin{subfigure}[t]{0.23\textwidth}
			\caption{\footnotesize $h=3$}
			\includegraphics[width=0.95\textwidth]{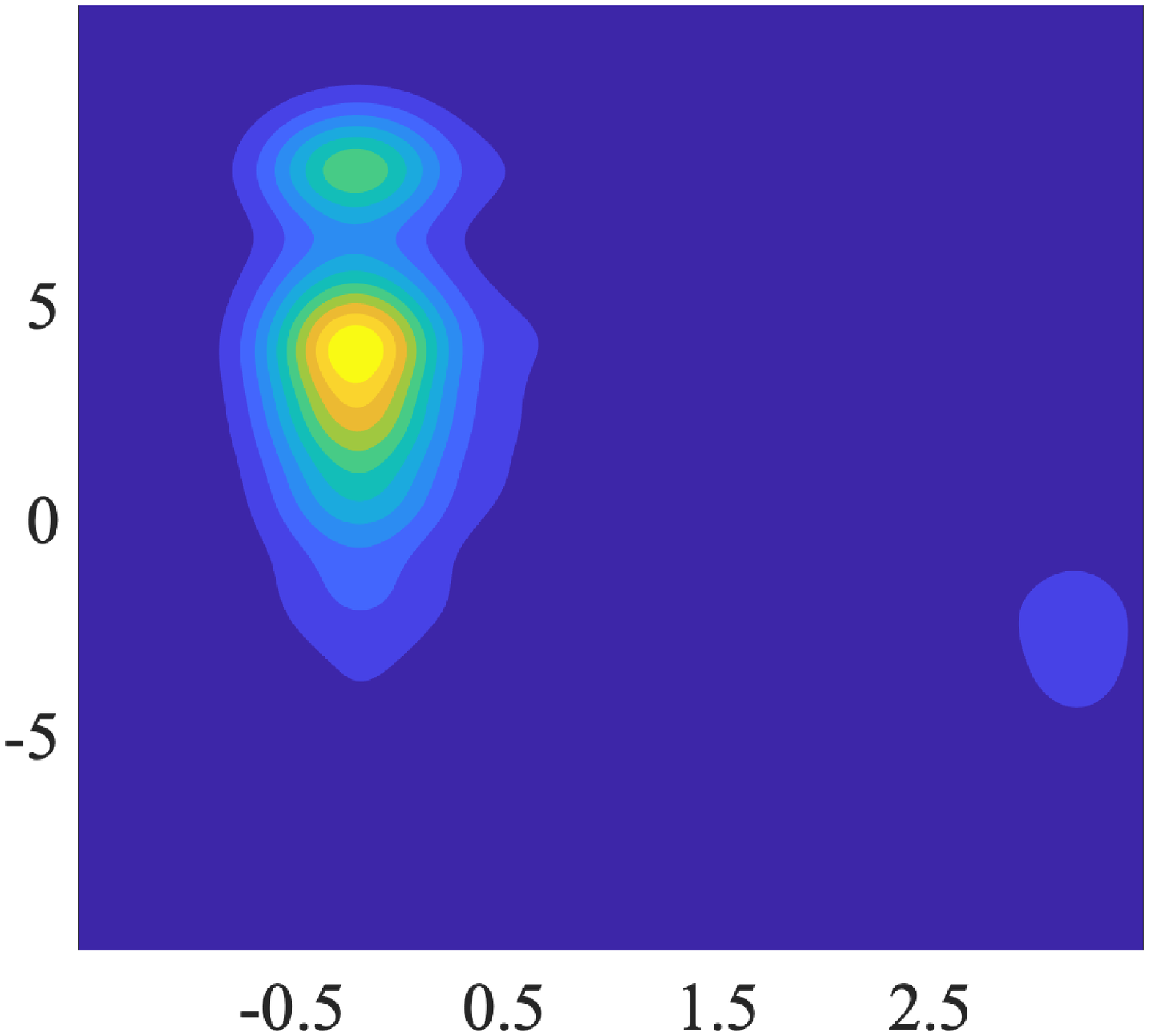}
		\end{subfigure}
		\begin{subfigure}[t]{0.23\textwidth}
			\caption{\footnotesize $h=4$}	
			\includegraphics[width=0.95\textwidth]{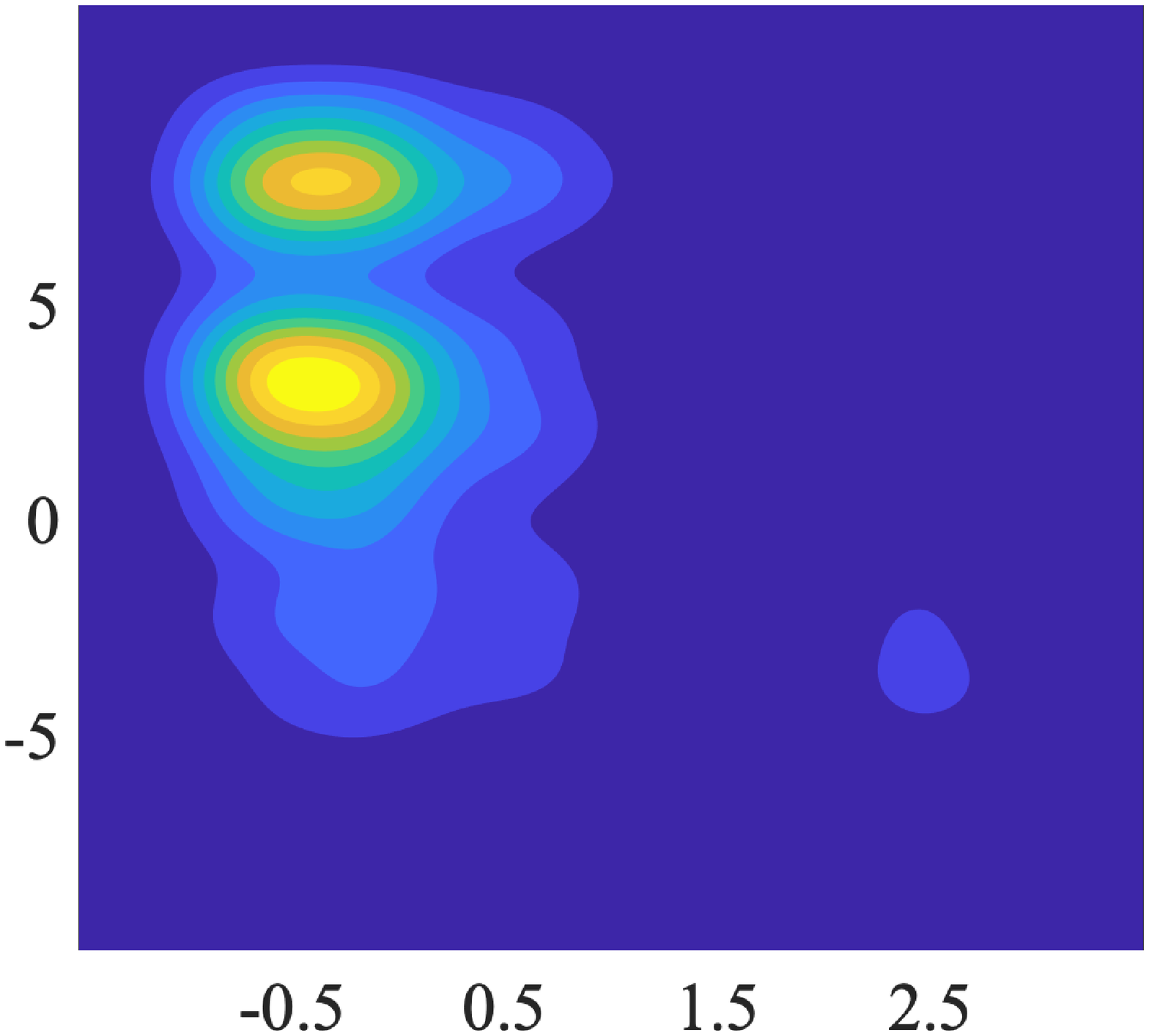}
		\end{subfigure} 
		\vspace{2mm}
	\end{minipage}
	\begin{minipage}[t]{\textwidth}
		\centering \footnotesize(e) $Y_{t+h}$ given $(Y_t,Z_t)$ for $t$=2009:Q3 \\		
		\begin{subfigure}[t]{0.23\textwidth}
			\caption{\footnotesize $h=1$}	
			\includegraphics[width=0.95\textwidth]{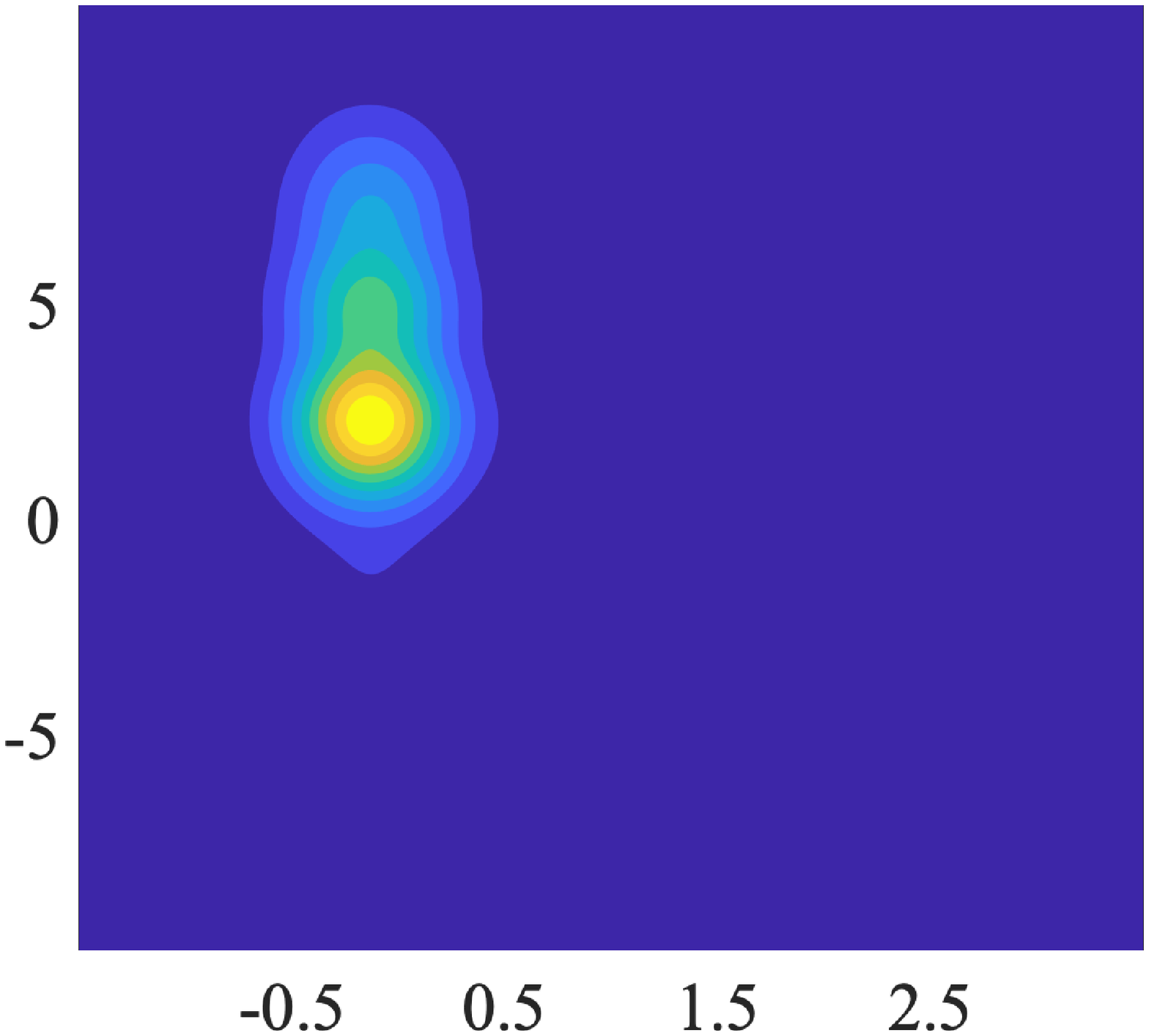}     		
		\end{subfigure}
		\begin{subfigure}[t]{0.23\textwidth}
			\caption{\footnotesize $h=2$}
			\includegraphics[width=0.95\textwidth]{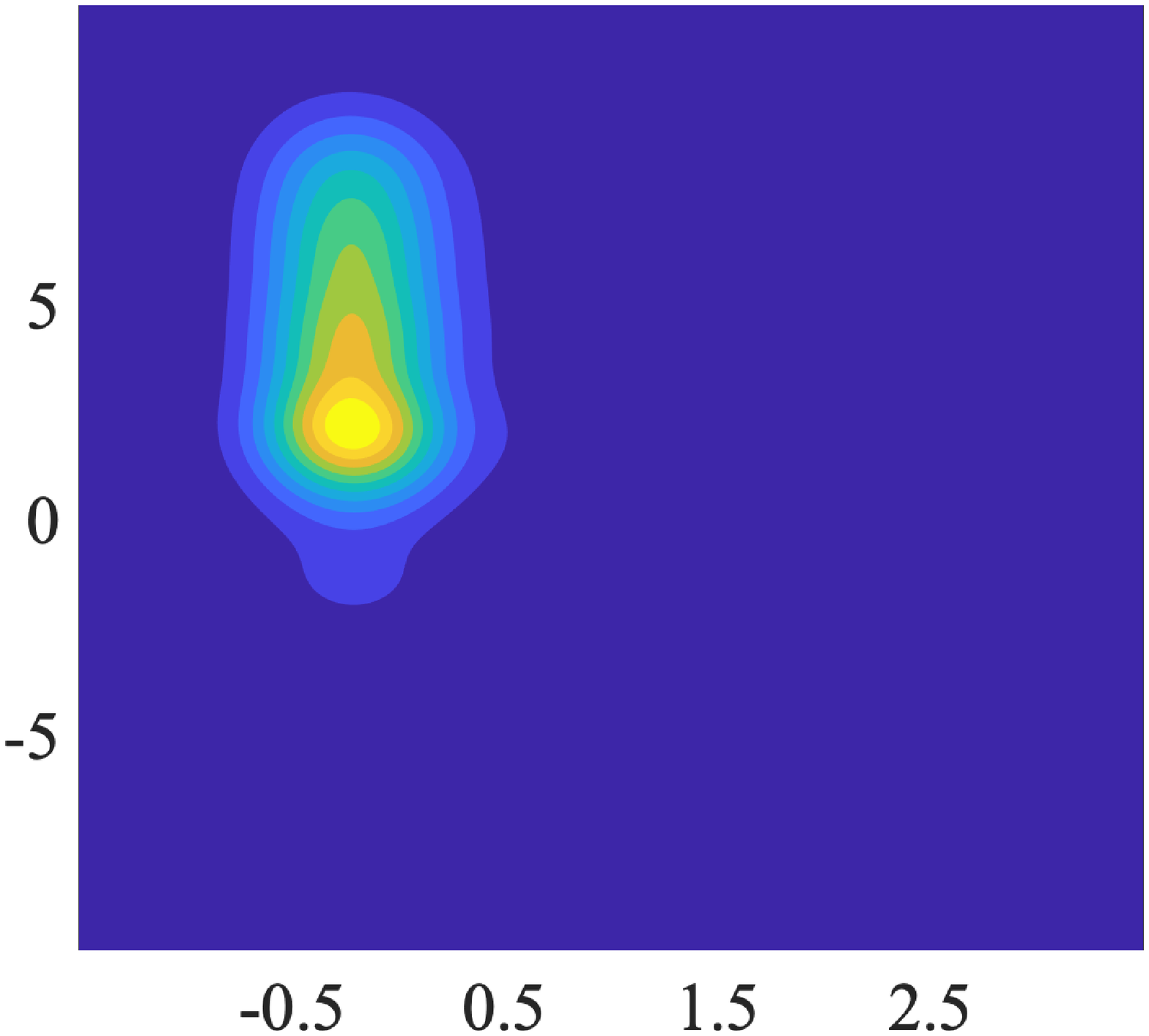}   
		\end{subfigure}
		\begin{subfigure}[t]{0.23\textwidth}
			\caption{\footnotesize $h=3$}
			\includegraphics[width=0.95\textwidth]{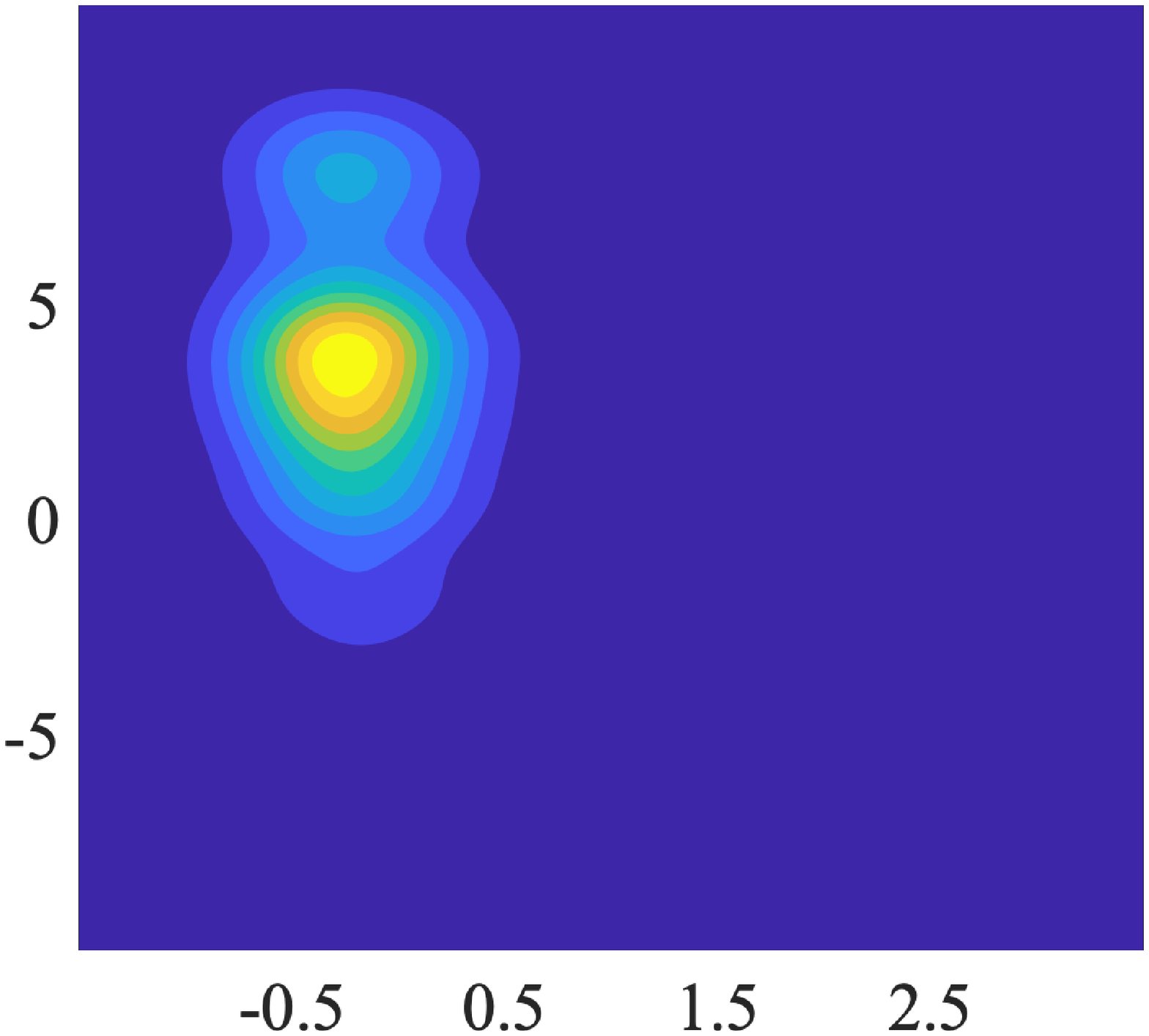}  			
		\end{subfigure}
		\begin{subfigure}[t]{0.23\textwidth}
			\caption{\footnotesize $h=4$}
			\includegraphics[width=0.95\textwidth]{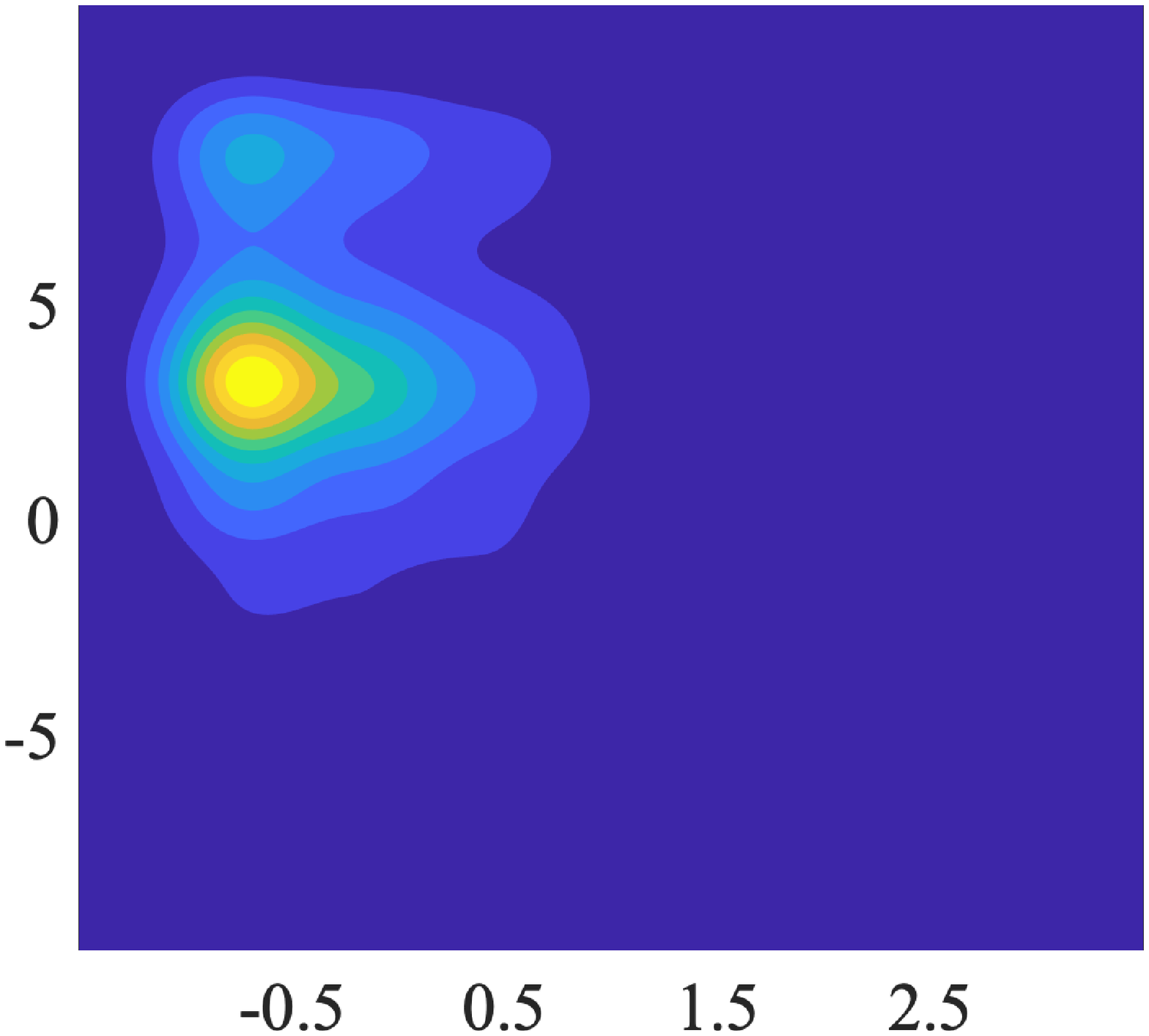}  			
		\end{subfigure} 
	\end{minipage}
   \vspace{0.2cm}
	\begin{minipage}{.93\linewidth} 
		\linespread{1}\footnotesize
		\textit{Notes}: Each contour plot (with NFCI in $x$-axis and  GDP in $y$-axis) is estimated using bivariate kernel density estimation with a bandwidth of $(0.25,0.8)$ based on samples generated from the multistep forecasting distribution. Different columns correspond to forecast horizons $h=1,2,3,4$, and different rows correspond to different conditioning information from 2008:Q1-Q3 to 2009:Q1-Q3.
  	\end{minipage}
\end{figure}

We further study how the joint distribution of the real GDP growth and NFCI evolved during the Great Recession. The joint distribution dynamics of the out-of-sample forecasting distributions, with different columns corresponding to forecast horizons from one (leftmost column, $h=1$) to four (rightmost column, $h=4$) quarters and rows corresponding to different conditioning information from 2008:Q1-Q3 (top row) to 2009:Q1-Q3 (bottom row), are illustrated in Figure \ref{fig: contour-plots}. 

The joint distributions of the real GDP growth and NFCI predicted using the data up to 2008:Q1-Q3, displayed in the top row of Figure \ref{fig: contour-plots}, are characterized by a single mode for all forecasting horizons. As the forecasting horizon increases, there is an increased likelihood of higher growth and looser financial conditions. However, with the inclusion of information from 2008:Q4, the predicted distributions exhibit multimodality. As depicted in the second row of Figure \ref{fig: contour-plots}, the one-quarter-ahead predicted distribution displays two distinct modes, both centered around low GDP growth of approximately $-2$ and tight financial conditions (one around 1 and the other around 2). As the forecasting horizon extends, the predicted distribution gradually shifts its weight towards higher GDP growth and looser financial conditions. Finally, the one-year-ahead distribution resolves into a unimodal distribution centered near 2 for real GDP growth and average financial conditions (the NFCI is approximately 0). As additional information from 2009:Q1 and 2009:Q2 becomes available, the predicted distributions in the third and fourth rows evolve similarly to those shown in the second row. However, given the more recent information, the multimodality resolves more quickly. Notably, the predicted distributions conditional on information as of 2009:Q1-Q3 are approximately unimodal for all horizons, as shown in the last row of Figure \ref{fig: contour-plots}. It is noteworthy that the multimodality in the distributions primarily stems from the shift in the distribution of the NFCI. 

Therefore, the distributional behavior depicted in Figure \ref{fig: contour-plots} implies that during normal periods, the joint distribution of the real GDP growth and NFCI is characterized by a single mode. During periods of tight financial conditions, however, there is a marked change in the distribution's shape, with the emergence of multiple modes. When financial conditions are less severe, the multimodality is observed only in short-term forecasts and is usually resolved within a couple of quarters. Additionally, the plots indicate that as the forecasting horizon extends, both variables become increasingly volatile, with the real GDP growth exhibiting a greater degree of uncertainty.

\subsection{Counterfactual Analysis During the Great Recession}\label{subsec: counterfactual}
Assuming that the information available about the economic and financial conditions is up to 2008:Q3, based on the out-of-sample forecasting distributions, we use the DIRF to explore the potential policy effects during the Great Recession. Specifically, we investigate the impact of the policy intervention aimed at limiting the possibility of tightening financial conditions or worsening GDP growth during 2008:Q4 on the predicted distributions in the following quarters. 
\subsubsection{Counterfactual Analysis of Distributional Impulse on the NFCI}

\begin{figure}[H]
	\captionsetup[subfigure]{aboveskip=-1pt,belowskip=0pt}
	\centering
	\caption{Distributional Impulse on NFCI in 2008:Q4} \label{fig: NFCI-Impulse}	
	\begin{minipage}[t]{\textwidth}	
		\centering Distribution of $Y_t$ given $Z_t$ for $t$ equals to 2008:Q4 ($h=0$)\\
		\begin{subfigure}[t]{0.32\textwidth}
			\centering 		
			\includegraphics[width=0.95\textwidth]{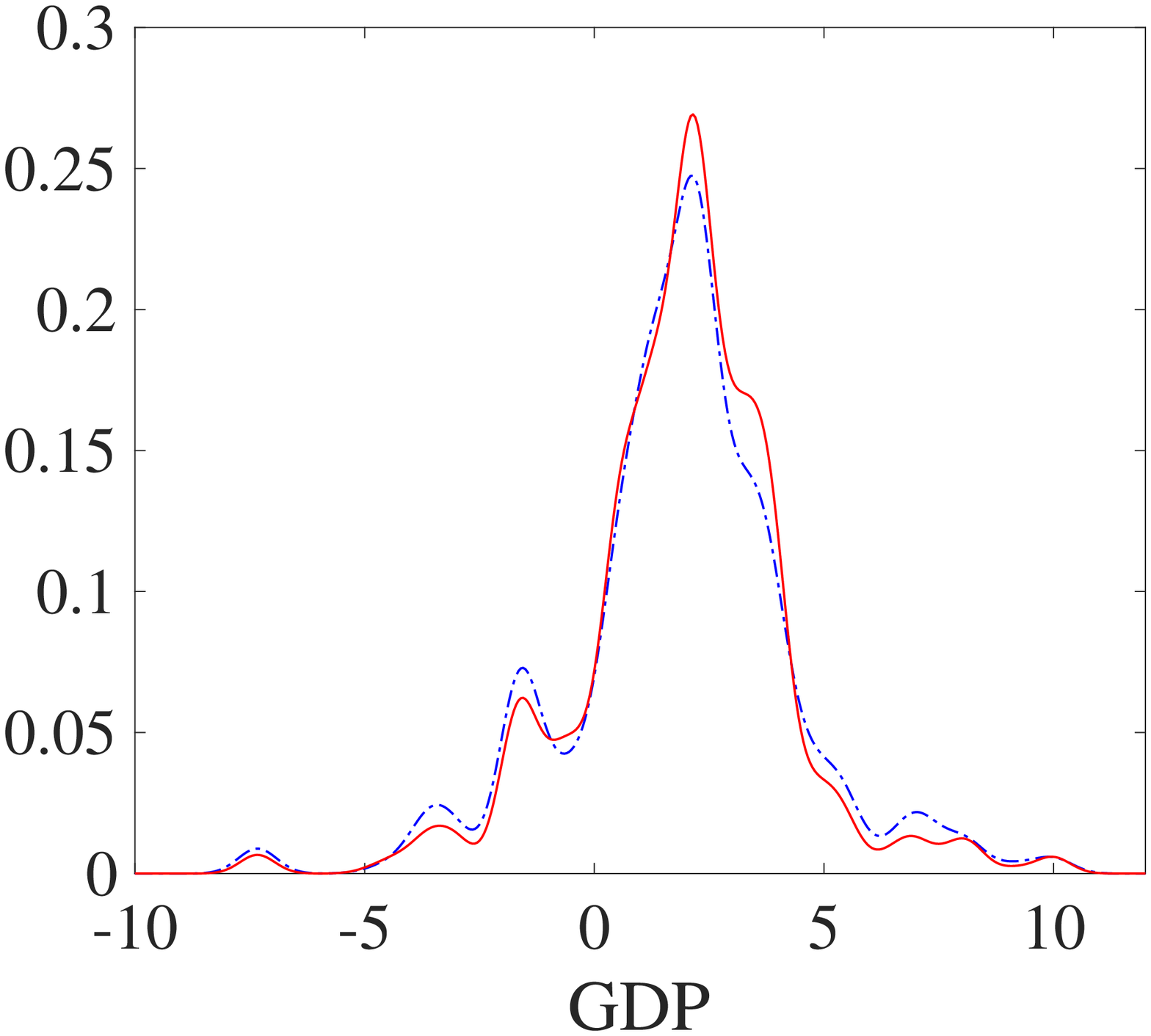}     	
			\label{fig:Q-NFCI-h0}
		\end{subfigure}
		\begin{subfigure}[t]{0.32\textwidth}
			\centering		
			\includegraphics[width=0.95\textwidth]{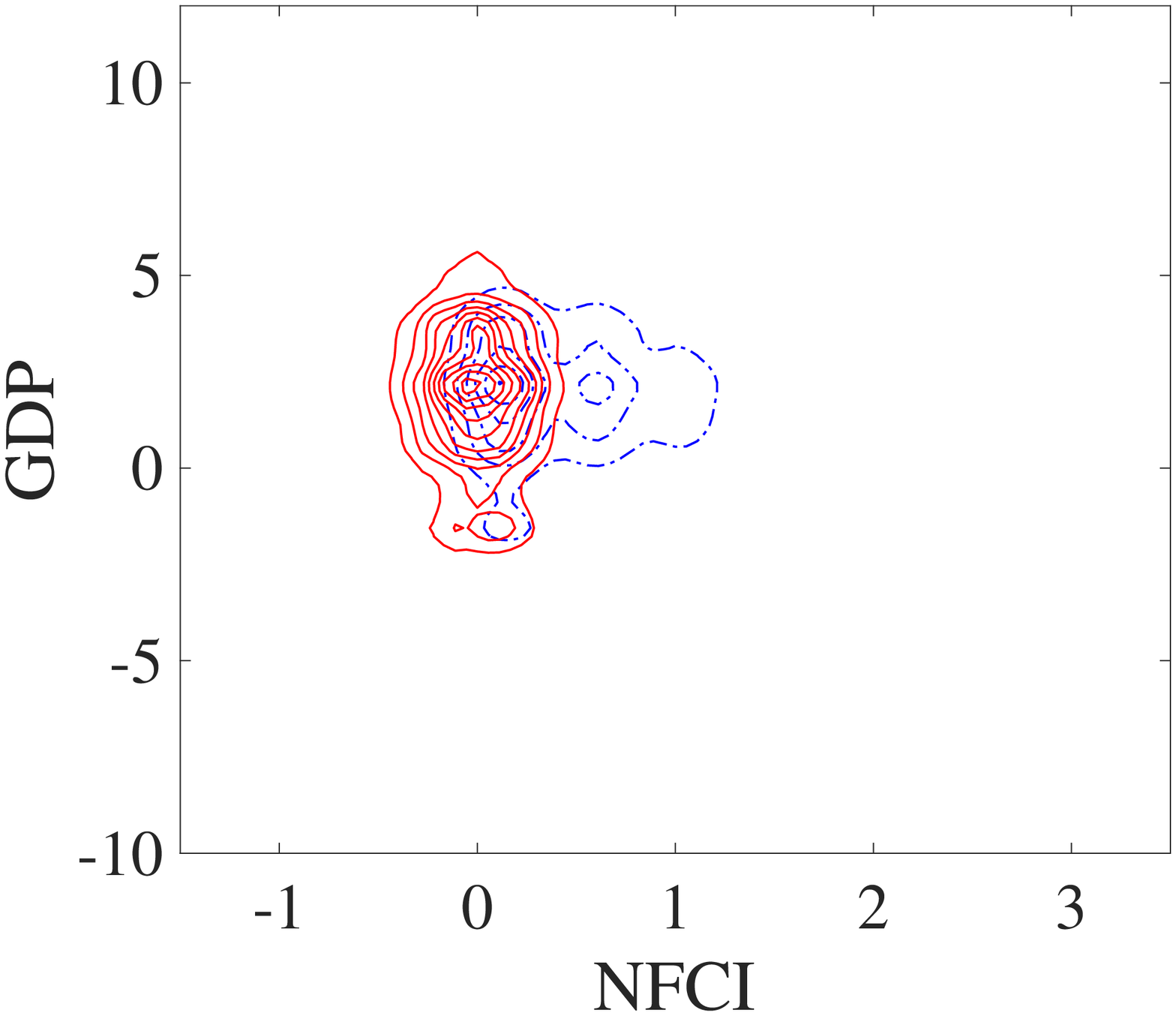}
			\label{fig:Q-joint-h0}
		\end{subfigure}
		\begin{subfigure}[t]{0.32\textwidth}
			\centering	
			\includegraphics[width=0.95\textwidth]{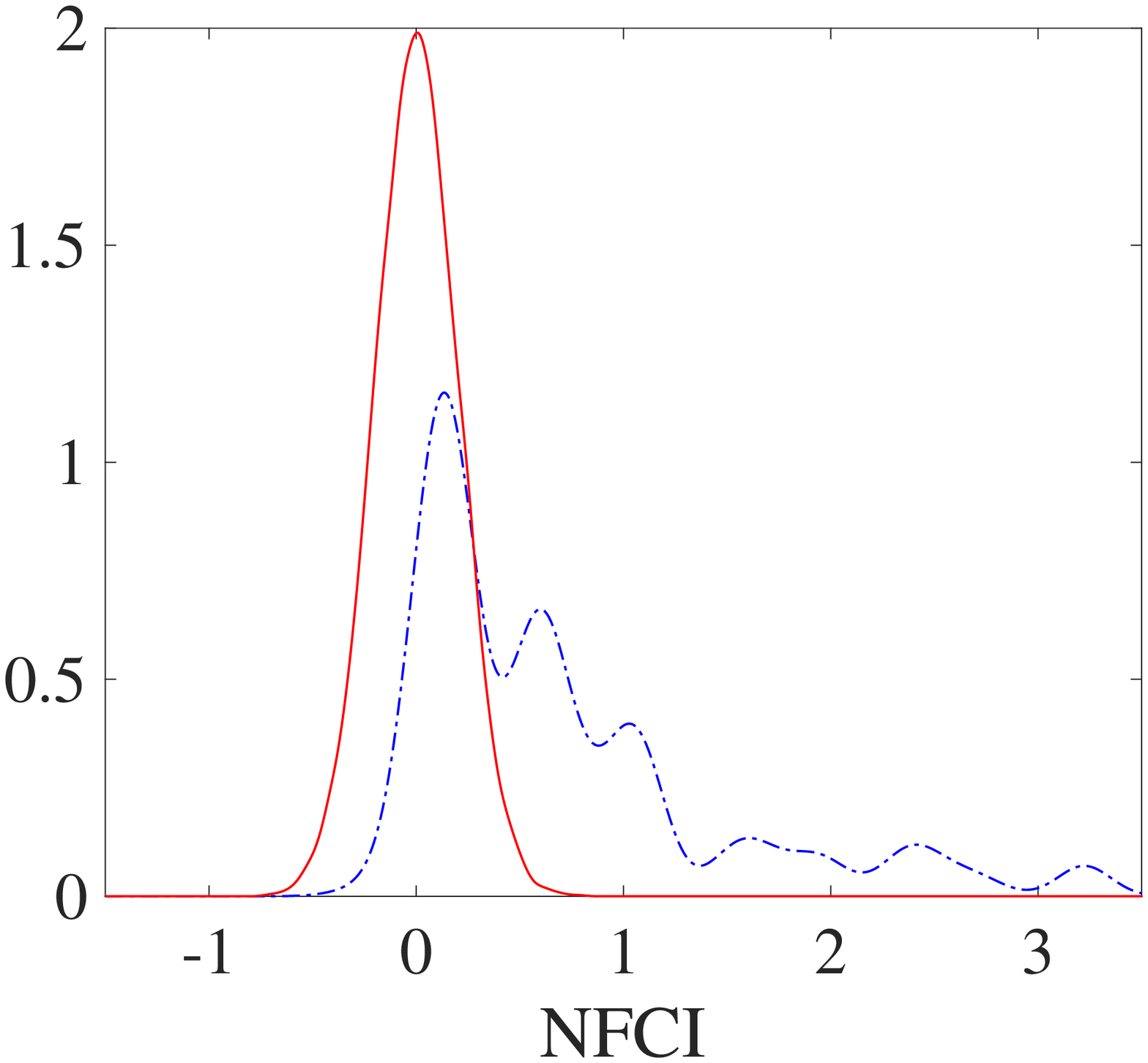}
			\label{fig:Q-GDP-h0}
		\end{subfigure}
	\end{minipage}
	\begin{minipage}{.95\linewidth} 
		\linespread{1}\footnotesize
		\textit{Notes}: Based on samples of the real GDP growth and NFCI generated from the joint distribution of 2008:Q4, conditional on 2008:Q1-Q3, the marginal densities and the joint contour plot are estimated using kernel density estimation with `rule-of-thumb' bandwidths. The baseline distributions are plotted with blue dotdash lines, and counterfactual distributions are plotted with red solid lines.
	\end{minipage}
\end{figure}

We first explore the effect of the policy intervention in 2008:Q3, which could limit the possibility of tightening financial conditions during 2008:Q4. A counterfactual distribution, truncated normal distribution with mean of $0$ and standard deviation of $0.2$ on $(-1.5,2)$, is considered for the NFCI in 2008:Q4. In Figure \ref{fig: NFCI-Impulse}, we provide the initial one-step-ahead (baseline) joint and marginal distributions in 2008:Q4, together with their counterfactual counterparts under the distributional impulse on $Y_{1t}$. Additionally, their differences in different quantiles over the distribution and moments, including the mean, standard deviation, skewness and kurtosis, are presented in Table \ref{tab: DIR-QN-MQ}, with $h=0$. The distributional impulse significantly reduces the 95\% quantiles, skewness, and kurtosis of the NFCI, thereby also lowering other quantiles and moments. Simultaneously, it increases the 95\% quantiles of GDP by approximately 1 and kurtosis by approximately 0.65; however, it has a minor influence on other quantiles and moments. The results suggest a minor contemporaneous effect of the NFCI shock on the GDP.

\begin{figure}[H]
	\captionsetup[subfigure]{aboveskip=-3pt,belowskip=0pt}
	\centering
	\caption{Distributional Response to the NFCI Impulse} \label{fig: NFCI-response}		
	\begin{minipage}[t]{\textwidth}
		\centering  Distributions of $Y_{t+h}$ given $Z_t$ for $t$=2008:Q4\\$h=1$\\
		\begin{subfigure}[t]{0.32\textwidth} 		
			\includegraphics[width=0.95\textwidth]{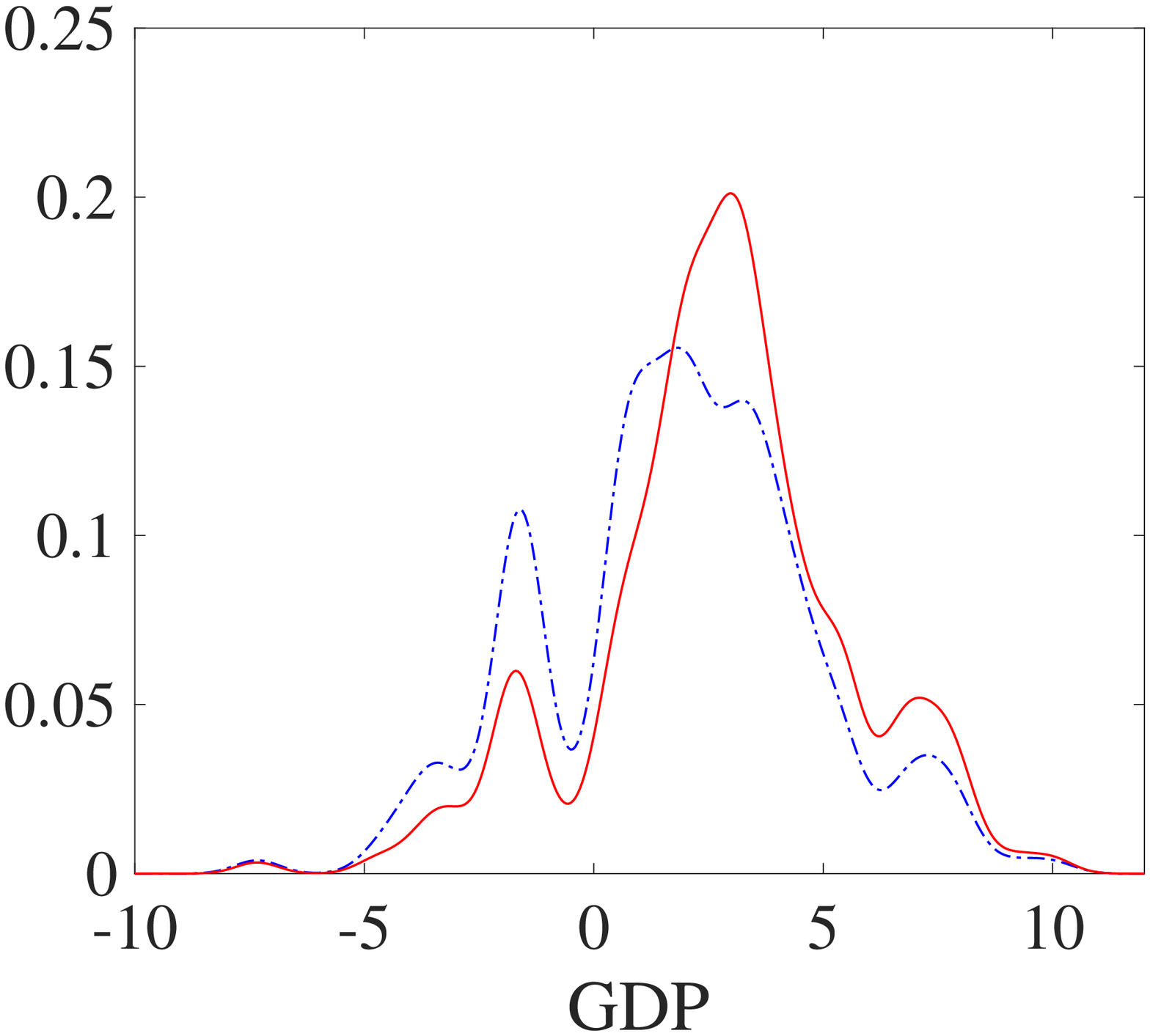}     		
		\end{subfigure}
		\begin{subfigure}[t]{0.32\textwidth}
			\includegraphics[width=0.95\textwidth]{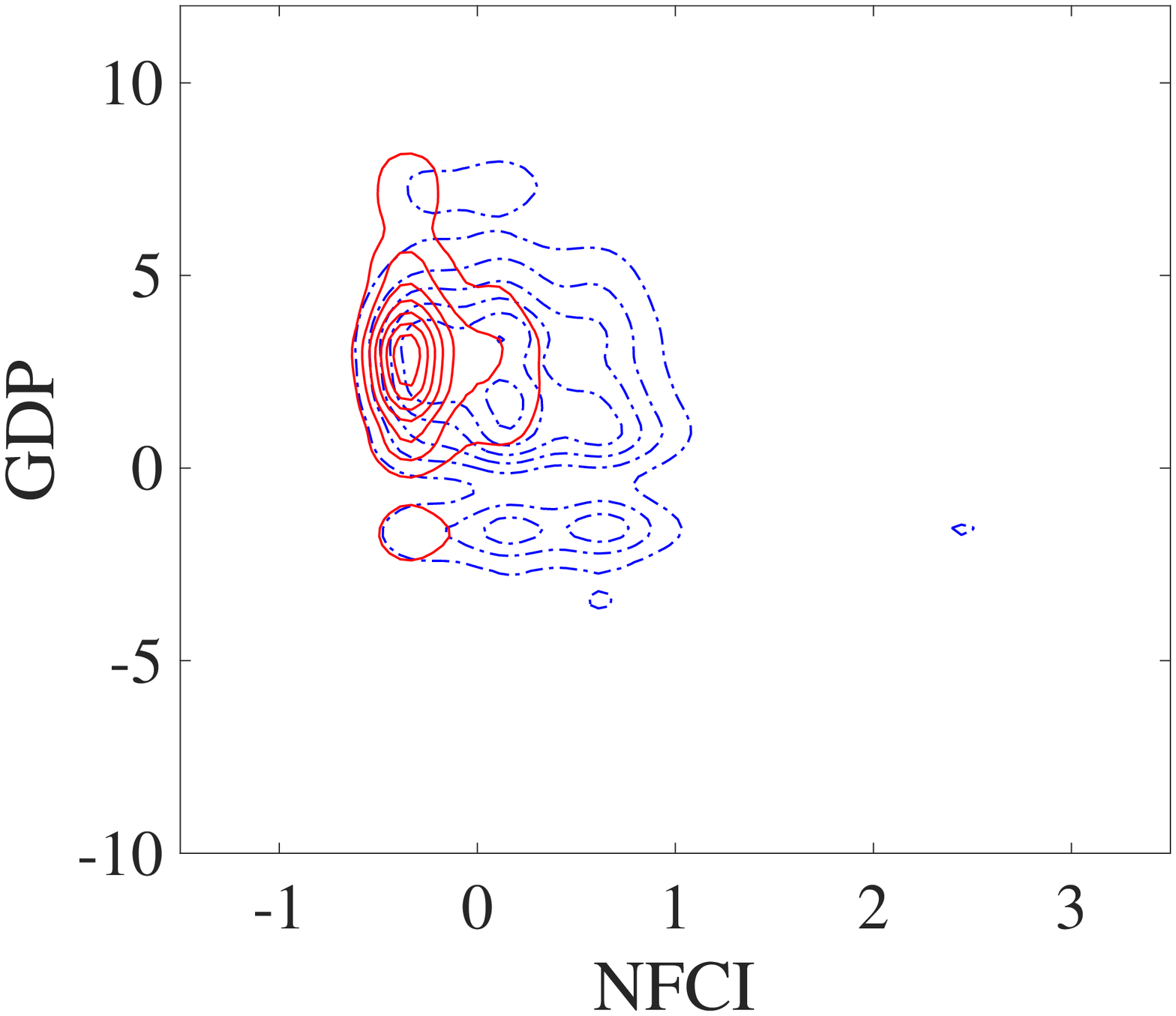}
		\end{subfigure}
		\begin{subfigure}[t]{0.32\textwidth}
			\includegraphics[width=0.95\textwidth]{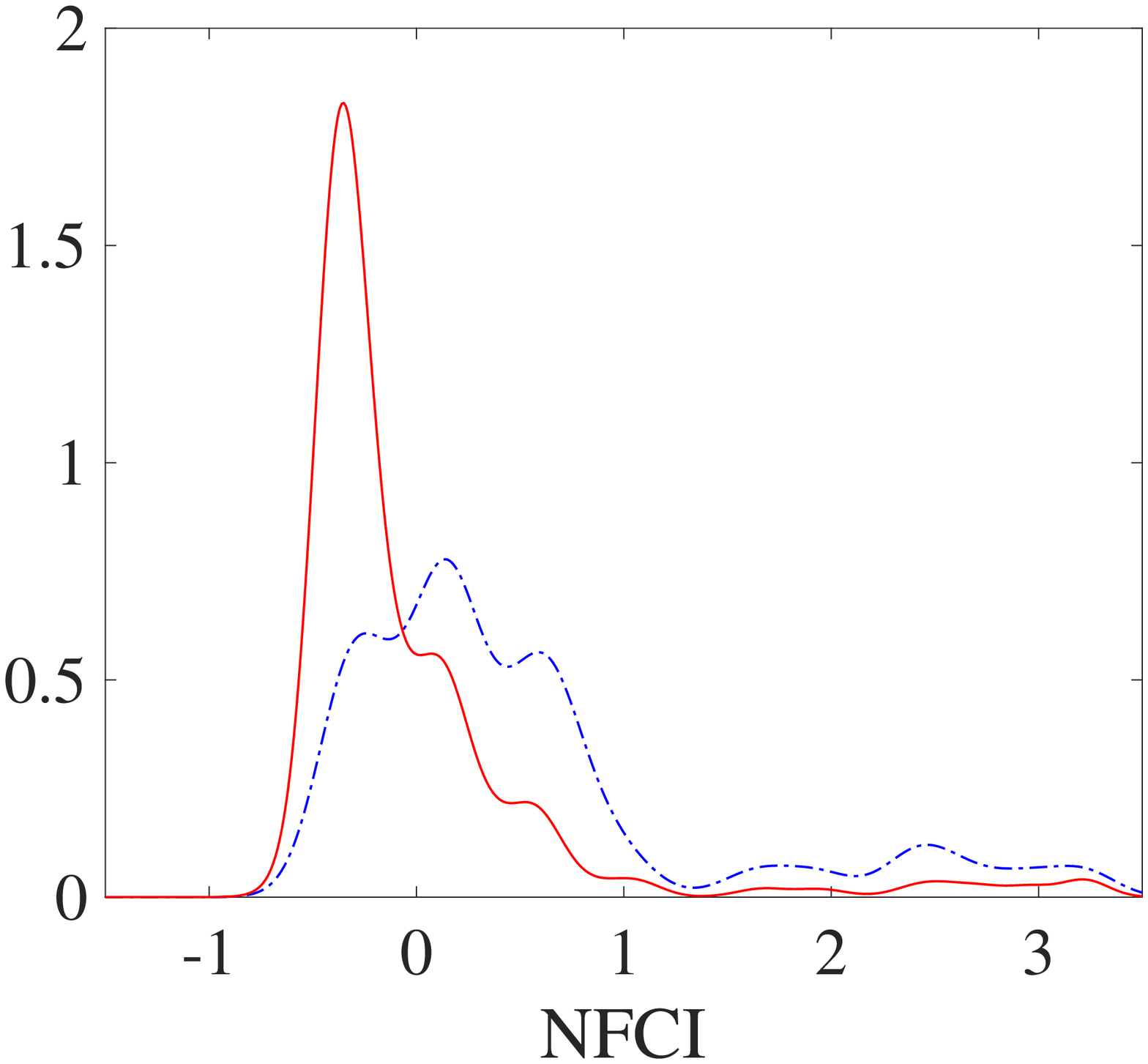}
		\end{subfigure}
		\vspace{0.1cm}
	\end{minipage}	
	\begin{minipage}[t]{\textwidth}
		\centering $h=2$\\
		\begin{subfigure}[t]{0.32\textwidth} 		
			\includegraphics[width=0.95\textwidth]{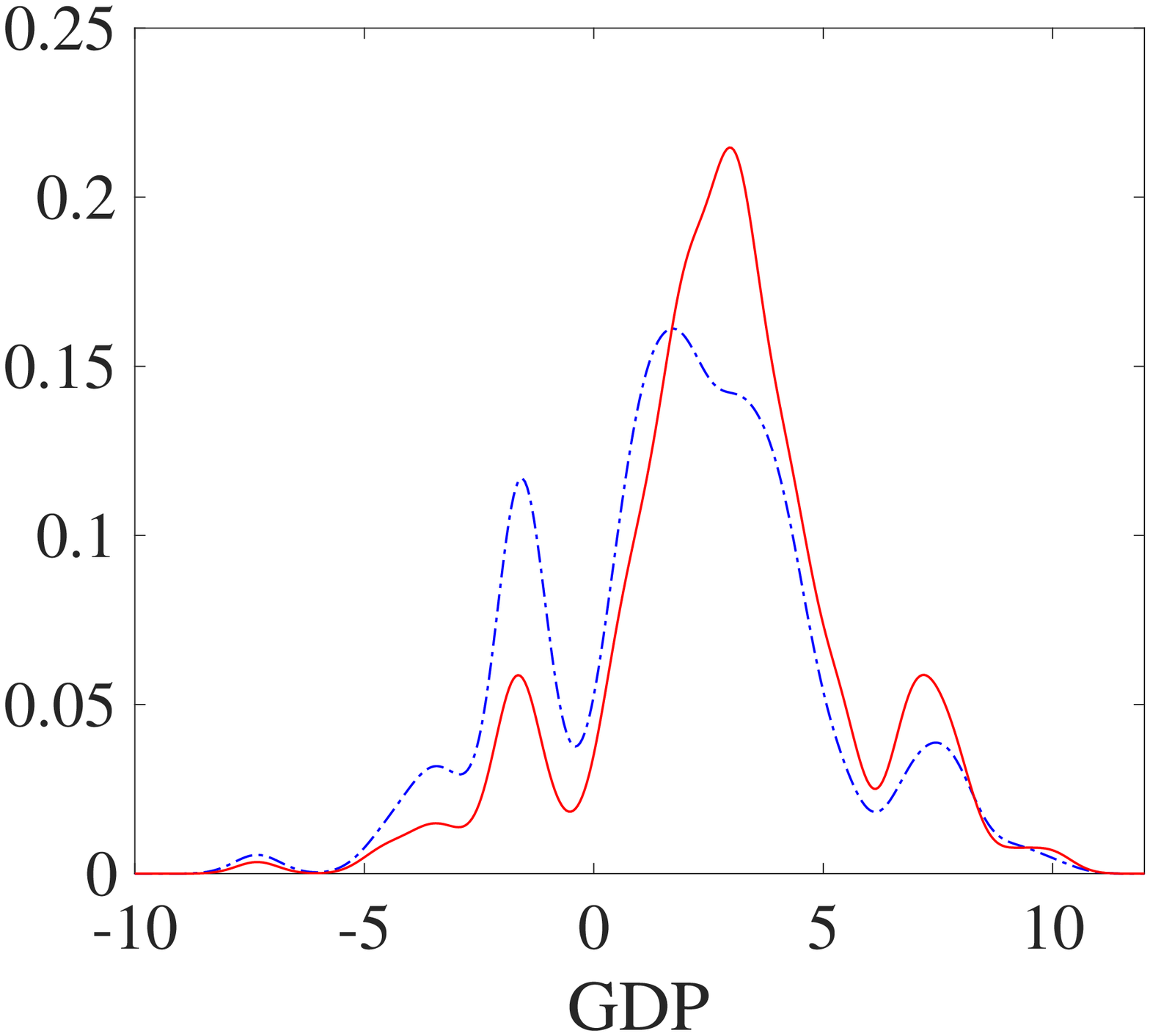}     		
		\end{subfigure}
		\begin{subfigure}[t]{0.32\textwidth}
			\includegraphics[width=0.95\textwidth]{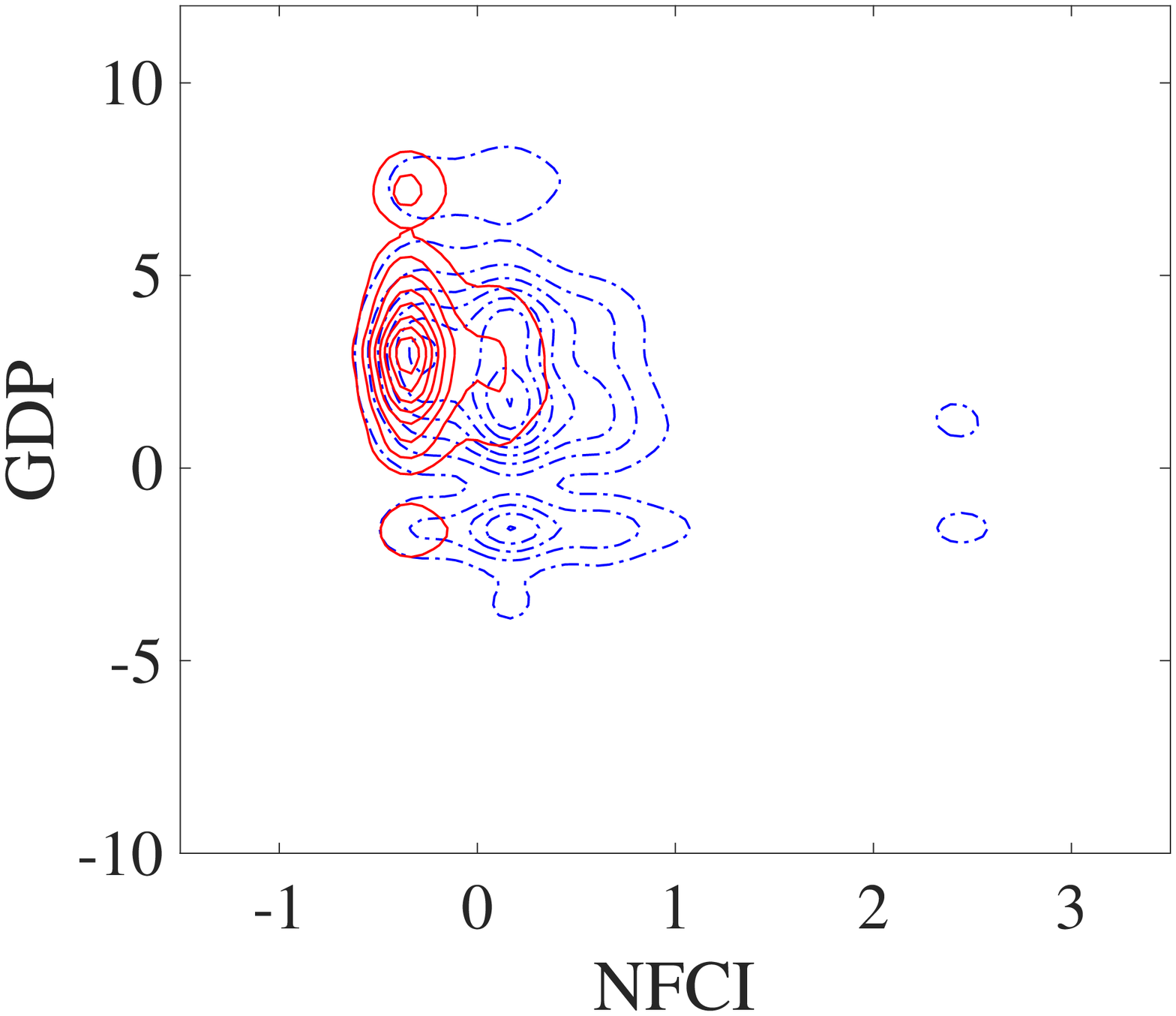}
		\end{subfigure}
		\begin{subfigure}[t]{0.32\textwidth}
			\includegraphics[width=0.95\textwidth]{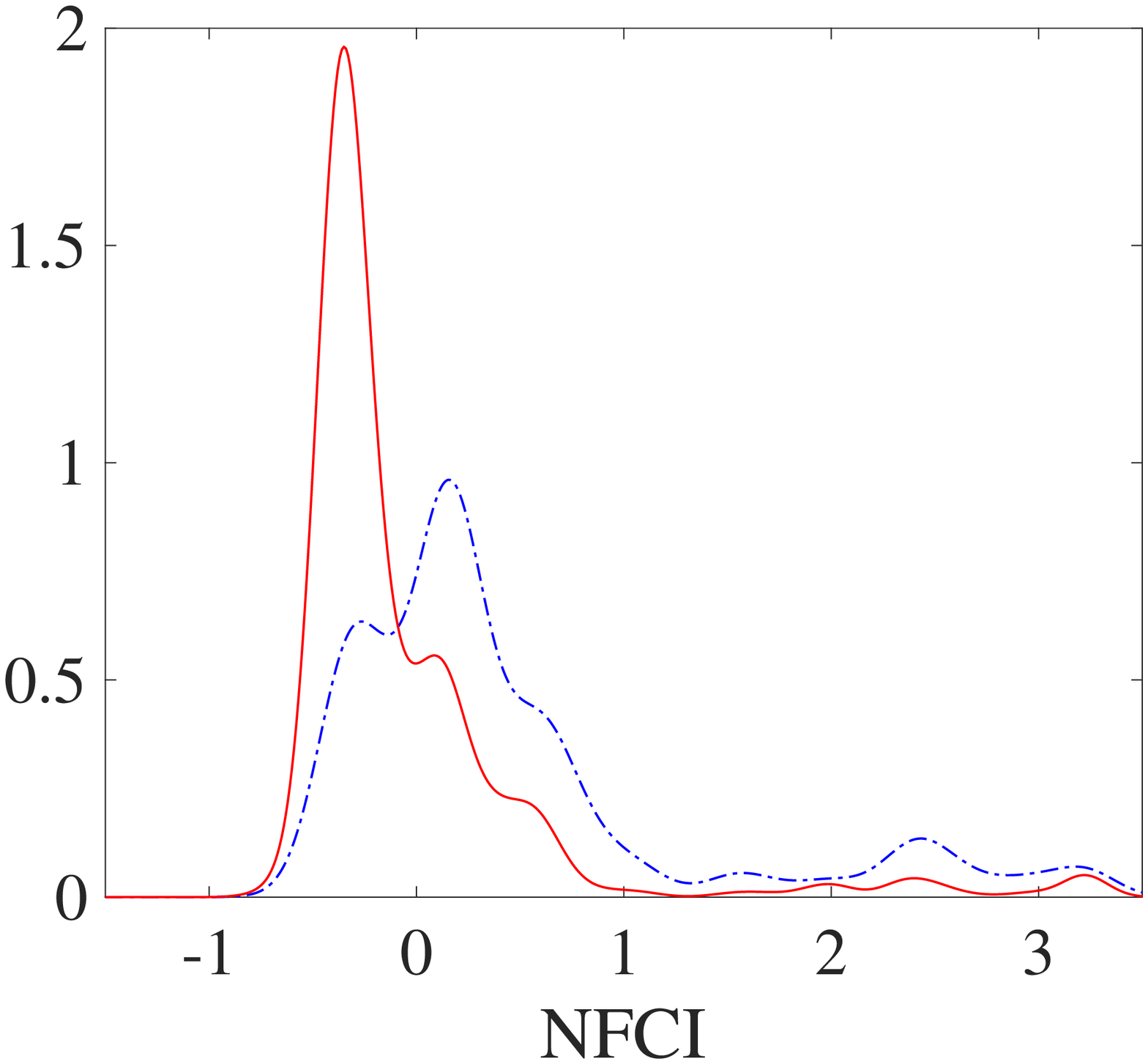}
		\end{subfigure}
		\vspace{0.1cm}
	\end{minipage}
	\begin{minipage}[t]{1\textwidth}
		\centering $h=3$\\
		\begin{subfigure}[t]{0.32\textwidth} 		
			\includegraphics[width=0.95\textwidth]{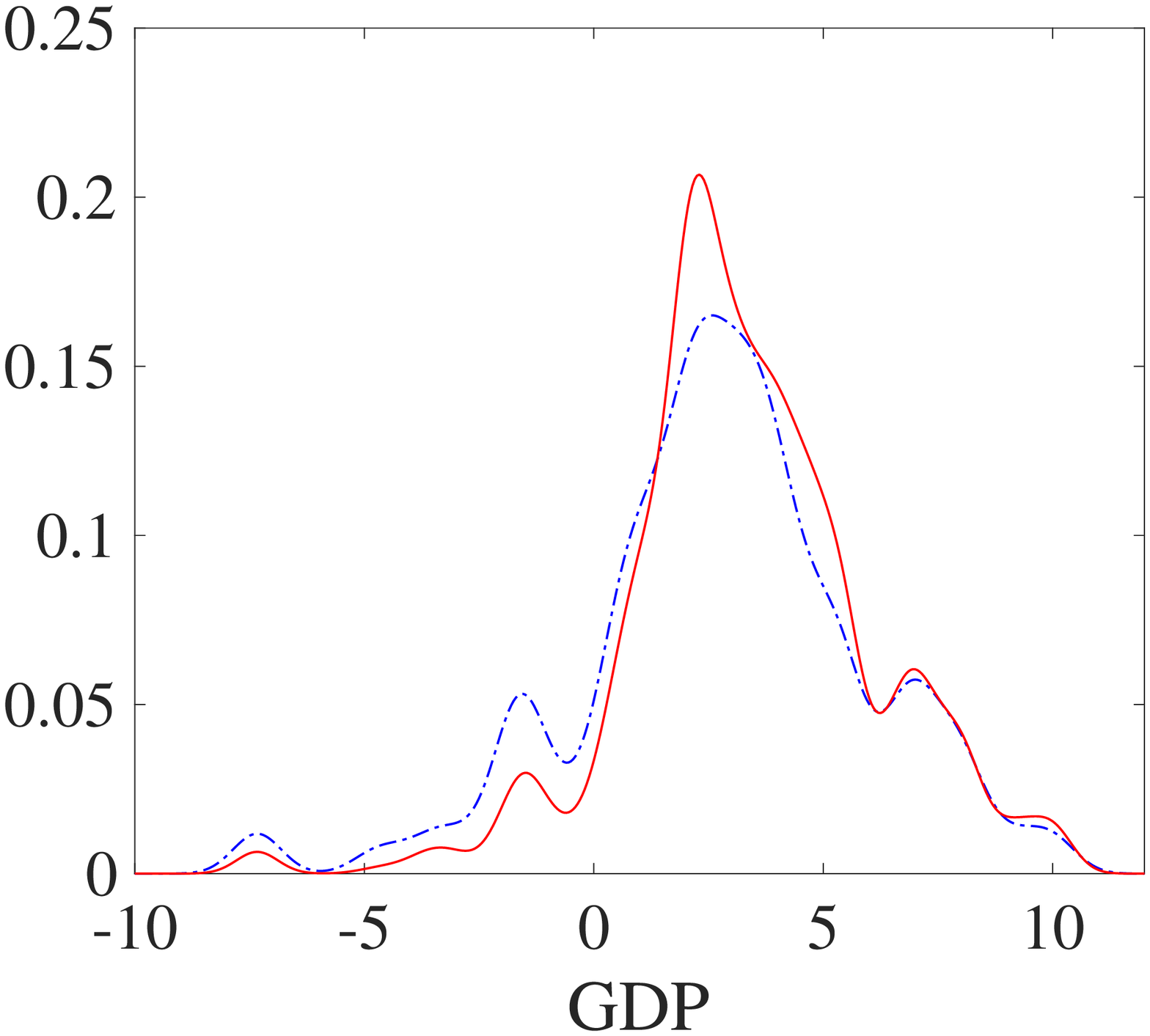}     		
		\end{subfigure}
		\begin{subfigure}[t]{0.32\textwidth}
			\includegraphics[width=0.95\textwidth]{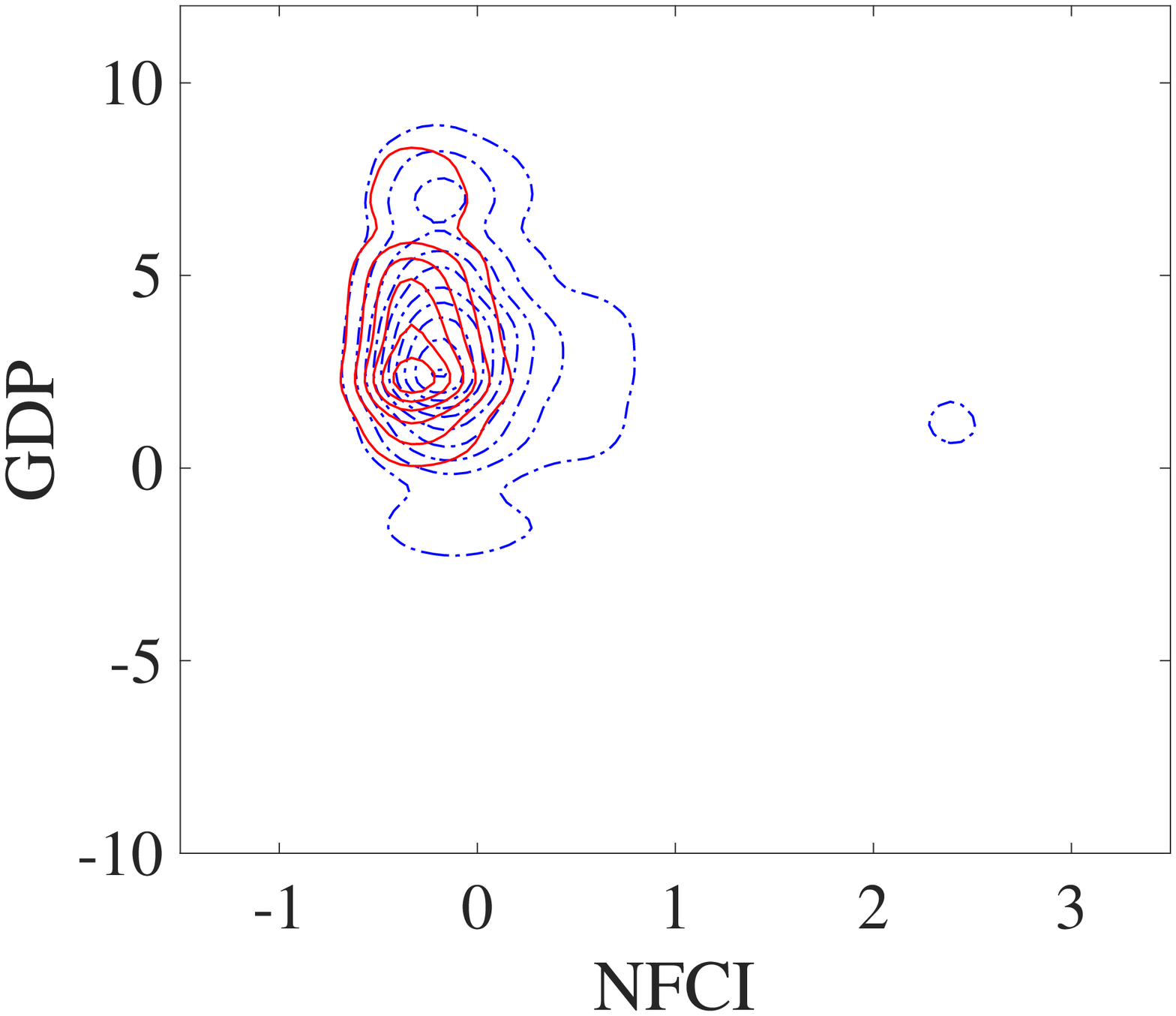}
		\end{subfigure}
		\begin{subfigure}[t]{0.32\textwidth}
			\includegraphics[width=0.95\textwidth]{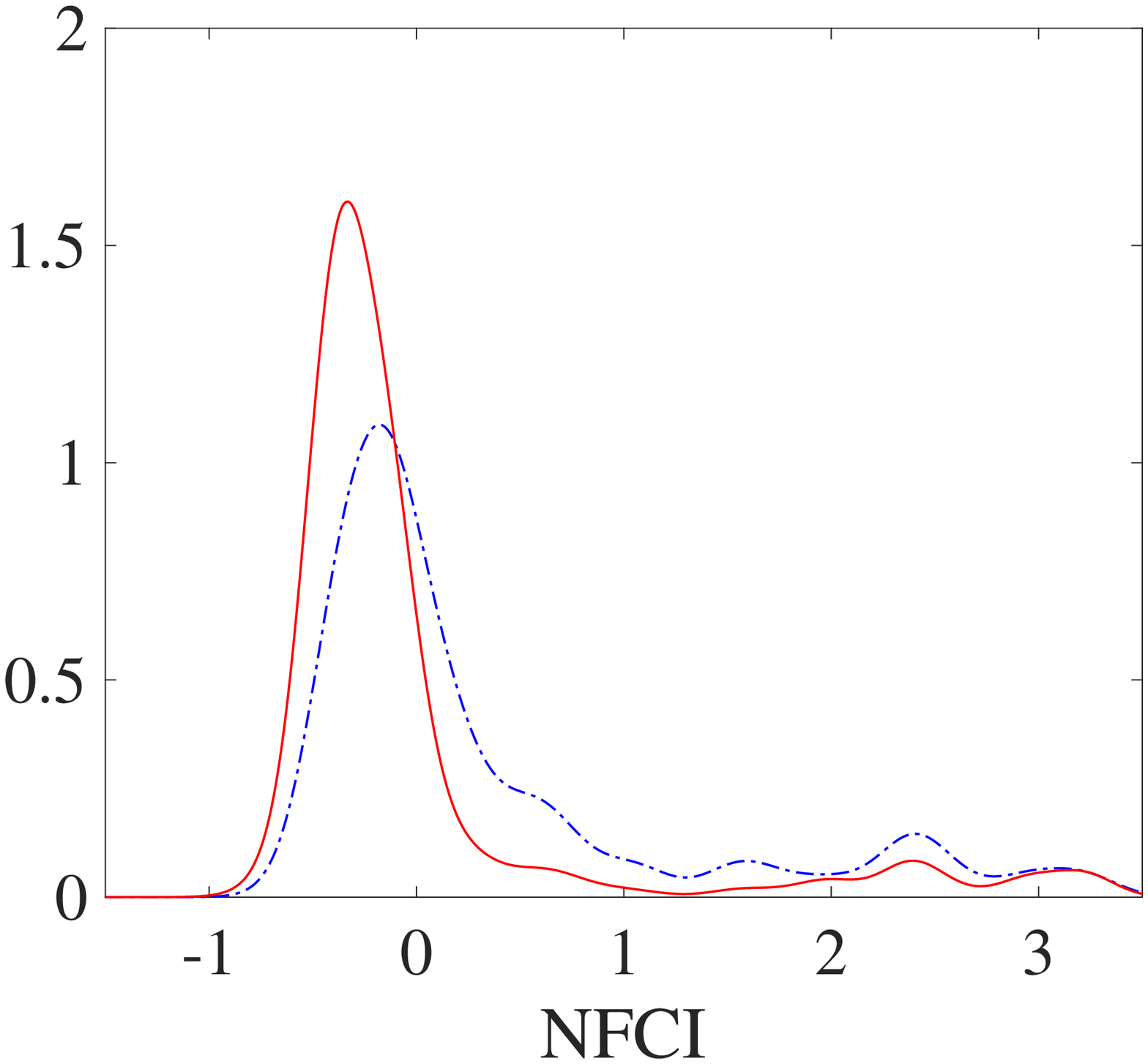}
		\end{subfigure}
		\vspace{0.1cm}
	\end{minipage}
	\begin{minipage}[t]{1\textwidth}
		\centering $h=4$\\
		\begin{subfigure}[t]{0.32\textwidth} 		
			\includegraphics[width=0.95\textwidth]{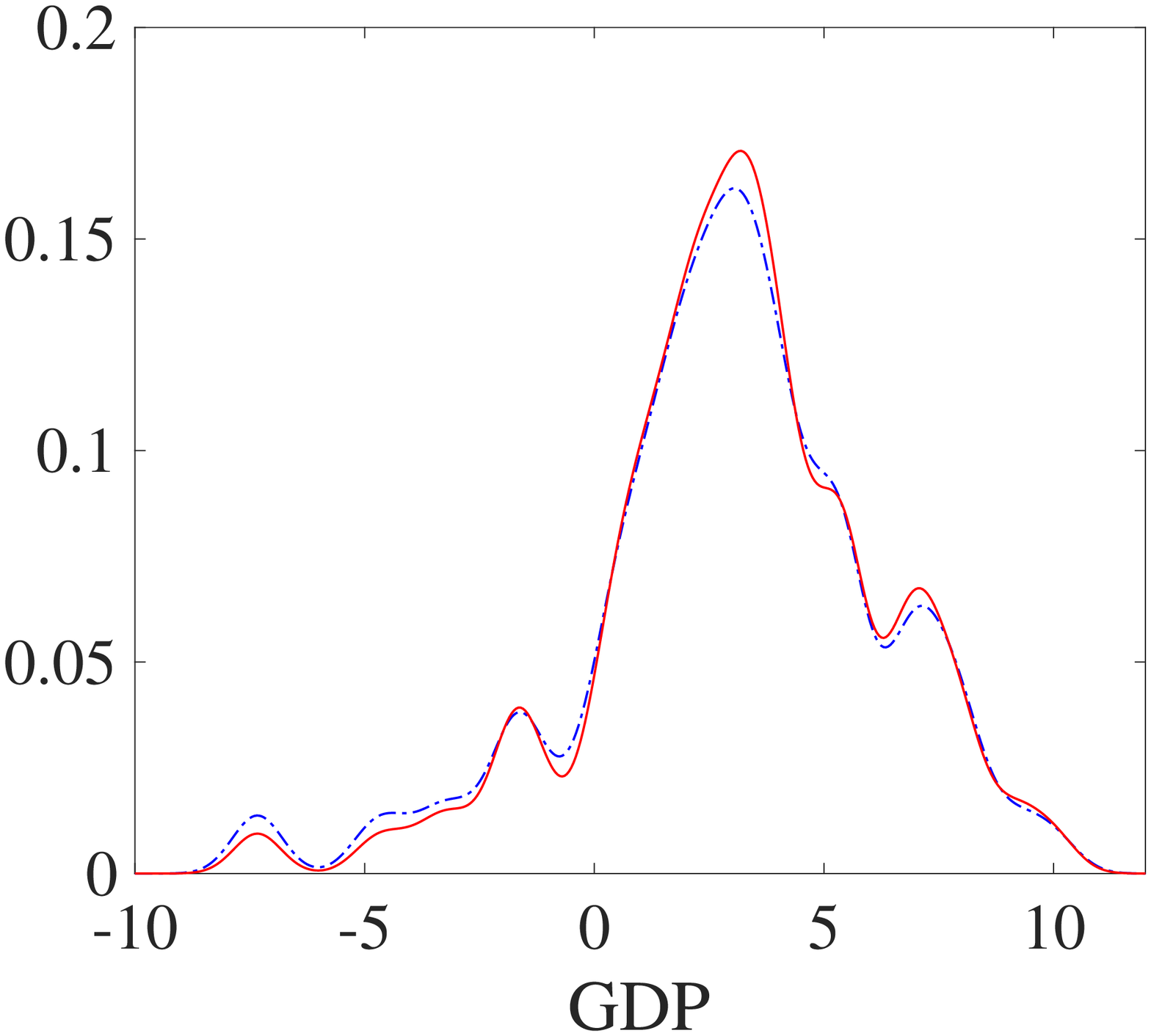}     		
			\label{fig:Q-NFCI-h1}
		\end{subfigure}
		\begin{subfigure}[t]{0.32\textwidth}
			\includegraphics[width=0.95\textwidth]{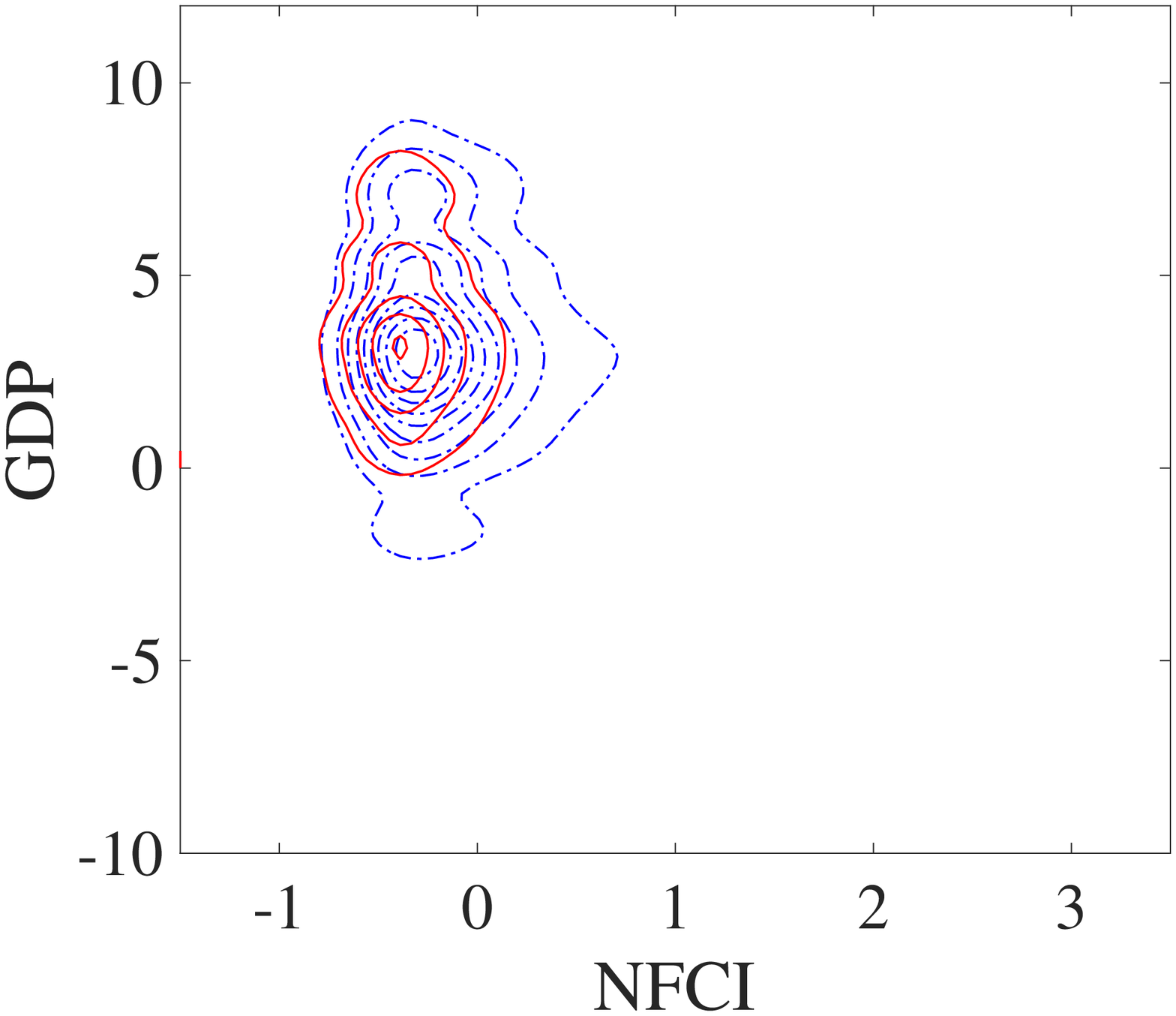}
			\label{fig:Q-joint-h1}
		\end{subfigure}
		\begin{subfigure}[t]{0.32\textwidth}
			\includegraphics[width=0.95\textwidth]{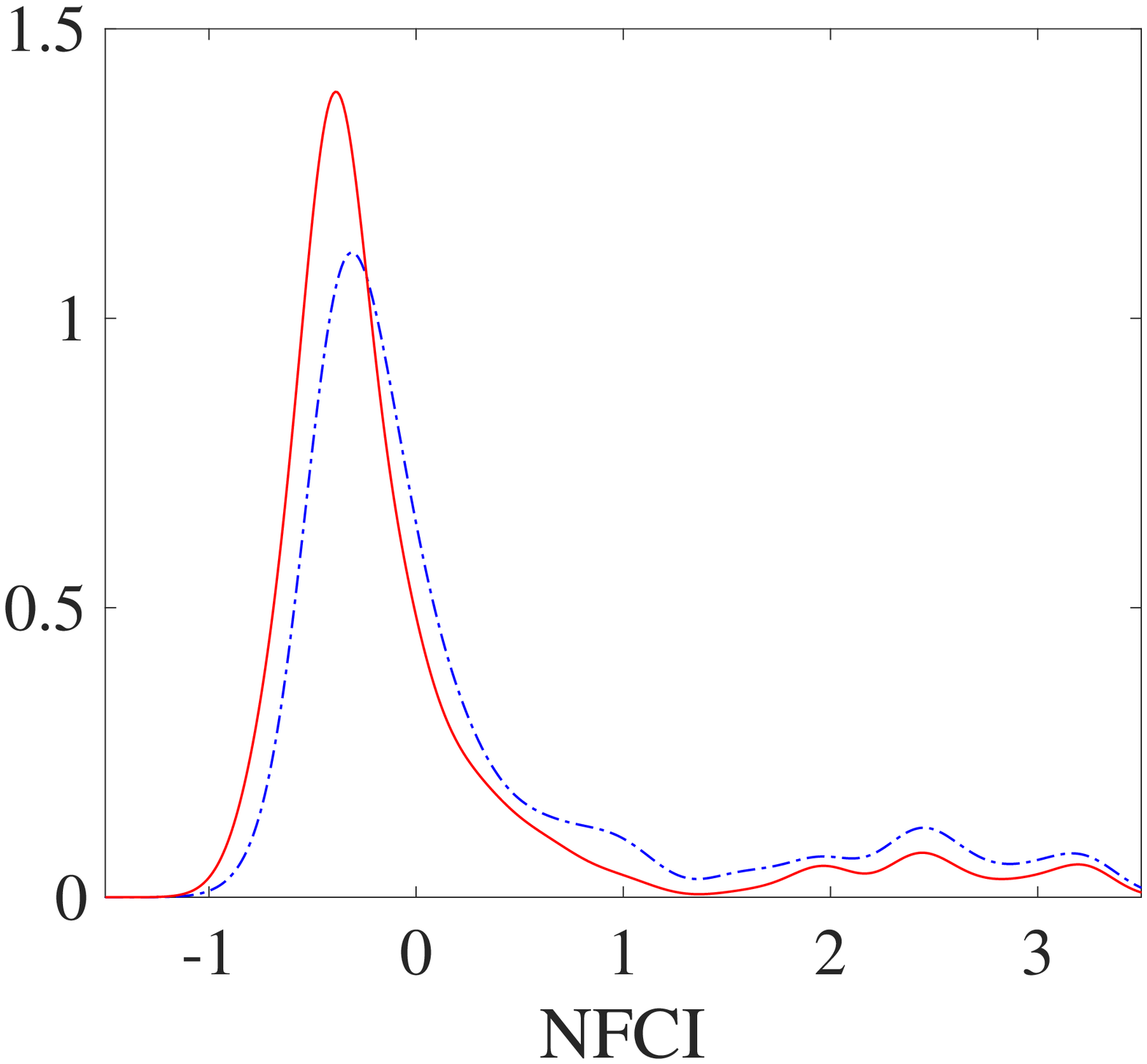}
			\label{fig:Q-GDP-h1}
		\end{subfigure}
	\end{minipage}
	\begin{minipage}{.95\linewidth} 
		\linespread{1}\footnotesize
		\textit{Notes}: Different rows corresponding to forecasting distributions for forecast horizons from one (first row, $h=1$) to four (last raw, $h=4$) quarters, conditional on 2008:Q2-Q3. Refer to Figure \ref{fig: NFCI-Impulse} for details of the plots in each row.
	\end{minipage}
\end{figure}

With the distributional impulse on the NFCI in 2008:Q4, we study the DIRFs for the following year, which ranges from 2009:Q1 to 2009:Q4 corresponding to the horizons $h=1,2,3$ and 4 quarters, respectively. Figure \ref{fig: NFCI-response} presents a complete picture of how the entire joint and marginal distributions of the real GDP growth and NFCI respond to the impulse on the NFCI in the following year. Table \ref{tab: DIR-QN-MQ} presents the results for the quantile and moment IRFs. From Figure \ref{fig: NFCI-response}, in the following two quarters, the right tail of the NFCI is greatly reduced, the left tail of the GDP becomes much thinner, and the left tail of the NFCI and right tail of the GDP become slightly fatter. Such an effect continues but decays at more distant horizons.

\begin{table}[H]
	\footnotesize
	\centering
	\caption{QIR and MIR to NFCI Impulse \label{tab: DIR-QN-MQ}}
	\scalebox{0.95}{
		\begin{tabular}{lllrrrrrrrrrrr}
			\hline \hline
			Variables                   & \multicolumn{1}{l}{}  &      & \multicolumn{5}{c}{NFCI}                                                                                              & \multicolumn{1}{l}{} & \multicolumn{5}{c}{GDP}                                                                                               \\ \cline{4-8} \cline{10-14} 
			& \multicolumn{1}{l}{}  & $h$  & \multicolumn{1}{c}{0} & \multicolumn{1}{c}{1} & \multicolumn{1}{c}{2} & \multicolumn{1}{c}{3} & \multicolumn{1}{c}{4} & \multicolumn{1}{c}{} & \multicolumn{1}{c}{0} & \multicolumn{1}{c}{1} & \multicolumn{1}{c}{2} & \multicolumn{1}{c}{3} & \multicolumn{1}{c}{4} \\ 
			\hline
			\multirow{10}{*}{Quantiles} & \multirow{2}{*}{0.05}       & Base & 0.01  & -0.36 & -0.36 & -0.41 & -0.49 &  & -1.77 & -3.03 & -3.03 & -2.10 & -3.03 \\
			&                             & Diff & -0.34 & -0.09 & -0.09 & -0.08 & -0.16 &  & 0.00  & 0.93  & 1.26  & 0.80  & 0.93  \\
			& \multirow{2}{*}{0.25}       & Base & 0.15  & -0.09 & -0.11 & -0.23 & -0.37 &  & 0.68  & 0.40  & 0.34  & 1.20  & 1.36  \\
			&                             & Diff & -0.28 & -0.29 & -0.25 & -0.16 & -0.05 &  & 0.02  & 0.96  & 1.11  & 0.80  & 0.09  \\
			& \multirow{2}{*}{0.5}        & Base & 0.55  & 0.18  & 0.17  & -0.09 & -0.14 &  & 2.03  & 1.96  & 1.96  & 2.90  & 3.00  \\
			&                             & Diff & -0.55 & -0.45 & -0.48 & -0.23 & -0.22 &  & 0.00  & 0.94  & 0.94  & 0.20  & 0.10  \\
			& \multirow{2}{*}{0.75}       & Base & 1.00  & 0.68  & 0.62  & 0.56  & 0.40  &  & 3.20  & 3.60  & 3.60  & 4.67  & 4.94  \\
			&                             & Diff & -0.87 & -0.57 & -0.51 & -0.69 & -0.42 &  & 0.00  & 0.60  & 0.60  & 0.17  & 0.17  \\
			& \multirow{2}{*}{0.95}       & Base & 2.40  & 2.72  & 2.55  & 2.47  & 2.72  &  & 6.26  & 7.06  & 7.44  & 8.10  & 8.10  \\
			&                             & Diff & -2.08 & -1.63 & -1.68 & -0.06 & -0.32 &  & -0.88 & 0.38  & 0.00  & 0.00  & 0.00       \\ 
			\hline
			\multirow{8}{*}{Moments} & \multirow{2}{*}{Mean}       & Base & 0.71  & 0.52  & 0.46  & 0.35  & 0.27  &  & 1.86  & 1.85  & 1.84  & 2.81  & 2.91  \\
			&                             & Diff & -0.71 & -0.53 & -0.50 & -0.34 & -0.28 &  & 0.05  & 0.88  & 0.92  & 0.54  & 0.17  \\
			& \multirow{2}{*}{Std}        & Base & 0.77  & 0.89  & 0.88  & 0.94  & 1.00  &  & 2.50  & 2.90  & 3.00  & 3.12  & 3.25  \\
			&                             & Diff & -0.56 & -0.22 & -0.23 & -0.10 & -0.14 &  & -0.23 & -0.16 & -0.30 & -0.43 & -0.18 \\
			& \multirow{2}{*}{Skewess}    & Base & 1.53  & 1.55  & 1.70  & 1.66  & 1.70  &  & -0.20 & -0.06 & -0.03 & -0.41 & -0.59 \\
			&                             & Diff & -1.52 & 1.52  & 1.54  & 1.00  & 0.71  &  & -0.02 & -0.22 & -0.22 & 0.16  & 0.10  \\
			& \multirow{2}{*}{Kurtosis}   & Base & 4.83  & 4.70  & 5.14  & 4.66  & 4.72  &  & 4.90  & 3.07  & 3.16  & 3.90  & 3.92  \\
			&                             & Diff & -1.79 & 8.49  & 9.43  & 4.40  & 3.44  &  & 0.65  & 0.53  & 0.72  & 0.59  & 0.03   \\ 
			\hline			
	\end{tabular}}
	\begin{minipage}{0.95\linewidth} 
		\linespread{1}\footnotesize
		\textit{Notes}: This table presents different quantiles and moments for the baseline distributions $F_{Y_{j,t+h}|Z_t}$ (Base) and the quantile and moments differences from the counterfactual distributions $F^*_{Y_{j,t+h}|Z_t}$ to the baseline distributions (Diff) of NFCI and real GDP growth from 2008:Q4 ($h=0$) to 2009:Q4 ($h=4$). 
	\end{minipage}
\end{table}

This observation is further confirmed by investigating quantile IRFs. All quantiles of the NFCI decreases; however, the differences lessen over time. For the GDP, the impulse effect on the 95\% quantile is negligible from 2009:Q2. All other quantiles, specifically the 5\%, 25\% and 75\% quantiles, increase significantly in the following two quarters. The difference becomes smaller at further horizons but remains significant.
Moreover, the exploration of the moments IRFs reveals that the mean of NFCI at all horizons decreases to approximately 0, driven by the significant impulse effect on the upper tail. A more substantial impact on the lower tail for GDP increases its mean to $2{\sim}3$ for all horizons. The impulse significantly changes the skewness and kurtosis of the NFCI, and the results for the other moments also demonstrate significant long-run effects of both variables. Finally, we conclude that if the policies in 2008:Q3 had been able to limit the possibility of financial tightening in 2008:Q4, the likelihood of adverse GDP growth (left tail) and tight financial conditions (right tail) would have been largely eliminated in 2009:Q1-Q2 and reduced in 2009:Q3-Q4.

\subsubsection{Counterfactual Analysis of Distributional Impulse on GDP}		

\begin{figure}[H]
	\captionsetup[subfigure]{aboveskip=-1pt,belowskip=0pt}
	\centering
	\caption{Distributional Impulse to real GDP growth in 2008:Q4} \label{fig: GDP-Impulse}		
	\begin{minipage}[t]{\textwidth}	
		\centering Distribution of $Y_t$ given $Z_t$ for $t$ equals to 2008:Q4 ($h=0$)\\
		\begin{subfigure}[t]{0.32\textwidth}
			\centering 		
			\includegraphics[width=0.95\textwidth]{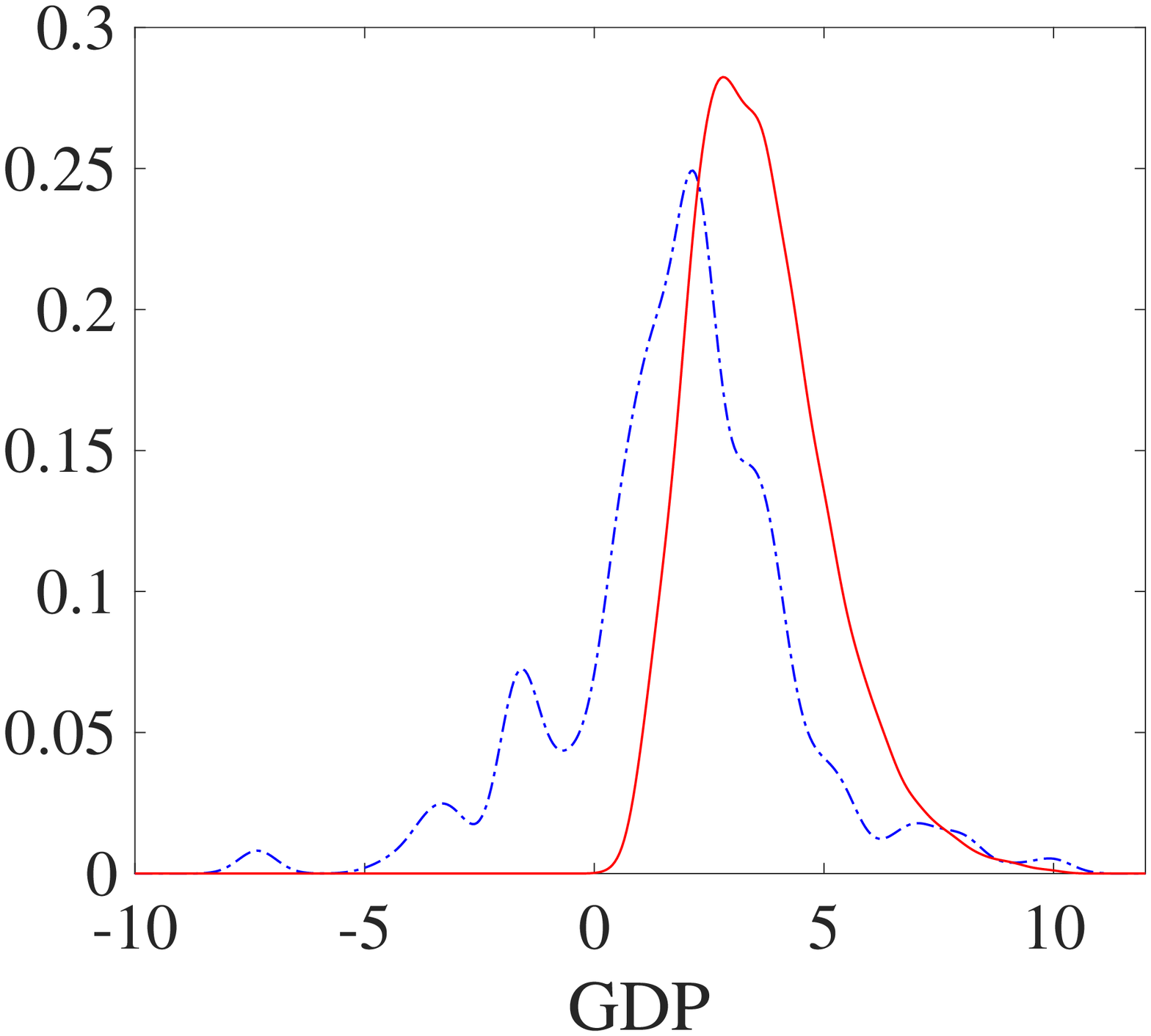}     		
		\end{subfigure}
		\begin{subfigure}[t]{0.32\textwidth}
			\centering		
			\includegraphics[width=0.95\textwidth]{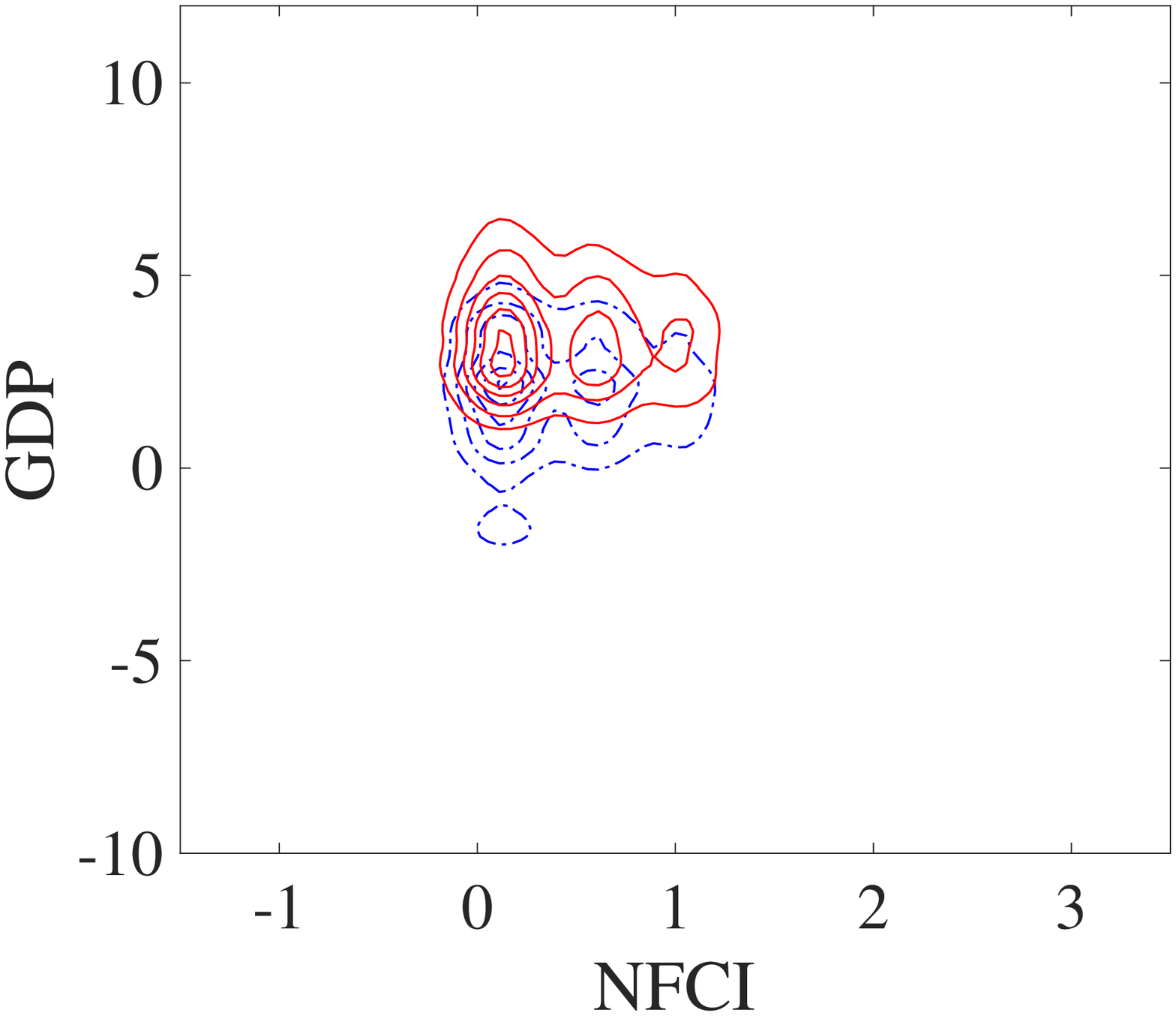}
		\end{subfigure}
		\begin{subfigure}[t]{0.32\textwidth}
			\centering	
			\includegraphics[width=0.95\textwidth]{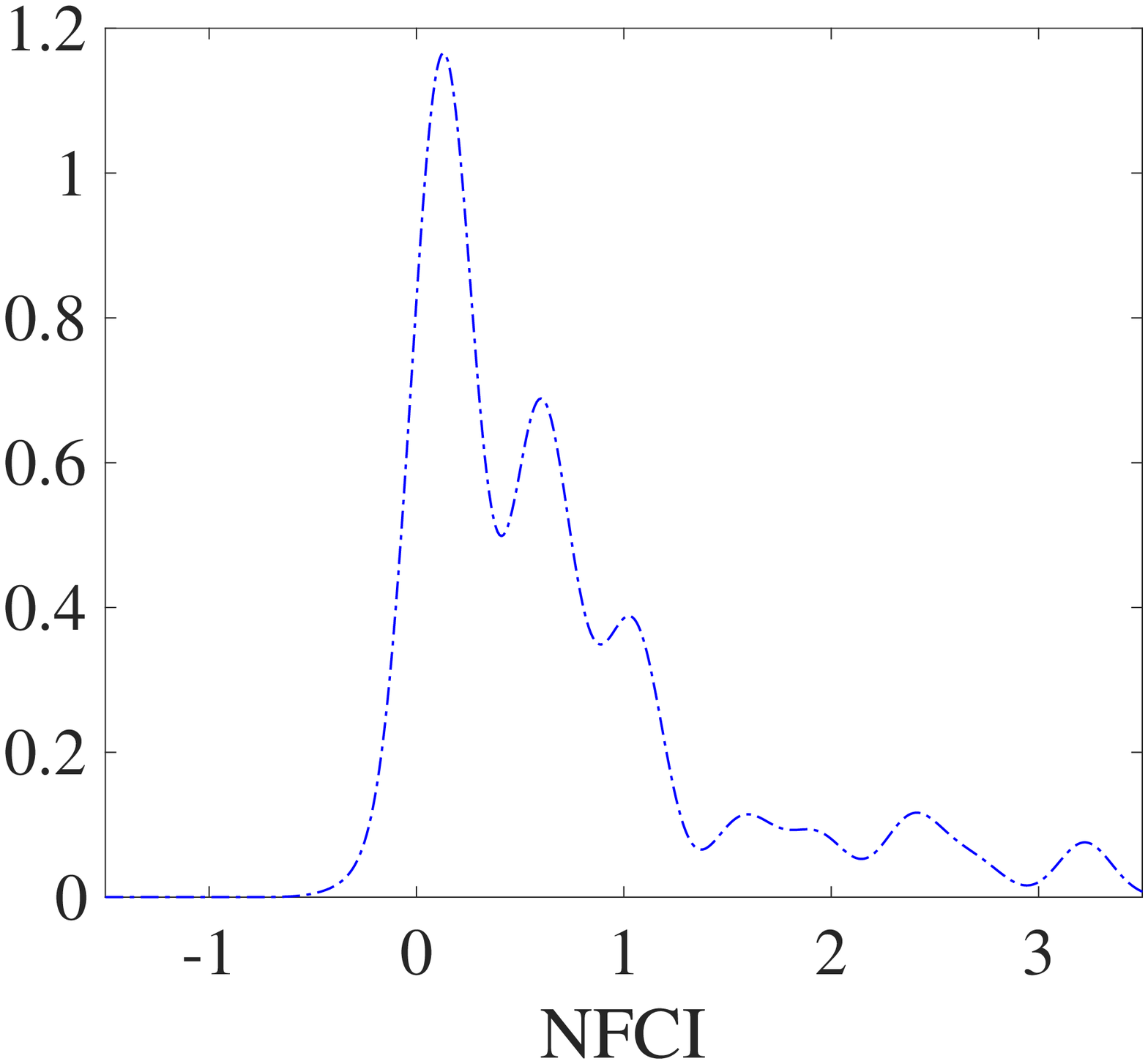}
		\end{subfigure}
	\end{minipage}
	\begin{minipage}{.9\linewidth} 
		\linespread{1}\footnotesize
		\textit{Notes}: Refer to Figure \ref{fig: NFCI-Impulse}.
	\end{minipage}
\end{figure}

\begin{figure}[H]
	\captionsetup[subfigure]{aboveskip=-1pt,belowskip=0pt}
	\centering
	\caption{Distributional Response to the GDP Impulse} \label{fig: GDP-response}		
	\begin{minipage}[t]{\textwidth}
		\centering  Distributions of $Y_{t+h}$ given $Z_{t}$ for $t$=2008:Q4\\$h=1$\\
		\begin{subfigure}[t]{0.32\textwidth} 		
			\includegraphics[width=0.95\textwidth]{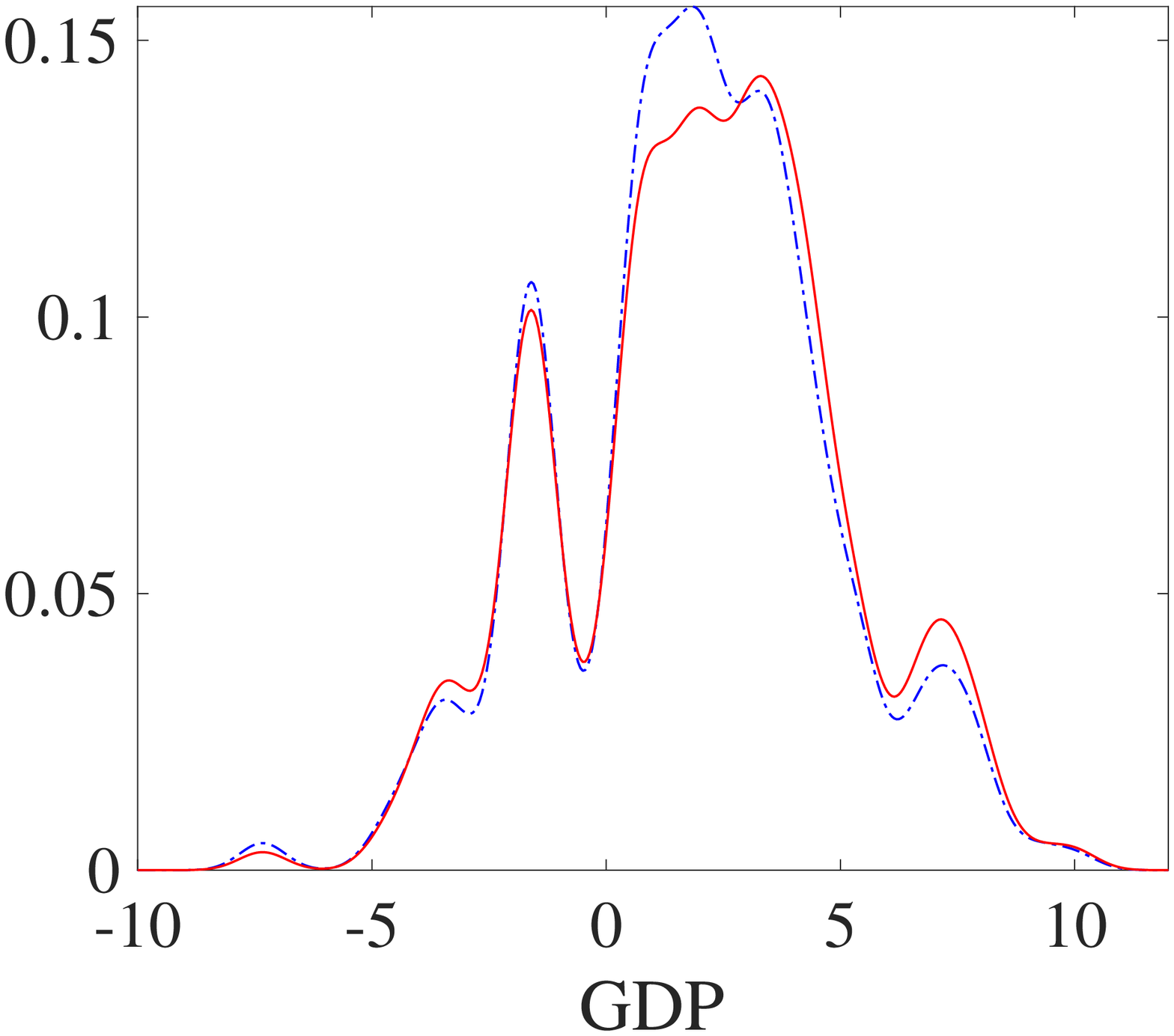}     		
		\end{subfigure}
		\begin{subfigure}[t]{0.32\textwidth}
			\includegraphics[width=0.95\textwidth]{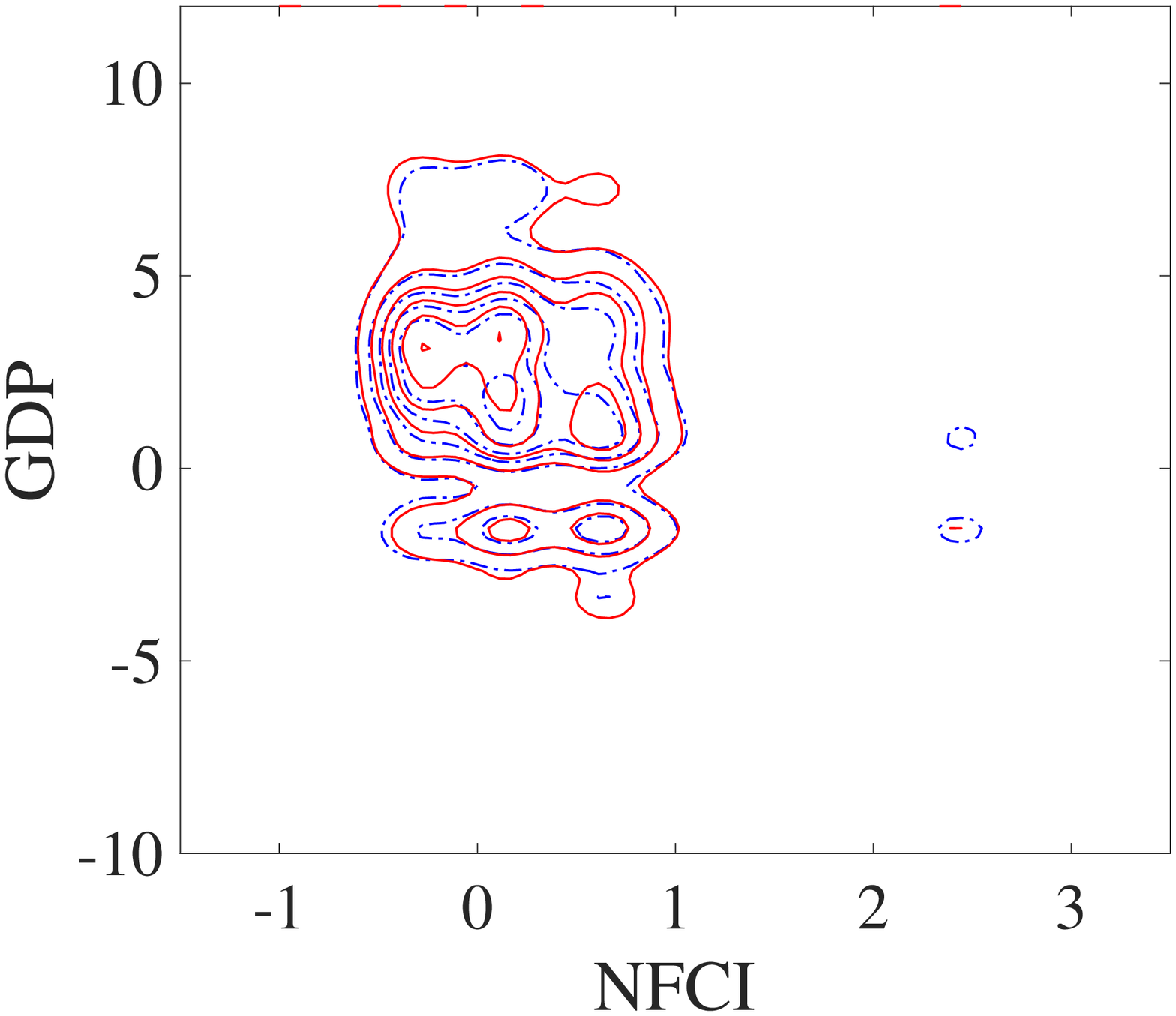}
		\end{subfigure}
		\begin{subfigure}[t]{0.32\textwidth}
			\includegraphics[width=0.95\textwidth]{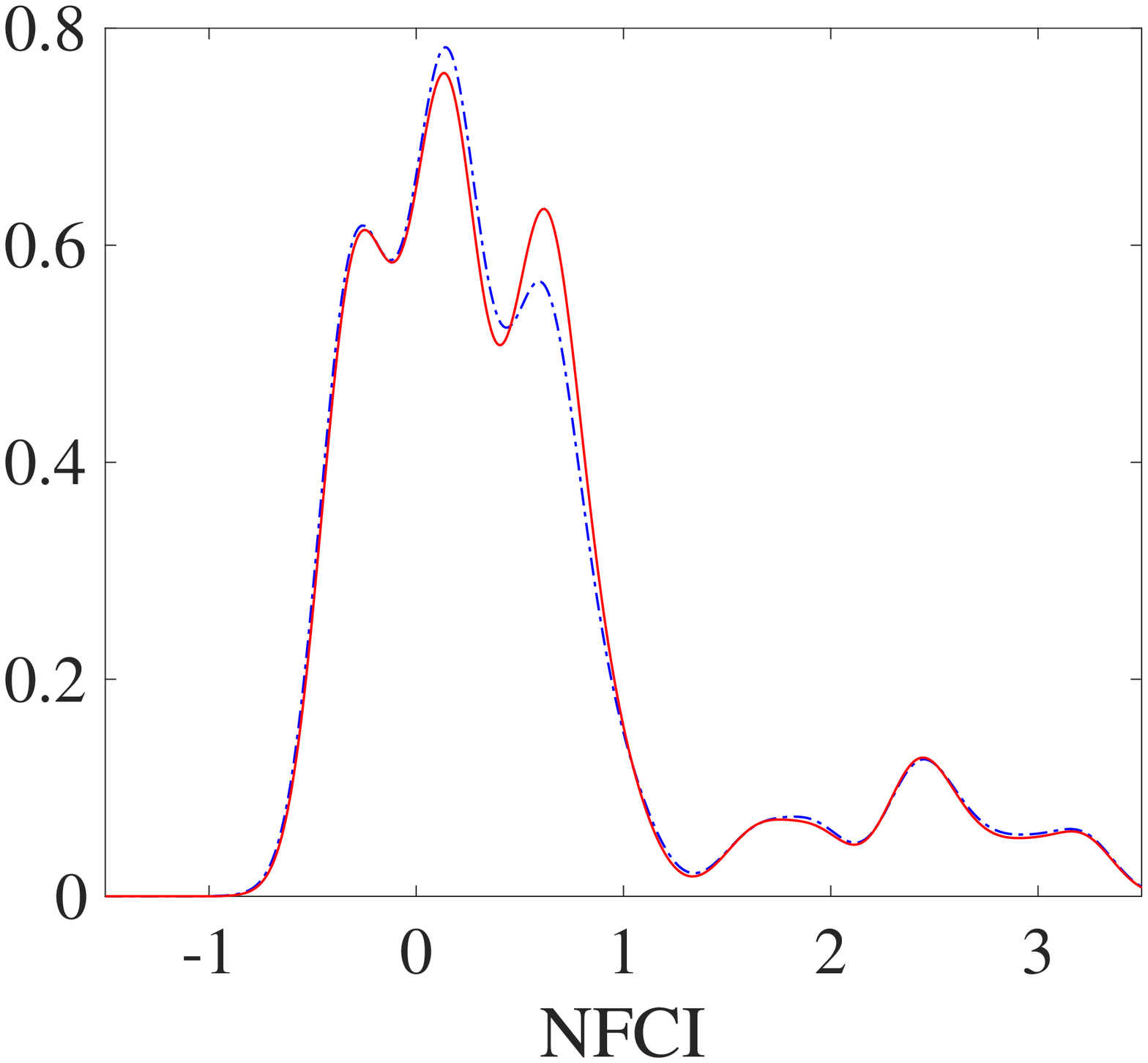}
		\end{subfigure}
		\vspace{0.1cm}
	\end{minipage}	
	\begin{minipage}[t]{1\textwidth}
		\centering $h=2$\\
		\begin{subfigure}[t]{0.32\textwidth} 		
			\includegraphics[width=0.95\textwidth]{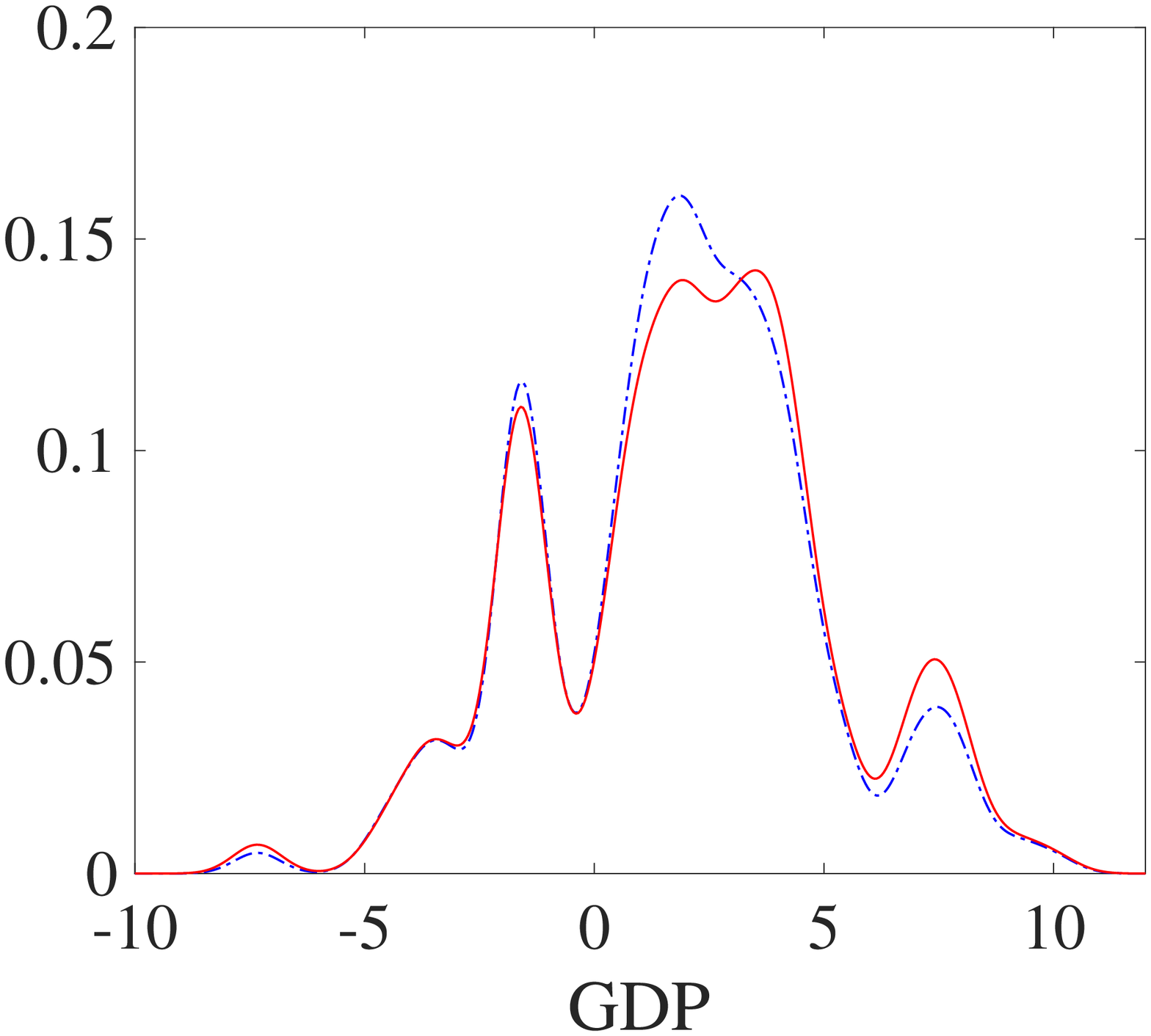}     		
		\end{subfigure}
		\begin{subfigure}[t]{0.32\textwidth}
			\includegraphics[width=0.95\textwidth]{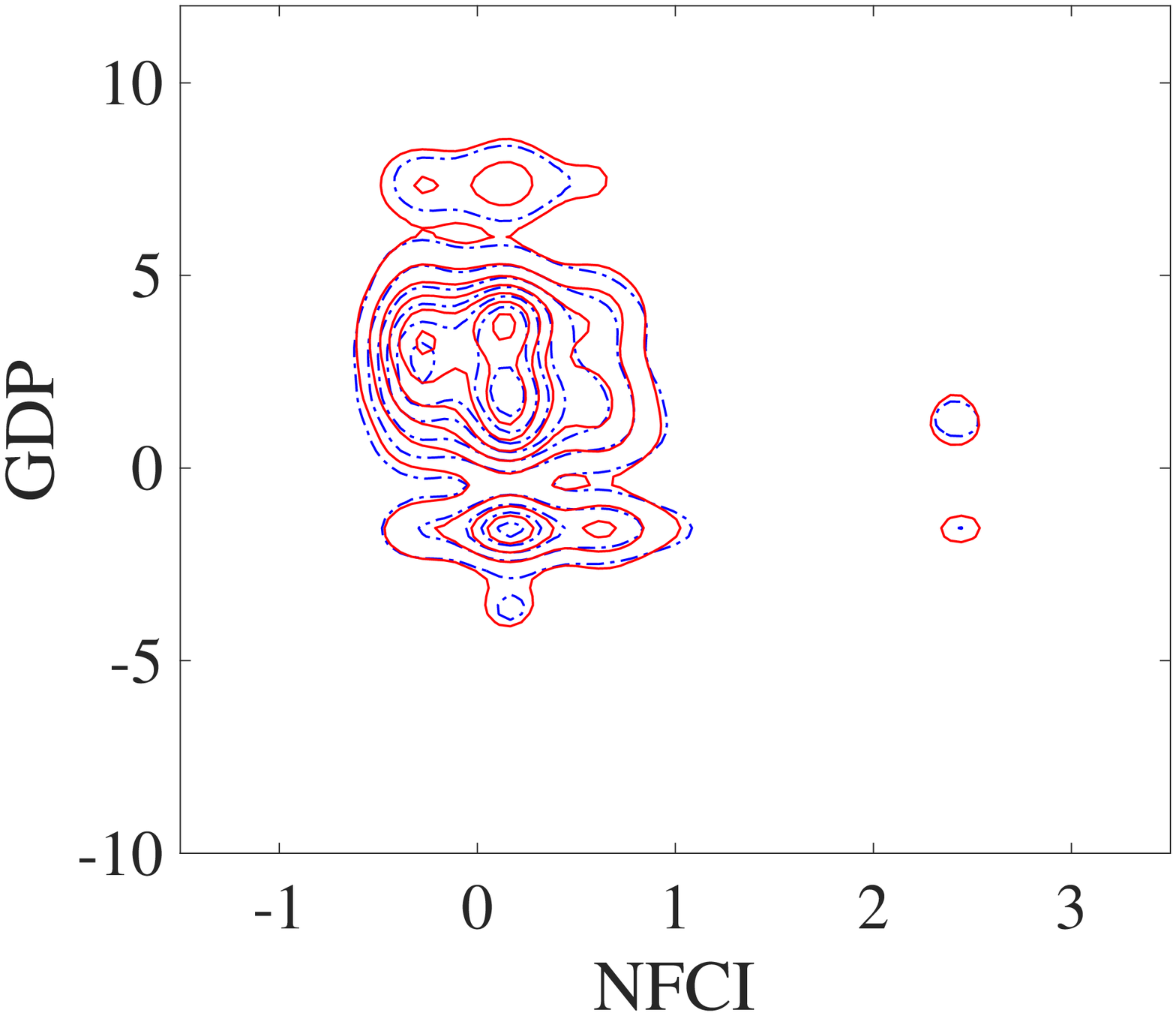}
		\end{subfigure}
		\begin{subfigure}[t]{0.32\textwidth}
			\includegraphics[width=0.95\textwidth]{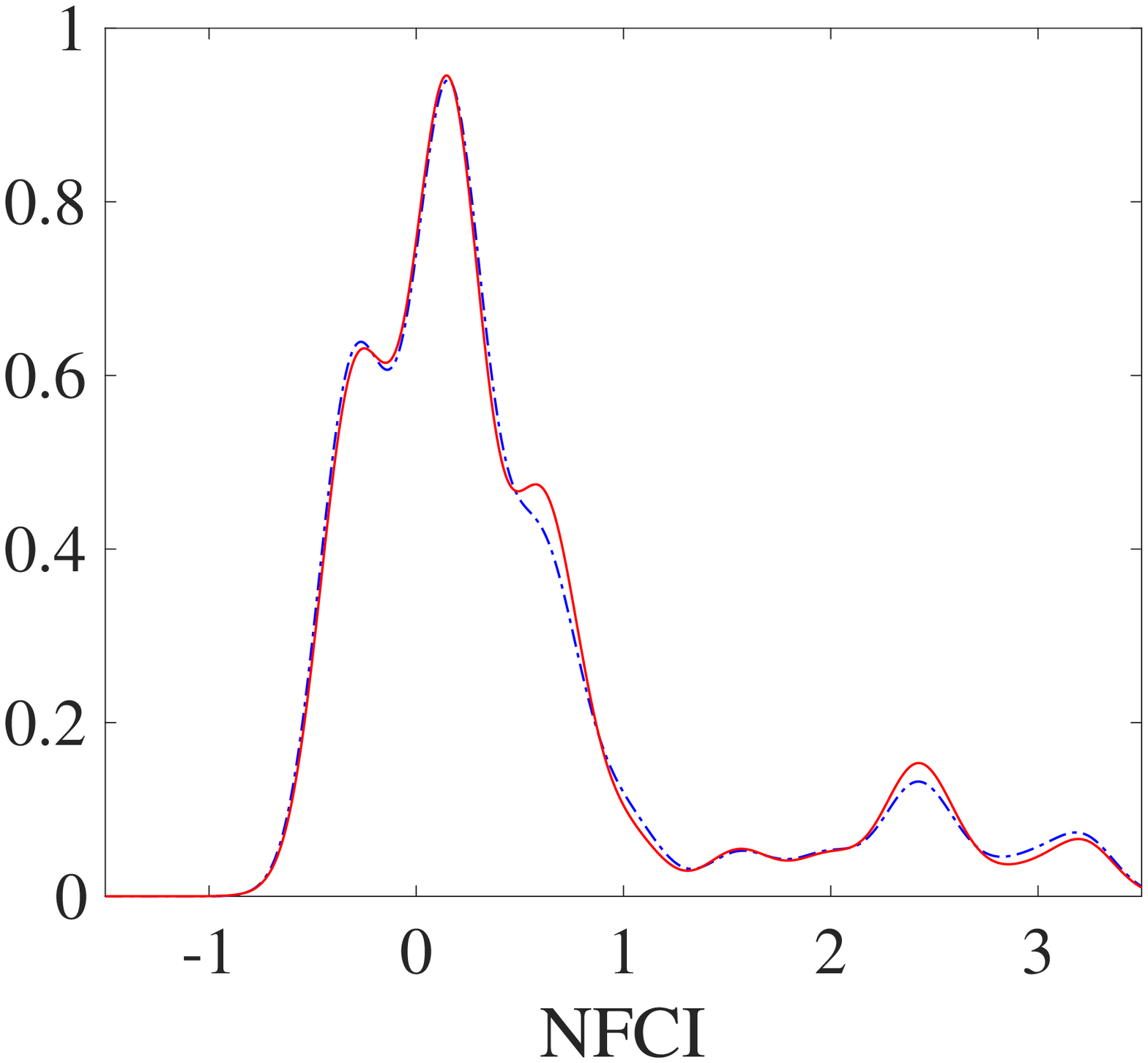}
		\end{subfigure}
		\vspace{0.1cm}
	\end{minipage}
	\begin{minipage}[t]{1\textwidth}
		\centering $h=3$\\
		\begin{subfigure}[t]{0.32\textwidth} 		
			\includegraphics[width=0.95\textwidth]{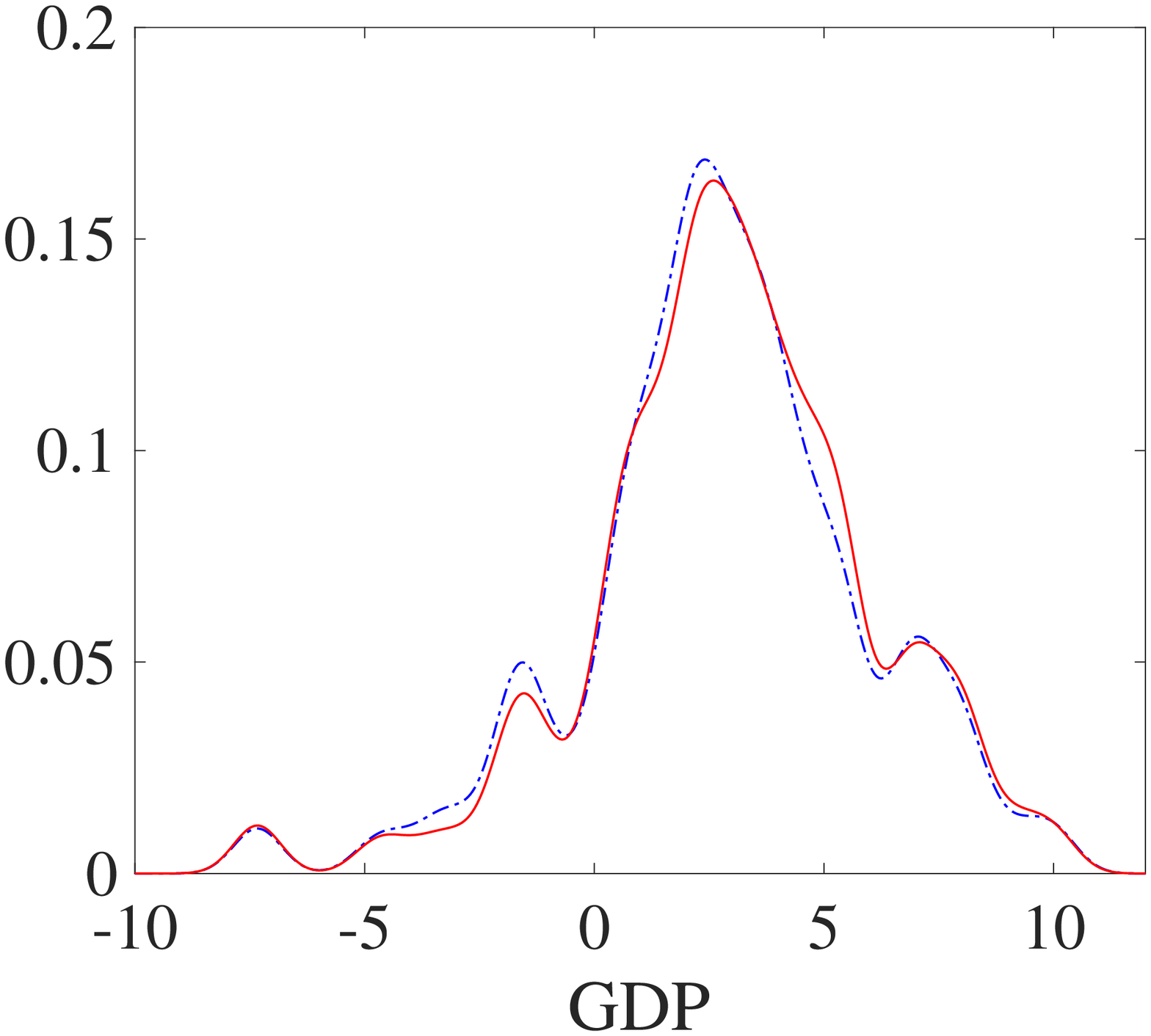}     		
		\end{subfigure}
		\begin{subfigure}[t]{0.32\textwidth}
			\includegraphics[width=0.95\textwidth]{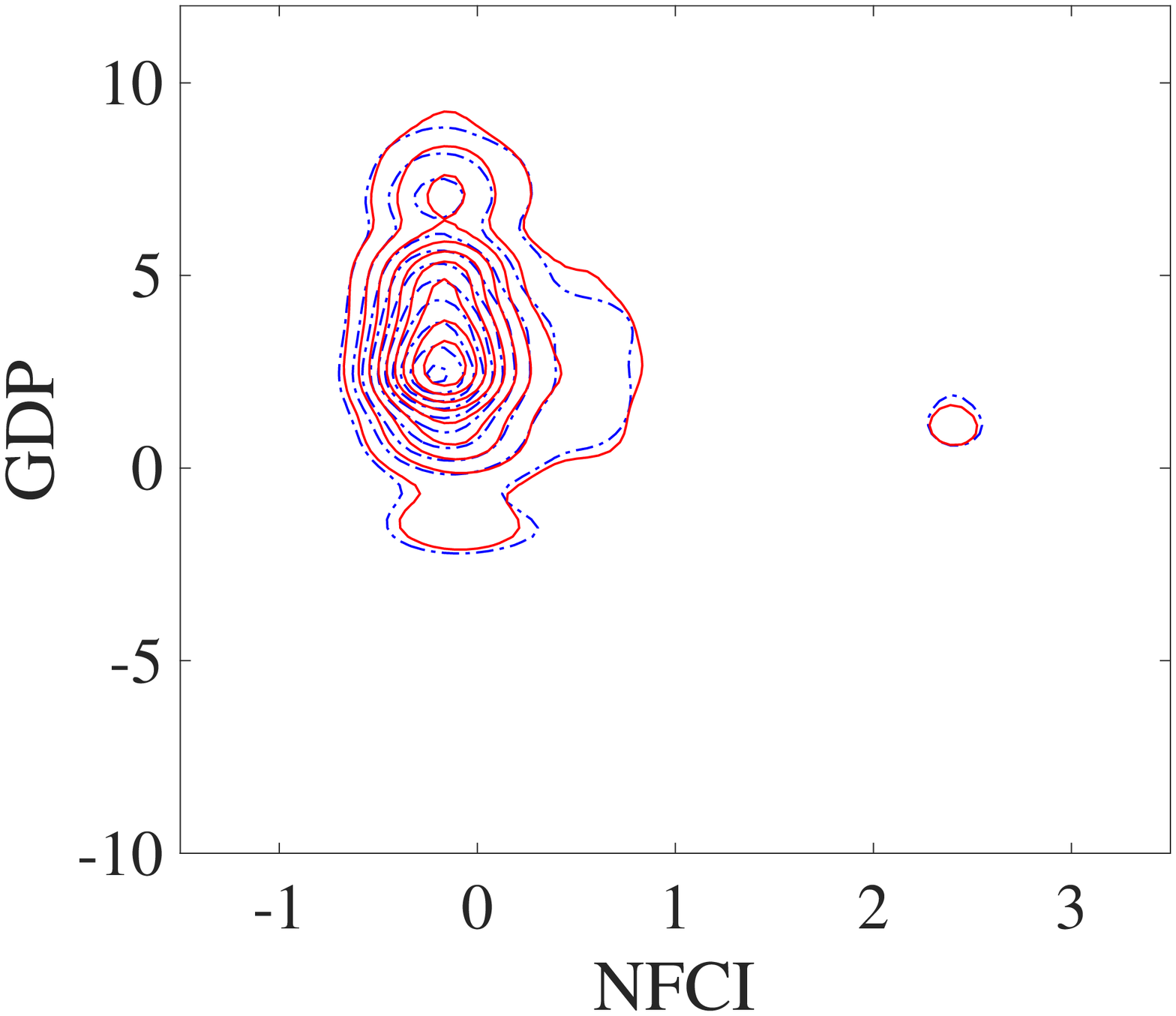}
		\end{subfigure}
		\begin{subfigure}[t]{0.32\textwidth}
			\includegraphics[width=0.95\textwidth]{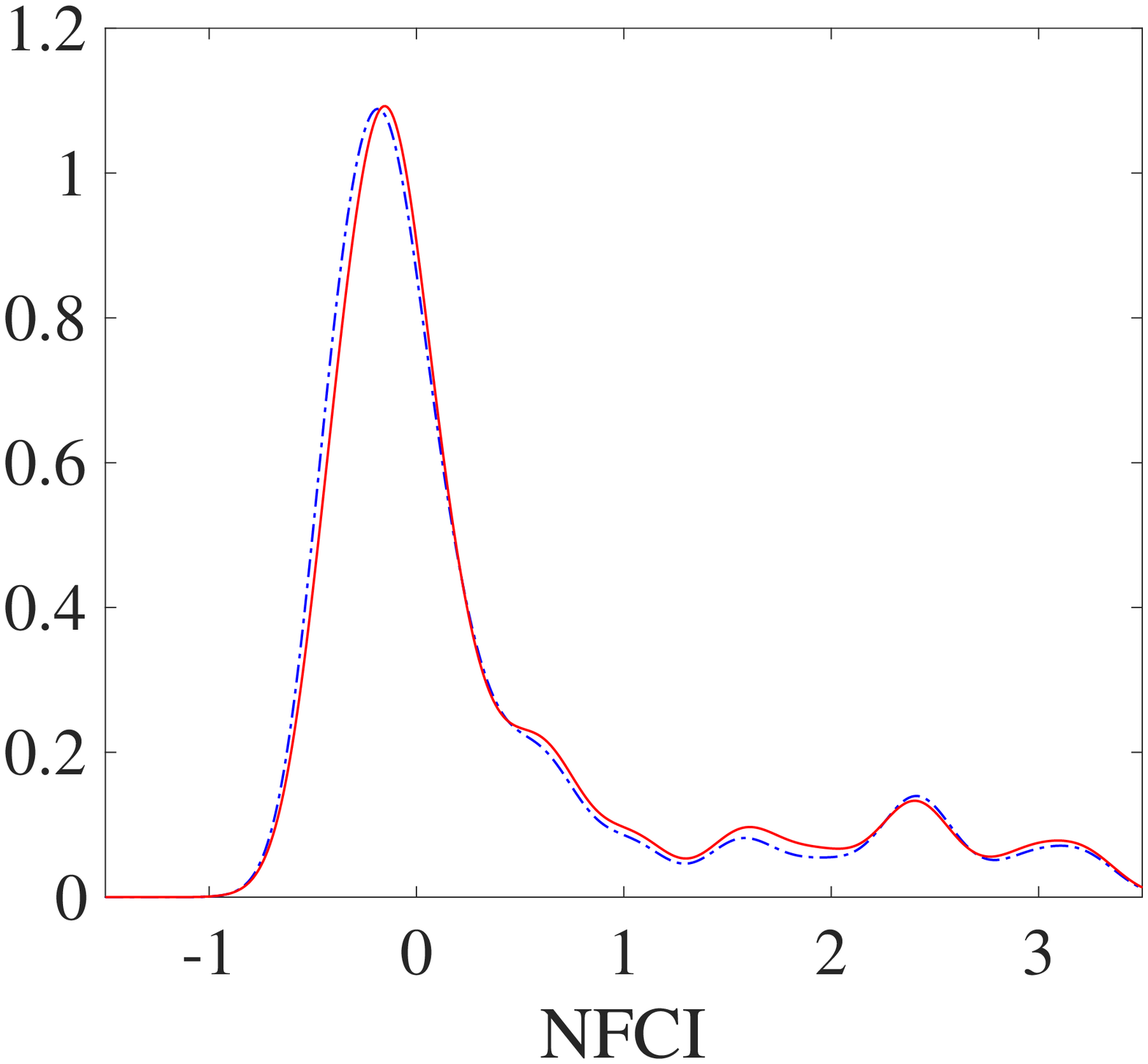}
		\end{subfigure}
		\vspace{0.1cm}
	\end{minipage}
	\begin{minipage}[t]{1\textwidth}
		\centering $h=4$\\
		\begin{subfigure}[t]{0.32\textwidth} 		
			\includegraphics[width=0.95\textwidth]{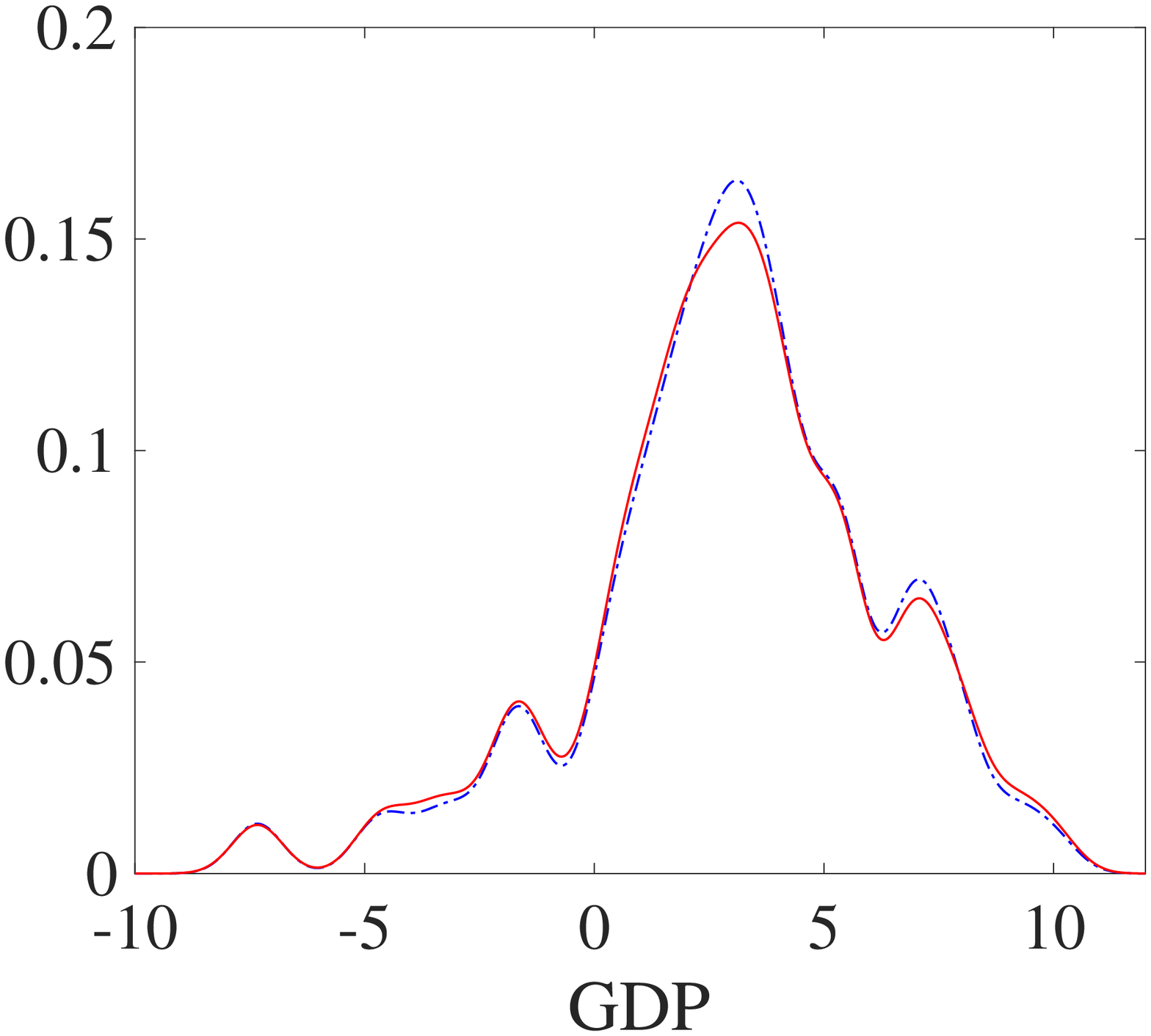}     		
		\end{subfigure}
		\begin{subfigure}[t]{0.32\textwidth}
			\includegraphics[width=0.95\textwidth]{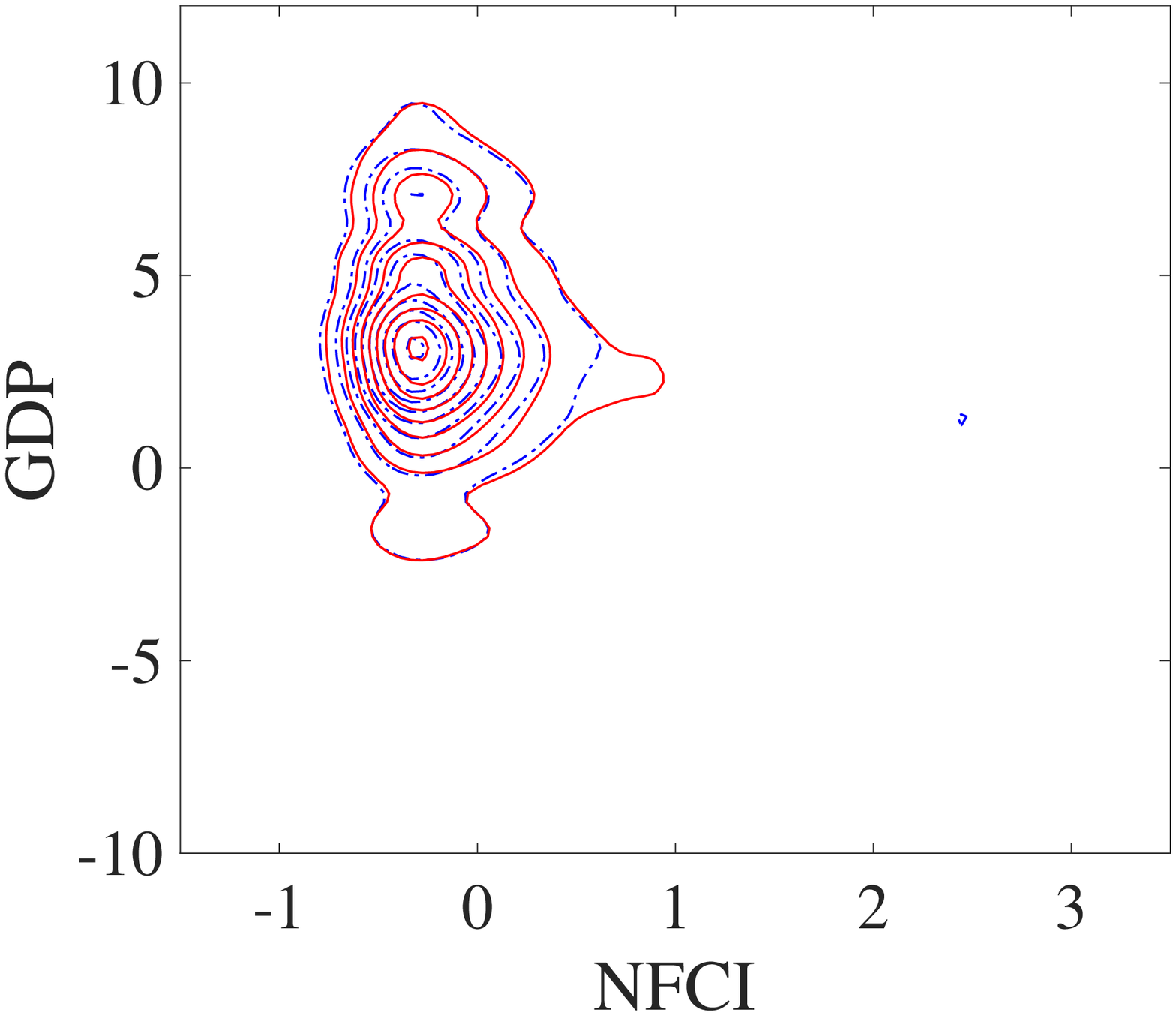}
		\end{subfigure}
		\begin{subfigure}[t]{0.32\textwidth}
			\includegraphics[width=0.95\textwidth]{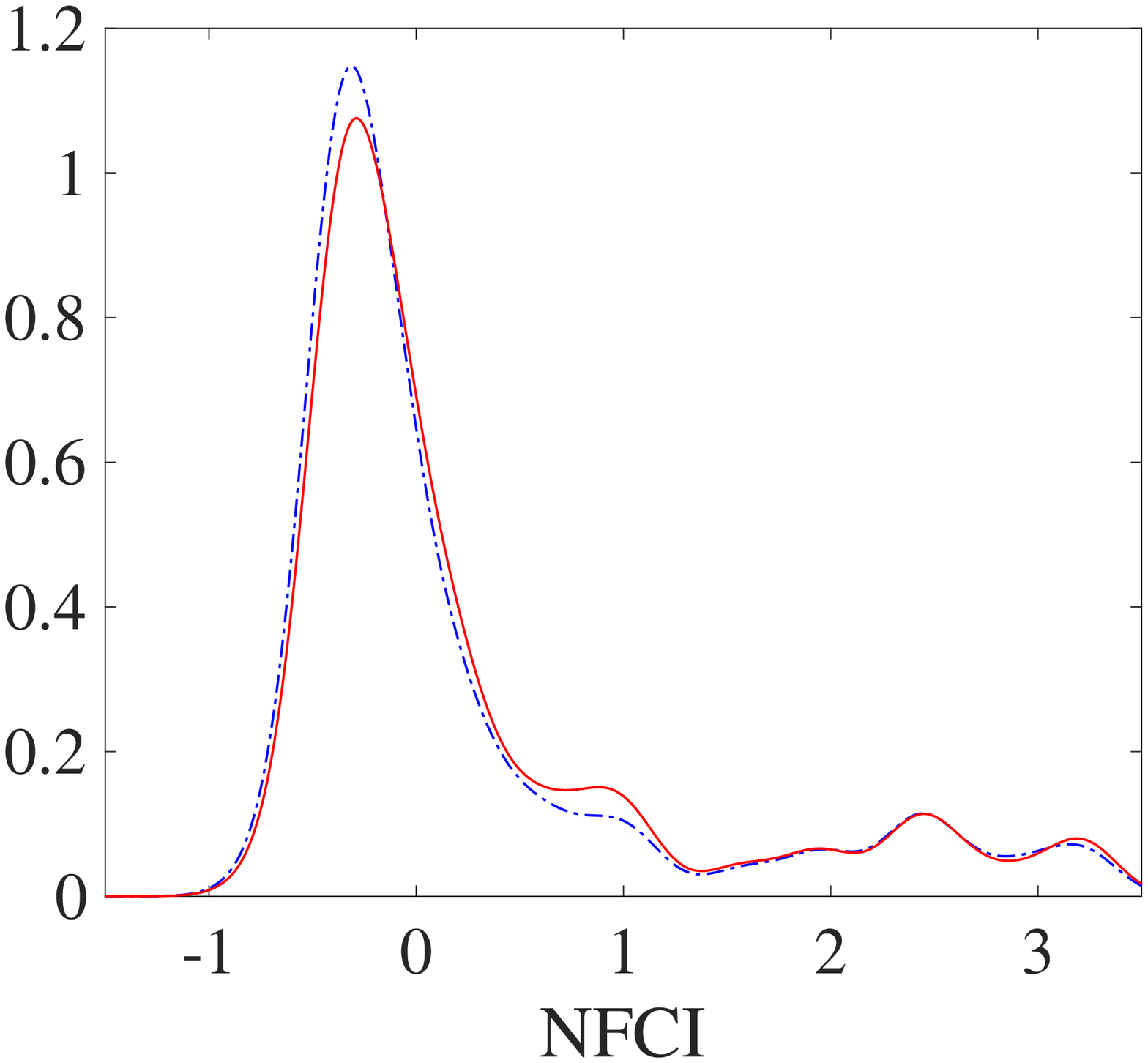}
		\end{subfigure}
	\end{minipage}	
	\begin{minipage}{.9\linewidth} 
		\linespread{1}\footnotesize
		\textit{Notes}: Refer to Figure \ref{fig: NFCI-response}.
	\end{minipage}
\end{figure}

We explore the effect of the policy in 2008:Q3 that could have limited the possibility of low economic activity during 2008:Q4. Specifically, as shown in Figure \ref{fig: GDP-Impulse}, we maintain the conditional distribution of the NFCI in 2008:Q4 as it is, and a counterfactual distribution, truncated gamma distribution on $(0,11)$ with scale parameter of $6$ and shape parameter of $0.6$ is considered for the GDP. Figure \ref{fig: GDP-response} shows the joint and marginal distributions of $Y_t$ given $Z_t$ and their counterfactual counterparts; their differences in quantiles and moments are presented in Table \ref{tab: DIR-QG-MQ}, with $h=0$. The distributional impulse on GDP increases the 5\% quantile significantly from -2.1 to 1.5, and the other quantiles increases by approximately $1{\sim}2$. This impulse also significantly changes different moments of the GDP.

Using the distributional impulse on real GDP growth in 2008:Q4, we study the DIRFs for the following year from 2009:Q1 to 2009:Q4, which corresponds to the horizons ${h=1,2,3}$, and 4 quarters, respectively. The dynamic effect of this impulse on the entire distributions of the NFCI and real GDP growth is illustrated in Figure \ref{fig: GDP-response}, where the baseline distributions and the counterfactual distributions are plotted in blue and red, respectively. A slight difference between the baseline and counterfactual distributions for the NFCI and real GDP growth is observed for all horizons.

\begin{table}[H]
	\centering
	\footnotesize
	\caption{QIR and MIR to GDP Impulse\label{tab: DIR-QG-MQ}}
	\scalebox{0.95}{
		\begin{tabular}{lllrrrrrlrrrrr}
			\hline \hline
			Variables                   & \multicolumn{1}{l}{}  &      & \multicolumn{5}{c}{NFCI}                                                                                              & \multicolumn{1}{l}{} & \multicolumn{5}{c}{GDP}                                                                                               \\ \cline{4-8} \cline{10-14} 
			& \multicolumn{1}{l}{}  & $h$  & \multicolumn{1}{c}{0} & \multicolumn{1}{c}{1} & \multicolumn{1}{c}{2} & \multicolumn{1}{c}{3} & \multicolumn{1}{c}{4} & \multicolumn{1}{c}{} & \multicolumn{1}{c}{0} & \multicolumn{1}{c}{1} & \multicolumn{1}{c}{2} & \multicolumn{1}{c}{3} & \multicolumn{1}{c}{4} \\ 
			\hline
			\multirow{10}{*}{Quantiles} & \multirow{2}{*}{0.05}       & Base & 0.01 & -0.37 & -0.36 & -0.41 & -0.50 &  & -2.10 & -3.03 & -3.03 & -2.10 & -3.03 \\
			&                             & Diff & 0.00 & 0.02  & 0.00  & 0.00  & 0.03  &  & 3.64  & 0.00  & -0.59 & 0.00  & 0.00  \\
			& \multirow{2}{*}{0.25}       & Base & 0.15 & -0.10 & -0.12 & -0.24 & -0.37 &  & 0.68  & 0.40  & 0.11  & 1.20  & 1.36  \\
			&                             & Diff & 0.00 & 0.01  & 0.02  & 0.08  & 0.02  &  & 1.84  & 0.11  & 0.00  & 0.10  & -0.06 \\
			& \multirow{2}{*}{0.5}        & Base & 0.55 & 0.18  & 0.17  & -0.10 & -0.14 &  & 2.03  & 2.00  & 2.00  & 2.90  & 3.04  \\
			&                             & Diff & 0.00 & 0.06  & 0.01  & 0.08  & 0.02  &  & 1.39  & 0.10  & 0.10  & 0.10  & 0.00  \\
			& \multirow{2}{*}{0.75}       & Base & 1.00 & 0.68  & 0.62  & 0.55  & 0.31  &  & 3.20  & 3.60  & 3.80  & 4.67  & 5.11  \\
			&                             & Diff & 0.00 & 0.00  & 0.00  & 0.07  & 0.09  &  & 1.25  & 0.30  & 0.10  & 0.17  & 0.00  \\
			& \multirow{2}{*}{0.95}       & Base & 2.47 & 2.55  & 2.55  & 2.55  & 2.55  &  & 5.50  & 7.06  & 7.44  & 8.10  & 7.78  \\
			&                             & Diff & 0.00 & 0.00  & -0.08 & 0.00  & 0.00  &  & 0.81  & 0.38  & 0.34  & 0.00  & 0.32    \\ 
			\cline{1-8} \cline{10-14} 
			\multirow{8}{*}{Moments} & \multirow{2}{*}{Mean}       & Base & 0.70 & 0.50  & 0.46  & 0.35  & 0.24  &  & 1.84  & 1.88  & 1.87  & 2.79  & 2.99  \\
			&                             & Diff & 0.00 & 0.01  & 0.00  & 0.05  & 0.05  &  & 1.77  & 0.17  & 0.16  & 0.17  & -0.05 \\
			& \multirow{2}{*}{Std}        & Base & 0.77 & 0.87  & 0.88  & 0.95  & 0.98  &  & 2.46  & 2.91  & 3.02  & 3.10  & 3.23  \\
			&                             & Diff & 0.00 & -0.01 & -0.01 & 0.01  & 0.00  &  & -0.99 & 0.08  & 0.15  & -0.03 & 0.06  \\
			& \multirow{2}{*}{Skewess}    & Base & 1.59 & 1.56  & 1.69  & 1.66  & 1.77  &  & -0.22 & -0.09 & -0.02 & -0.39 & -0.58 \\
			&                             & Diff & 0.00 & 0.00  & -0.02 & -0.10 & -0.07 &  & 1.00  & 0.00  & -0.09 & -0.07 & 0.06  \\
			& \multirow{2}{*}{Kurtosis}   & Base & 5.03 & 4.77  & 5.10  & 4.64  & 5.03  &  & 4.87  & 3.14  & 3.11  & 3.85  & 3.86  \\
			&                             & Diff & 0.00 & 0.07  & -0.07 & -0.33 & -0.18 &  & -1.11 & -0.27 & -0.13 & 0.20  & -0.17 \\
			\hline
	\end{tabular}}
	\begin{minipage}{0.95\linewidth} 
		\linespread{1}\footnotesize
		\textit{Notes}: Refer to Table \ref{tab: DIR-QN-MQ}.
	\end{minipage}
\end{table}

We further explore the quantile and moment IRFs presented in Table \ref{tab: DIR-QG-MQ} to draw a more concrete comparison. First, the quantile and moment IRFs of the NFCI at one- to four-quarters ahead are all close to 0. Regarding real GDP growth, the one- and two-quarters-ahead counterfactual distributions have a slightly fatter right tail and a larger mean than the baseline distributions. However, the counterfactual and baseline distributions are almost identical for more distant horizons. Thus, in the absence of a corresponding improvement in the distribution of NFCI, limiting the likelihood of negative real GDP growth in 2008:Q4 only increases the likelihood of positive economic activity in the short run and does not significantly affect the NFCI, even in the subsequent quarter.

We provide more exploration of this empirical application in Appendix \ref{sec: appendix-B}. Section \ref{sec: appendix-B1} introduces the simulation of samples of the outcome variables from the multiperiod forecasting distributions. The same model specification but with an alternative order is considered in Section \ref{sec: appendix-B2}, where $Y_{1t}$ represents the real GDP growth and $Y_{2t}$ represents the quarterly NFCI. The results suggest that the order does not significantly affect the estimation of multiperiod forecasting distributions. We compare the proposed approach with the kernel regression method introduced by \cite{adrian2021multimodality} in Section \ref{sec: appendix-B3}. The results indicate that the performance of the kernel regression approach is sensitive to bandwidths, whereas the proposed DR approach demonstrats superior performance in eliciting this conditional distribution, specifically for the NFCI distribution. In Section \ref{sec: appendix-B4}, we explore the use of the monthly NFCI in the system and a four-dimensional mixed-frequency model conditional on two lags by treating the three monthly NFCI series as separate observations within a quarter. Based on this model, we revisit the distribution forecasting and counterfactual analysis. For each scenario, $95\%$ confidence bands for the entire distribution as well as density impulse responses are estimated using the moving block bootstrap approach, the results of which are presented in Section \ref{sec: appendix-B5}.

\section{Conclusion}\label{sec: con}
This study develops a flexible  semiparametric approach for characterizing the conditional joint distribution of multivariate time series. The resulting DIRF provides a more comprehensive picture of the dynamic heterogeneity. The asymptotic properties of the conditional distribution estimators and their transformations are also derived. Based on an analysis of the real GDP growth and NFCI in U.S., the empirical results confirm some existing findings in the literature: First, the tight financial conditions create multimodality in the conditional joint distribution. Second, restricting the upper tail of financial conditions has a noticeable impact on long-term GDP growth. However, with the inclusion of additional lag information, the extracted results of the proposed model on the effect of restricting the lower tail of the GDP during the global financial crisis suggest a negligible impact on the financial conditions.

\clearpage
\setstretch{0.5}
\bibliographystyle{chicago}    
\bibliography{Distributional_VAR_ref}

\newpage    
\setstretch{1.3}
\begin{center}
  \part*{Appendix}
\end{center}

\renewcommand{\qedsymbol}{$\blacksquare$}

 \setcounter{section}{0} 
\setcounter{figure}{0} 
\setcounter{table}{0} 
\setcounter{equation}{0} 
\setcounter{lemma}{0}\setcounter{page}{1}
\setcounter{proposition}{0} %
\renewcommand{\thepage}{A-\arabic{page}}
\renewcommand{\theequation}{A.\arabic{equation}}
\renewcommand{\thelemma}{A.\arabic{lemma}} 
\renewcommand{\theproposition}{A.\arabic{proposition}} 
\renewcommand\thesection{\Alph{section}} 
\renewcommand\thesubsection{A.{subsection}}
\renewcommand\thefigure{\thesection.\arabic{figure}}
\renewcommand\thetable{\thesection.\arabic{table}}

	\section{Theoretical Results}\label{sec: appendix-A}

	We will use some notations and results from the literature on empirical process.
	For more details we refer to \cite{van1994weak} and
        \cite{PollardDavid1984Cosp}. 
	Let $\mathcal{F}$ denote a class of real-valued (measurable) functions 
	with envelope $F$. 
	Given $\epsilon \in (0,1)$ and $p \ge 1$,
	the covering number 
	$N
	\big(\epsilon\|F\|_{Q,p}, \mathcal{F}, L_p(Q) \big)$
	of $\mathcal{F}$ with respect to some probability measure $Q$
	is defined as the smallest cardinality of 
	$\epsilon\|F\|_{Q,p}$-cover of $\mathcal{F}$
	with respect to the $L_{p}(Q)$-norm $\| \cdot \|_{Q,p}:= (Q|\cdot|^p)^{1/p}$.
	The class $\mathcal{F}$ is said to be Euclidean 
	for the envelop $F$
	if there exist constants  
	$A$ and $V$ such that
	\begin{eqnarray*}
		N\big(
		\epsilon\|F\|_{Q,1}, \mathcal{F}, L_1(Q)
		\big ) \le A \epsilon^{-V}, 
	\end{eqnarray*}
	for all $\epsilon \in (0,1]$
	and all measures $Q$
	whenever $Q F \in (0, \infty)$. 
	It can be shown that 
	if $\mathcal{F}$ is Euclidean, then for each $p > 1$,
	\begin{eqnarray}
		\label{eq:eudp}
		N
		\big (
		\epsilon\|F\|_{Q,p}, \mathcal{F}, L_p(Q)
		\big )
		\le 
		A 2^{pV}\epsilon^{-pV},  
	\end{eqnarray}
	whenever $Q F^p \in (0, \infty)$.
        See \cite{nolan1987u} for instance.

	We define
	$
	\psi_{y}(\theta)
	:=
	\big [
	\psi_{y, 1}(\theta_{1})^{\top},
	\dots,
	\psi_{y, J}(\theta_{J})^{\top}
	\big ]^{\top}
	$
	for
	$(\theta,y) \in \Theta \times \mathcal{Y}$,
	where 
	\[
	\psi_{y, j}(\theta_{j})
	:=
	\big[\Lambda\big(\phi_{j}(X_{jt})^{\top}\theta_j\big)-\1\{Y_{jt}\leq y_j\}\big]R\big(
	\phi_{j} (X_{jt})^{\top}\theta_j\big) \phi_{j}(X_{jt}).
	\]
The lemma blow shows 
the Donskerness
of
the class of functions
$\{ \psi_{y}\big(\theta(y) \big):y \in {\cal Y} \}$

	\vspace{0.5cm}
	\begin{lemma}
		\label{lemma:donsker}
		Suppose that Assumptions A1-A2 hold. Then,  
		the function class 
		$\big\{ 
		\psi_{y}\big( \theta(y) \big):
		y
		\in \mathcal{Y} \big\}$ is Donsker
		with a square-integrable envelope.    
	\end{lemma}
	\begin{proof}
		We define
		the function classes, for each $j=1, \dots, J$,
		\begin{eqnarray*}
			\mathcal{H}_{j}:=\big\{ 
			x_{j} \mapsto \phi_{j}(x_{j})^{\top}\theta_{j}:  \theta_{j} \in \Theta_{j} \big\}
			\ \ \ \mathrm{and} \ \ \
			\mathcal{I}_{j}:=\big\{v_{j} \mapsto \1\{v_{j} \le y_{j}\}: y_{j} \in \mathcal{Y}_{j}\big\}.
		\end{eqnarray*}
		Lemma 2.6.15 of \cite{van1996weak}
		shows that 
		$\mathcal{H}_{j}$ 
		and 
		$\mathcal{I}_{j}$
		are VC-subgraph classes.
		Letting 
		$
		\mathcal{G}_{j}
		:=\big \{x_{j} \mapsto \phi_{j}(x_{j})^{\top}\theta_{j}(y_{j}):
		y_{j} \in \mathcal{Y}_{j} \big\}
		$ for $j=1, \dots, J$,
		we can write
		$\mathcal{G}_{j} = \mathcal{H}_{j} \circ \theta(\cdot)$.
		Then,
		the class 
		$\mathcal{G}_{j}$
		and
		its monotonic transformation 
		$\Lambda(\mathcal{G}_{j})$
		are VC-subgraph classes
		by 
		Lemma 2.6.18~(vii) and (viii)
		of \cite{van1996weak}, respectively.
		Given that
		the transformation $\phi_{j}: \mathcal{X}_{j} \to \R^{d_{j}}$,
		we can write 
		$\phi_{j}(x_{j}) \equiv[
		\phi_{j}^{(1)}(x_{j}), \dots, \phi_{j}^{(d_{j})}(x_{j})]^{ \top}$.
		Lemma 2.6.18~(vi)
		of \cite{van1996weak}
		shows that 
		$\Lambda(\mathcal{G}_{j}) \cdot \phi_{j}^{(\ell)}$
		and
		$\mathcal{I}_{j} \cdot \phi_{j}^{(\ell)}$
		are VC-subgraph.

		Because all VC-classes are Euclidean by Lemma II.25 of \cite{PollardDavid1984Cosp},
		the classes 
		$\Lambda(\mathcal{G}_{j}) \cdot \phi_{j}^{(\ell)}$
		and
		$\mathcal{I}_{j} \cdot \phi_{j}^{(\ell)}$
		are Euclidean 
		with envelop $|\phi_{j}^{(\ell)}|$.
		Also, under Assumption A2,
		$R(\cdot)$ is continuously differentiable and has uniformly bounded 
		derivative, so that
		$|R'(\cdot)| \le M$ uniformly
		for some constant $M$. It follows that,
		for any $g_{1}, g_{2} \in \mathcal{G}_{j}$
		with $\|g_{2} - g_{1}\|_{Q,1} < \epsilon / M$,
		we can show that 
		\begin{eqnarray*}
			\|R(g_{2}) - R(g_{1}) \|_{Q, 1}
			\le 
			M \|
			g_{2} 
			- 
			g_{1} 
			\|_{Q,1}
			\le 
			\epsilon.
		\end{eqnarray*}
		Since $\mathcal{G}_{j}$ is Euclidean,
		we have that, for some constants $A_{j}$ and $V_{j}$,
		\begin{eqnarray*}
			N
			\big(
			\epsilon \|\bar{R}_{j} \|_{Q,1}, 
			R(\mathcal{G}_{j}), 
			L_{1}(Q)
			\big)
			\le 
			N
			\big(
			(\epsilon/M) \|G_{j} \|_{Q,1}, 
			\mathcal{G}_{j}, 
			L_{1}(Q)
			\big)
			\le
			(A_{j} M^{V_{j}}) \epsilon^{-V_{j}},
		\end{eqnarray*}
		Thus, $R(G_{j})$ is Euclidean with envelop 
		$\bar{R}_{j} := R(g_{j, 0}) + c (|g_{j, 0}| + G_{j})$ 
		for some $g_{j,0} \in \mathcal{G}_{j}$. 
		Lemma 19 and Corollary 17 of \cite{nolan1987u} show that 
		addition of Euclidean classes is Euclidean
		and 
		Lemma 2.14~(ii) of \cite{pakes1989simulation} shows that
		multiplication of Euclidean classes is Euclidean.
		Thus, 
		function class 
		$
		\mathcal{F}_{j}^{(\ell)}
		:=
		\{
		\big\{[\Lambda(\mathcal{G}_{j}) - \mathcal{I}_{j}]R(\mathcal{G}_{j})
		\phi_{j}^{(\ell)}
		\}
		$
		is Euclidean
		with
		envelop
		$F_{j}^{(\ell)}:=|\phi_{j}^{(\ell)} |\cdot \bar{R}_{j}$
		for each $j=1, \dots, J$
		and 
		$\ell = 1 ,\dots, d_{j}$.
		Then,
		each entry of vector $\psi_{y}\big(\theta(y) \big)$
		is an element 
		in the following Euclidean classes of functions:
		\begin{eqnarray*}
			\mathcal{F}:=
			\cup_{j=1}^{J}
			\cup_{\ell=1}^{d_{j}}
			\mathcal{F}_{j}^{(\ell)}.
		\end{eqnarray*}

		It follows from (\ref{eq:eudp})
		that,
		for some constants
		$A_{j}^{(\ell)}$
		and 
		$V_{j}^{(\ell)}$,
		\begin{eqnarray*}
			N
			\big(
			\epsilon \|\bar{F}_{j}^{(\ell)} \|_{Q,2}, 
			\mathcal{F}_{j}^{(\ell)}, 
			L_{2}(Q)
			\big)
			\le
			A_{j}^{(\ell)}2^{2V_{j}^{(\ell)}}\epsilon^{-2 V_{j}^{(\ell)}},
		\end{eqnarray*}
		for every $j=1, \dots, J$
		and 
		$\ell = 1 ,\dots, d_{j}$,
		and thus we can show that 
		\begin{eqnarray*}
			\int_{0}^{1}
			\bigg ( 
			\sup_{Q} \log
			N(\epsilon \|F_{j}^{(\ell)}\|_{Q,2},
			\mathcal{F}_{j}^{(\ell)}, L_{2}(Q) )
			\log(1/\epsilon)
			\bigg )^{1/2}
			d \epsilon
			< \infty.
		\end{eqnarray*}
		This together with Theorem 1 of \cite{rio1998processus}
		implies that the class $\mathcal{F}$
		is Donsker. 
	\end{proof}
	\vspace{0.5cm}


	\vspace{0.5cm}
	\begin{proof}[\textbf{Proof of Theorem \ref{theorem:beta}}]

          Fist, we shall obtain the asymptotic linear representation
          for the DR estimator
          under the $\beta$-mixing condition. 
          That is, we will show that, 
          uniformly in $y \in \mathcal{Y}$,
          \begin{eqnarray}
            \label{eq:asym-lin}
            \sqrt{T} 
            \big \{
            \widehat{\theta}(y)
            -
            \theta(y)
            \big \}
            =
            -
            \big (
            H(y)
            \big )^{-1}
            \widehat{\Psi}_{y}
            \big (\theta(y)\big) 
            + o_p(1). 
          \end{eqnarray}
          Since
          $
          \widehat{\theta}(y)
          $
          and 
          $
          \theta(y)
          $ respectively 
          consists of
          subvectors 
          $\{\widehat{\theta}_{j}(y_{j})\}_{j=1}^{J}$
          and 
          $\{\theta_{j}(y_{j})\}_{j=1}^{J}$
          with finite $J$,
          it suffices to prove
          that, 
          for each $j = 1, \dots, J$,
          uniformly in $y_j \in \mathcal{Y}_j$,
          \begin{eqnarray} 
            \label{eq:asym-lin-j}
            \sqrt{T} 
            \big \{
            \widehat{\theta}_{j}(y_{j})
            -
            \theta_{j}(y_{j})
            \big \}
            =
            -
            \big (
            H_{j}(y_{j})
            \big )^{-1}
            \widehat{\Psi}_{y, j}
            \big (\theta_{j}(y_{j})\big) 
            + o_p(1). 
          \end{eqnarray}
          Fix $j \in \{1, \dots, J\}$
		and 
		let
		$M$ be a finite positive constant.
		For notational simplicity,   
		we define a localized objective function
		given $\delta_{j} \in \R^{d_{j}}$
		\begin{eqnarray*}
			\widehat{Q}_{y, j}(\delta_{j})
			:=
			\widehat{\ell}_{y,j}
			\big (
			\theta_{j}(y_{j}) + T^{-1/2}\delta_{j}  
			\big)
			-
			\widehat{\ell}_{y,j}
			\big (
			\theta_{j}(y_{j}) 
			\big).
		\end{eqnarray*}
		Then, the estimator 
		$\widehat{\delta}_{j}(y_{j}) :=
		\sqrt{T}
		\big (
		\widehat{\theta}_{j}(y_{j}) - \theta_{j}(y_{j})
		\big)$
		is the solution to 
		$  \max_{\delta_{j} \in \R^{d_j}}
		\widehat{Q}_{y, j}
		(\delta_{j})
		$.

		Under Assumption A.2,
		the map 
		$\delta_{j} \mapsto \widehat{Q}_{y, j}(\delta_{j})$
		is twice continuously differentiable
                and thus we can show that
		\begin{eqnarray*}
			T
			\widehat{Q}_{y, j}(\delta_{j})
			=
			\delta_{j}^{\top}
			\widehat{\Psi}_{y, j}
			\big(
			\theta_{j}(y_{j})
			\big)             
			+
			\frac{1}{2}
			\delta_{j}^{\top}
			\widehat{H}_{j}(y_{j})
			\big (
			\theta_{j}(y_{j})
			\big)
			\delta_{j}
			+
			o(T^{-1}\|\delta_{j}\|^2),
		\end{eqnarray*}
		uniformly
		in $y_{j} \in \mathcal{Y}_{j}$,
		for each fixed $\delta_{j}$
		with $\|\delta_{j} \| \le M$.
		Also,
		we can show that 
		$
		\widehat{H}_{j}
		(y_{j})
		\to^p
		H_{j}(y_{j})
		$
		uniformly in $y_{j} \in \mathcal{Y}_{j}$,
		by the uniform law of large numbers.
		Thus, for each $\delta_{j}$
		with $\|\delta_{j} \| \le M$,
		we can show that 
		$
		\sup_{y_{j} \in \mathcal{Y}_{j}}
		\big |
		\widehat{Q}_{y, j}(\delta_{j}) 
		-
		\widetilde{Q}_{y, j}(\delta_{j})
		\big |
		=
		o_p(T^{-1})
		$,
		where 
		\begin{eqnarray*}
			T 
			\widetilde{Q}_{y,j}(\delta_{j})
			:=
			\delta_{j}^{\top}
			\widehat{\Psi}_{y, j}
			\big( \theta_{j}(y_{j}) \big)
			+
			\frac{1}{2}
			\delta_{j}^{\top}
			H_{j}(y_{j})
			\delta_{j}.
		\end{eqnarray*} 
		The convexity lemma
		\citep[see][]{pollard1991asymptotics, kato2009asymptotics}
		extends
		the point-wise convergence 
		with respect $\delta_{j}$
		to the uniform converges and thus,
		under Assumption A2,
		\begin{eqnarray}
			\label{eq:Q2}
			\sup_{y_{j} \in \mathcal{Y}_{j}}
			\sup_{\delta_{j} : \|\delta_{j} \| \le M}
			\big |
			\widehat{Q}_{y, j}(\delta_{j})
			-
			\widetilde{Q}_{y, j}(\delta_{j})
			\big |
			=
			o_p(T^{-1}).
		\end{eqnarray}

		Let 
		$\widetilde{\delta}_{j}(y_{j})
		:=
		-
		\big (H_{j} (y_{j}) \big )^{-1}
		\widehat{\Psi}_{y, j}
		\big (\theta_{j}(y_{j})\big) 
		$,
		which maximizes
		$\widetilde{Q}_{y, j}(\delta_{j})$.
		Then,  
		simple algebra can show that,
		for any $\delta_{j}$ and for some constant $c>0$,
		\begin{eqnarray}
			\label{eq:lbd1}
			\widetilde{Q}_{y, j}
			\big(\tilde{\delta}_{j}(y_{j}) \big)
			-
			\widetilde{Q}_{y, j}
			\big (\delta_{j} \big)
			=
			-
			\frac{1}{2 T}
			\big (
			\widetilde{\delta}_{j}(y_{j}) - \delta_{j}  
			\big )^{\top}
			H_{j}(y_{j})
			\big (
			\widetilde{\delta}_{j}(y_{j}) - \delta_{j}  
			\big )
			\ge 
			\frac{c}{2 T}
			\| 
			\widetilde{\delta}_{j}(y_{j}) - \delta_{j}  
			\|^2,
		\end{eqnarray}
		where
		the last inequality is
		due to that
		$H_{j}(y_{j})$
		is negative definite
		under Assumption A2.  
		For any subset $D \subseteq \R^{d_{j}}$ including $\widetilde{\delta}_{j}(y_{j})$,
		an application of the triangle inequality obtains 
		\begin{eqnarray}
			2
			\sup_{\delta_{j} \in D}
			|
			\widehat{Q}_{y,j}
			(\delta_{j})
			-
			\widetilde{Q}_{y, j}
			(\delta_{j})
			|
			&\ge& \notag
			\sup_{\delta_{j} \in D}
			\big \{
			\widetilde{Q}_{y, j}\big(\widetilde{\delta}_{j}(y_{j}) \big)
			-
			\widetilde{Q}_{y, j}\big (\delta_{j} \big)
			\big \}\\
			&& - \label{eq:tri5}
			\sup_{\delta_{j} \in D}
			\big \{
			\widehat{Q}_{y,j}
			\big (\widetilde{\delta}_{j}(y_{j})\big)
			-
			\widehat{Q}_{y, j}
			(\delta_{j})
			\big\}.
		\end{eqnarray}
		
		Let $\eta>0$ be an arbitrary constant.
		Because of the concavity under Assumption A2,
		difference quotients
		satisfy that,
		for any $\zeta > \eta$
		and
		for any $v \in S^{d_1}$
		with the unit sphere  
		$S^{d_1}$
		in $\R^{d_1}$,
		\begin{eqnarray*}
			\frac{
				\widehat{Q}_{y, j}
				\big (
				\widetilde{\delta}_{j}(y_{j})
				+
				\eta v
				\big)
				-
				\widehat{Q}_{y, j}
				\big(
				\widetilde{\delta}_{j}(y_{j})
				\big)
			}{
				\eta
			}
			\ge
			\frac{
				\widehat{Q}_{y, j}
				\big(
				\widetilde{\delta}_{j}(y_{j})
				+
				\zeta v
				\big)
				-
				\widehat{Q}_{y,j}
				\big(
				\widetilde{\delta}_{j}(y_{j})
				\big)
			}{
				\zeta
			}.
		\end{eqnarray*}
		This inequality with
		a set $\widetilde{D}_{j}(\eta):=
		\big\{\delta_{j} \in \R^{d_j}: \|\delta_{j} - \widetilde{\delta}_{j}(y_{j}) \| \le \eta \big\}$
		implies
		that, 
		given the event 
		$\big\{
		\sup_{y \in \mathcal{Y}_{j}}\| \widehat{\delta}_{j}(y_{j}) - \widetilde{\delta}_{j}(y_{j}) \|
		\ge \eta
		\big\}$,
		we have,
		for any $y_{j} \in \mathcal{Y}_{j}$,
		\begin{eqnarray} 
			\label{eq:Q1}
			\sup_{\delta_{j} \in \widetilde{D}_{j}(\eta)}
			\widehat{Q}_{y, j}
			(\delta_{j})
			-
			\widehat{Q}_{y, j}
			\big (\widetilde{\delta}_{j}(y_{j})\big) \ge 0,
		\end{eqnarray}
		where
		the last inequality is due to that 
		$\widehat{Q}_{y, j} 
		\big (\widehat{\delta}_{j}(y_{j})\big)
		-
		\widehat{Q}_{y, j}
		\big ( \widetilde{\delta}_{j}(y_{j})\big)
		\ge  0
		$,
		by definition of
		$\widehat{\delta}_{j}(y_{j})$.
		It follows from
		(\ref{eq:lbd1})-(\ref{eq:Q1}) that,
		given the event 
		$\big\{
		\sup_{y \in \mathcal{Y}_{j}}\| \widehat{\delta}_{j}(y_{j}) - \widetilde{\delta}_{j}(y_{j}) \|
		\ge \eta
		\big\}$,
		\begin{eqnarray*}
			\sup_{\delta_{j} \in \widetilde{D}_{j}(\eta)}
			\big|
			\widehat{Q}_{y, j}
			(\delta_{j})
			-
			\widetilde{Q}_{y, j}
			(\delta_{j})
			\big|
			\ge
			\frac{c}{4 T} \eta^{2}. 
		\end{eqnarray*}
		Because
		$\widehat{\Psi}_{y, j}
		\big (\theta(y_{j}) \big)$  
		is Donsker by Lemma \ref{lemma:donsker},
		we can show that,
		for any $\xi>0$,
		there exists a constant $C$
		such that 
		$\Pr\big(\sup_{y_{j} \in \mathcal{Y}_{j}}\| \widetilde{\delta}_{j}(y_{j}) \| \ge C\big)
		\le \xi
		$ 
		for a sufficiently large $T$. 
		Thus, the above display implies that 
		\begin{eqnarray*}
			\Pr
			\bigg (
			\sup_{y_{j} \in \mathcal{Y}_{j}}\| \widehat{\delta}_{j}(y_{j}) - \widetilde{\delta}_{j}(y_{j}) \|
			\ge \eta 
			\bigg )
			\le
			\Pr 
			\bigg (
			\sup_{y_{j} \in \mathcal{Y}_{j}}
			\sup_{\delta_{j}: \|\delta_{j} \| \le \eta + C}
			\big |
			\widehat{Q}_{y, j}
			(\delta_{j})
			-
			\widetilde{Q}_{y, j}
			(\delta_{j})
			\big |
			>
			\frac{c }{4 T} \eta^2
			\bigg )
			+
			\xi,
		\end{eqnarray*} 
		for a sufficiently large $T$.
		It follows from (\ref{eq:Q2}) that
		the first term on the right side of the above equation
		converges to 0 as $T \to \infty$.
		Thus, we obtain
                (\ref{eq:asym-lin-j}),
                which implies
                (\ref{eq:asym-lin}).

                

		Now, Lemma \ref{lemma:donsker} shows that 
		the empirical process
		$\widehat{\Psi}_{y}\big( \theta(y) \big)$ is 
		stochastically equicontinuous over $\mathcal{Y}$.
		The finite dimensional convergence follows from 
		the central limit theorem for $\beta$-mixing processes \citep[Theorem 4,][]{bradley1985central}.
		This with
		the stochastic equicontinuity of the map $y \mapsto \widehat{\Psi}\big( \theta(y) \big)$
		implies that 
		$
		\widehat{\Psi}(\cdot)
		\rightsquigarrow
		\mathbb{B}(\cdot)
		$
		in
		$\times_{j=1}^{J} \ell^{\infty}(\mathcal{Y}_{j})^{d_j}$
		where $\mathbb{B}(\cdot)$ is
		a zero-mean Gaussian process with covariance function
		defined in Theorem \ref{theorem:beta}.
	\end{proof}
	\vspace{0.5cm}

	\begin{proof}[\textbf{Proof of Theorem \ref{theorem: joint}}]
		
		\textbf{(a)}
		Under Assumptions A1-A3, 
		the map $\varphi(\cdot)$
		is shown to be Hadamard differentiable
		at
		$
		\theta(\cdot)
		$
		tangentially to
		$\mathbb{D}$
		with the derivative map
		$b:=(b_{1}, \dots, b_{J})
		\mapsto 
		\varphi_{\theta(\cdot)}'(b)$,
		given by 
		\begin{eqnarray*}
			\varphi_{\theta(\cdot)}'(b)(x,y)
			=
			\left [
			\begin{array}{c}
				\lambda
				\big ( \phi_1(x_{1})^{\top} \theta_{1}(y_{1})\big)
				\phi_1(x_{1})^{\top} b_{1}(y_{1}) \\  
				\vdots \\ 
				\lambda
				\big ( \phi_J(x_{1})^{\top} \theta_{J}(y_{J})\big)
				\phi_J(x_{J})^{\top} b_{J}(y_{J})
			\end{array}
			\right ].
		\end{eqnarray*}
		Then,
		we can write 
		$
		\big (
		\widehat{F}_{Y_{1t}|X_{1t}},
		\dots,
		\widehat{F}_{Y_{Jt}|X_{Jt}}
		\big )^{\top}
		=
		\varphi\big(\widehat{\theta}(\cdot) \big)
		$
		and
		$
		\big (
		F_{Y_{1t}|X_{1t}},
		\dots,
		F_{Y_{Jt}|X_{Jt}}
		\big )^{\top}
		=
		\varphi\big(\theta(\cdot) \big)
		$.
		Applying the functional delta method
		with
		the result in Theorem \ref{theorem:beta},  
		we can show that
		\begin{eqnarray*}
			\sqrt{T}\left(
			\begin{array}{c}
				\widehat{F}_{Y_{1t}|X_{1t}}-F_{Y_{1t}|X_{1t}}\\
				\vdots\\
				\widehat{F}_{Y_{Jt}|X_{Jt}}-F_{Y_{Jt}|X_{Jt}}
			\end{array}
			\right)
			\rightsquigarrow
			\varphi_{\theta(\cdot)}'
			(
			\mathbb{B}
			)
			\ \ \mathrm{in} \ \
			\times_{j=1}^{J}    
			\ell^{\infty}(\mathcal{X}_{j}{\times}\mathcal{Y}_{j}).
		\end{eqnarray*}
		
		\noindent 
		\textbf{(b)}
		The chain rule for Hadamard differentiable maps
		\citep[Lemma 3.9.3,][]{van1996weak} shows that
		$\nu \circ \varphi: \mathbb{D}_{\varphi} \to
		\ell^{\infty}(\mathcal{Z}{\times}\mathcal{Y})$
		is Hadamard differntiable
		at $\theta$ tangentially to $\mathbb{D}$
		with
		derivative
		$\nu'_{\varphi(\theta(\cdot) )} \circ \varphi_{\theta(\cdot)}'$.
		An application of the functional delta method yields
		the desired conclusion.
              \end{proof}

              \newpage

	\setcounter{equation}{0} 
	\setcounter{lemma}{0}\setcounter{page}{1}\setcounter{proposition}{0} %
	\renewcommand{\thepage}{B-\arabic{page}} \renewcommand{\theequation}{B.%
		\arabic{equation}}\renewcommand{\thelemma}{B.\arabic{lemma}} 
	\renewcommand{\theproposition}{B.\arabic{proposition}} 
        \renewcommand\thesubsection{B.\arabic{subsection}}
  
	\section{Additional Results for Empirical Application}\label{sec: appendix-B}
	We provide more results of the empirical application. Section \ref{sec: appendix-B1} introduces a simulation of samples of the outcome variables from the multiperiod forecasting distributions. In Section \ref{sec: appendix-B2}, we consider the same model specification as in Section \ref{sec: empirical}, but with an alternative order where $Y_{1t}$ represents the real GDP growth and $Y_{2t}$ represents the quarterly NFCI. We compare the proposed approach with the kernel regression method introduced by \cite{adrian2021multimodality} in Section \ref{sec: appendix-B3}. In Section \ref{sec: appendix-B4}, we explore the use of the monthly NFCI in the system and a four-dimensional mixed-frequency model conditional on two lags by treating the three monthly NFCI series as separate observations within a quarter. Based on this model, we revisit the distribution forecasting and counterfactual analysis. Finally, in Section \ref{sec: appendix-B5}, we estimate $95\%$ confidence bands for the distribution as well as density impulse responses using the moving block bootstrap approach for each scenario, and present the results.

\subsection{Sample from Joint Conditional Distributions}\label{sec: appendix-B1}
We explain how to generate random numbers of $Y_t$ from the estimated joint conditional distribution $\widehat{F}_{Y_t|Z_{t}}(\cdot|z)$ based on Model (\ref{eq: estimate-joint}). We assume that for the $j$-th element $Y_{jt}$, the conditional CDF $\widehat{F}_{Y_{jt}|X_{jt}}(\cdot|x_{j})$ is estimated by DR approach over a set of finite points
$\widetilde{\cal Y}_{j} \subseteq \mathcal{Y}_j$
for each $j =1, \dots, J$.  
   
   \begin{algorithm}
   \caption{An algorithm for generating $S$ random samples of $Y_t$ from $\widehat{F}_{Y_t|Z_{t}}(\cdot|z)$}\label{alg:sample}
   For $s=1,...,S$, do
   \begin{enumerate}
   \item Let $j=1$ and set the covariate $x_{1}^{(s)}=(1, z^\top)^\top$. 
   \item Generate a standard uniform random number: $u_{j}^{(s)}\sim U(0,1)$.
   \item Obtain
     \[
       y_{j}^{(s)}=\min
       \Big \{ 
       y_{j} \in \widetilde{\cal Y}_{j}:
       \widehat{F}_{ Y_{jt}|X_{jt} }(y_{j}|x_{j}^{(s)})<u_{j}^{(s)}
       \Big \}.
     \]
    \item Let $j=j+1$ and set $x_{j+1}^{(s)}= \big (x_{j}^\top,y_{j}^{(s)} \big)^\top$.  
    \item Repeat Steps 2-4 untill $j=J$. 
    \item Output: $\big (y_{1}^{(s)}, \dots, y_{J}^{(s)} \big)$, which is
      a vector of realized values of $Y_t$.
    
	\end{enumerate}
   \end{algorithm}

	\subsection{Joint Conditional Distribution with Alternative order}\label{sec: appendix-B2}	
	\begin{figure}[H]
		\captionsetup[subfigure]{aboveskip=-3pt,belowskip=-10pt, labelformat=empty}
		\centering 
		\caption{Contour Plots of the Joint Distribution during the Great Recession} \label{fig: contour-plots-2}	
		\begin{minipage}[t]{0.95\textwidth}
			\centering \footnotesize (a) $Y_{t+h}$ given $(Y_t, Z_t)$ for $t$=2008:Q3\\
			\begin{subfigure}[t]{0.24\textwidth} 	
				\caption{\footnotesize $h=1$}
				\includegraphics[width=0.95\textwidth]{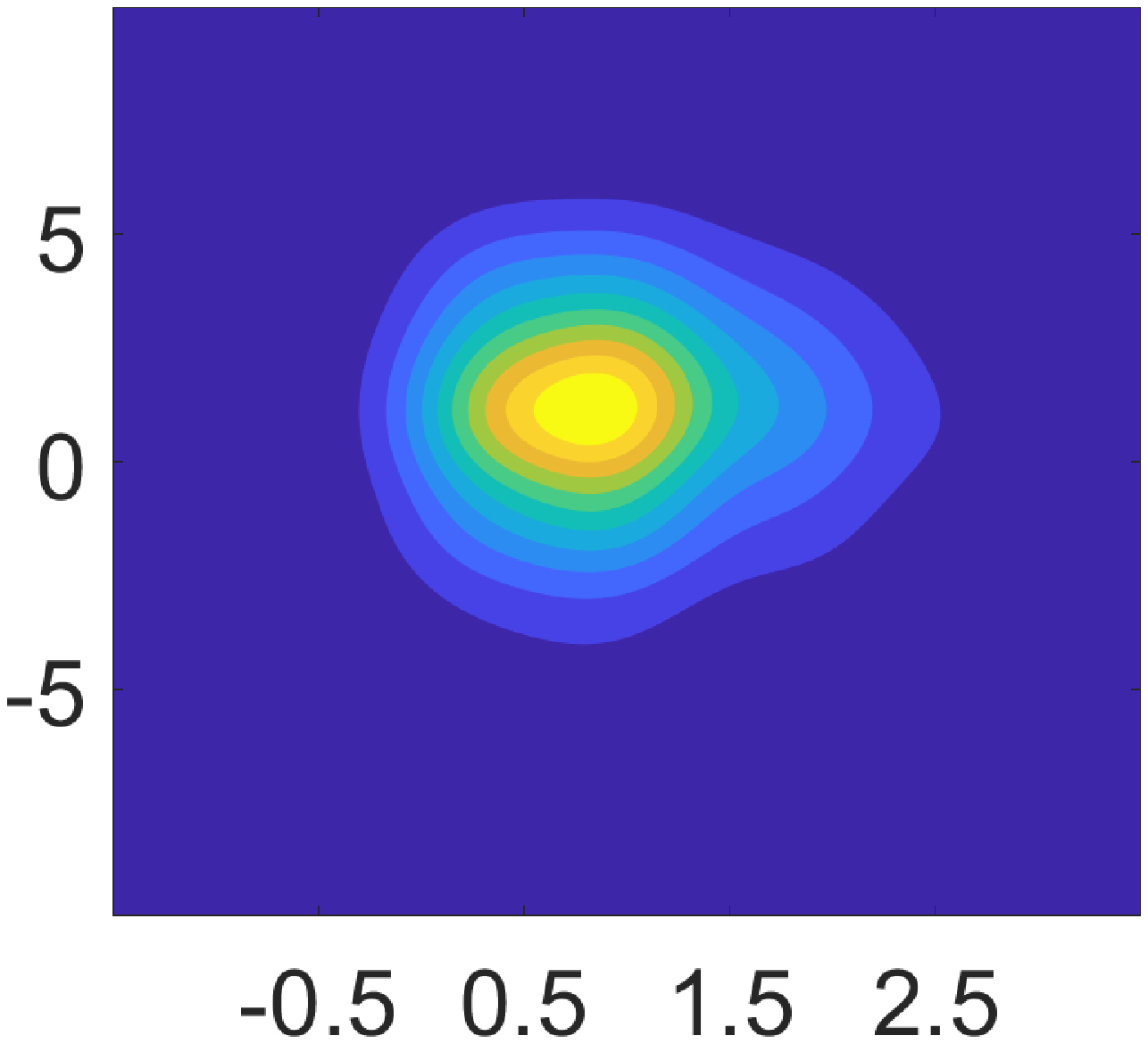}	
			\end{subfigure}
			\begin{subfigure}[t]{0.23\textwidth}
				\caption{\footnotesize $h=2$}	
				\includegraphics[width=0.95\textwidth]{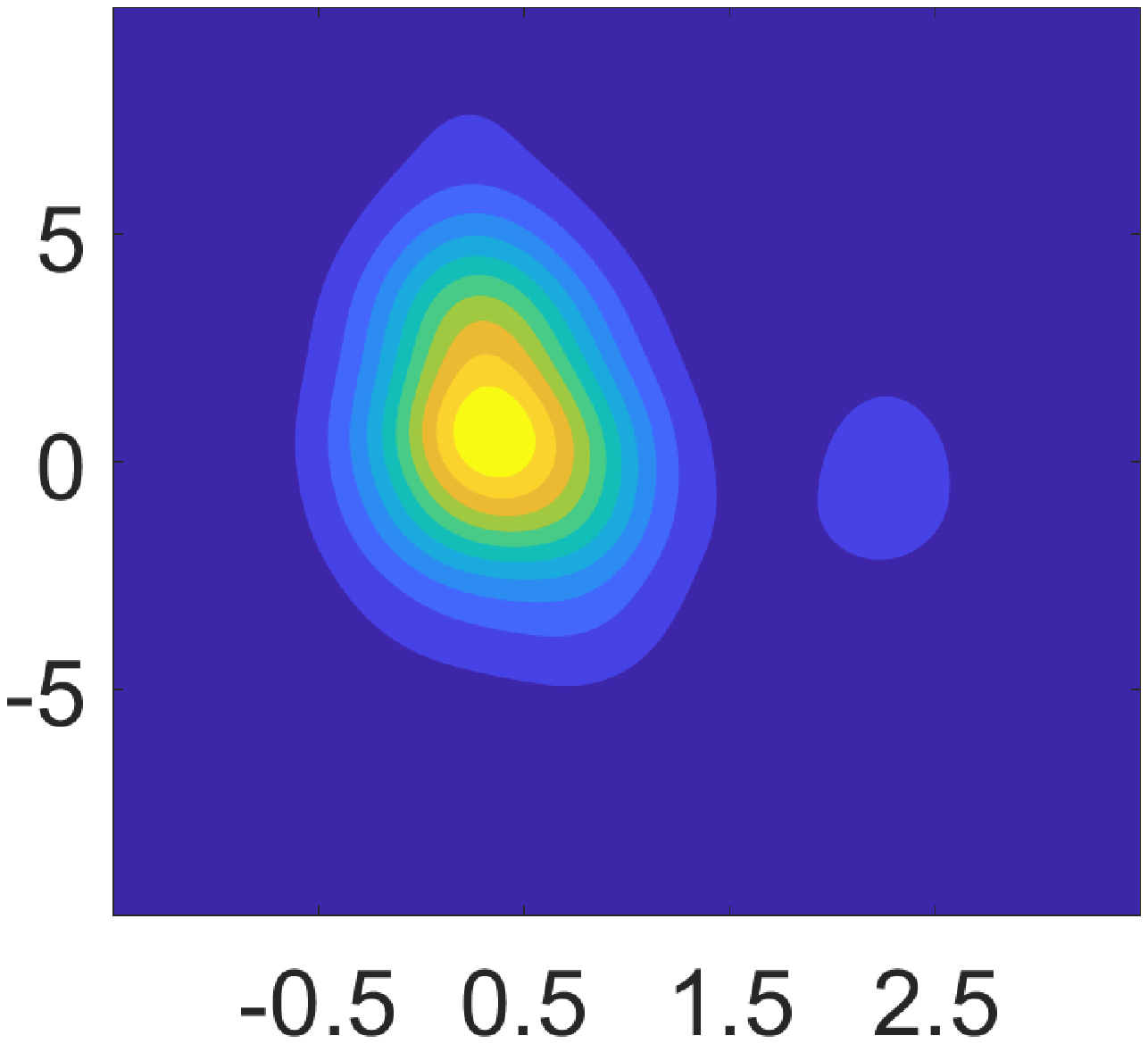}
			\end{subfigure}
			\begin{subfigure}[t]{0.23\textwidth}
				\caption{\footnotesize $h=3$}	
				\includegraphics[width=0.95\textwidth]{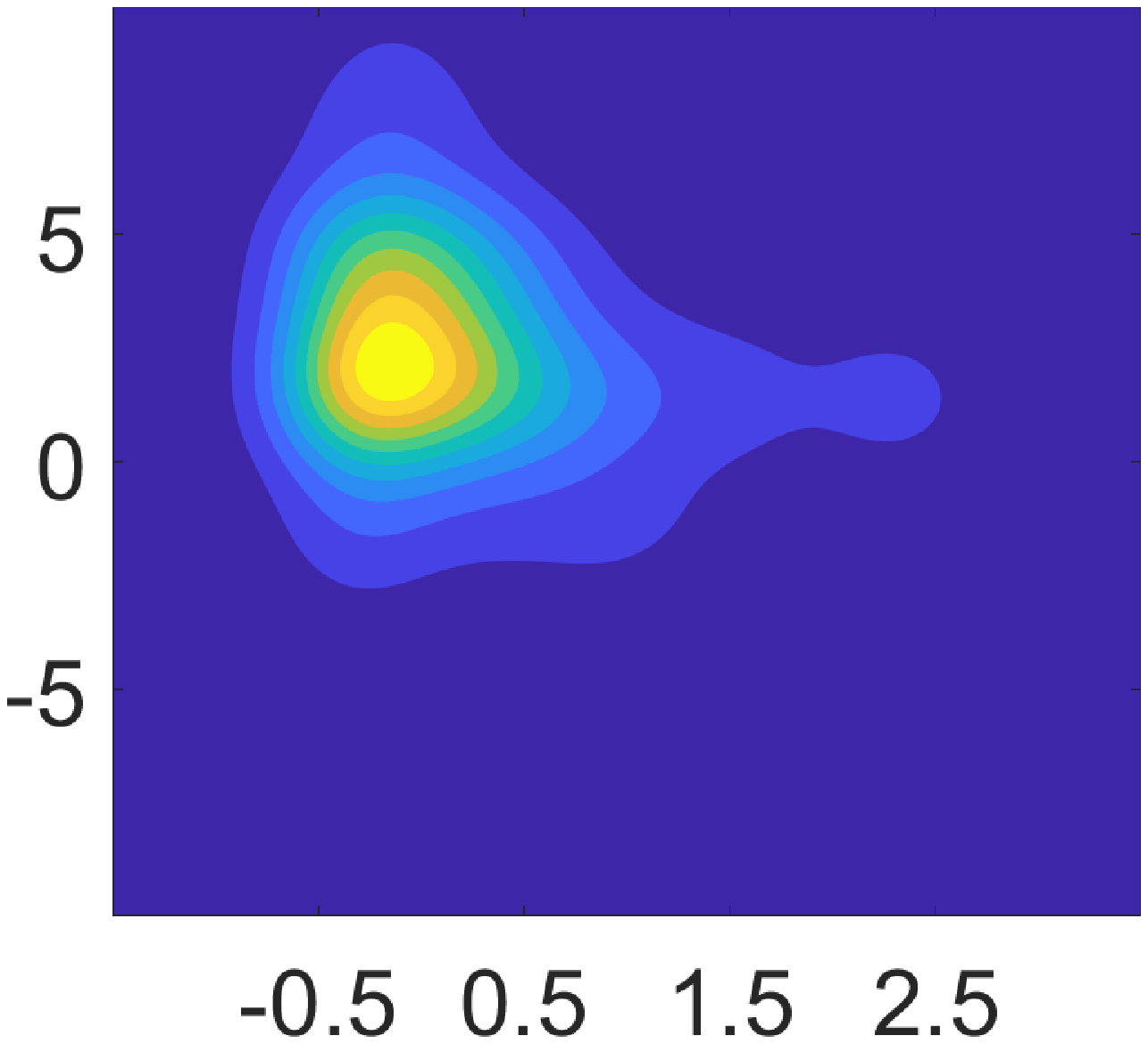}
			\end{subfigure}
			\begin{subfigure}[t]{0.23\textwidth}
				\caption{\footnotesize $h=4$}
				\includegraphics[width=0.95\textwidth]{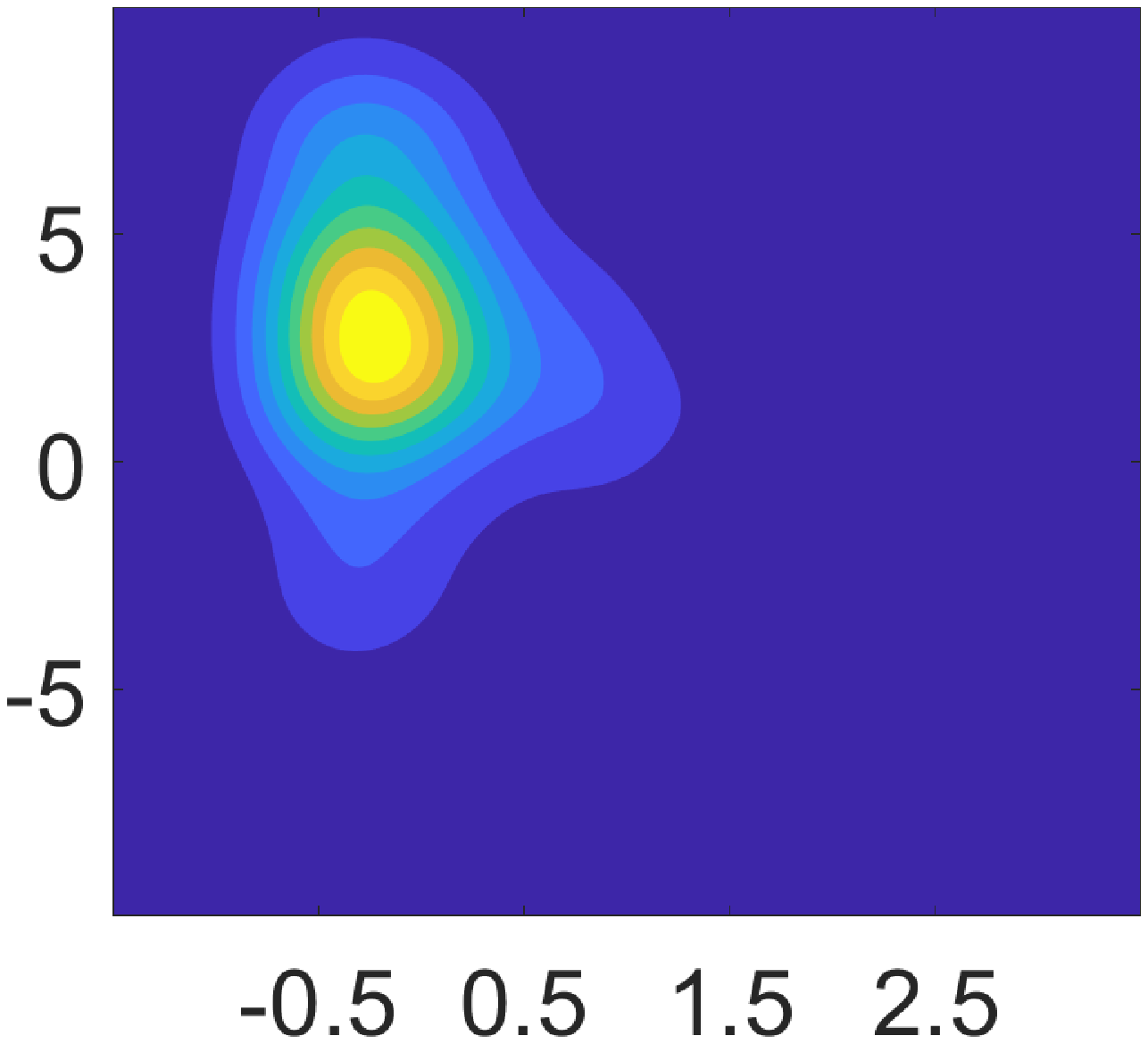}
			\end{subfigure}
			\vspace{3mm}
		\end{minipage}
		\begin{minipage}[t]{0.95\textwidth}
			\centering \footnotesize (b) $Y_{t+h}$ given $(Y_t, Z_t)$ for $t$=2008:Q4 \\
			\begin{subfigure}[t]{0.23\textwidth}
				\caption{\footnotesize $h=1$}	
				\includegraphics[width=0.95\textwidth]{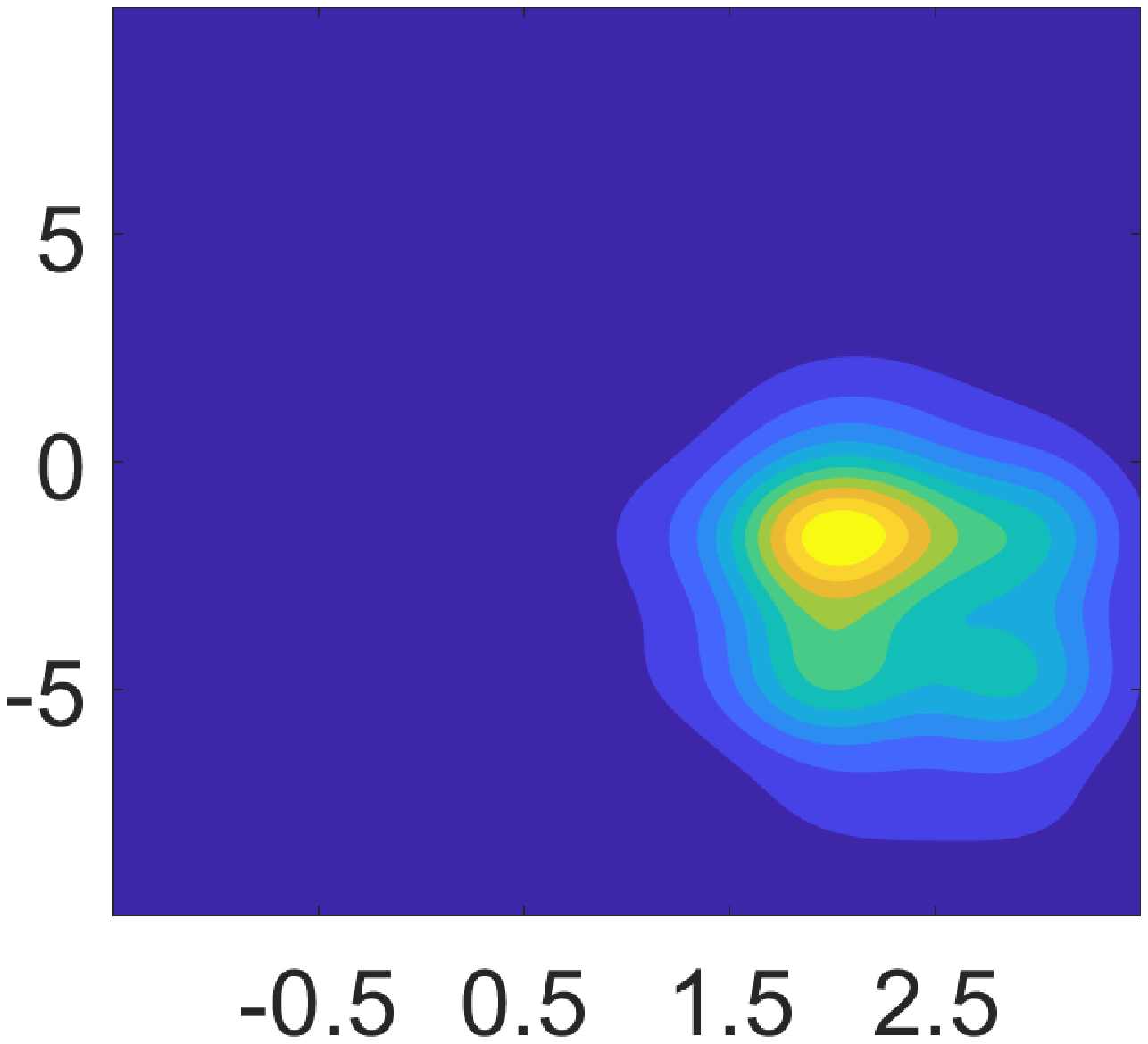}     		
			\end{subfigure}
			\begin{subfigure}[t]{0.23\textwidth}
				\caption{\footnotesize $h=2$}
				\includegraphics[width=0.95\textwidth]{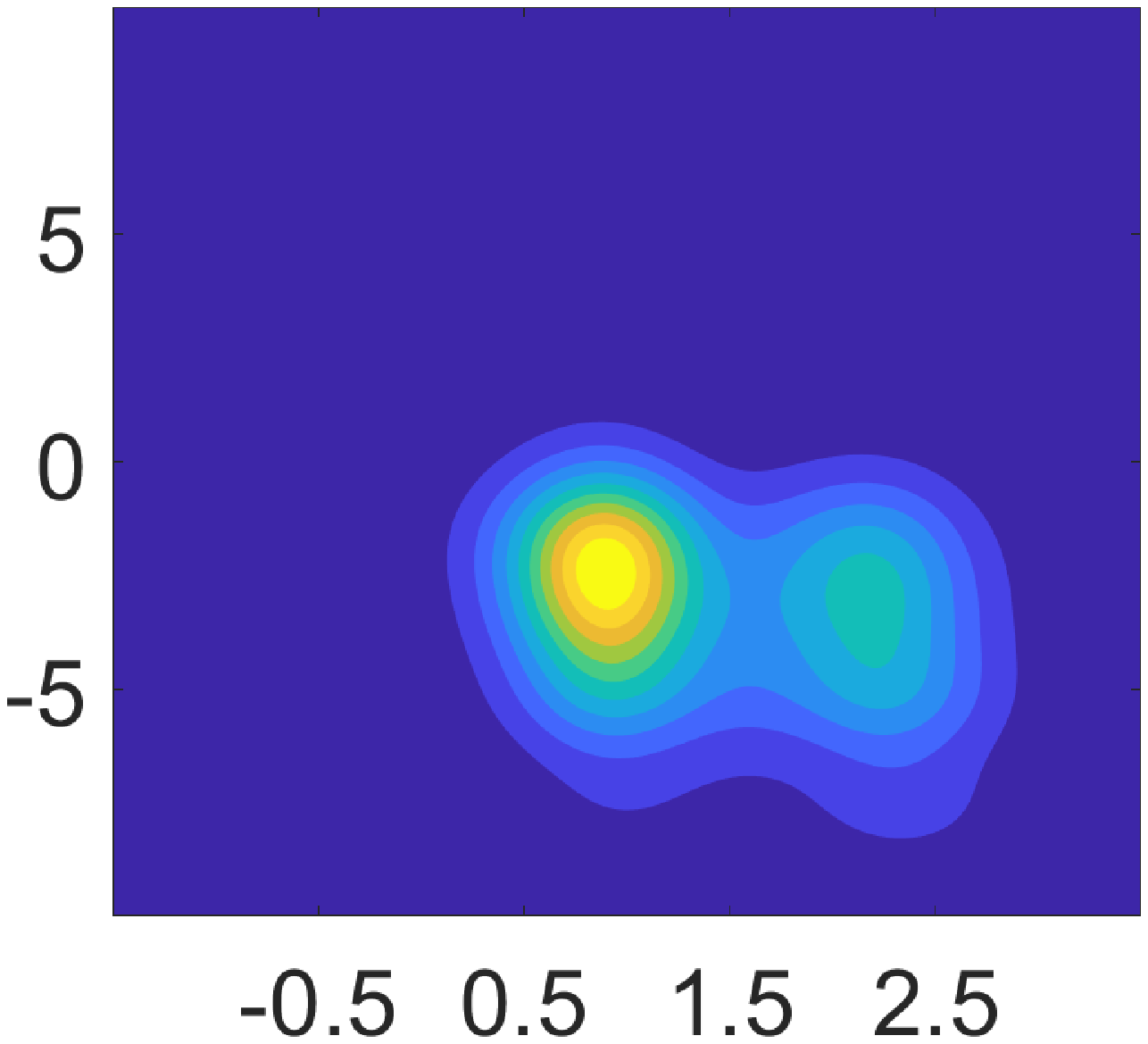}
			\end{subfigure}
			\begin{subfigure}[t]{0.23\textwidth}
				\caption{\footnotesize $h=3$}
				\includegraphics[width=0.95\textwidth]{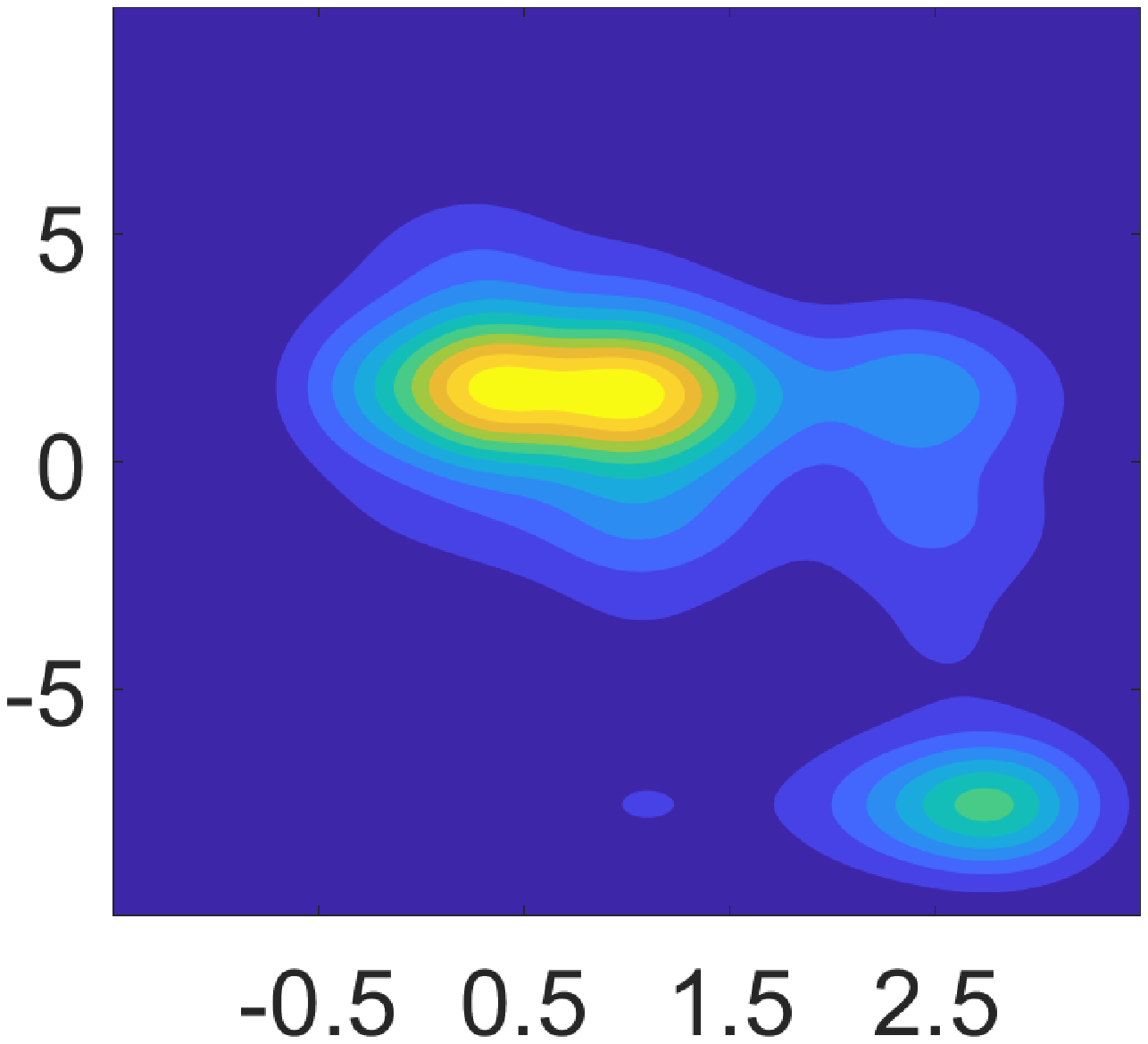}
			\end{subfigure}
			\begin{subfigure}[t]{0.23\textwidth}
				\caption{\footnotesize $h=4$}
				\includegraphics[width=0.95\textwidth]{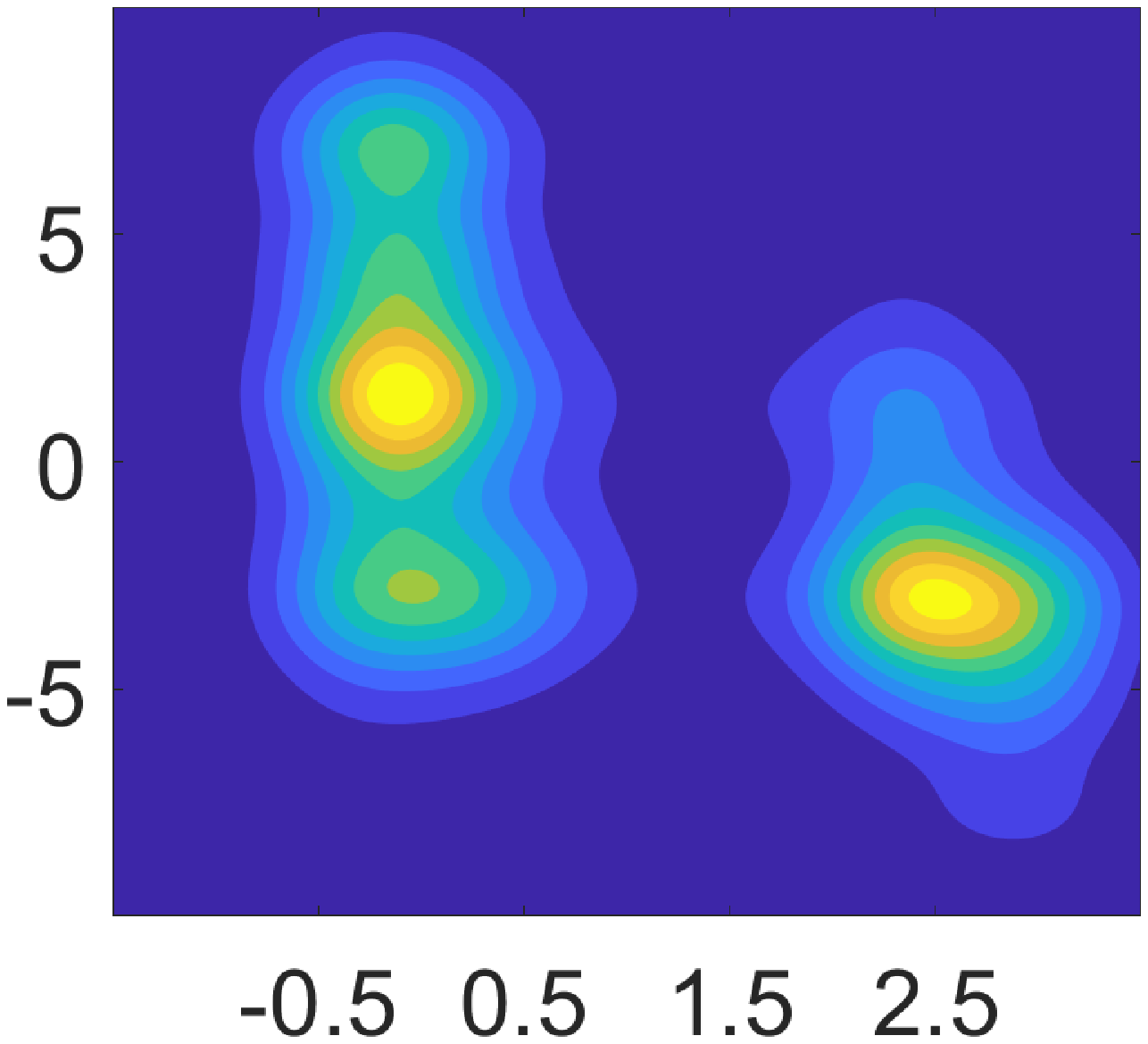}
			\end{subfigure}   
			\vspace{3mm}
		\end{minipage}
		\begin{minipage}[t]{0.95\textwidth}
			\centering \footnotesize (c) $Y_{t+h}$ given $(Y_t, Z_t)$ for $t$=2009:Q1 \\		
			\begin{subfigure}[t]{0.23\textwidth}
				\caption{\footnotesize $h=1$}	
				\includegraphics[width=0.95\textwidth]{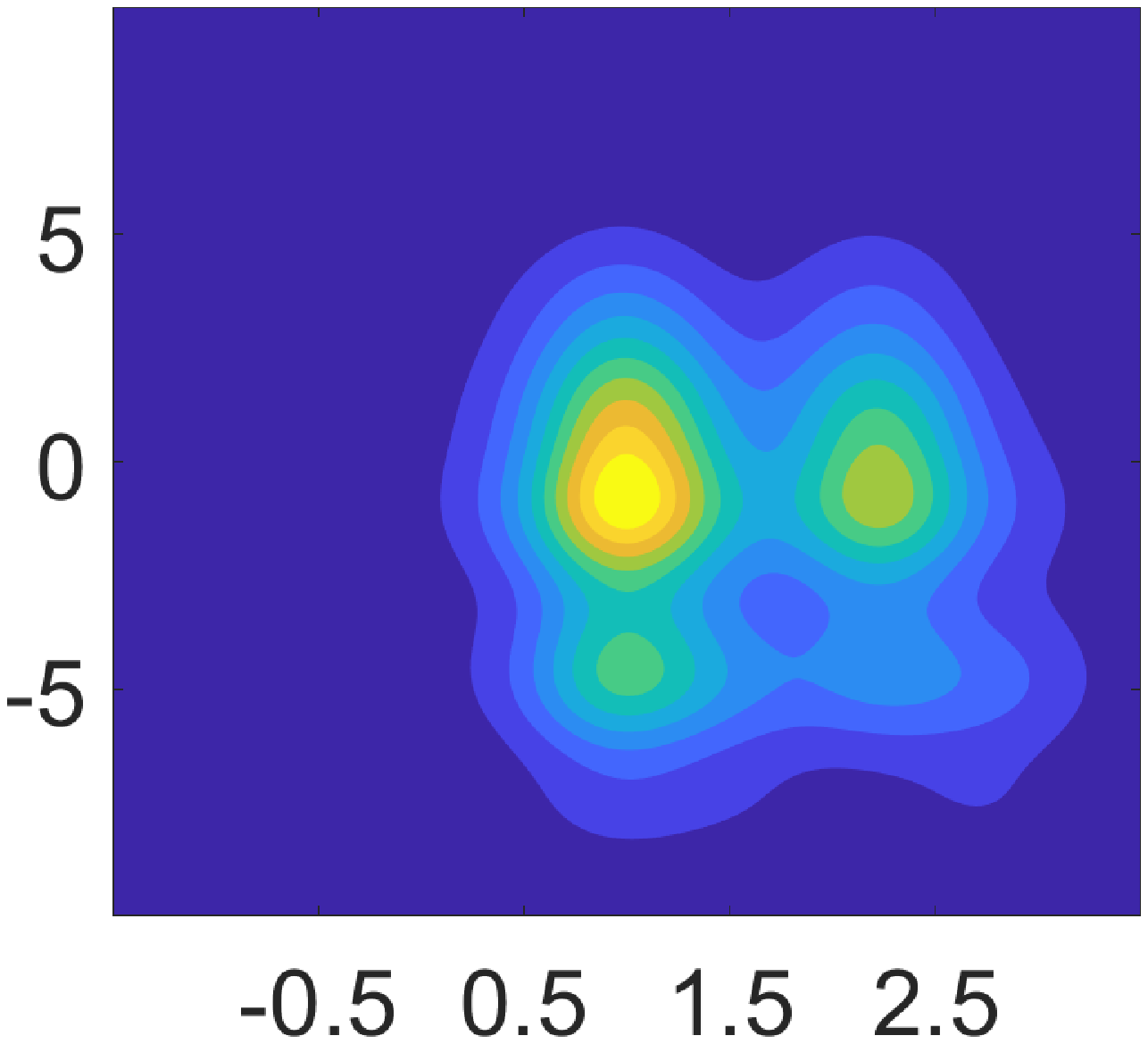}     		
			\end{subfigure}
			\begin{subfigure}[t]{0.23\textwidth}
				\caption{\footnotesize $h=2$}	
				\includegraphics[width=0.95\textwidth]{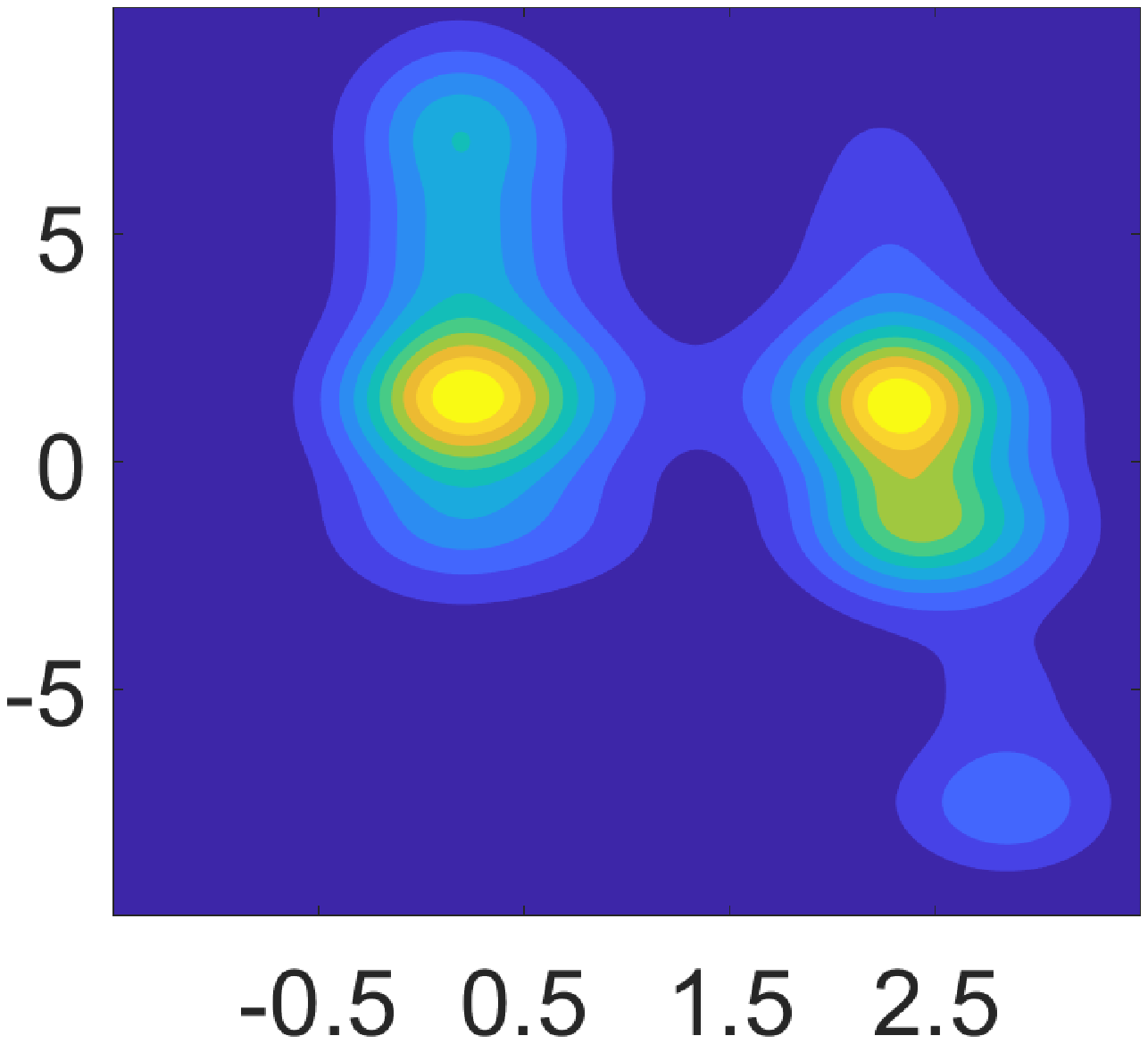}
			\end{subfigure}
			\begin{subfigure}[t]{0.23\textwidth}
				\caption{\footnotesize $h=3$}
				\includegraphics[width=0.95\textwidth]{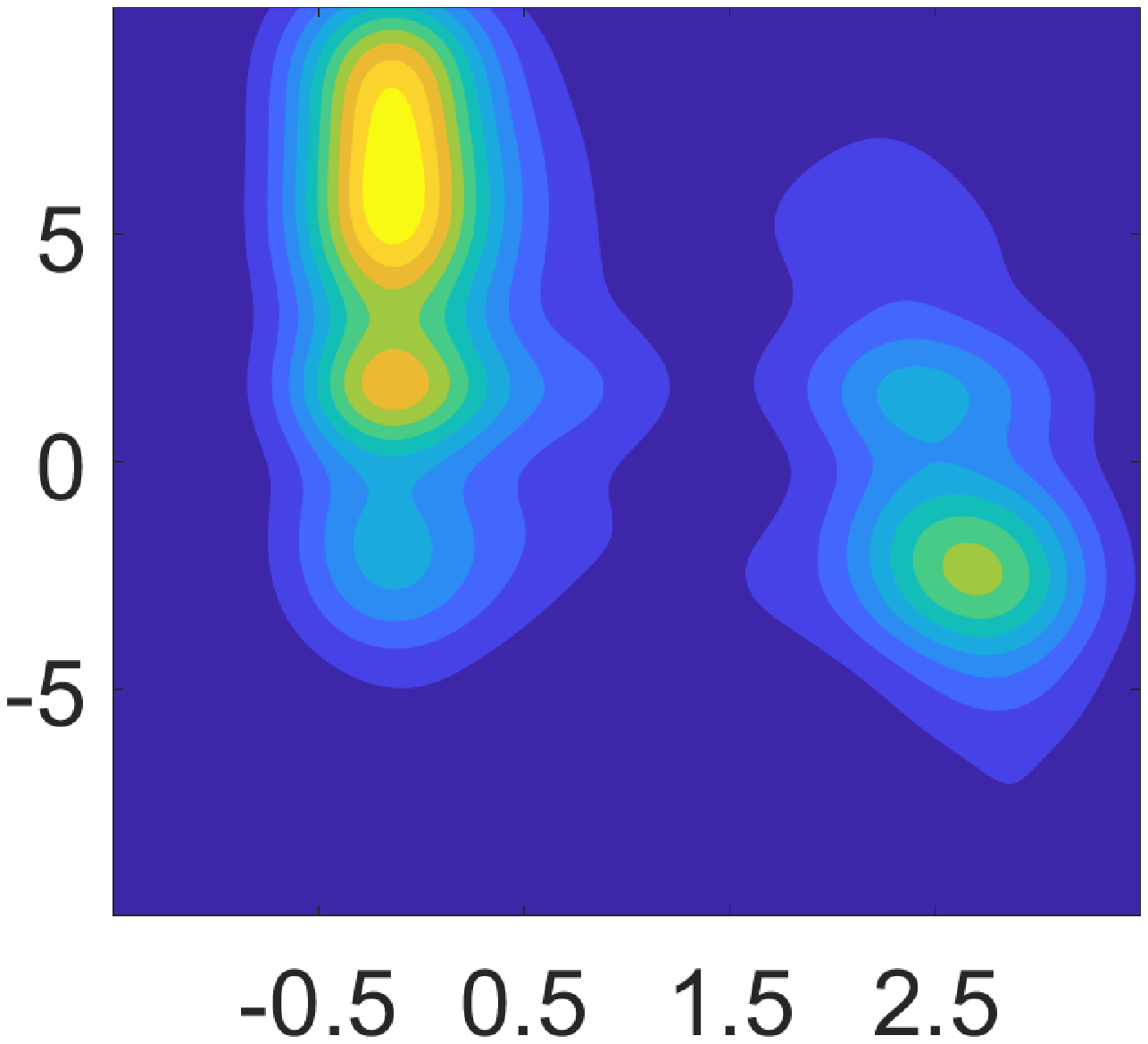}
			\end{subfigure}
			\begin{subfigure}[t]{0.23\textwidth}
				\caption{\footnotesize $h=4$}
				\includegraphics[width=0.95\textwidth]{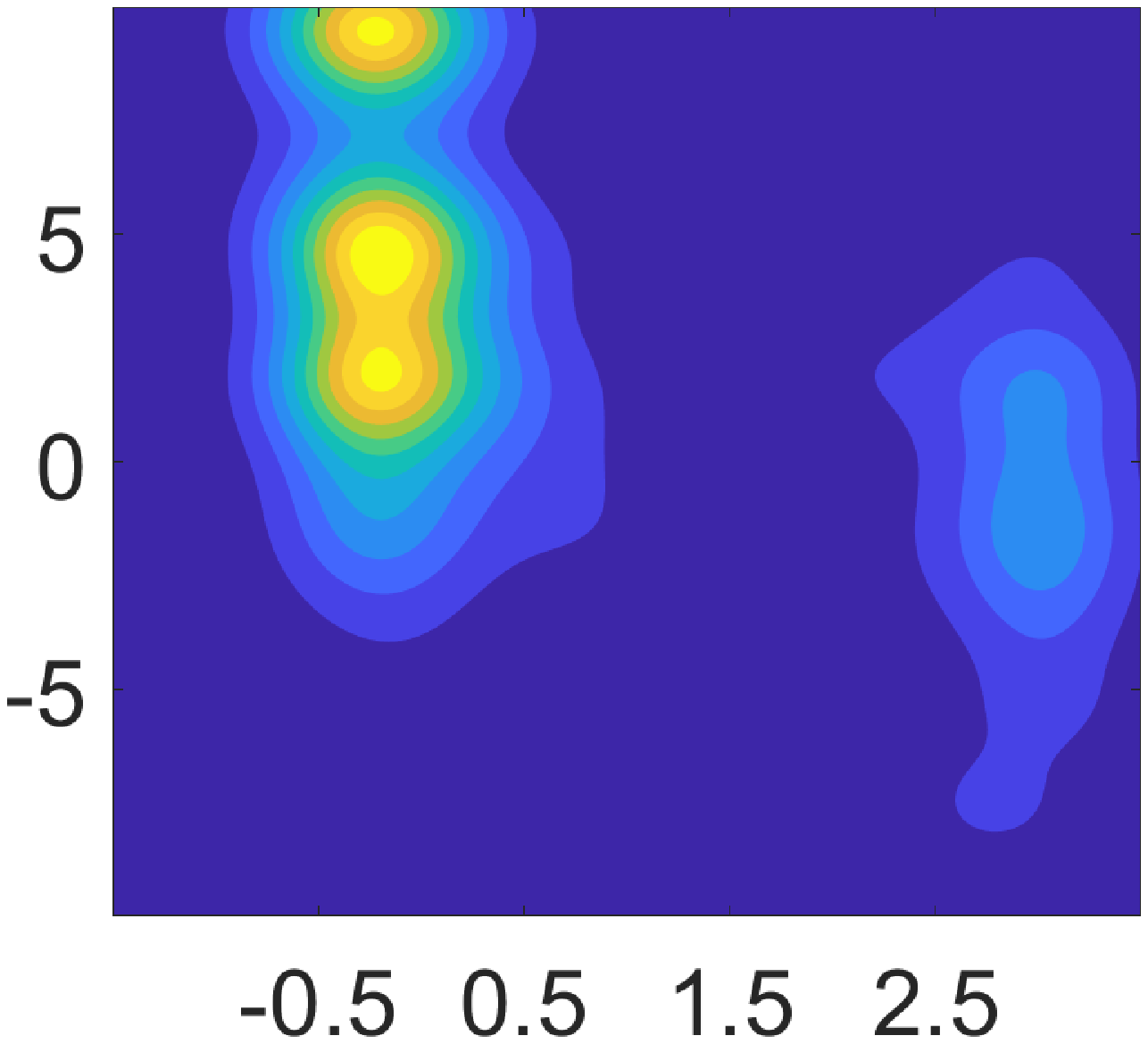}
			\end{subfigure} 
			\vspace{3mm}
		\end{minipage}
		\begin{minipage}[t]{0.95\textwidth}
			\centering \footnotesize (d) $Y_{t+h}$ given $(Y_t, Z_t)$ for $t$=2009:Q2 \\		
			\begin{subfigure}[t]{0.23\textwidth}
				\caption{\footnotesize $h=1$}	
				\includegraphics[width=0.95\textwidth]{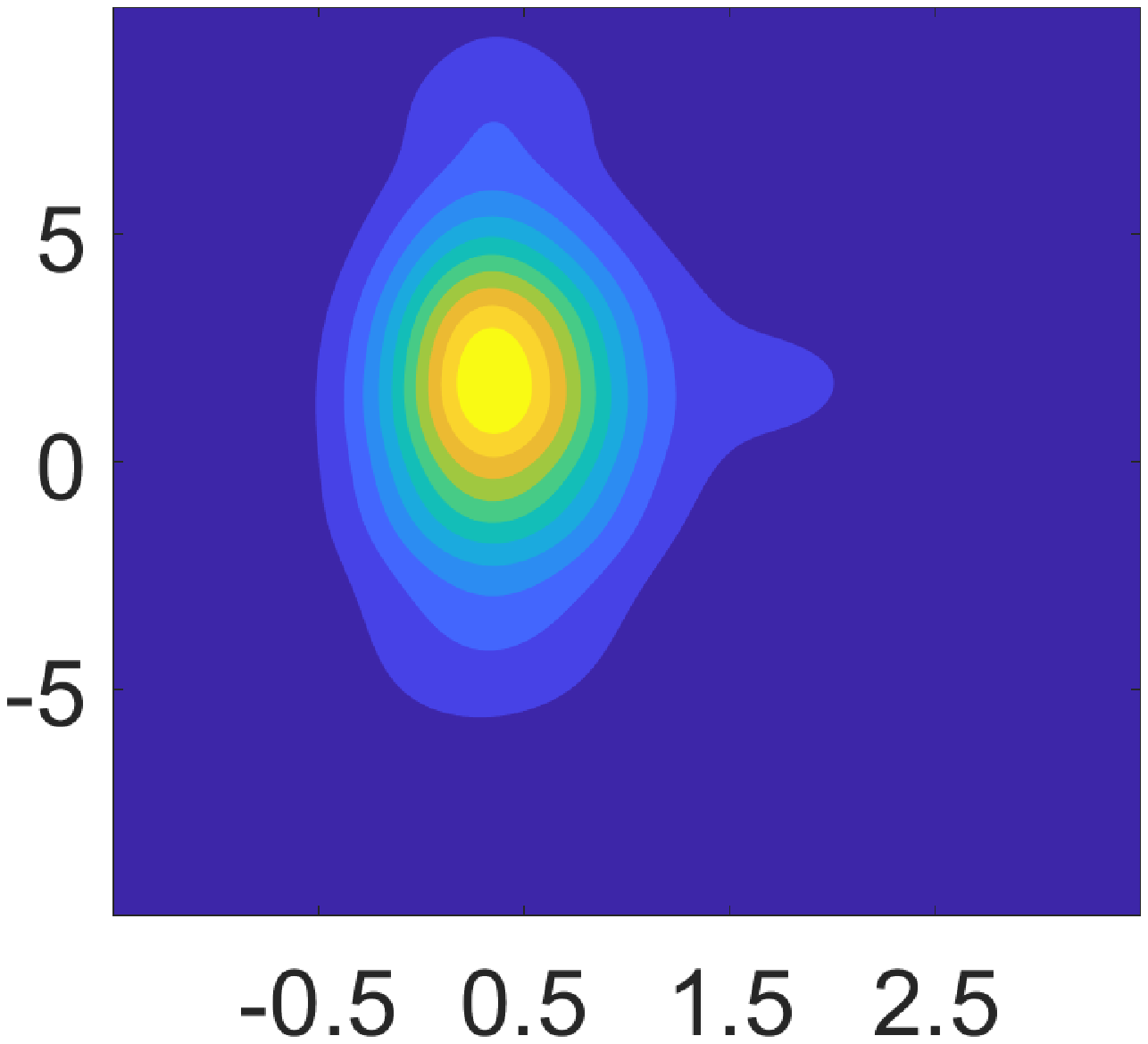}     		
			\end{subfigure}
			\begin{subfigure}[t]{0.23\textwidth}
				\caption{\footnotesize $h=2$}
				\includegraphics[width=0.95\textwidth]{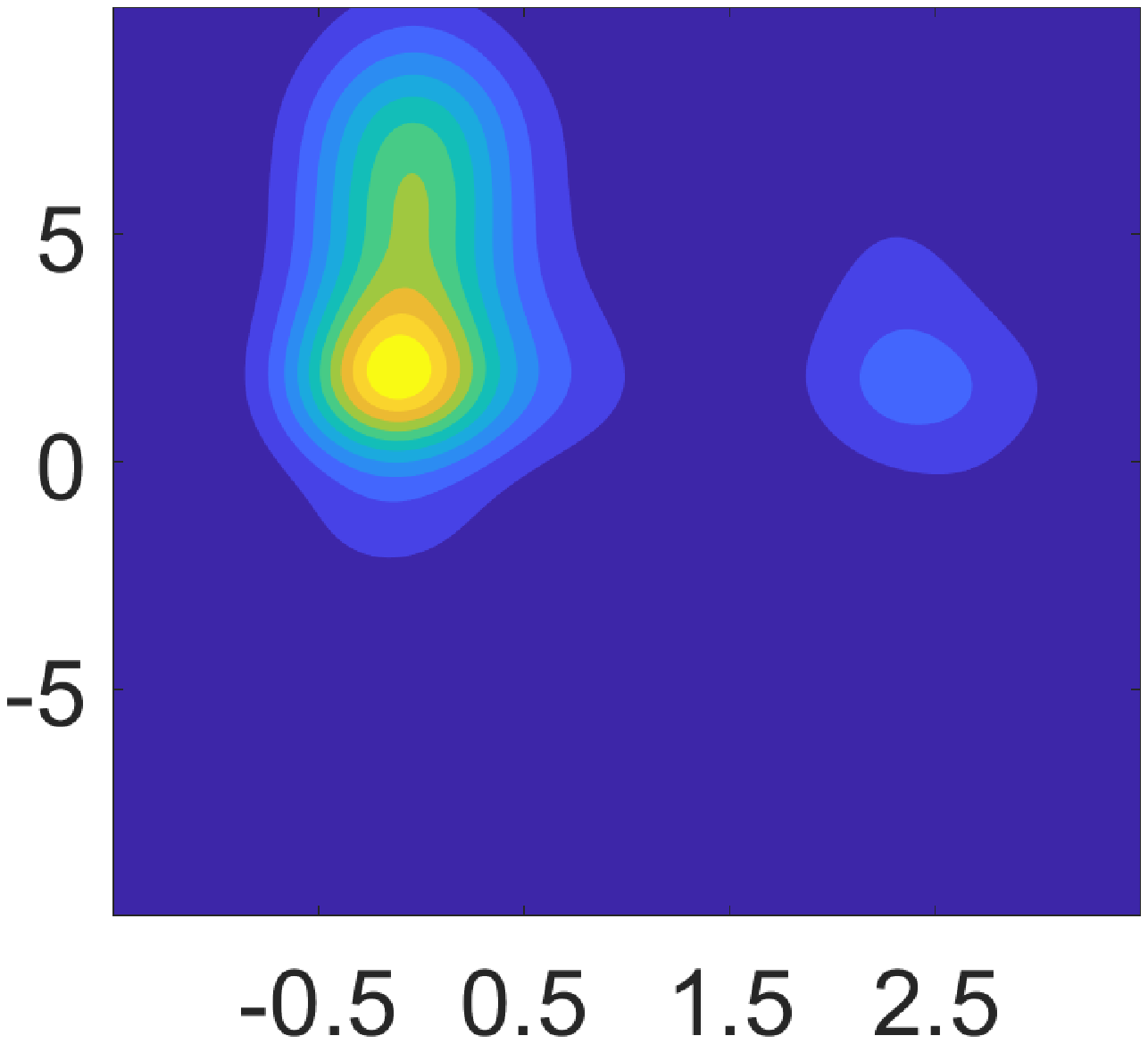}
			\end{subfigure}
			\begin{subfigure}[t]{0.23\textwidth}
				\caption{\footnotesize $h=3$}
				\includegraphics[width=0.95\textwidth]{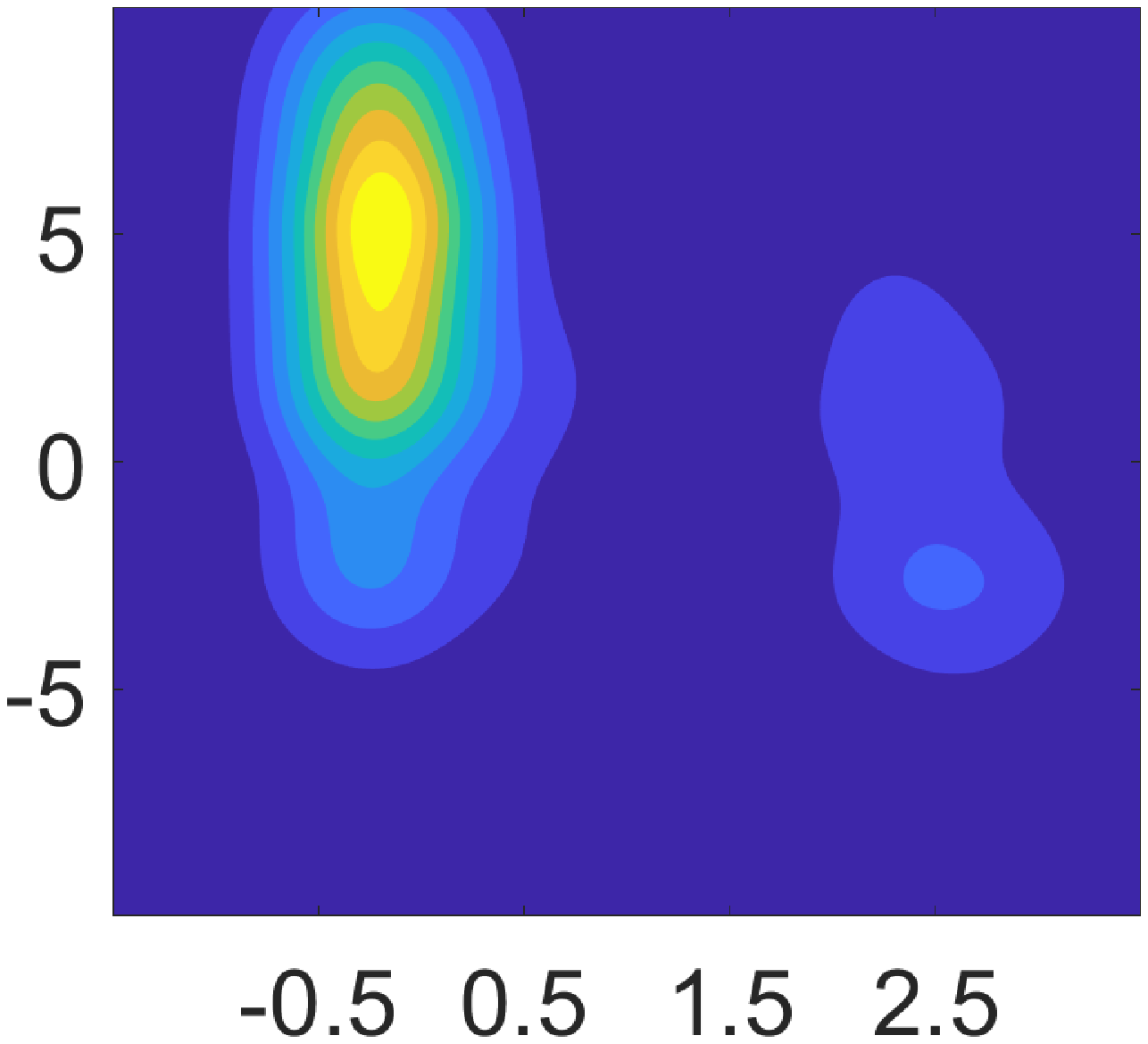}
			\end{subfigure}
			\begin{subfigure}[t]{0.23\textwidth}
				\caption{\footnotesize $h=4$}	
				\includegraphics[width=0.95\textwidth]{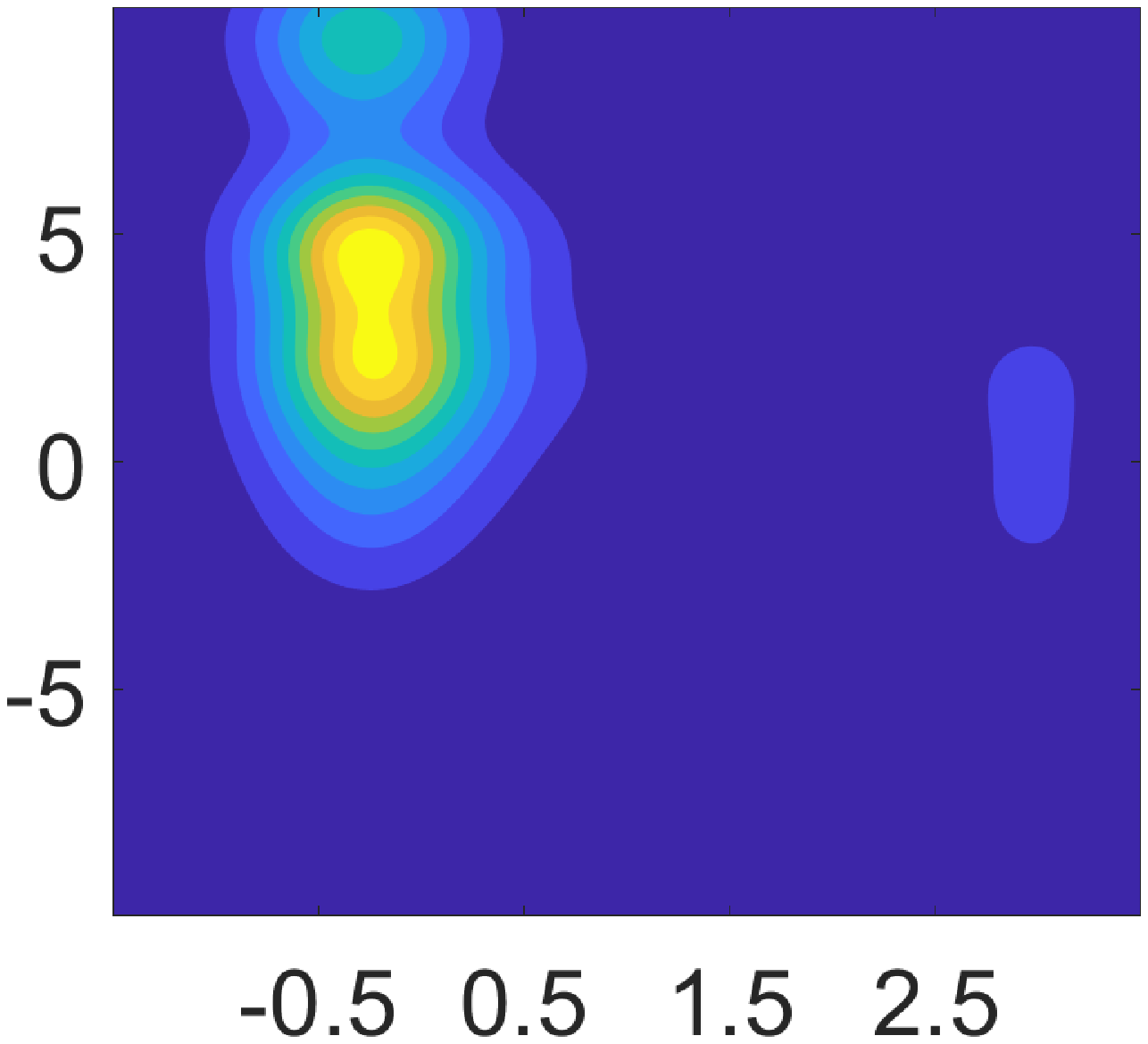}
			\end{subfigure} 
			\vspace{3mm}
		\end{minipage}
		\begin{minipage}[t]{0.95\textwidth}
			\centering \footnotesize (e) $Y_{t+h}$ given $(Y_t, Z_t)$ for $t$=2009:Q3 \\		
			\begin{subfigure}[t]{0.23\textwidth}
				\caption{\footnotesize $h=1$}	
				\includegraphics[width=0.95\textwidth]{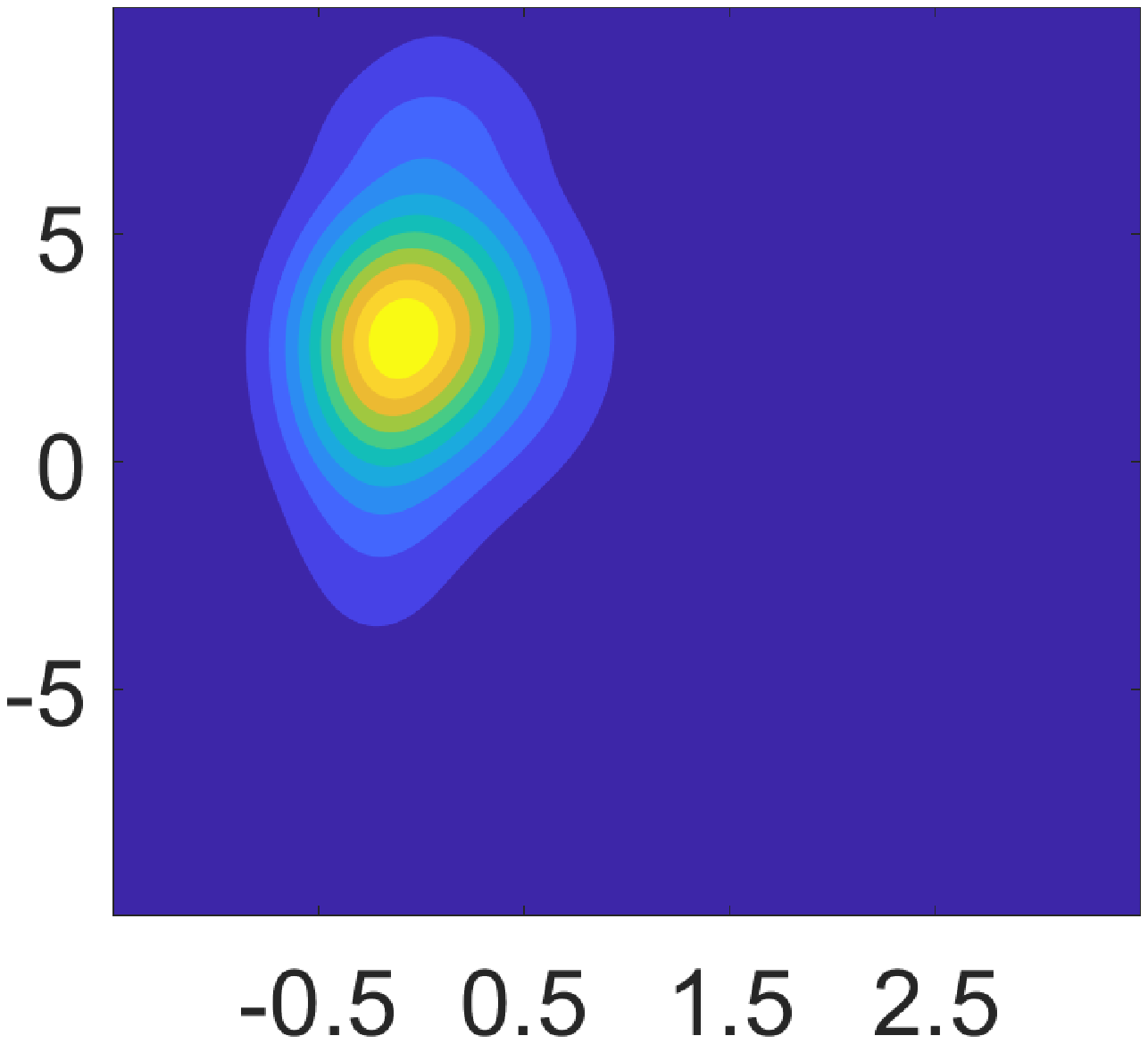}     		
			\end{subfigure}
			\begin{subfigure}[t]{0.23\textwidth}
				\caption{\footnotesize $h=2$}
				\includegraphics[width=0.95\textwidth]{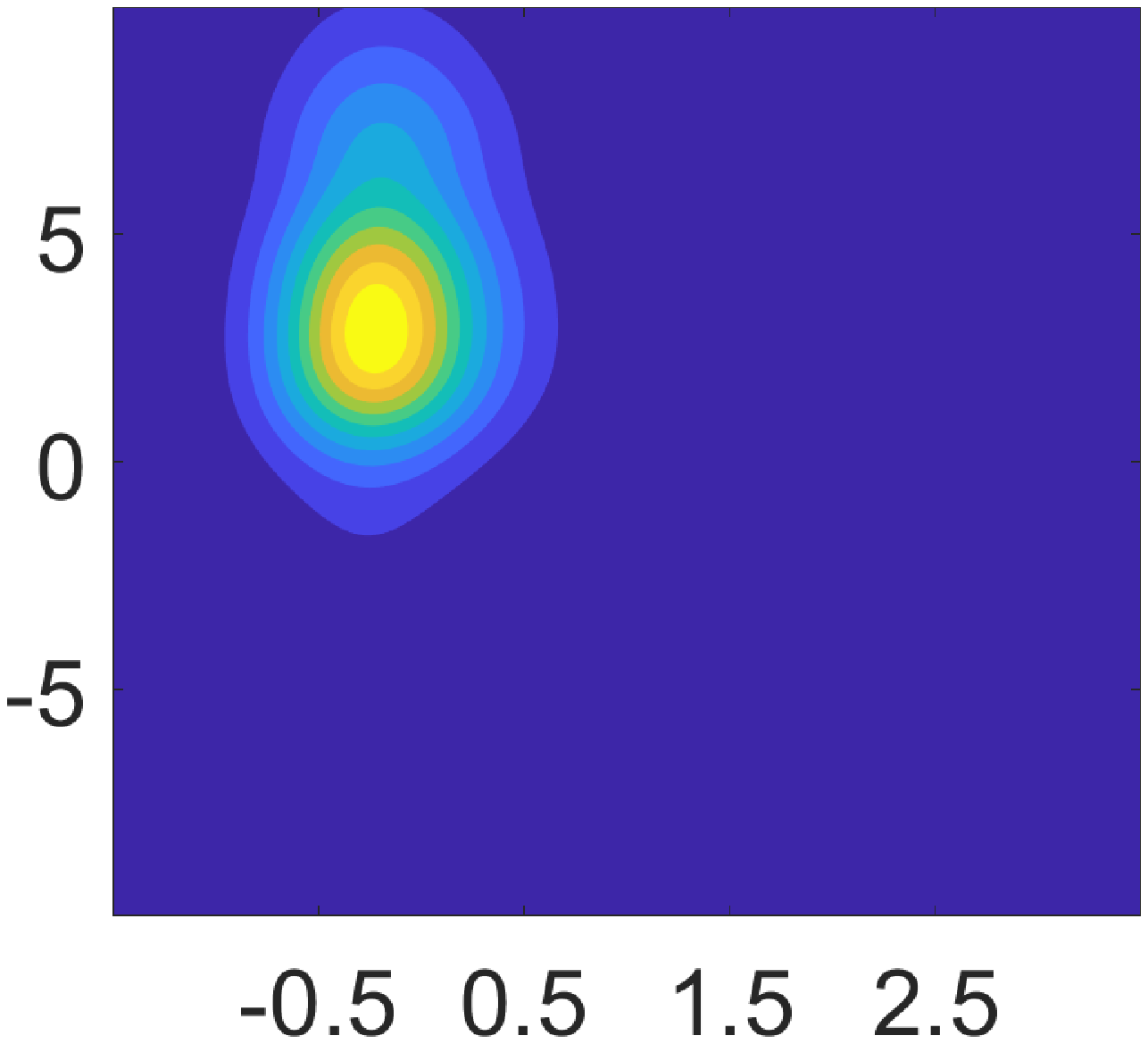}   
			\end{subfigure}
			\begin{subfigure}[t]{0.23\textwidth}
				\caption{\footnotesize $h=3$}
				\includegraphics[width=0.95\textwidth]{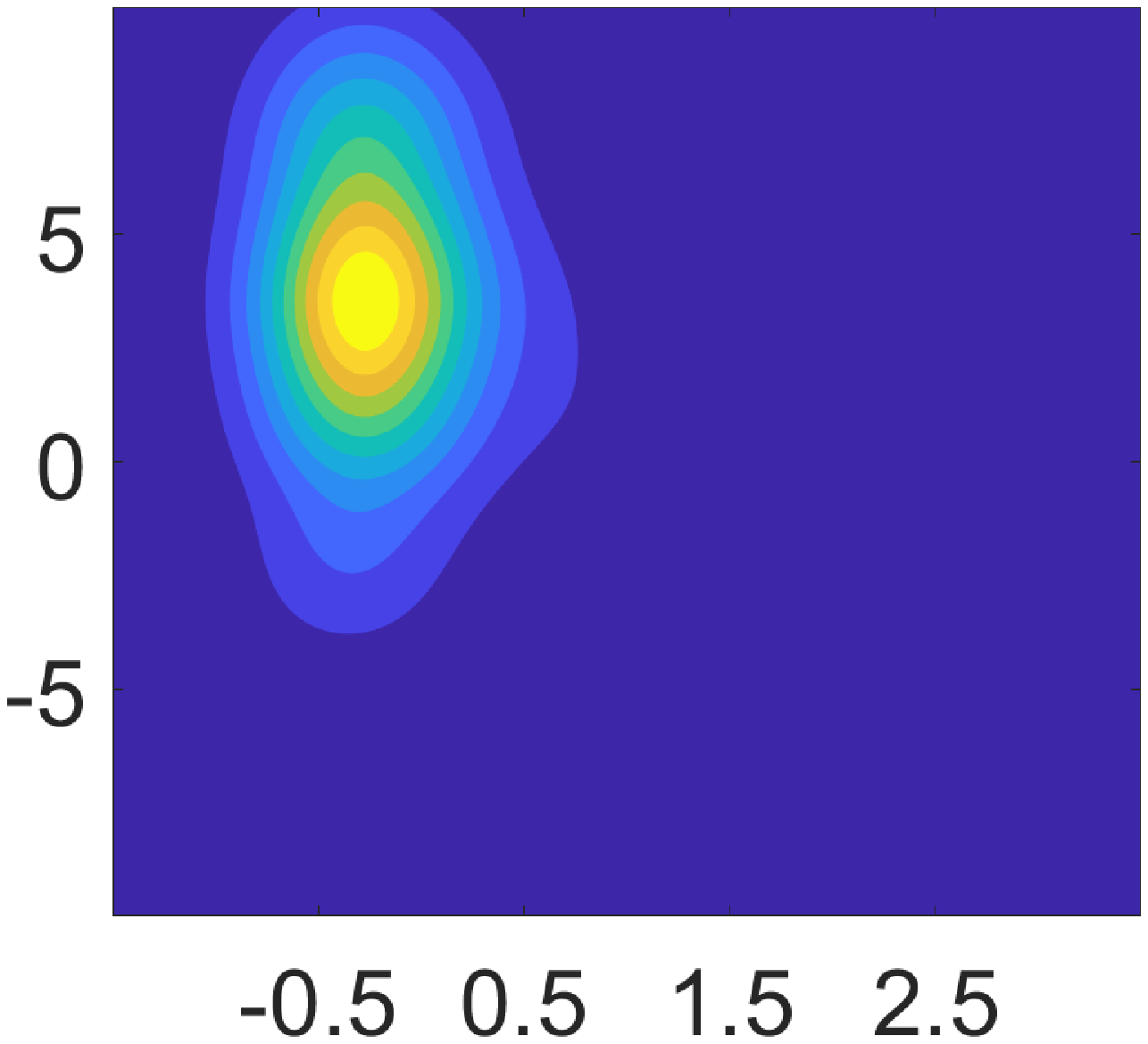}  			
			\end{subfigure}
			\begin{subfigure}[t]{0.23\textwidth}
				\caption{\footnotesize $h=4$}
				\includegraphics[width=0.95\textwidth]{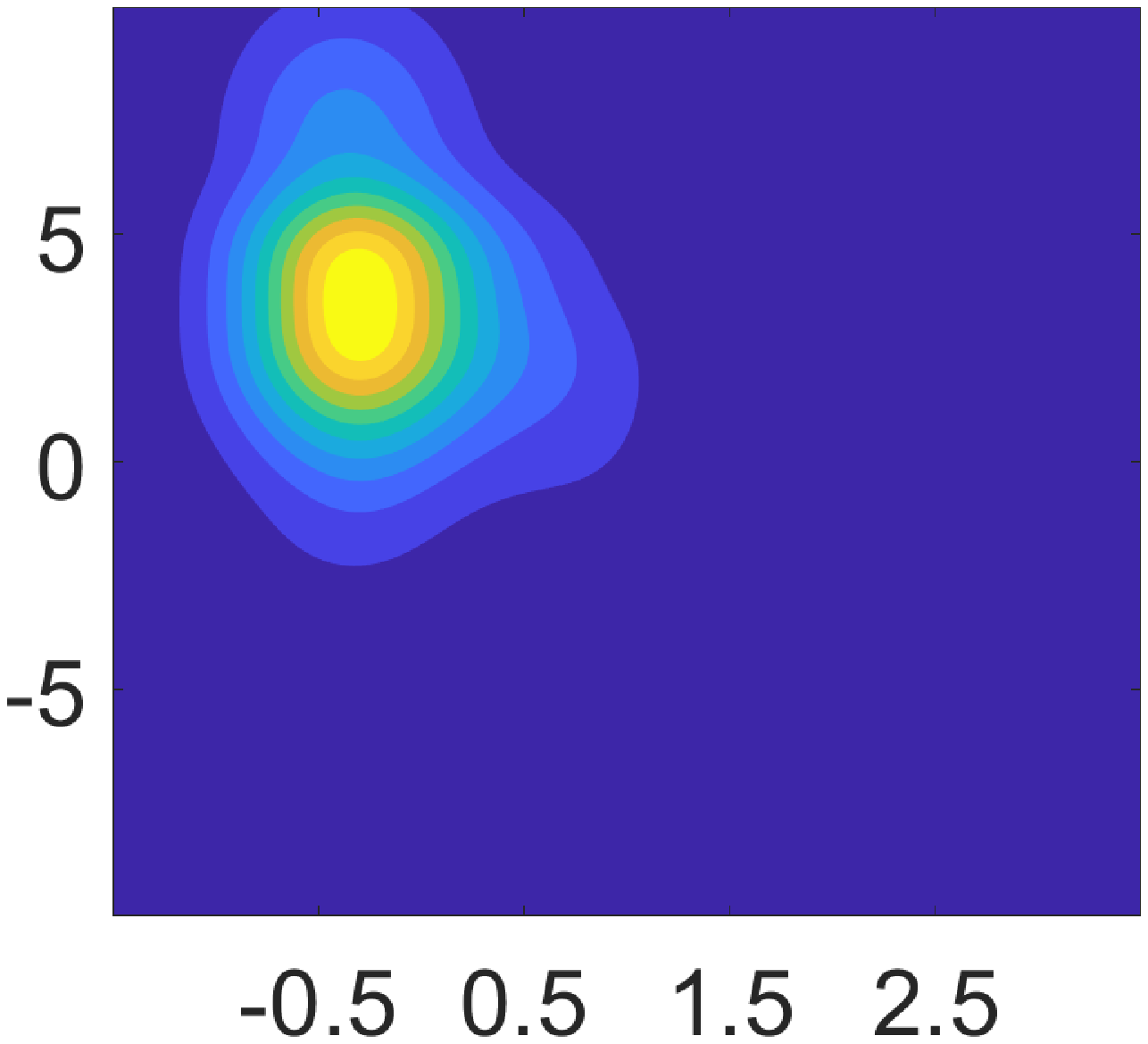}  			
			\end{subfigure} 
			\vspace{1.5mm}
		\end{minipage}
		\begin{minipage}{.9\linewidth} 
			\linespread{1}\footnotesize
			\textit{Notes}: Refer to Figure \ref{fig: contour-plots-2}. These plots are constructed based on the same model in Section \ref{sec: empirical} but with $Y_{1t}$ representing GDP and $Y_{2t}$ representing NFCI.
		\end{minipage}
	\end{figure}

	\subsection{Comparison with Kernel Regression}\label{sec: appendix-B3}
	Recently, \cite{adrian2021multimodality} attempted to construct the joint distribution of the financial conditions and real GDP growth by the kernel regression method. We compare our DR approach with the kernel regression on estimating the forecasting distributions.
	Following their work, we parameterize the bandwidths as being proportional to the in-sample unconditional standard deviation of the corresponding variable with the single proportionally constant $c$.
		
	\begin{figure}[H]
		\captionsetup[subfigure]{aboveskip=-2pt,belowskip=0pt}
		\centering
		\caption{Empirical CDF of the Out-of-sample PITs\label{fig: PITs-KD}}
		\begin{subfigure}[b]{0.45\textwidth}
			\centering
			\caption{GDP: One quarter ahead}     		
			\includegraphics[width=0.95\textwidth]{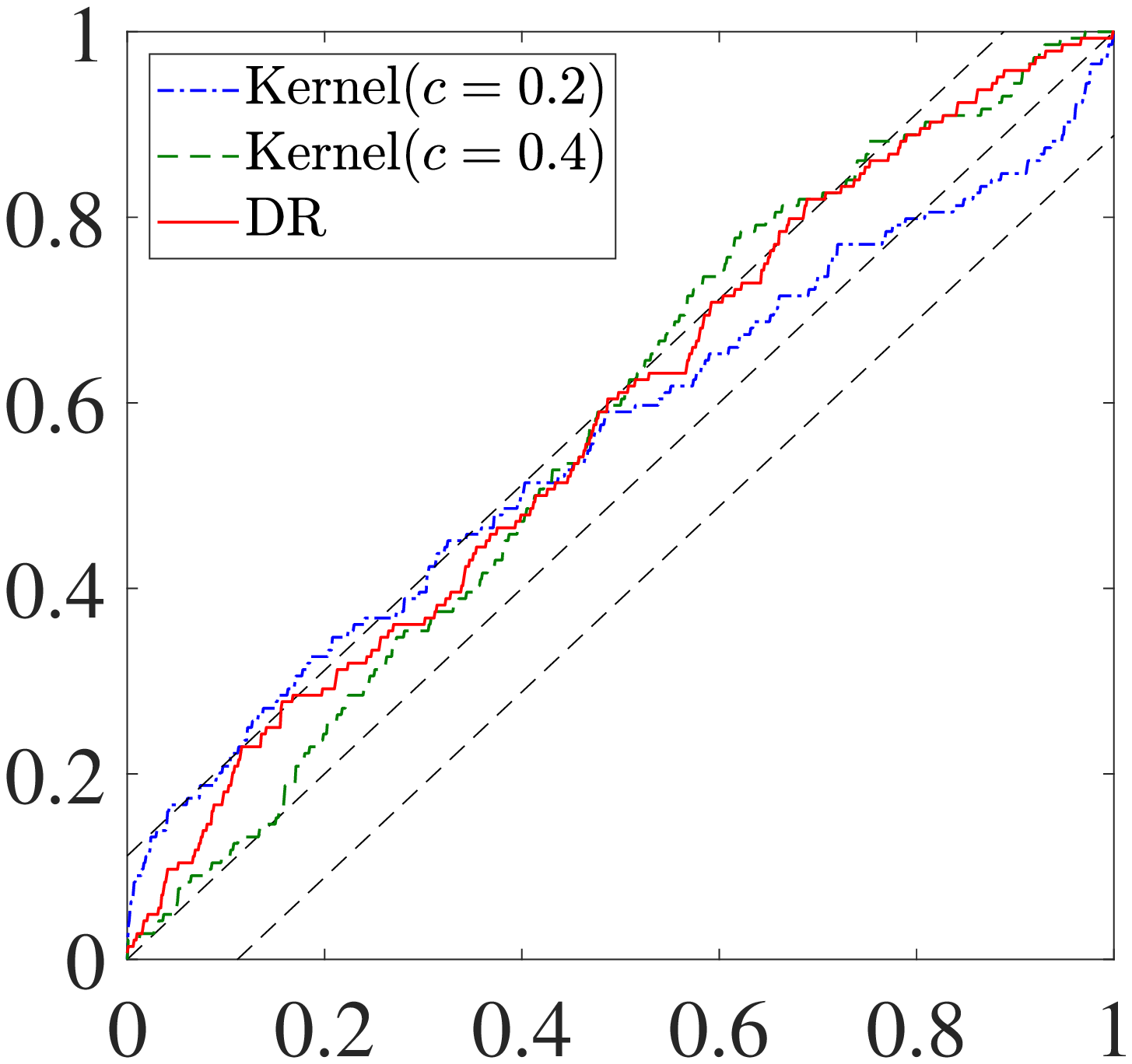}      		
			\label{GDP-h1}
		\end{subfigure}
		\begin{subfigure}[b]{0.45\textwidth}
			\centering
			\caption{GDP: One year ahead}		
			\includegraphics[width=0.95\textwidth]{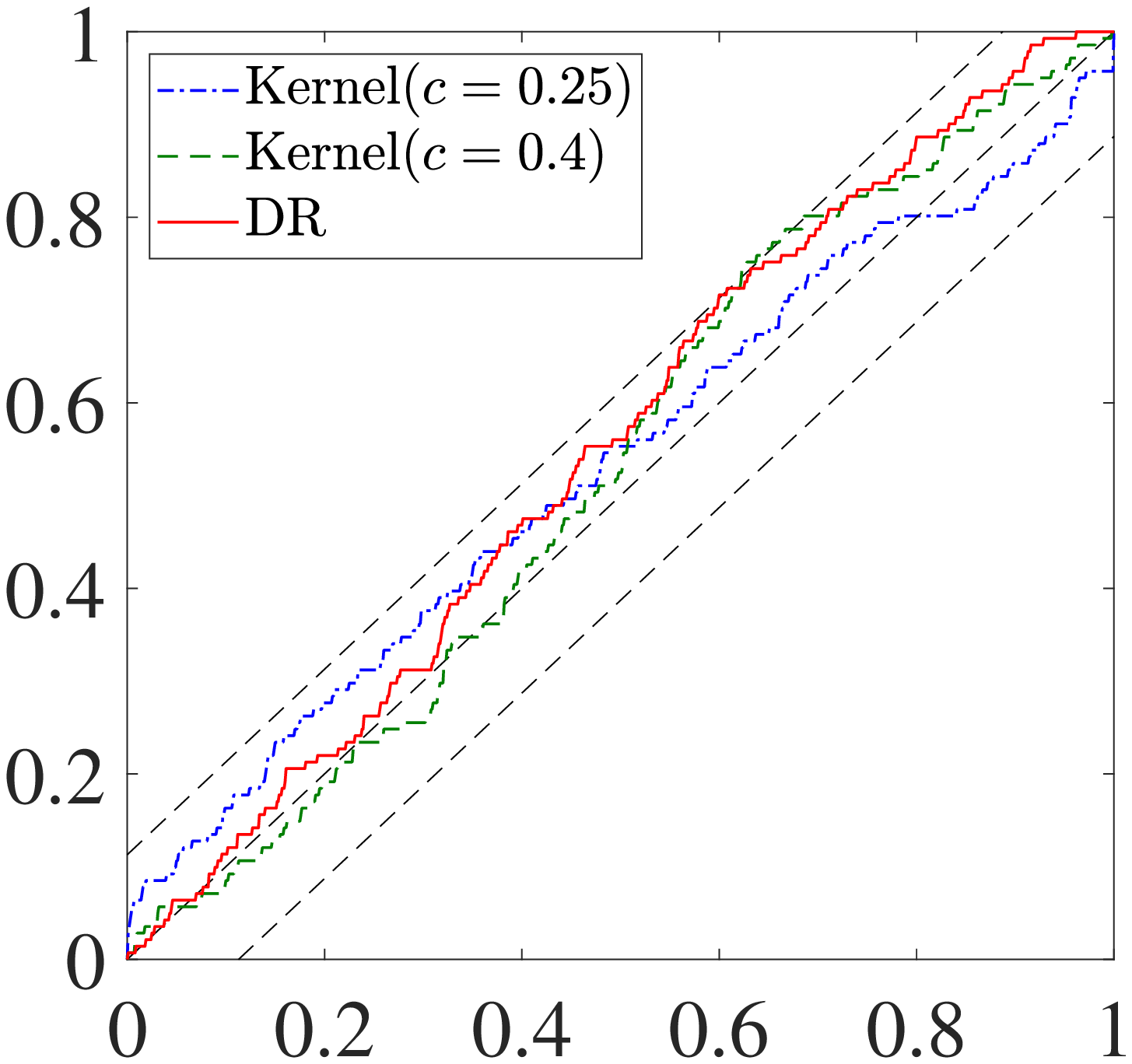}
			\label{GDP-h4}
		\end{subfigure}
		\hfill
		\begin{subfigure}[b]{0.45\textwidth}
			\centering
			\caption{NFCI: One quarter ahead}		
			\includegraphics[width=0.95\textwidth]{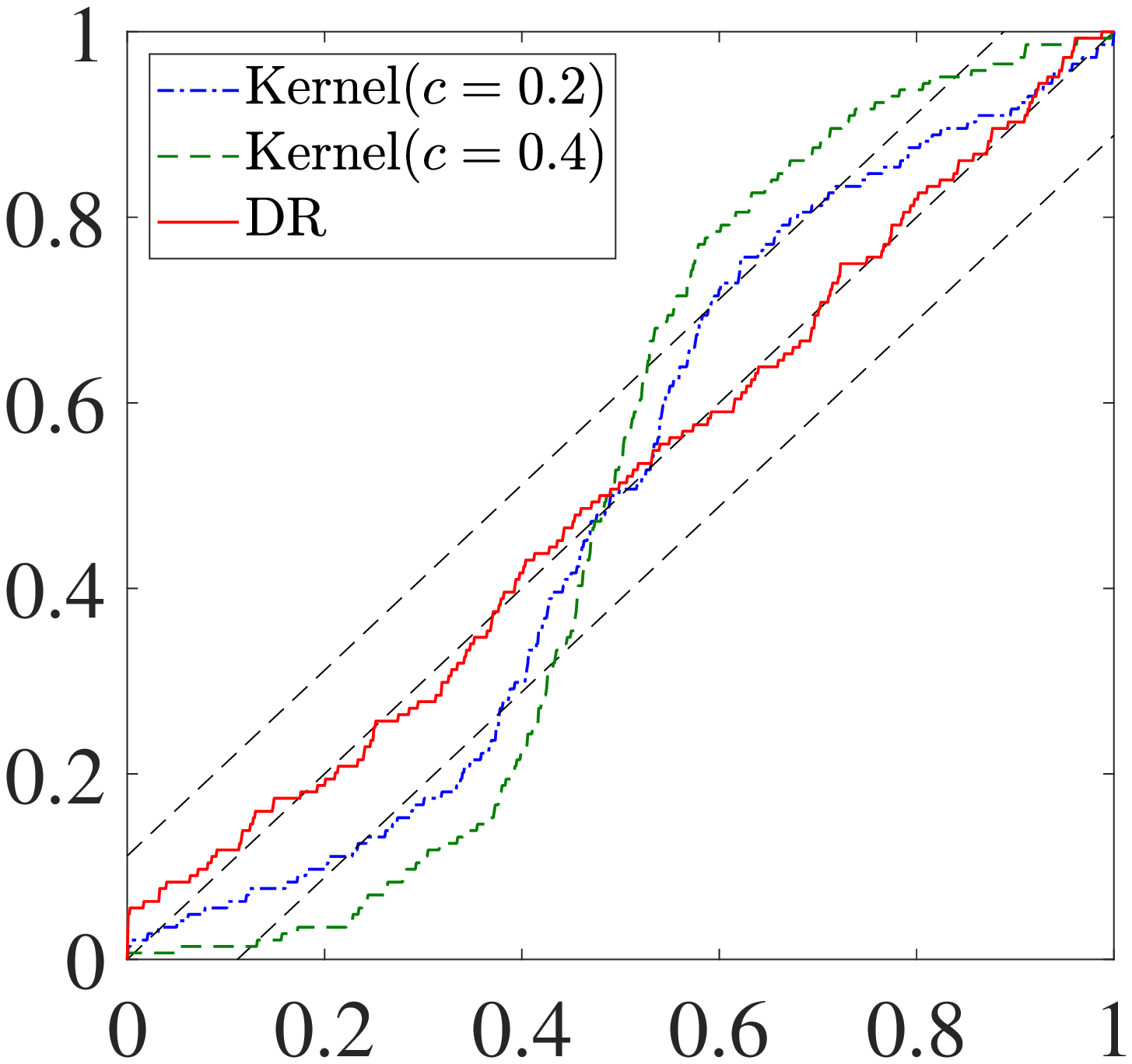}
			\label{NFCI-h1}
		\end{subfigure}          
		\begin{subfigure}[b]{0.45\textwidth}
			\centering
			\caption{NFCI: One year ahead}
			\includegraphics[width=0.95\textwidth]{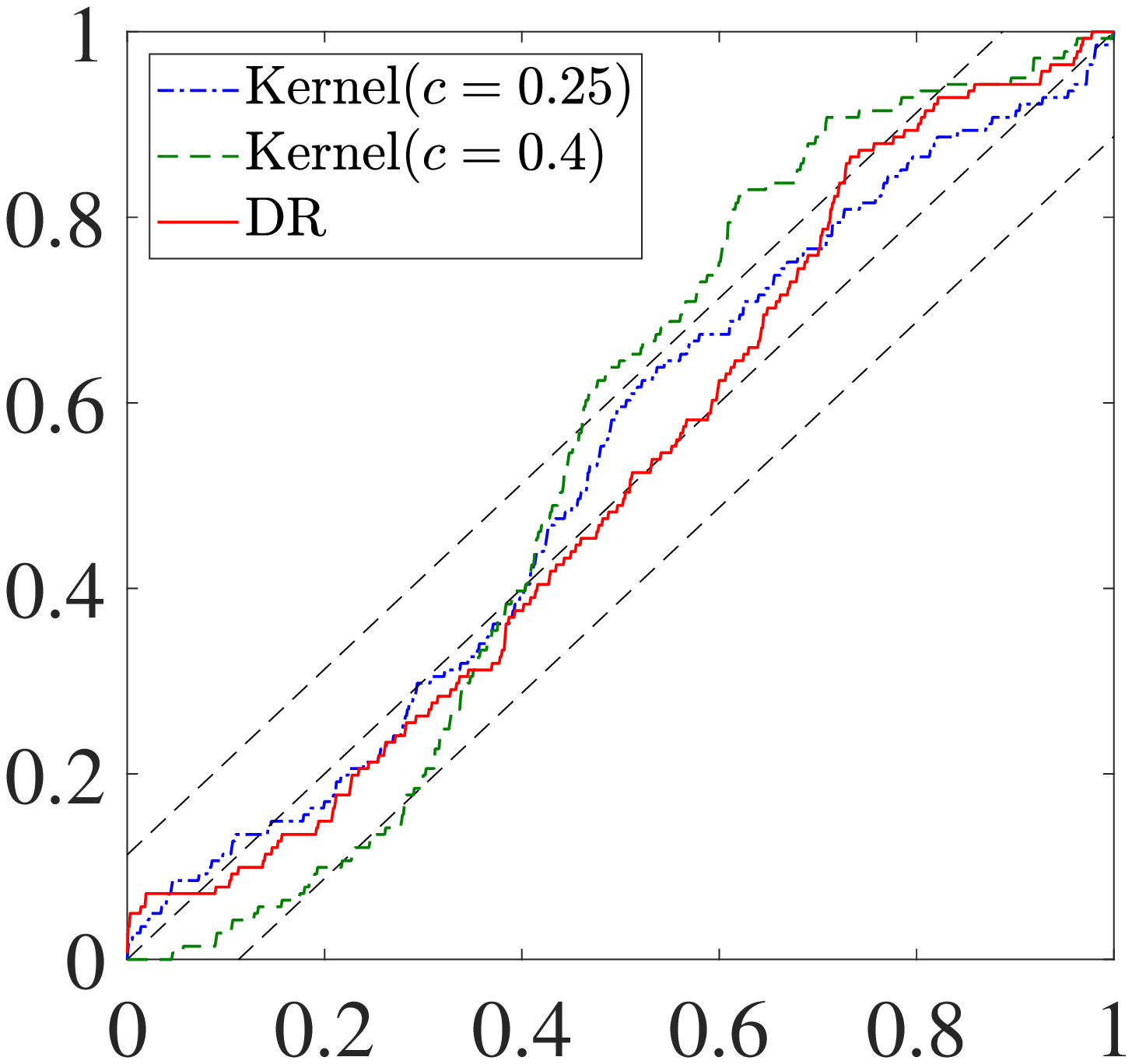}
			\label{NFCI-h4}
		\end{subfigure}
		\begin{minipage}{.8\linewidth} 
			\linespread{1}\footnotesize
			\textit{Notes}: This figure reports the empirical CDF of the PITs by the DR approach (red), the Kernel Regression approach with optimal multiperiod forecasing performance (green) and optimal model calibration (blue), plus the CDF of the PITs under the null hypothesis of correct calibration (the 45-degree line) and the 95\% confidence bands (dashed-line) of the \cite{rossi2019alternative} PITs test. 
		\end{minipage}
	\end{figure}
	Given a sequence of values of $c$, we figure out the one that maximizes the predictive accuracy of the resultant multiperiod conditional joint distribution (measured by log predictive score) and the one that optimizes the out-of-sample calibration of the model (measured by PITs).  We compare the PITs of kernel regression in these two cases with our approach in Figure \ref{fig: PITs-KD}. The results show that the kernel regression model can be a correct specification with appropriate bandwidths, while the bandwidths that maximize the predictive accuracy are likely to result in miss-specification.

	\subsection{Mixed Frequency Model}\label{sec: appendix-B4}
	We further explore the use of monthly financial conditions in the system considering that converting the monthly available financial conditions into quarterly series often results in information loss. 
	
	Spefically, we convert NFCI into monthly frequency by averaging weekly observations. We use three monthly NFCI data instead of the aggregate quarterly NFCI data used before to provide an updated view of the real GDP growth. A 4-dimensional mixed frequency model is developed by treating the three monthly NFCI series as separate observations within a quarter. We let $Y_{jt}, j=1,2,3$ be the NFCI of the $j$-th month in quarter $t$, and $Y_{4t}$ be the real GDP growth in this model. As in Section 4, we consider $Z_{t}=(Y_{t-1}^{\top},Y_{t-2}^{\top})^{\top}$ as the covariates to characterise the joint conditional distribution $F_{Y_{t}|Z_{t}}$ and $F_{Y_{t+h}|Y_{t}, Z_{t}}$. With such model specification, there are 8 conditional variables for the joint distribution. It is known that kernel-based methods suffer from the so-called `curse-of-dimensionality' when applied to multivariate data. However, taking into account overparameterization and collinearity among the covariates, the DR approach allows us to adapt popular regularization techniques easily to enhance the prediction accuracy. In this application, we apply the lasso regression when we estimate the model (\ref{model-DR}). 
	
\begin{figure}[H]
		\captionsetup[subfigure]{aboveskip=-3pt,belowskip=0pt}
		\centering
		\caption{Empirical CDF of the Out-of-sample PITs\label{fig: PITs-M}}
		\begin{subfigure}[b]{0.45\textwidth}
			\centering
			\caption{1st Month NFCI}     		
			\includegraphics[width=0.95\textwidth]{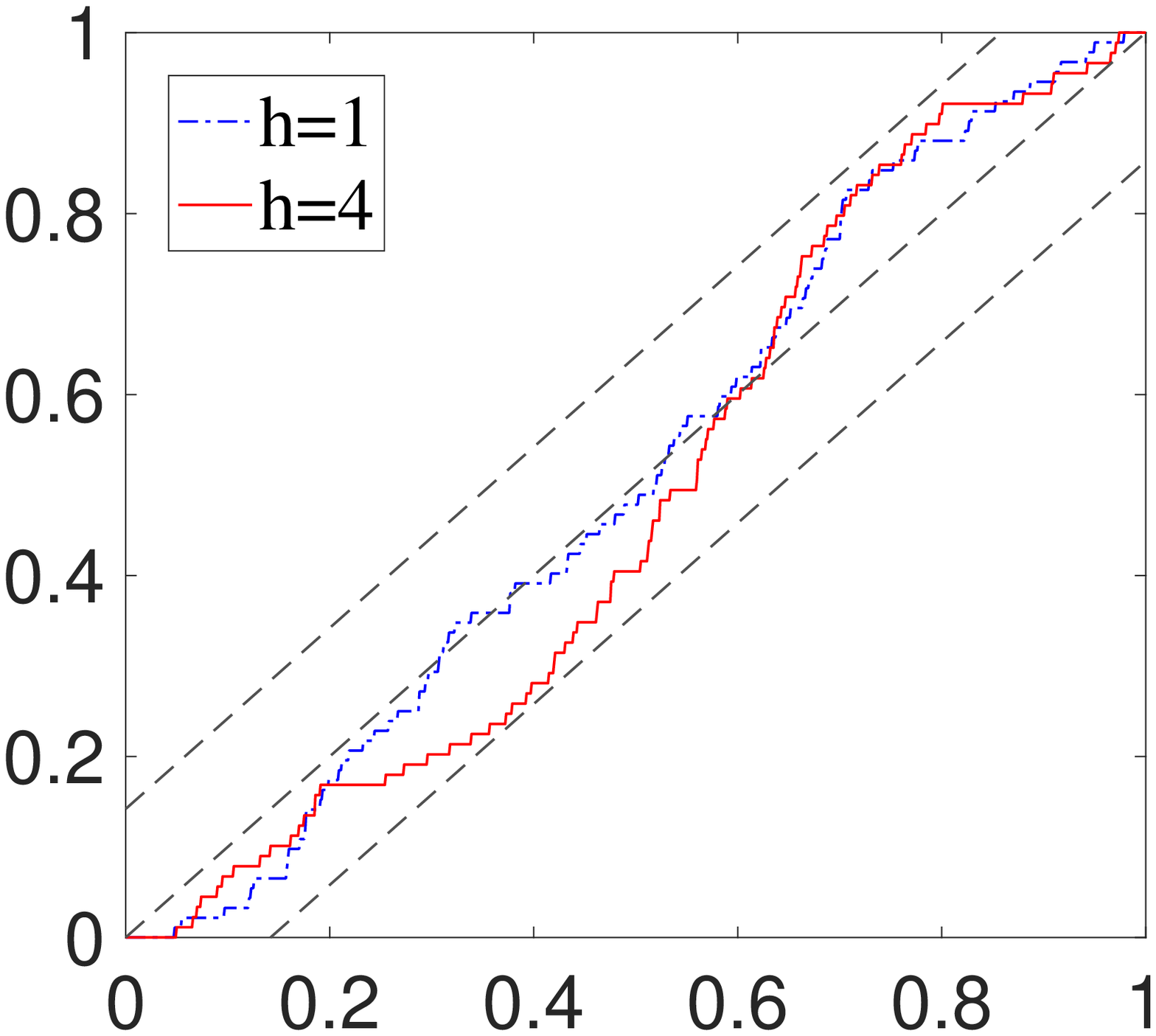}      		
			\label{M-lag2-NFCI1}
		\end{subfigure}
		\begin{subfigure}[b]{0.45\textwidth}
			\centering
			\caption{2nd Month NFCI}		
			\includegraphics[width=0.95\textwidth]{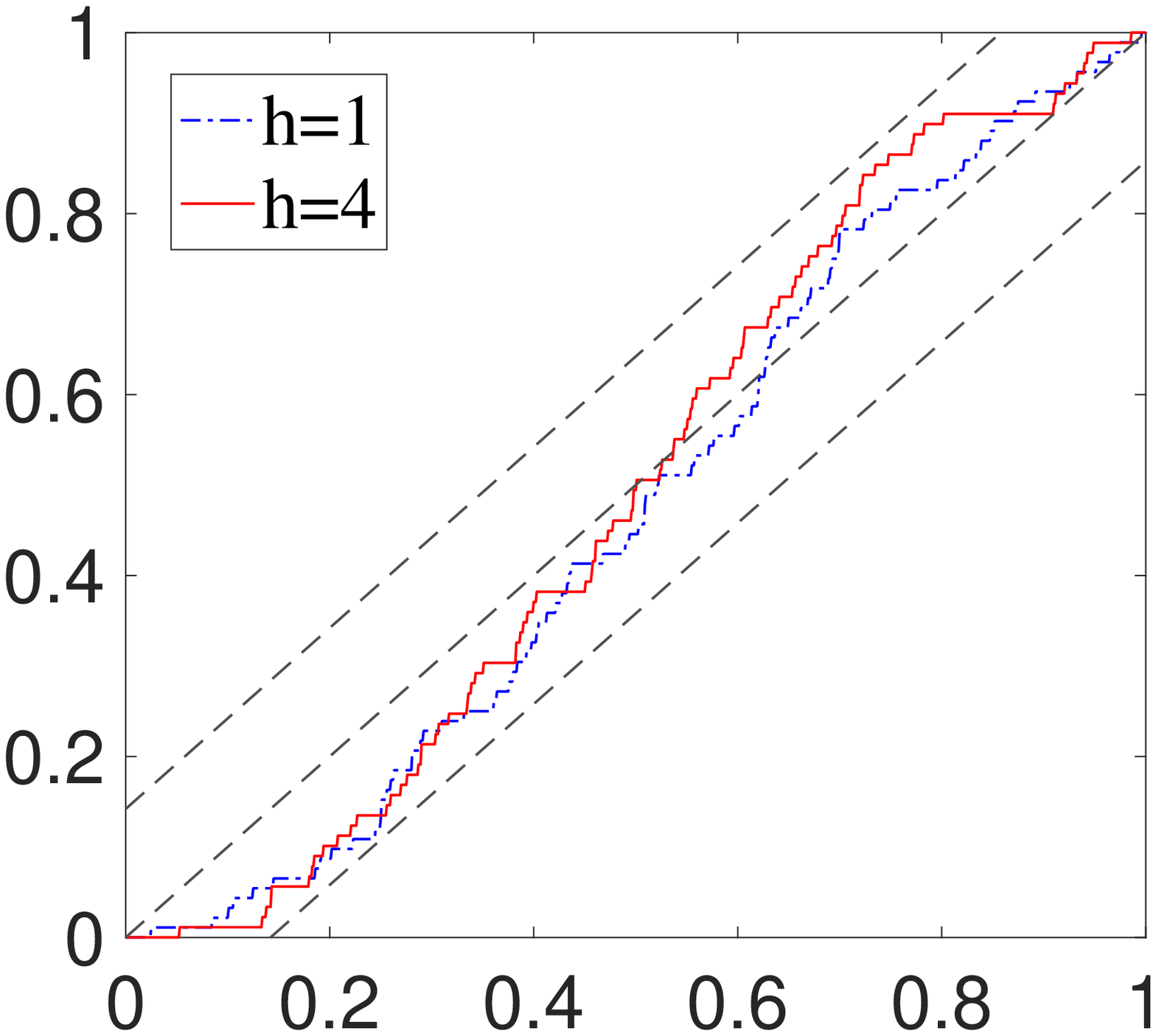}
			\label{M-lag2-NFCI2}
		\end{subfigure}
		\hfill
		\begin{subfigure}[b]{0.45\textwidth}
			\centering
			\caption{3rd Month NFCI}		
			\includegraphics[width=0.95\textwidth]{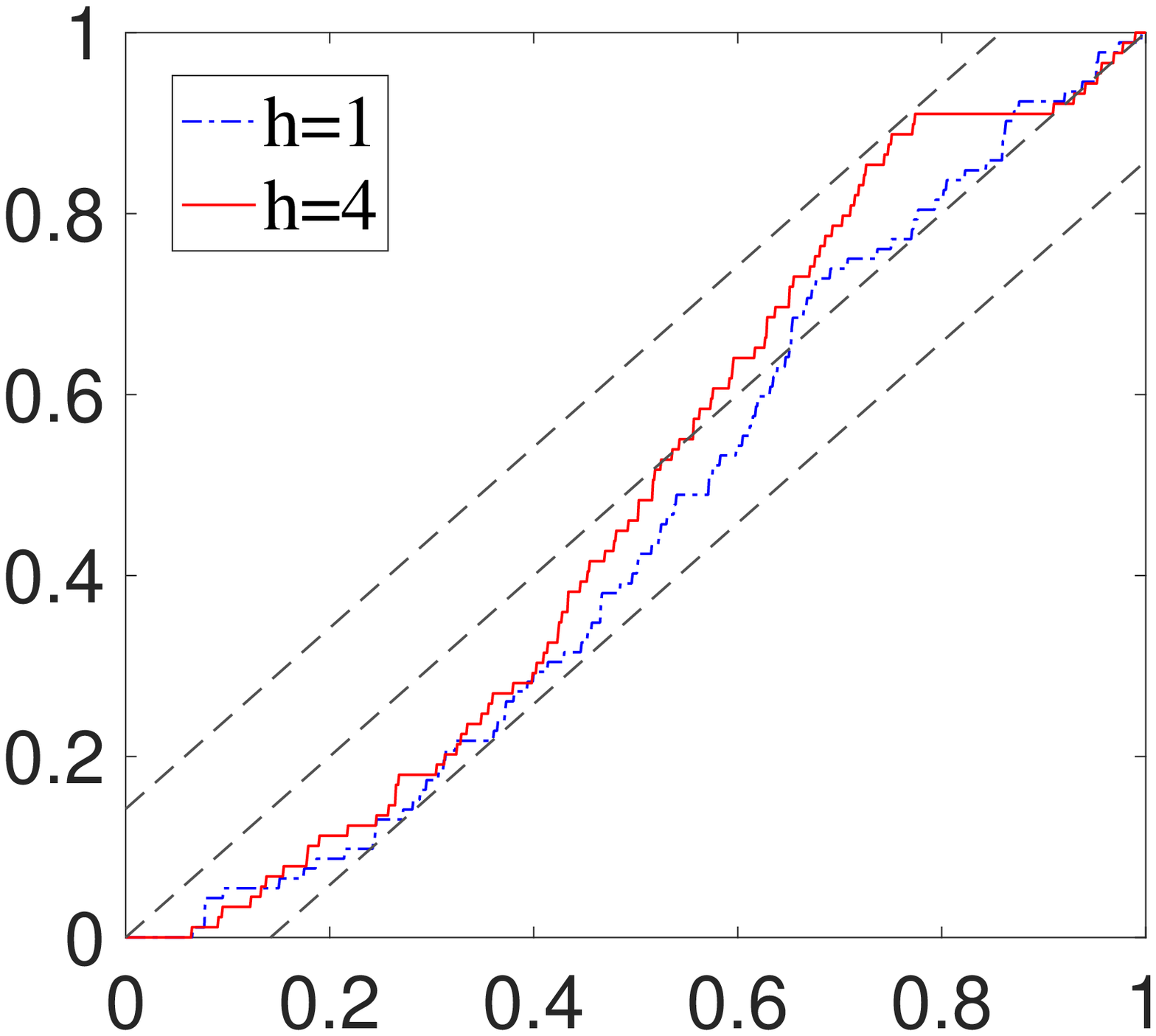}
			\label{M-lag2-NFCI3}
		\end{subfigure}          
		\begin{subfigure}[b]{0.45\textwidth}
			\centering
			\caption{Real GDP Growth}
			\includegraphics[width=0.95\textwidth]{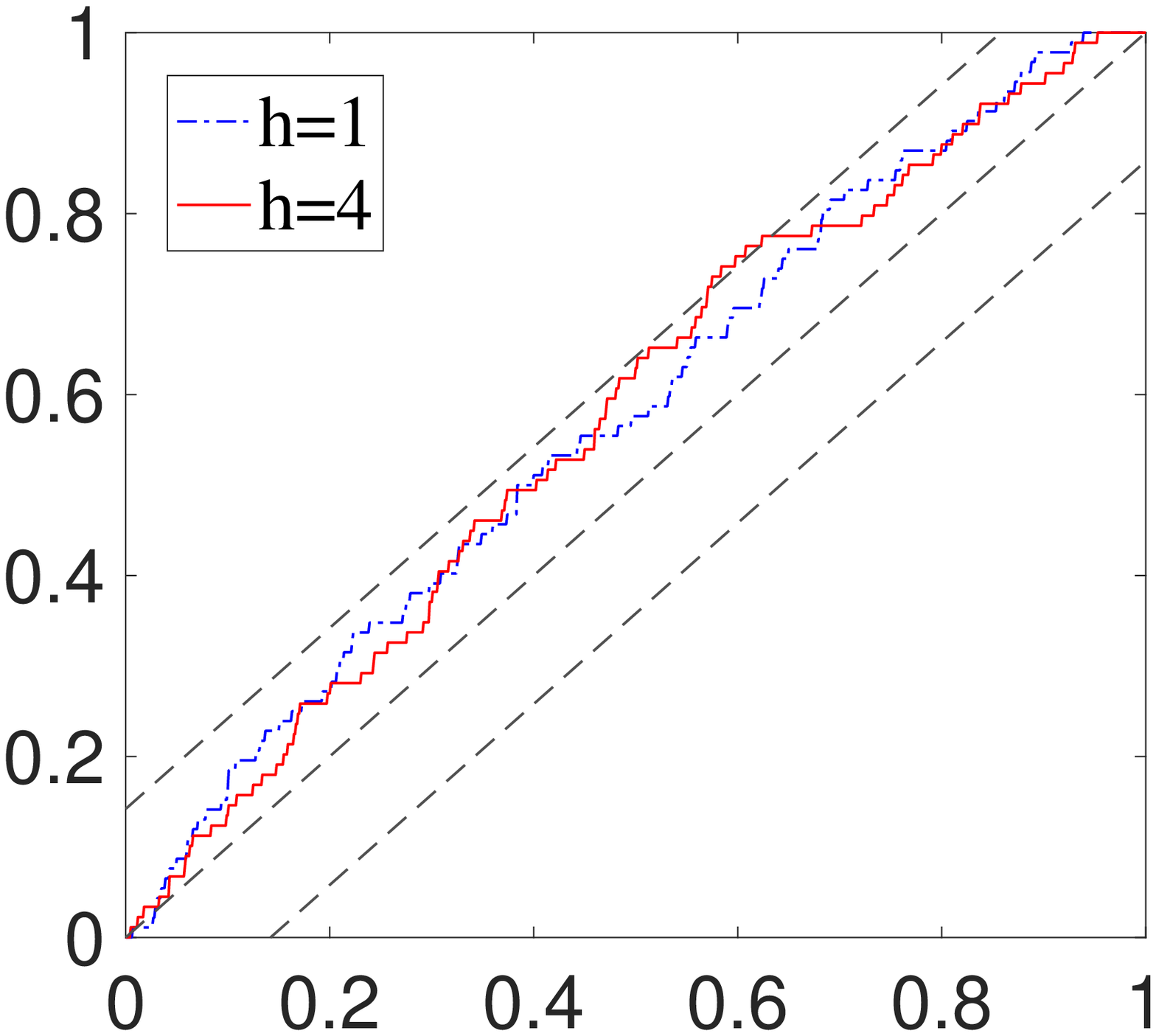}
			\label{M-lag2-GDP}
		\end{subfigure}
		\begin{minipage}{.8\linewidth} 
			\linespread{1}\footnotesize
			\textit{Notes}: This figure reports the empirical CDF of the PITs for the three monthly financial conditions and the real GDP growth. In each panel specific to different variables, the empirical CDF of the PITs for both one-quarter-ahead (red) and one-year-ahead (blue), plus the CDF of the PITs under the null hypothesis of correct calibration (the 45-degree line) and the 95\% confidence bands (dashed-line) of the \cite{rossi2019alternative} PITs test are plotted.
		\end{minipage}
	\end{figure}

	The out-of-sample calibration of the distribution forecasts is evaluated via the PITs test and presented in Figure \ref{fig: PITs-M}. The empirical CDF of PITs for all variables at both the one-quarter and one-year ahead horizons are all well within the 95\% confidence bands. Comparing the results of real GDP growth with Model 1 (see Figure \ref{fig: PITs-Q}) shows that using higher-frequency financial conditions data leads to improvement, especially for one quarter ahead of forecast.
	
\begin{figure}[H]
		\captionsetup[subfigure]{aboveskip=-3pt,belowskip=0pt, labelformat=empty}
		\centering
		\caption{Distributional Response to NFCI Impulse} \label{fig: M-NFCI-response}	
		\begin{minipage}[t]{\textwidth}	
			\centering  Distributions of $Y_{t+h}$ given $Z_{t}$ for $t$=2008:Q4\\$h=1$\\
			\begin{subfigure}[t]{0.24\textwidth}
				\centering 		
				\includegraphics[width=0.95\textwidth]{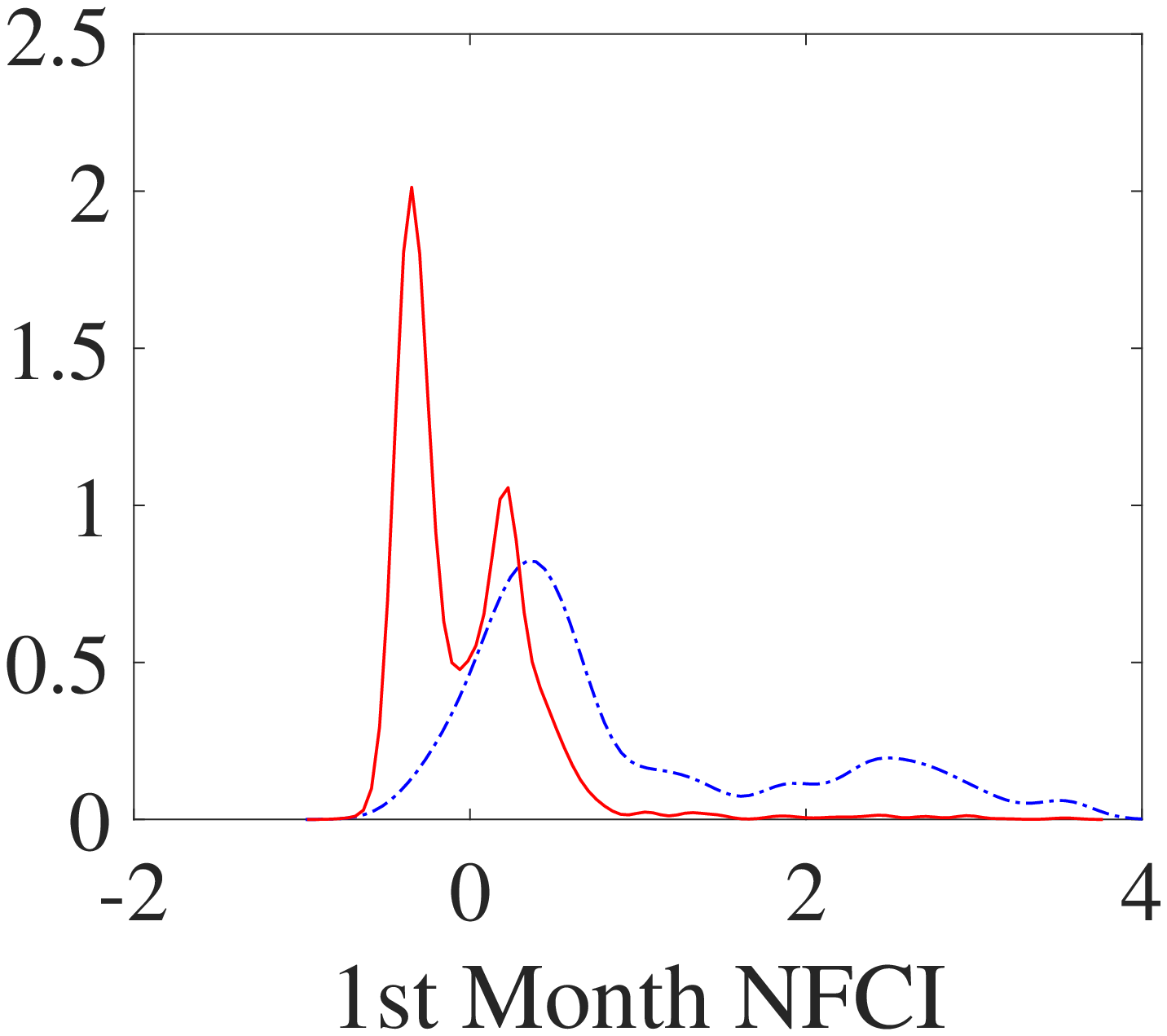}     		
			\end{subfigure}
			\begin{subfigure}[t]{0.24\textwidth}
				\centering	
				\includegraphics[width=0.95\textwidth]{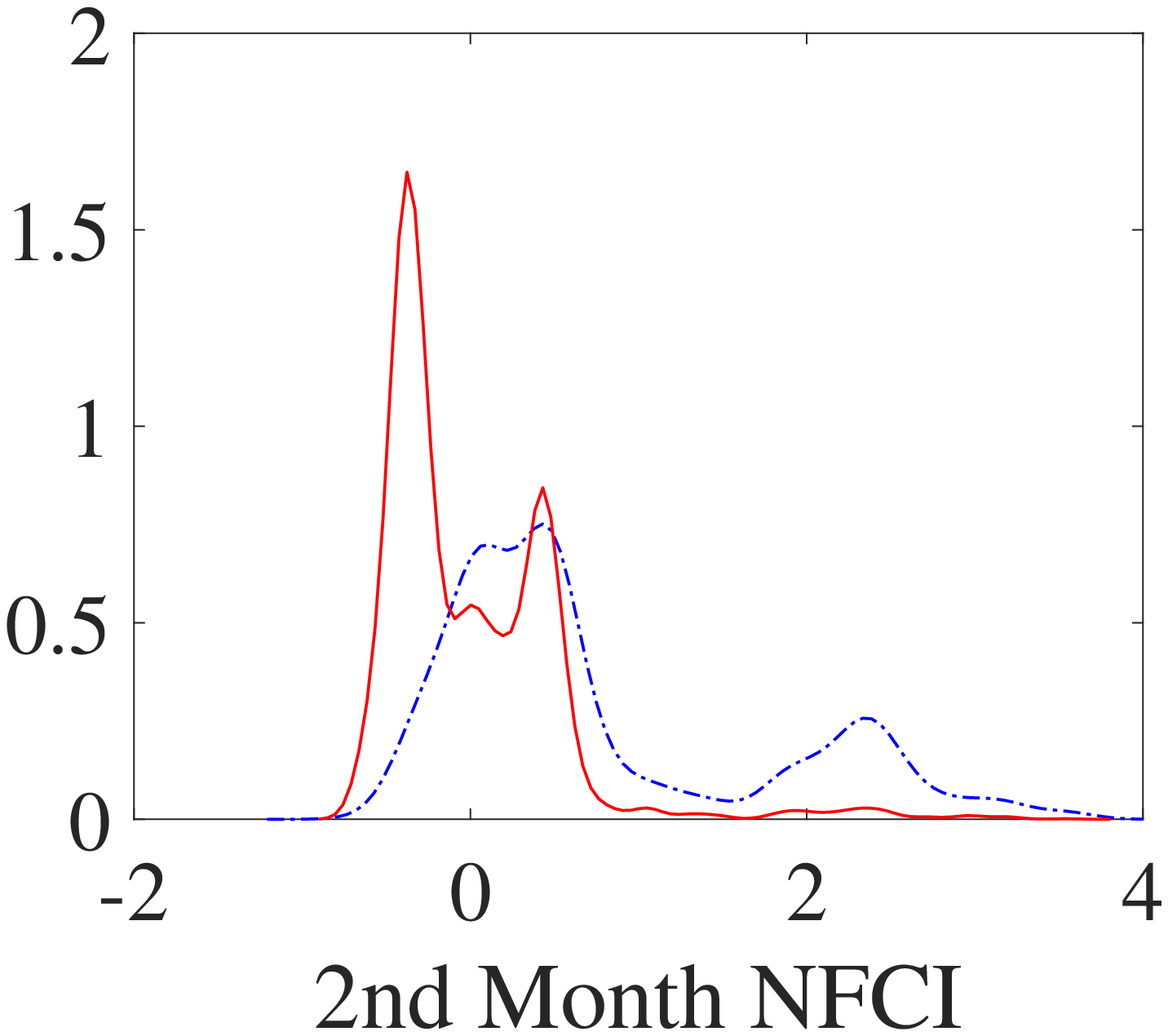}
			\end{subfigure}
			\begin{subfigure}[t]{0.24\textwidth}
				\centering	
				\includegraphics[width=0.95\textwidth]{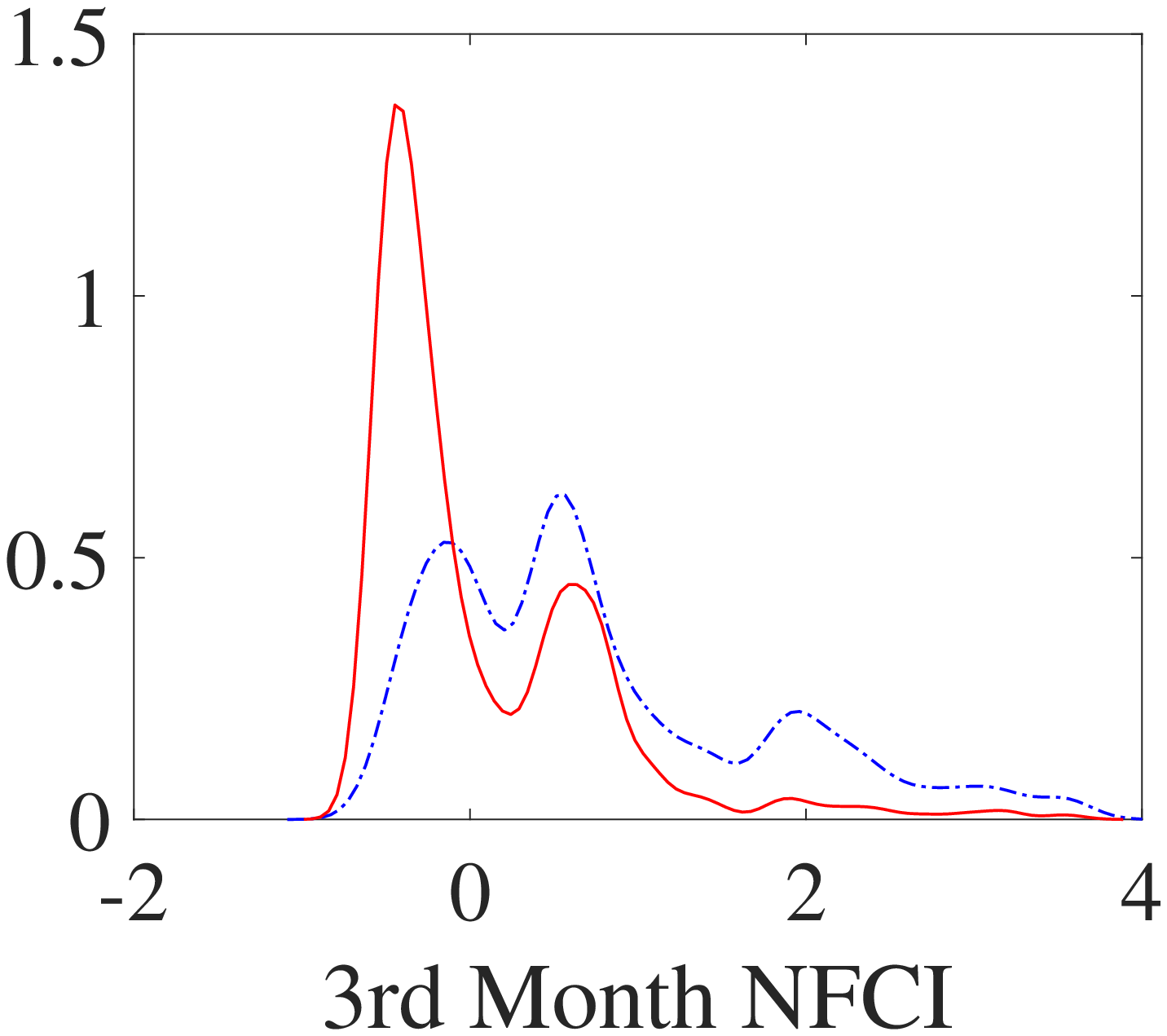}
			\end{subfigure}
			\begin{subfigure}[t]{0.24\textwidth}
				\centering	
				\includegraphics[width=0.95\textwidth]{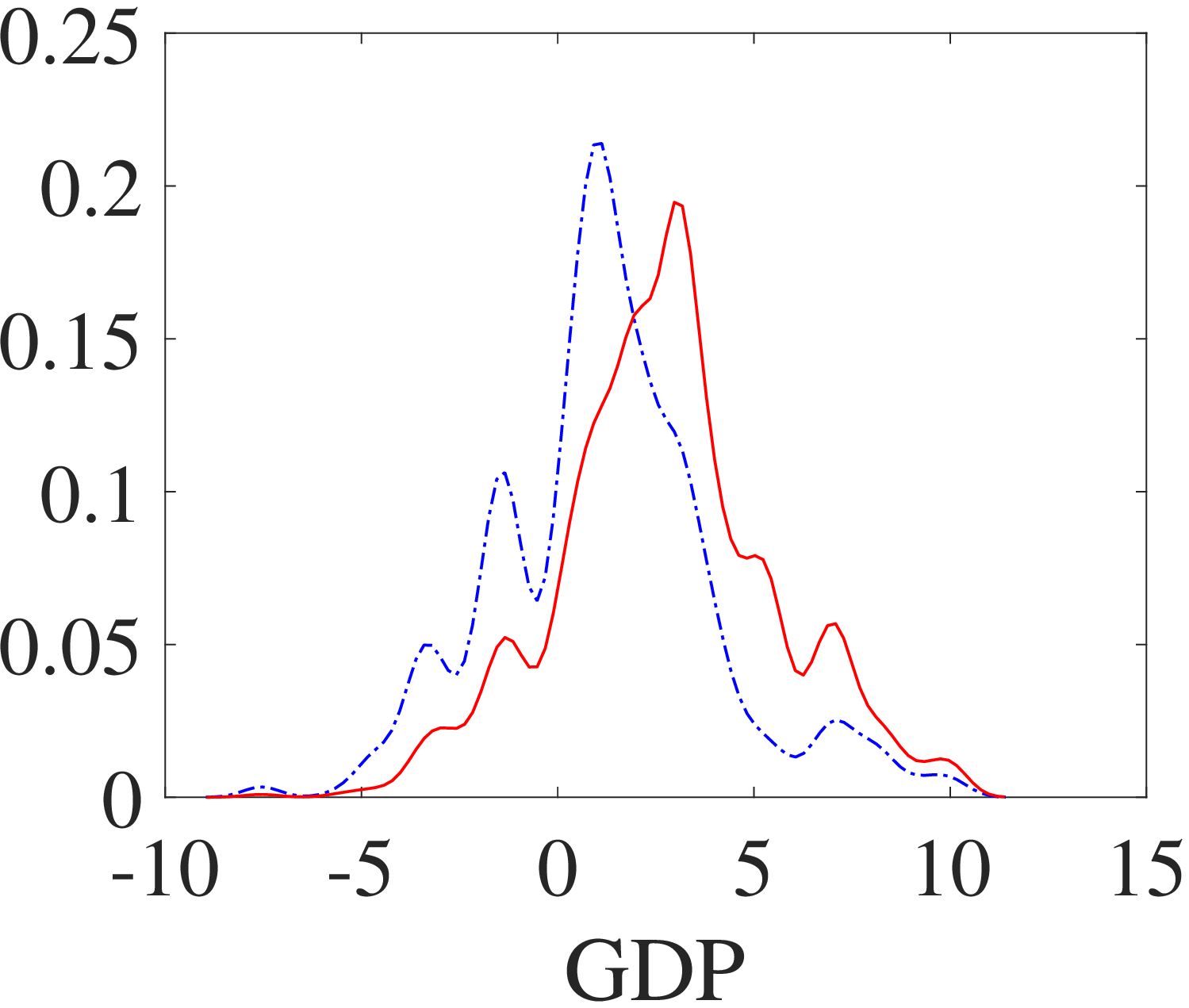}
			\end{subfigure}
			\vspace{0.2cm}
		\end{minipage}
		\begin{minipage}[t]{\textwidth}	
			\centering $h=2$\\
			\begin{subfigure}[t]{0.24\textwidth}
				\centering 		
				\includegraphics[width=0.95\textwidth]{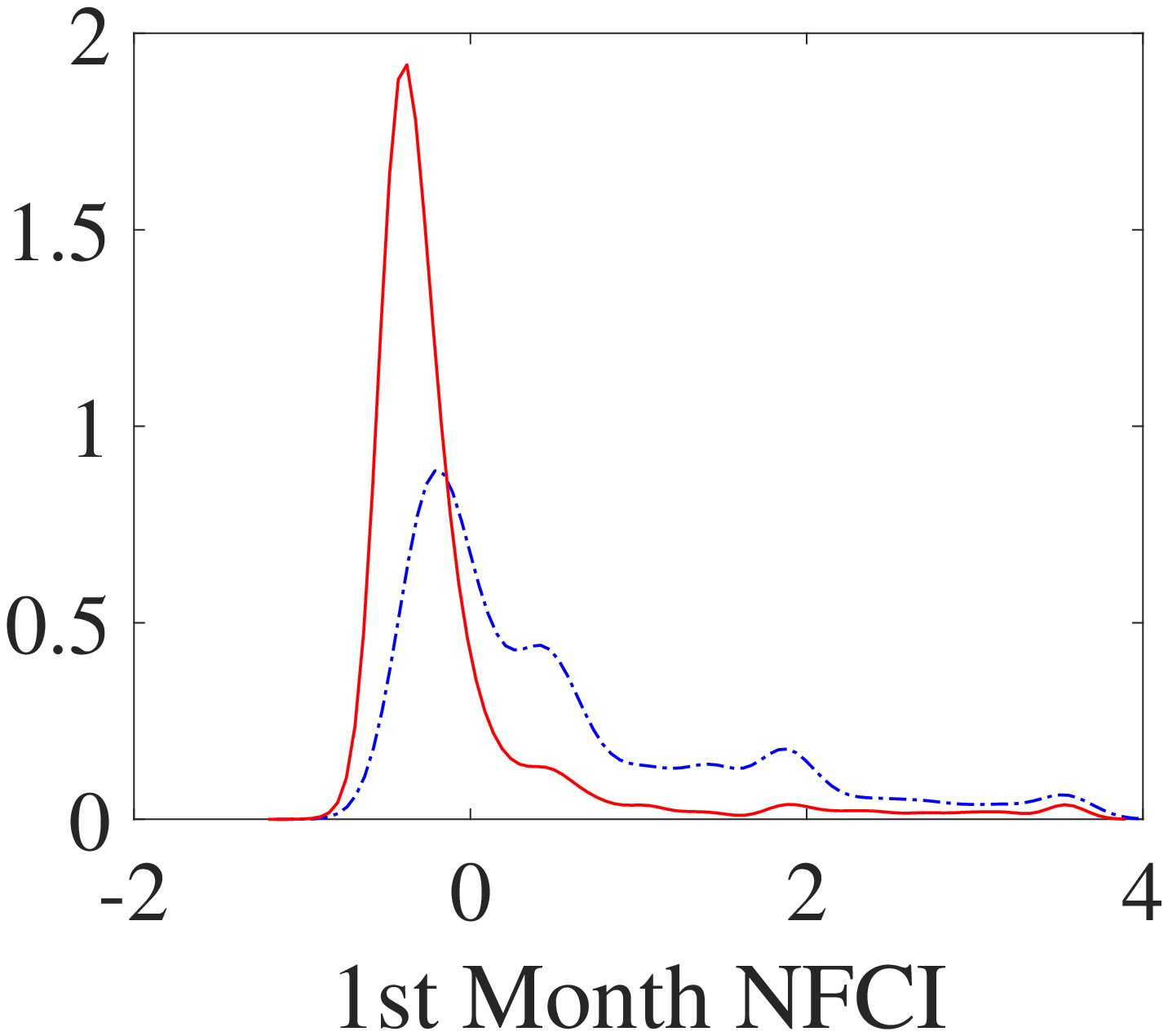}     		
			\end{subfigure}
			\begin{subfigure}[t]{0.24\textwidth}
				\centering	
				\includegraphics[width=0.95\textwidth]{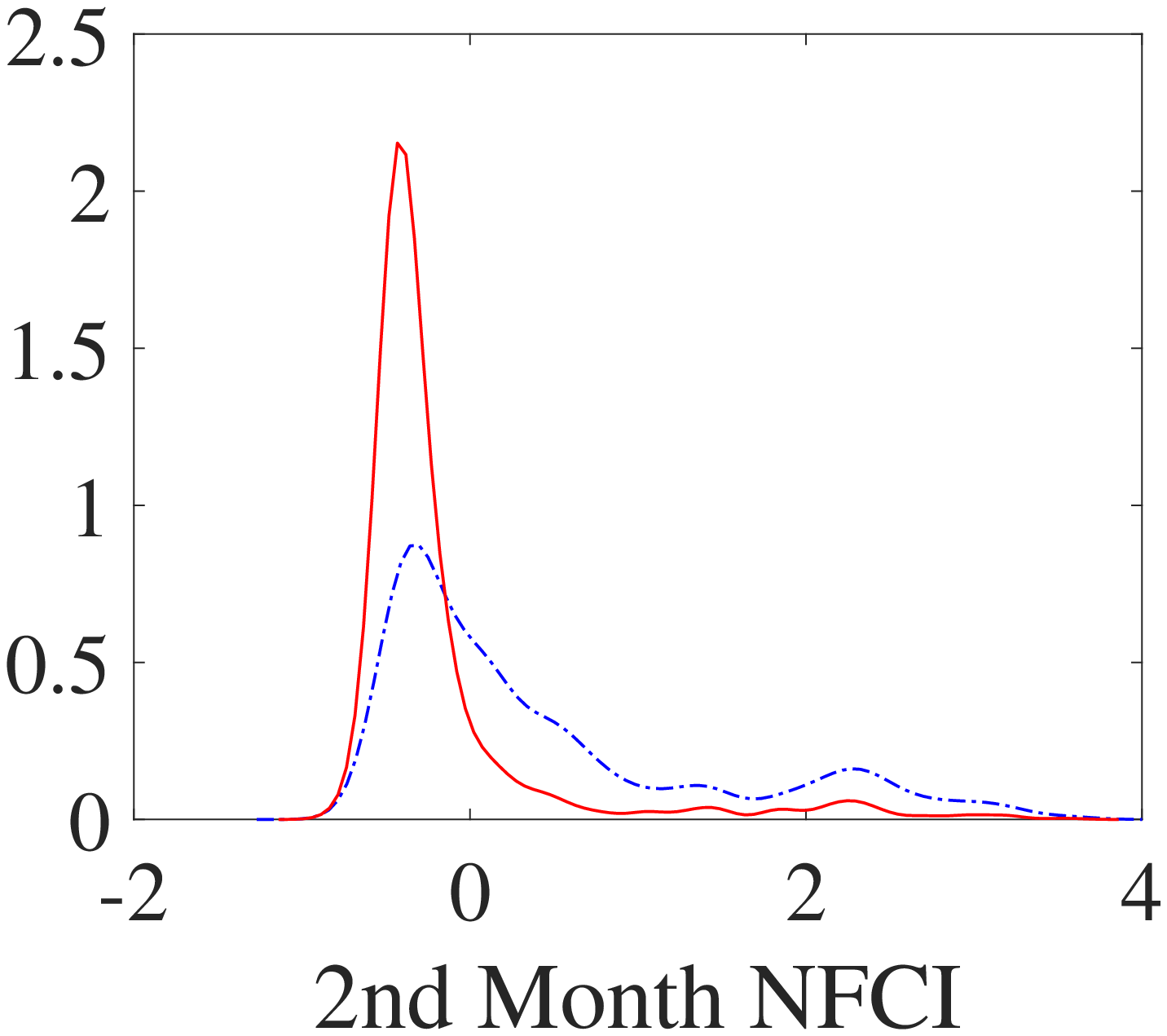}
			\end{subfigure}
			\begin{subfigure}[t]{0.24\textwidth}
				\centering	
				\includegraphics[width=0.95\textwidth]{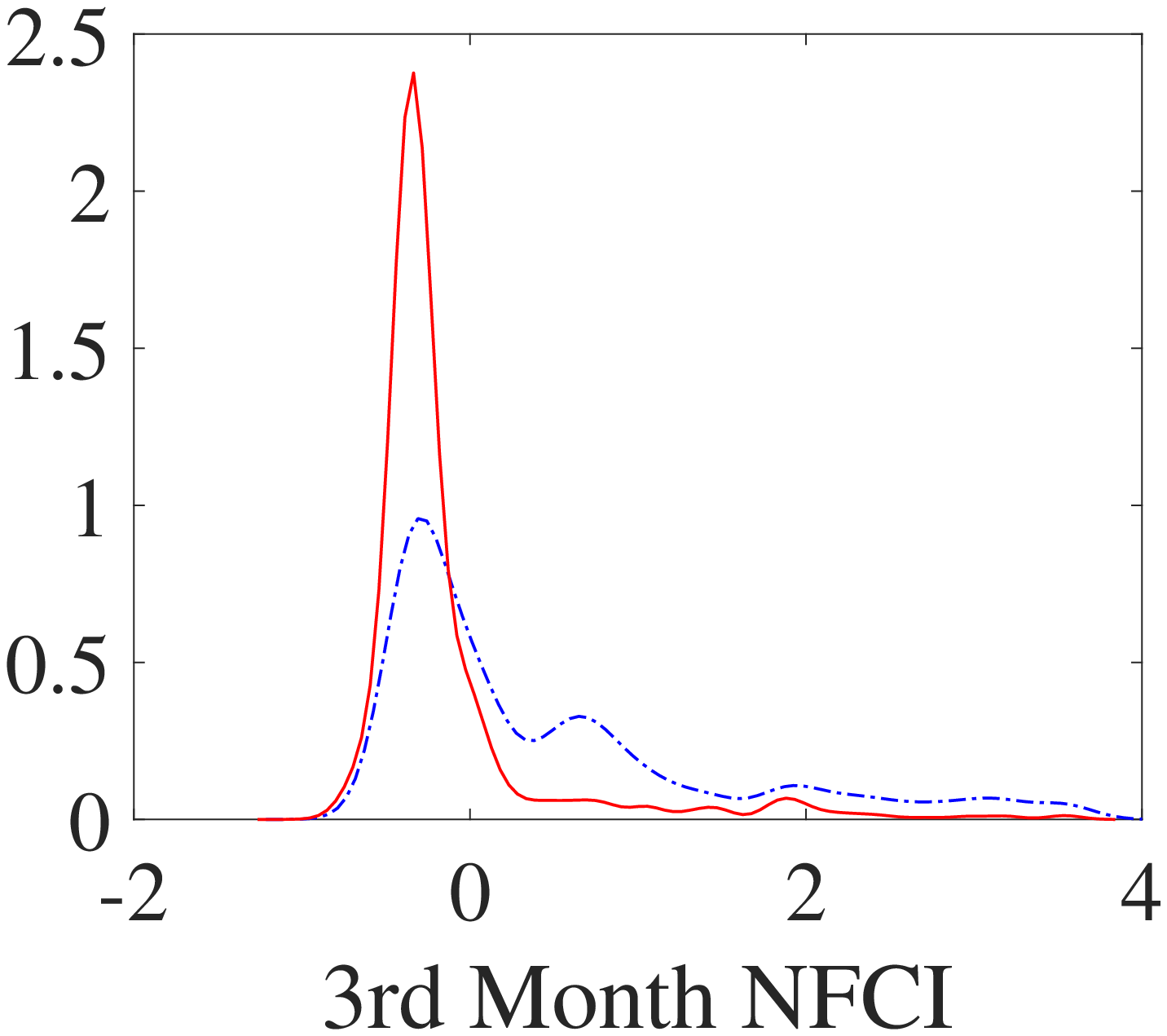}
			\end{subfigure}
			\begin{subfigure}[t]{0.24\textwidth}
				\centering	
				\includegraphics[width=0.95\textwidth]{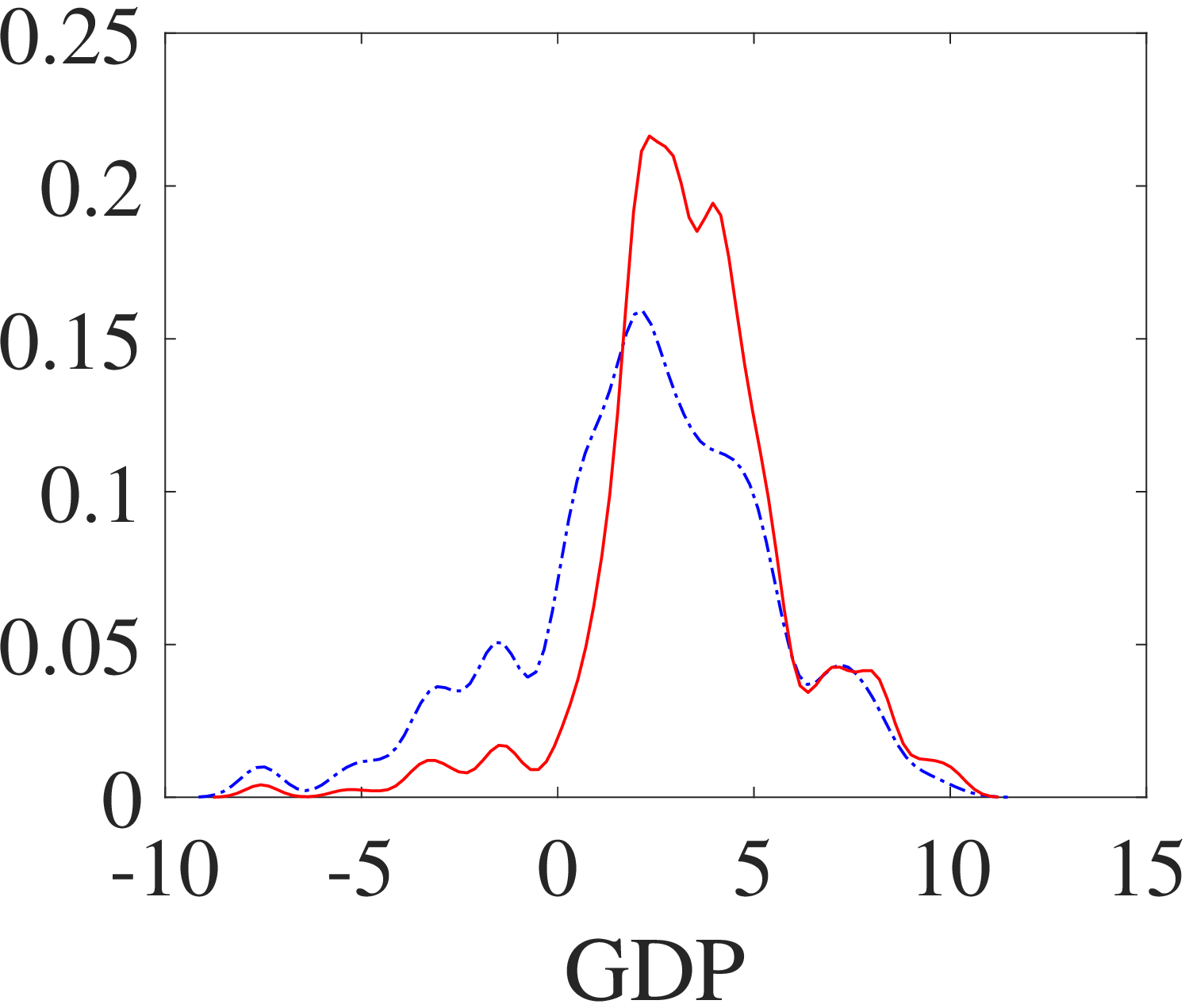}
			\end{subfigure}
			\vspace{0.2cm}
		\end{minipage}
		\begin{minipage}[t]{\textwidth}	
			\centering $h=3$\\
			\begin{subfigure}[t]{0.24\textwidth}
				\centering  		
				\includegraphics[width=0.95\textwidth]{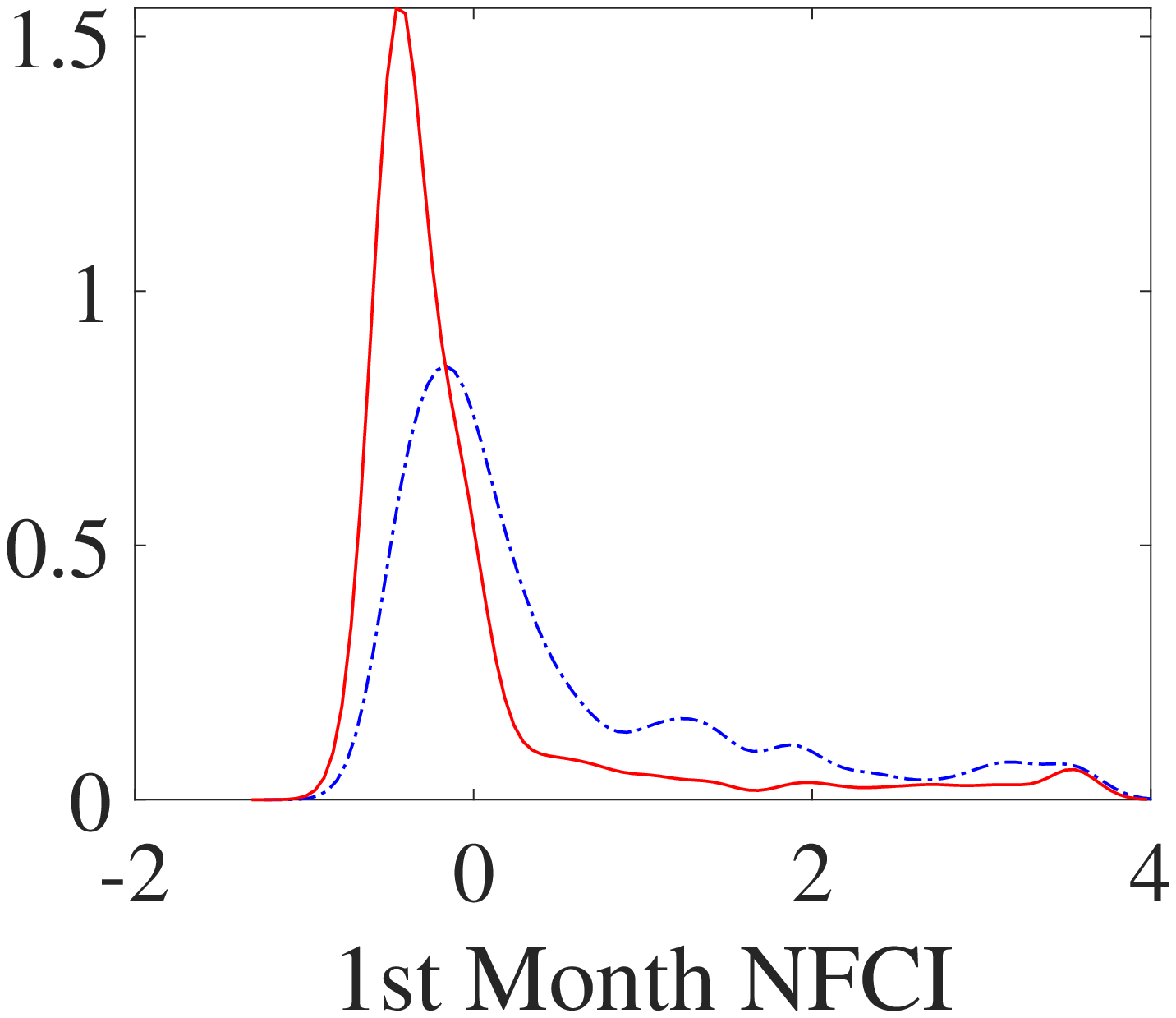}     		
			\end{subfigure}
			\begin{subfigure}[t]{0.24\textwidth}
				\centering	
				\includegraphics[width=0.95\textwidth]{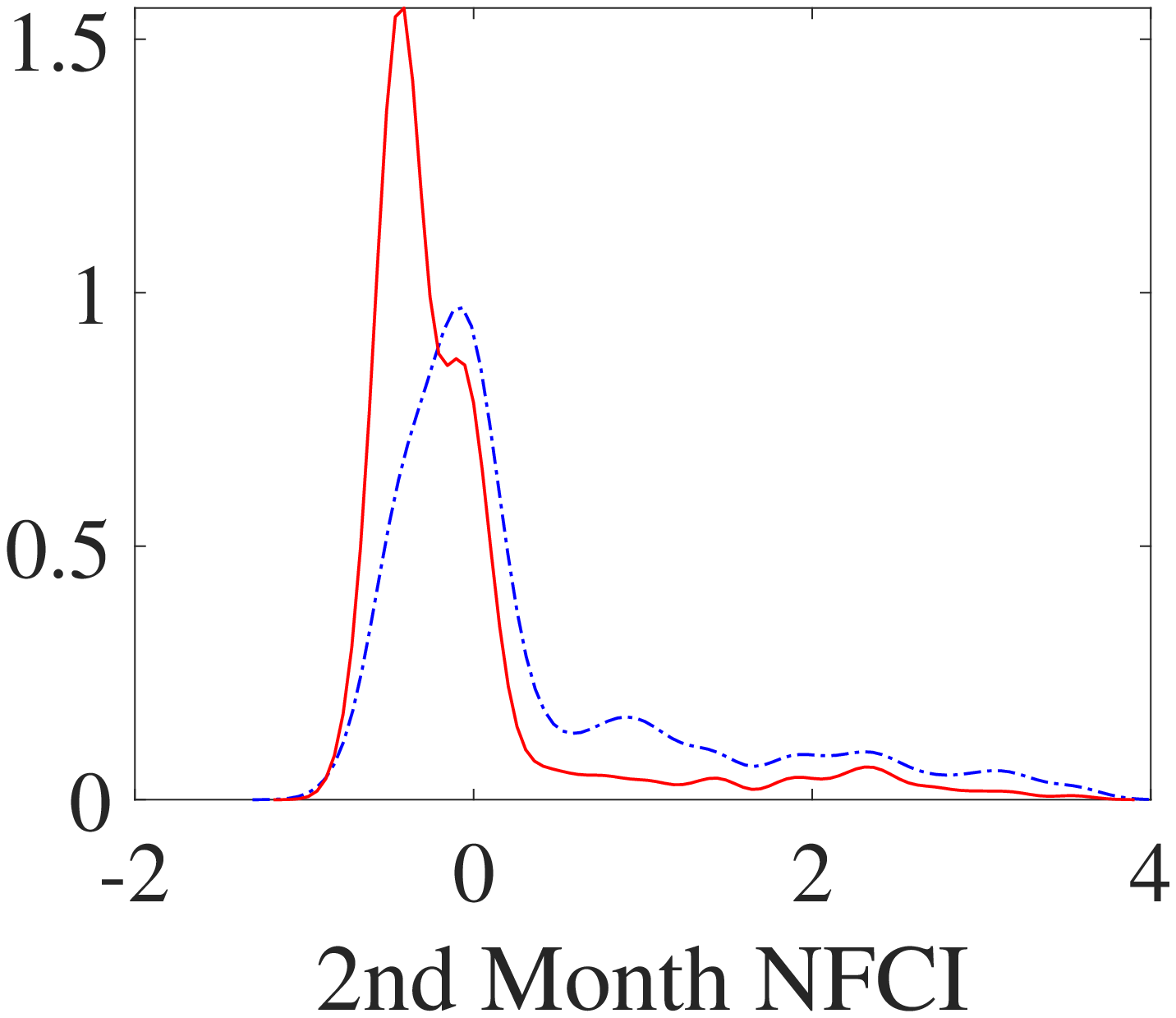}
			\end{subfigure}
			\begin{subfigure}[t]{0.24\textwidth}
				\centering	
				\includegraphics[width=0.95\textwidth]{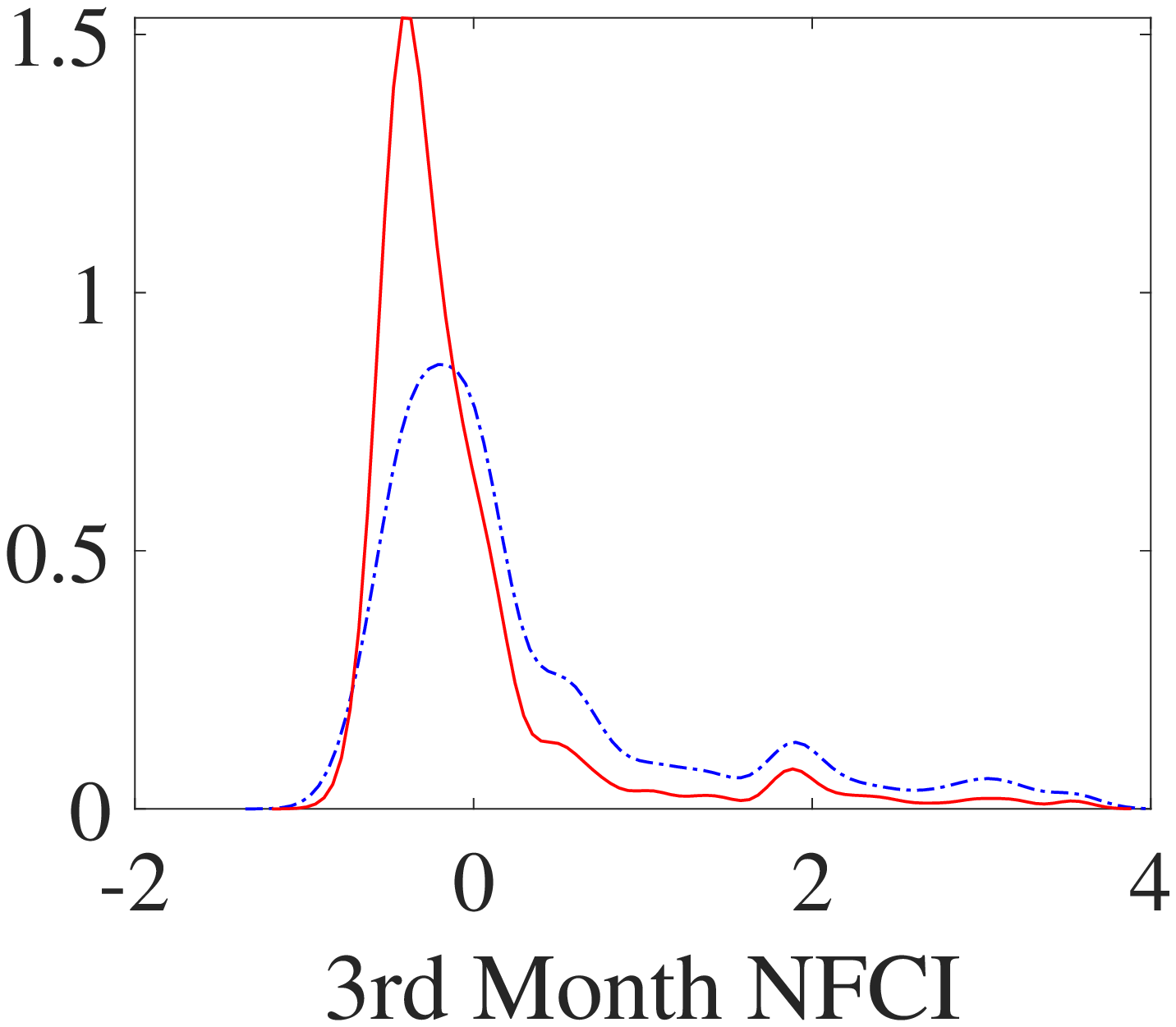}
			\end{subfigure}
			\begin{subfigure}[t]{0.24\textwidth}
				\centering	
				\includegraphics[width=0.95\textwidth]{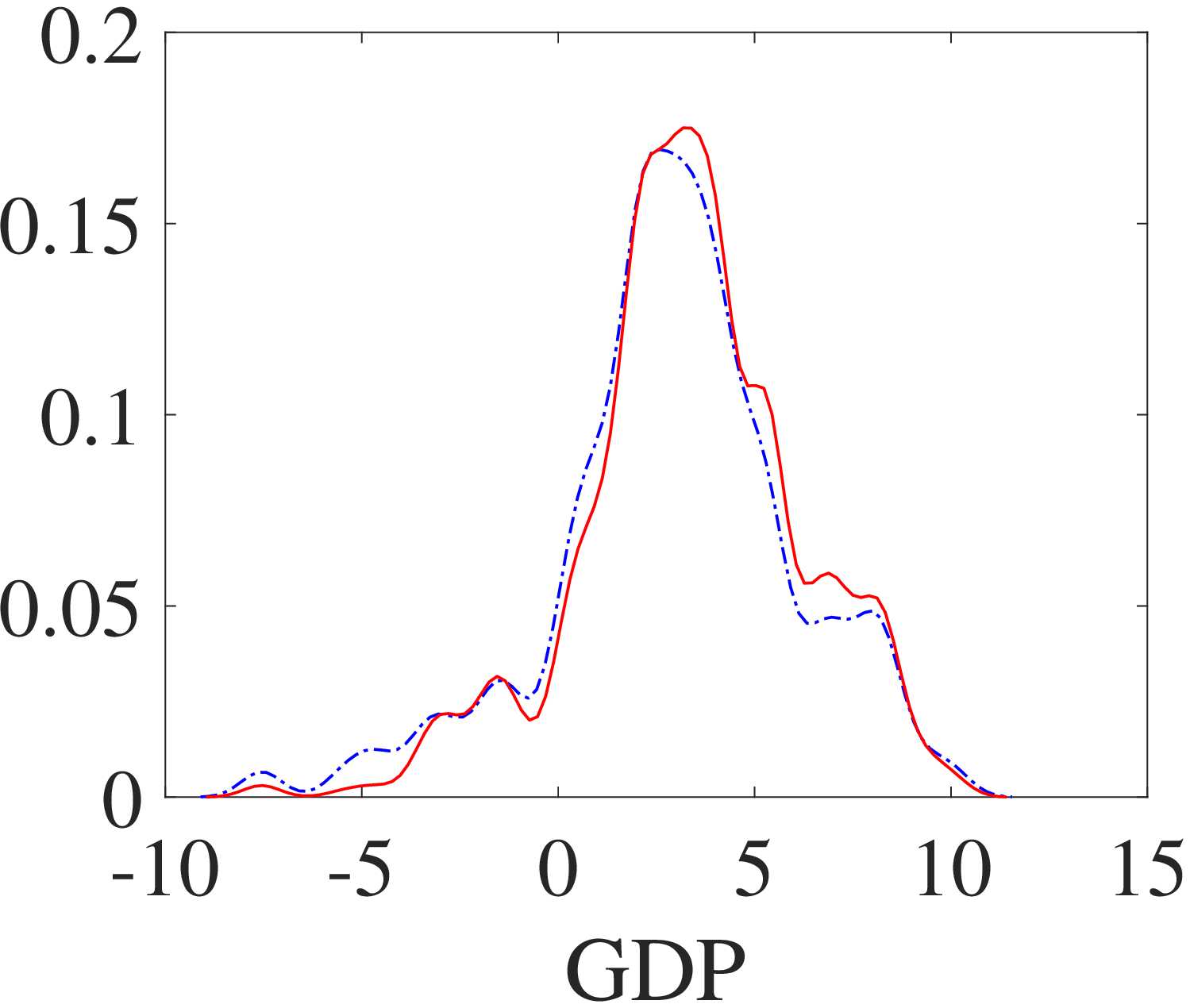}
			\end{subfigure}
			\vspace{0.2cm}
		\end{minipage}
		\begin{minipage}[t]{\textwidth}	
			\centering $h=4$\\
			\begin{subfigure}[t]{0.24\textwidth}
				\centering 		
				\includegraphics[width=0.95\textwidth]{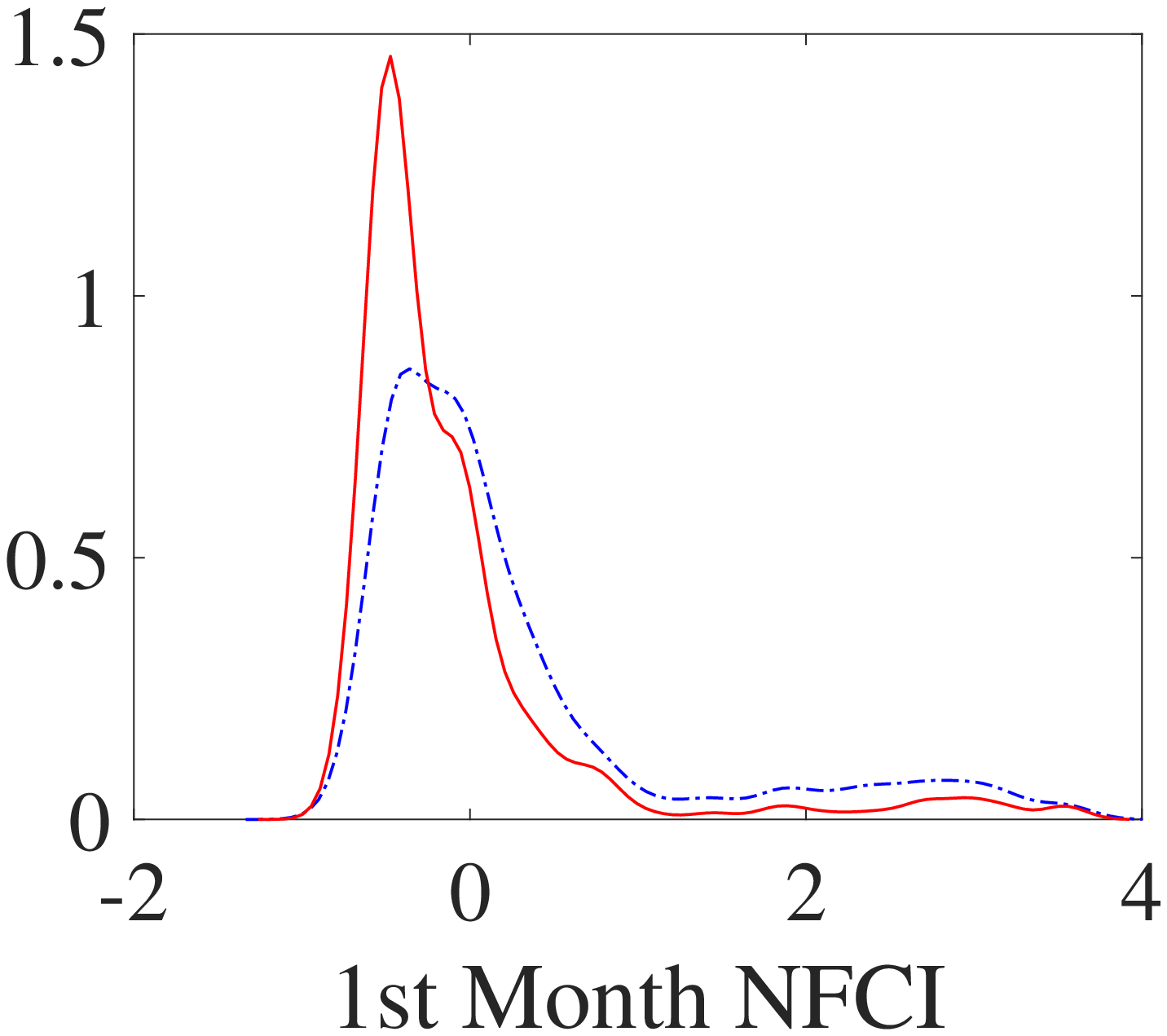}     		
			\end{subfigure}
			\begin{subfigure}[t]{0.24\textwidth}
				\centering	
				\includegraphics[width=0.95\textwidth]{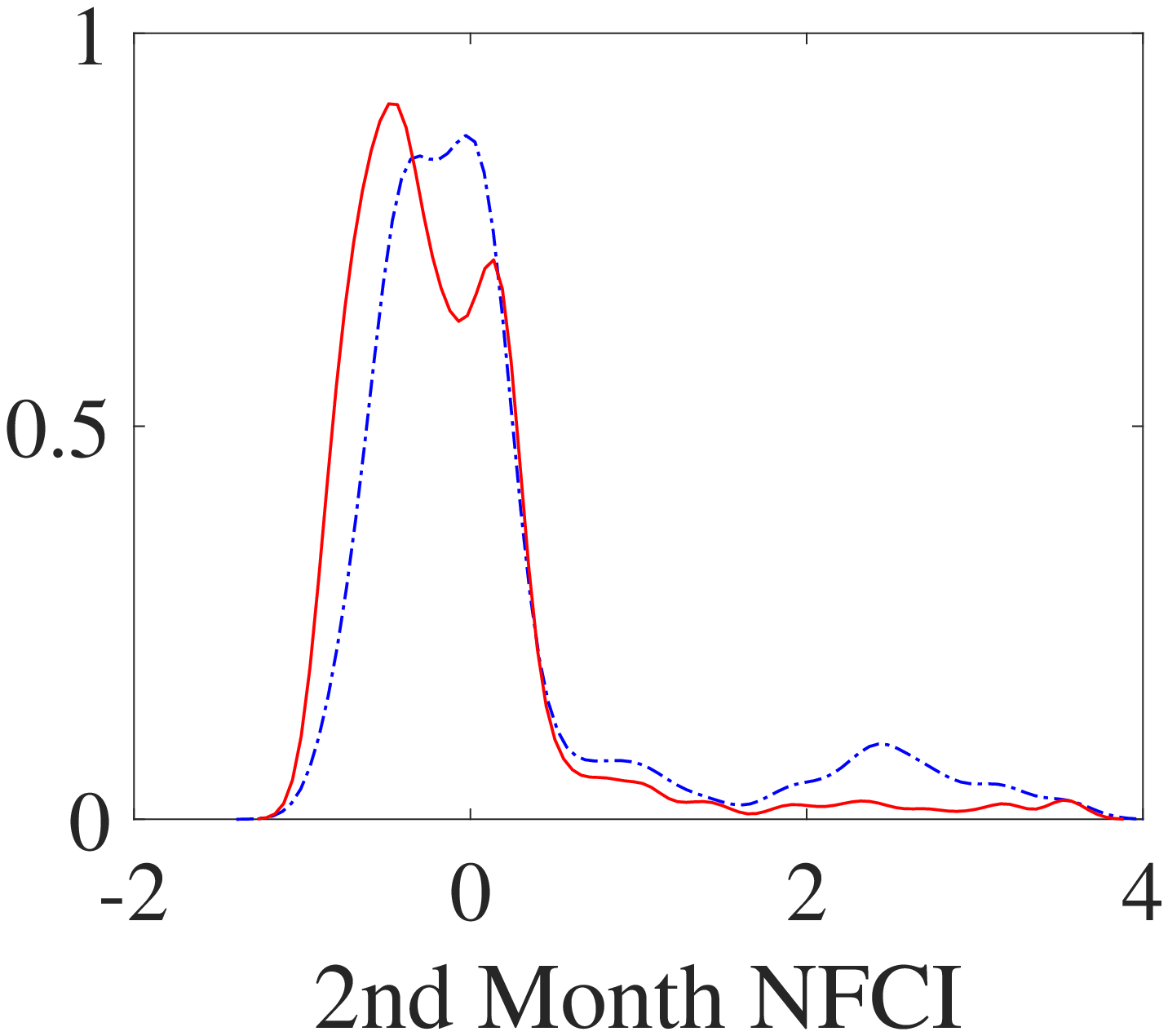}
			\end{subfigure}
			\begin{subfigure}[t]{0.24\textwidth}
				\centering	
				\includegraphics[width=0.95\textwidth]{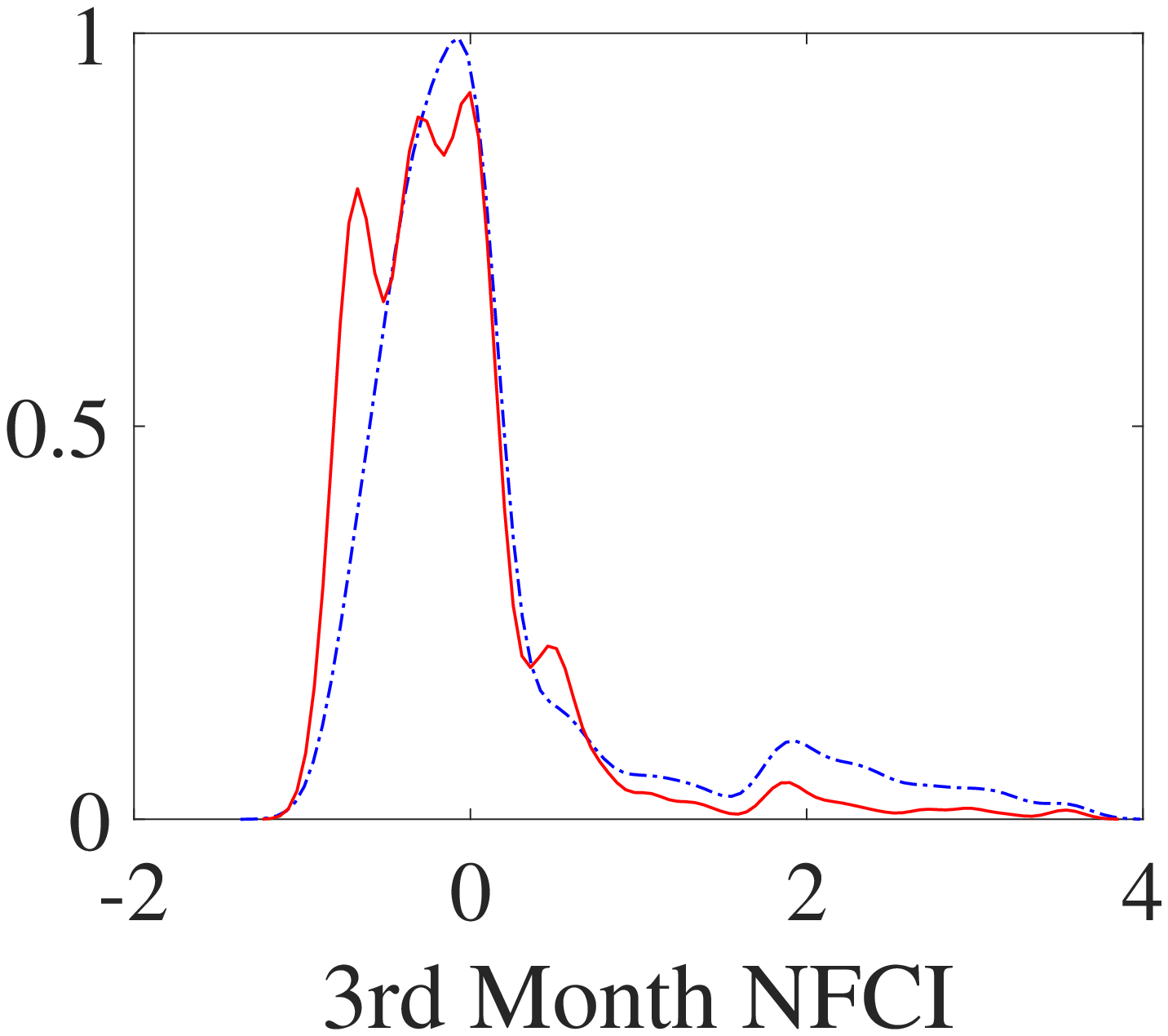}
			\end{subfigure}
			\begin{subfigure}[t]{0.24\textwidth}
				\centering	
				\includegraphics[width=0.95\textwidth]{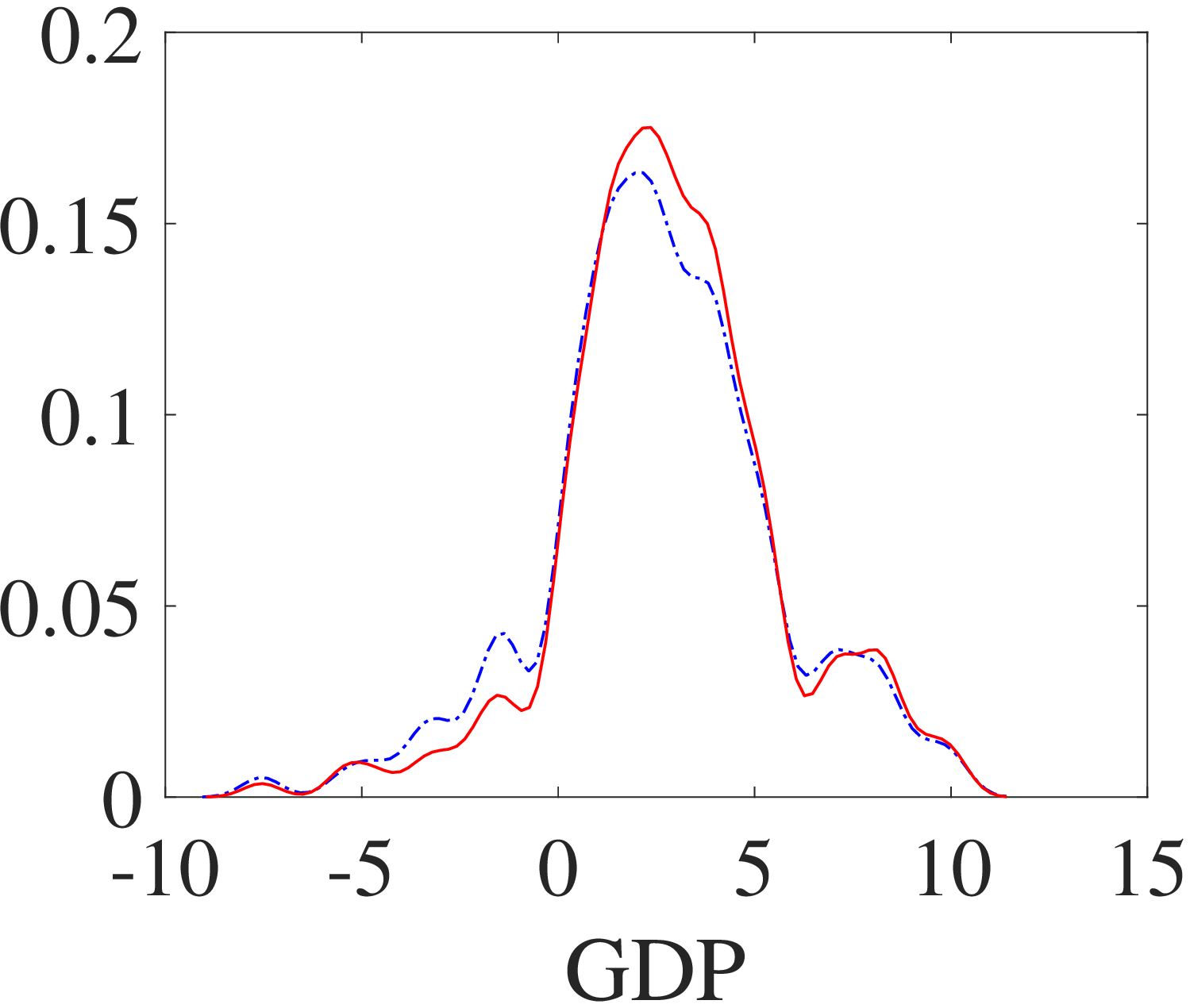}
			\end{subfigure}
		\end{minipage}
	    \vspace{0.5cm}
		\begin{minipage}{.95\linewidth} 
			\linespread{1}\footnotesize
			\textit{Notes}: Different rows correspond to different variables, and for each variable, different columns correspond to forecast horizons from one (first row, $h=1$) to four (last raw, $h=4$) quarters, given all the information available in 2008:Q4. The marginal densities are constructed using kernel density estimation with bandwidths of $0.25$ for the three monthly NFCI and $0.8$ for GDP. The baseline distributions are in blue, and counterfactual distributions are in red.		
		\end{minipage}
	\end{figure}

	Now, we reinvestigate the counterfactual analysis under this model specification. First, we consider the same counterfactual distribution for $Y_{1t}$ in the first month of 2008:Q4. We provide the impulse responses results on the entire distributions of each variable in the following one year in Figure \ref{fig: M-NFCI-response}. While the same conclusion can be reached as in Section 4, there is more evidence for the persistence of the impulse effect that the perturbation in the first month of 2008:Q4 still affects the distributions of both the financial conditions and real GDP growth in 2009:Q4.
	
    \begin{table}[H]
    	\centering
    	\footnotesize
    	\caption{QIR to NFCI Impulse \label{tab: DIR-MN-Quantile}}
    	\begin{tabular}{llllllllllll}
    		\hline \hline
    		Quantiles ($\tau$) & & \multicolumn{2}{c}{0.05} & \multicolumn{2}{c}{0.25} &  \multicolumn{2}{c}{0.5} &  \multicolumn{2}{c}{0.75} & \multicolumn{2}{c}{0.95}\\
    		& $h$ & Base & Diff & Base & Diff & Base & Diff & Base & Diff & Base & Diff   \\
    		\hline
    		NFCI1 
    		& 0 & 0.62  & -0.96 & 0.82  & -0.96 & 1.03  & -1.03 & 1.08 & -0.96 & 2.31 & -1.94 \\
    		& 1 & -0.16 & -0.26 & 0.25  & -0.60 & 0.52  & -0.70 & 1.47 & -1.26 & 2.96 & -2.41 \\
    		& 2 & -0.36 & -0.11 & -0.19 & -0.26 & 0.18  & -0.51 & 1.03 & -1.20 & 2.71 & -1.24 \\
    		& 3 & -0.45 & -0.08 & -0.21 & -0.25 & 0.00  & -0.41 & 0.82 & -0.88 & 3.18 & -0.87 \\
    		& 4 & -0.51 & -0.10 & -0.41 & -0.09 & -0.06 & -0.31 & 0.37 & -0.39 & 2.71 & -0.89\\
    		\hline
    		NFCI2 
    		& 0 & 0.20  & -0.58 & 0.45  & -0.72 & 0.82  & -0.96 & 1.89 & -1.91 & 2.46 & -2.04 \\
    		& 1 & -0.30 & -0.22 & 0.07  & -0.45 & 0.45  & -0.64 & 1.08 & -0.71 & 2.71 & -2.08 \\
    		& 2 & -0.46 & -0.07 & -0.36 & -0.11 & 0.07  & -0.45 & 0.72 & -0.93 & 2.46 & -0.99 \\
    		& 3 & -0.50 & -0.09 & -0.22 & -0.21 & -0.02 & -0.34 & 0.72 & -0.74 & 2.46 & -0.57 \\
    		& 4 & -0.66 & -0.16 & -0.41 & -0.16 & -0.08 & -0.22 & 0.18 & -0.06 & 2.46 & -1.43\\
    		\hline
    		NFCI3 
    		& 0 & -0.02 & -0.43 & 0.42  & -0.76 & 0.55  & -0.73 & 1.89 & -1.90 & 3.18 & -2.73 \\
    		& 1 & -0.41 & -0.13 & -0.02 & -0.43 & 0.55  & -0.80 & 1.31 & -0.79 & 2.71 & -1.40 \\
    		& 2 & -0.38 & -0.16 & -0.32 & -0.06 & 0.00  & -0.34 & 0.82 & -1.01 & 2.96 & -1.66 \\
    		& 3 & -0.56 & 0.00  & -0.32 & -0.10 & -0.02 & -0.28 & 0.52 & -0.52 & 2.46 & -0.64 \\
    		& 4 & -0.67 & -0.09 & -0.36 & -0.18 & -0.06 & -0.16 & 0.08 & -0.01 & 2.31 & -1.49\\
    		\hline
    		GDP 
    		& 0 & -3.20 & 1.45 & 0.50  & 0.30 & 1.90 & 0.10 & 3.10 & 0.00 & 6.48 & -0.98 \\
    		& 1 & -3.60 & 1.85 & -0.55 & 1.63 & 1.15 & 1.75 & 2.80 & 1.30 & 6.88 & 0.35  \\
    		& 2 & -3.20 & 2.65 & 0.50  & 1.60 & 2.30 & 0.90 & 4.50 & 0.18 & 7.60 & 0.42  \\
    		& 3 & -3.20 & 1.45 & 1.40  & 0.60 & 3.10 & 0.10 & 4.75 & 0.43 & 8.15 & 0.00  \\
    		& 4 & -2.70 & 0.95 & 1.15  & 0.15 & 2.45 & 0.25 & 4.10 & 0.20 & 8.15 & 0.00 \\
    		\hline         
    	\end{tabular}
    	\begin{minipage}{0.9\linewidth} 
    		\linespread{1}\footnotesize
    		\textit{Notes}: This table presents different quantiles for the baseline distributions $F_{Y_{j,t+h}|Z_t}$ (Base) and the quantile and moments differences from the counterfactual distributions $F^*_{Y_{j,t+h}|Z_t}$ to the baseline distributions (Diff) of three monthly financial conditions within each quarter (NFCI1, NFCI2, NFCI3) and real GDP growth from 2008:Q4 ($h=0$) to 2009:Q4 ($h=4$). 
    	\end{minipage}
    \end{table}

    \begin{table}[H]
    	\centering
    	\footnotesize
    	\caption{MIR to NFCI Impulse\label{tab: DIR-MN-Moment}}
    	\begin{tabular}{llllllllll}
    		\hline \hline
    		Moments& & \multicolumn{2}{c}{Mean} & \multicolumn{2}{c}{Std} &  \multicolumn{2}{c}{Skewess} &  \multicolumn{2}{c}{Kurtosis} \\
    		& $h$ & Base & Diff & Base & Diff & Base & Diff & Base & Diff   \\
    		\hline
    		NFCI1 
    		& 0 & 1.17 & -1.17 & 0.48 & -0.28 & 1.62 & -1.62 & 5.60 & -2.59 \\
    		& 1 & 0.94 & -0.97 & 1.01 & -0.56 & 1.05 & 1.82  & 2.86 & 14.39 \\
    		& 2 & 0.53 & -0.64 & 0.97 & -0.27 & 1.40 & 2.02  & 4.24 & 11.00 \\
    		& 3 & 0.47 & -0.51 & 1.03 & -0.16 & 1.52 & 1.30  & 4.44 & 5.99  \\
    		& 4 & 0.26 & -0.35 & 0.97 & -0.20 & 1.85 & 1.12  & 5.56 & 6.64 \\
    		\hline
    		NFCI2 
    		& 0 & 1.09 & -1.17 & 0.78 & -0.46 & 0.76 & 2.73 & 2.42 & 19.61 \\
    		& 1 & 0.75 & -0.76 & 0.97 & -0.44 & 1.08 & 1.25 & 2.95 & 8.62  \\
    		& 2 & 0.43 & -0.60 & 0.99 & -0.36 & 1.28 & 1.98 & 3.51 & 10.28 \\
    		& 3 & 0.35 & -0.43 & 0.97 & -0.25 & 1.57 & 1.17 & 4.60 & 5.80  \\
    		& 4 & 0.17 & -0.31 & 0.93 & -0.25 & 1.99 & 0.78 & 6.23 & 7.04 \\
    		\hline
    		NFCI3 
    		& 0 & 1.05 & -1.17 & 0.98 & -0.59 & 0.87 & 3.14 & 2.68 & 24.43 \\
    		& 1 & 0.76 & -0.70 & 0.97 & -0.28 & 0.97 & 0.90 & 3.21 & 4.28  \\
    		& 2 & 0.46 & -0.59 & 1.02 & -0.42 & 1.44 & 1.94 & 4.20 & 11.23 \\
    		& 3 & 0.29 & -0.36 & 0.94 & -0.26 & 1.69 & 1.21 & 5.22 & 6.89  \\
    		& 4 & 0.15 & -0.28 & 0.88 & -0.26 & 1.97 & 0.53 & 6.32 & 5.66 \\
    		\hline
    		GDP 
    		& 0 & 1.77 & 0.24 & 2.68 & -0.49 & 0.06  & 0.05  & 4.33 & 1.10  \\
    		& 1 & 1.22 & 1.52 & 2.81 & -0.03 & 0.33  & -0.25 & 3.83 & -0.56 \\
    		& 2 & 2.27 & 1.17 & 3.17 & -0.73 & -0.43 & 0.04  & 3.48 & 1.95  \\
    		& 3 & 2.95 & 0.39 & 3.08 & -0.27 & -0.49 & 0.14  & 3.92 & -0.21 \\
    		& 4 & 2.61 & 0.26 & 2.99 & -0.18 & -0.16 & 0.05  & 3.76 & 0.39 \\
    		\hline         
    	\end{tabular}
    	\begin{minipage}{0.75\linewidth} 
    		\linespread{1}\footnotesize
    		\textit{Notes}: This table presents moments for the baseline distributions (Base) and the moments differences from the counterfactual distributions to the baseline distributions (Diff) of three monthly financial conditions within each quarter (NFCI1, NFCI2, NFCI3) and real GDP growth from 2008:Q4 ($h=0$) to 2009:Q4 ($h=4$).
    	\end{minipage}
    \end{table}

\begin{figure}[H]
	\captionsetup[subfigure]{aboveskip=-3pt,belowskip=0pt, labelformat=empty}
	\centering
	\caption{Distributional Response to GDP Impulse} \label{fig: M-GDP-response}	
	\begin{minipage}[t]{\textwidth}	
		\centering  Distributions of $Y_{t+h}$ given $Z_{t}$ for $t$=2008:Q4\\$h=1$\\
         \begin{subfigure}[t]{0.24\textwidth}
         	\centering 		
         	\includegraphics[width=0.95\textwidth]{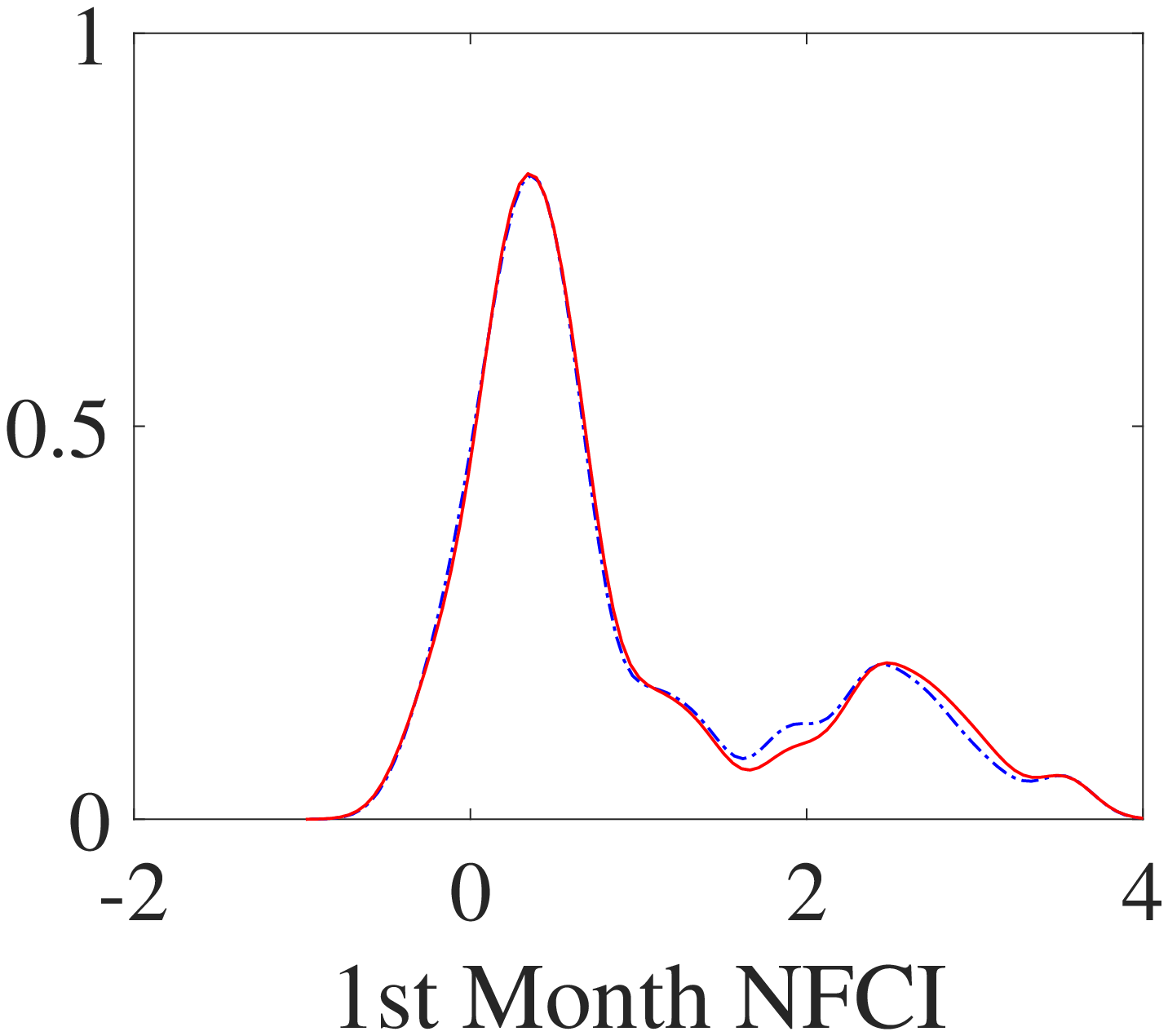}     		
         \end{subfigure}
         \begin{subfigure}[t]{0.24\textwidth}
         	\centering	
         	\includegraphics[width=0.95\textwidth]{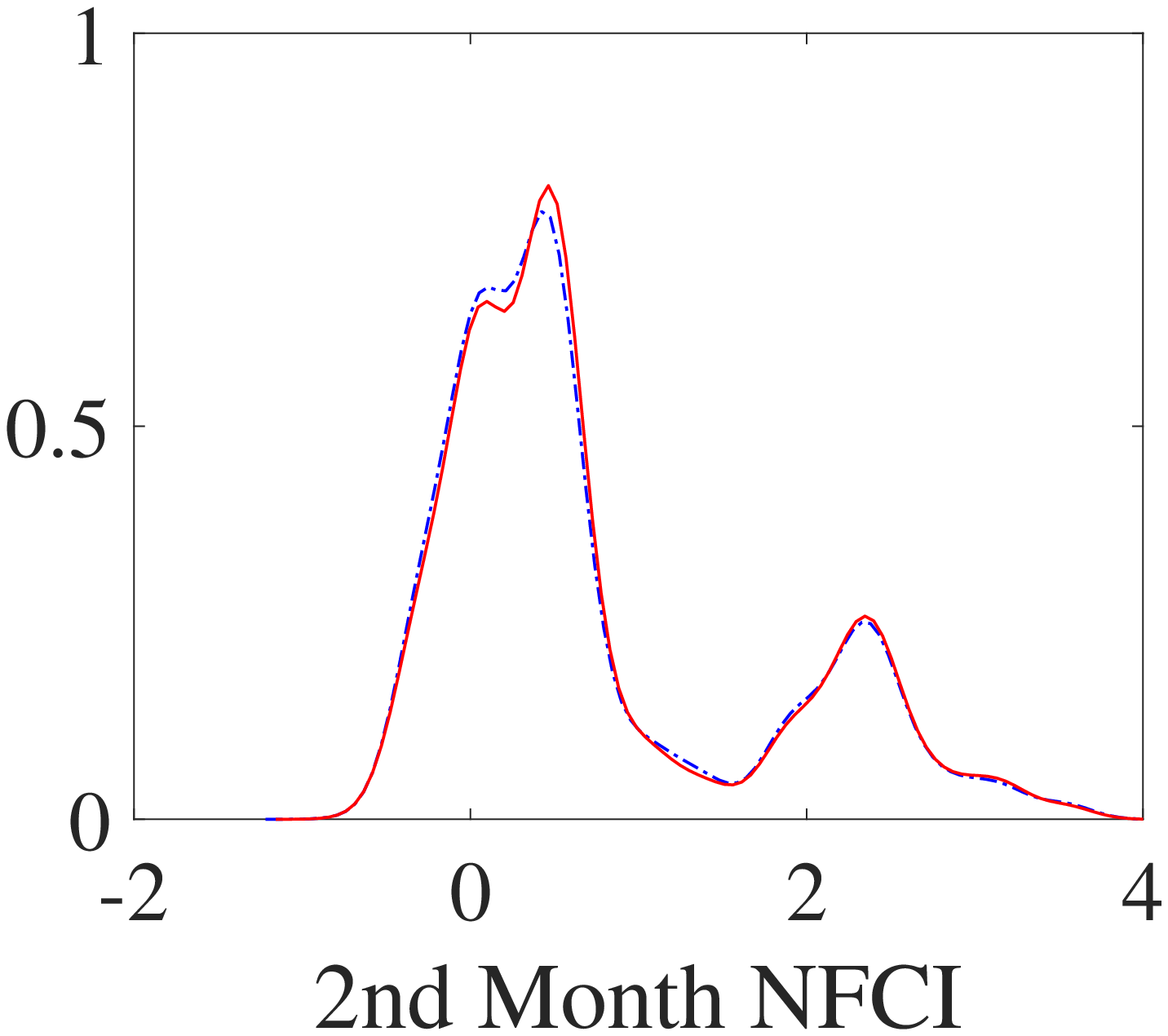}
         \end{subfigure}
         \begin{subfigure}[t]{0.24\textwidth}
         	\centering	
         	\includegraphics[width=0.95\textwidth]{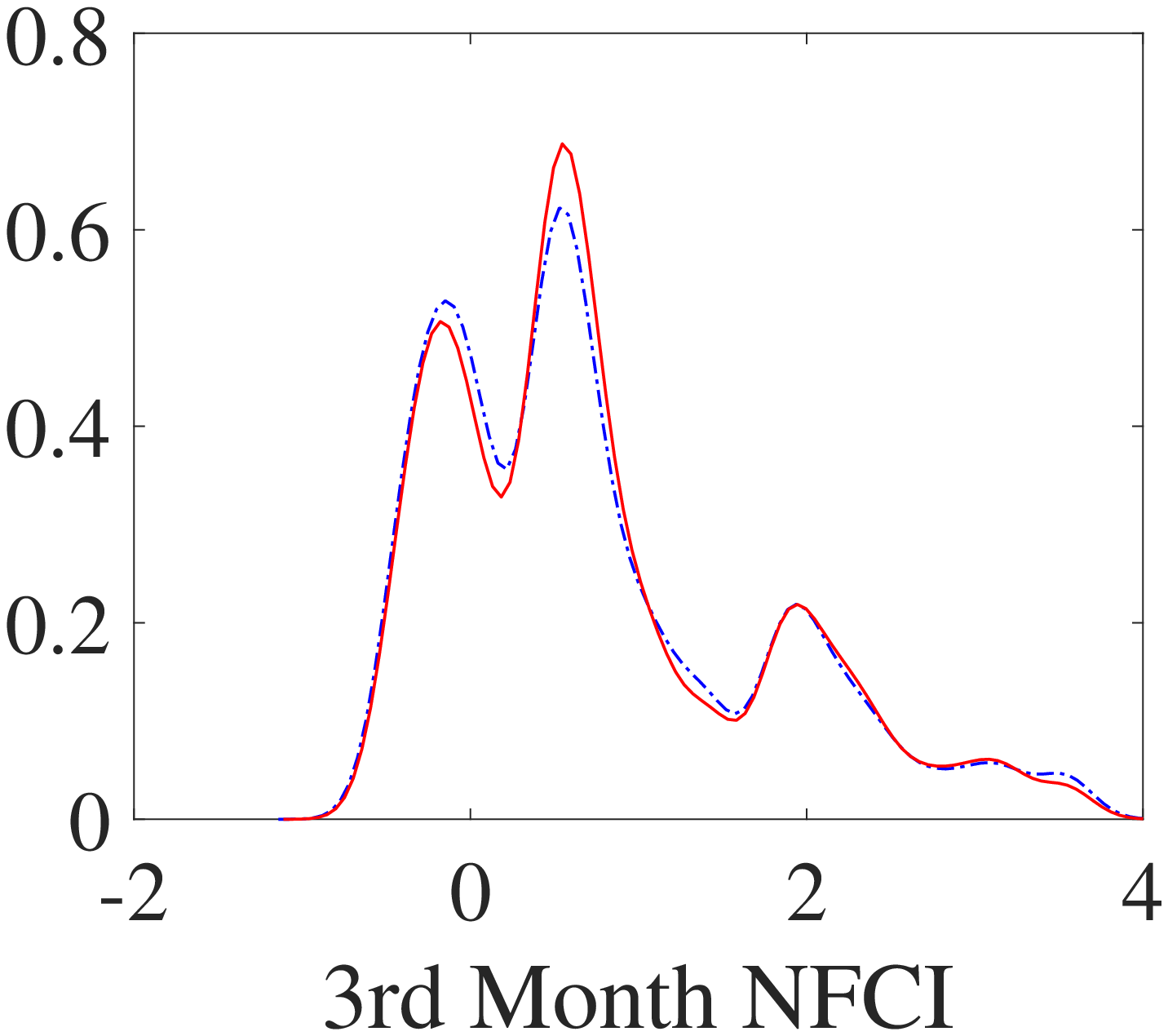}
         \end{subfigure}
         \begin{subfigure}[t]{0.24\textwidth}
         	\centering	
         	\includegraphics[width=0.95\textwidth]{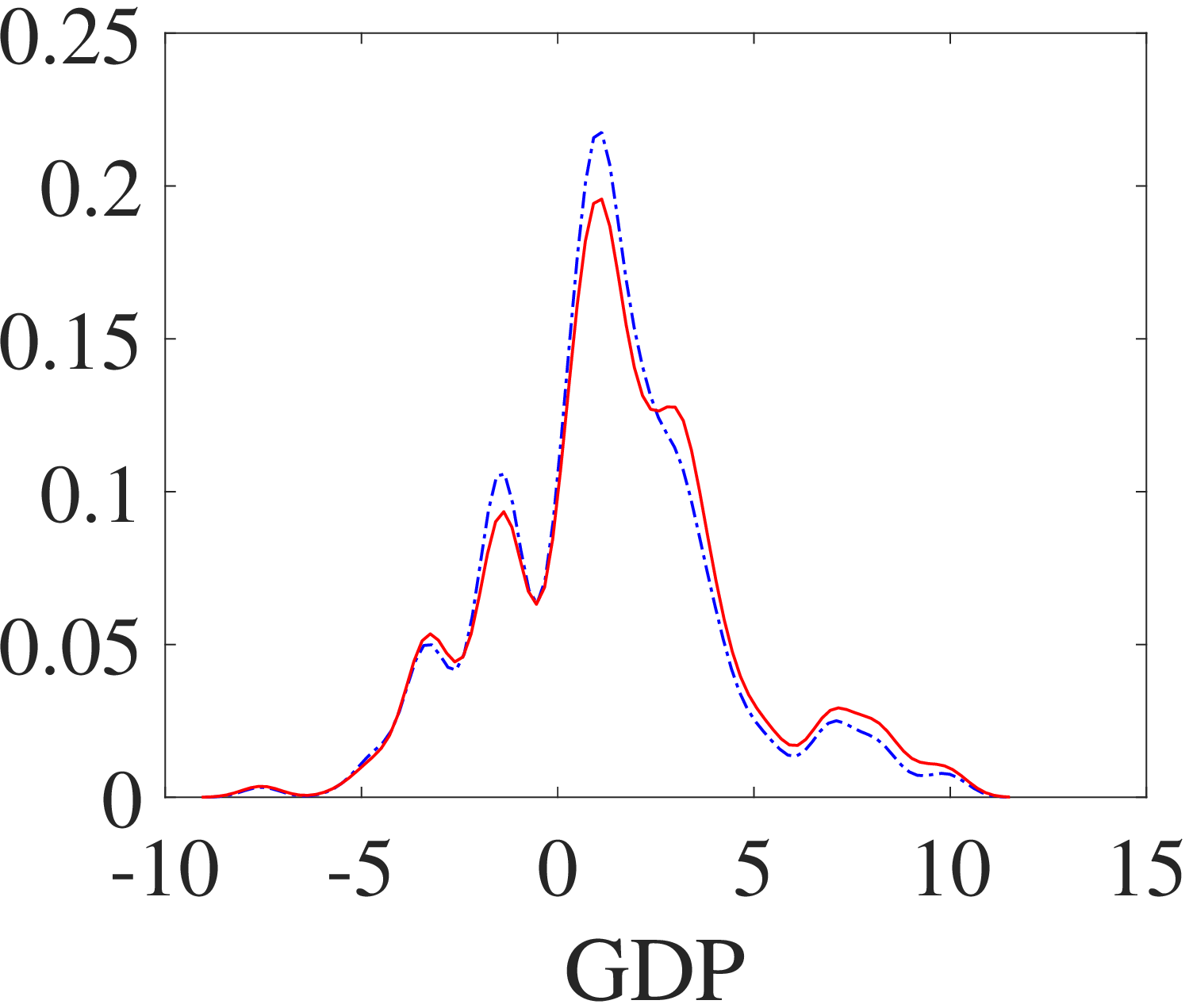}
         \end{subfigure}
     \vspace{0.2cm}
     \end{minipage}
		\begin{minipage}[t]{\textwidth}	
		\centering $h=2$\\
		\begin{subfigure}[t]{0.24\textwidth}
			\centering 		
			\includegraphics[width=0.95\textwidth]{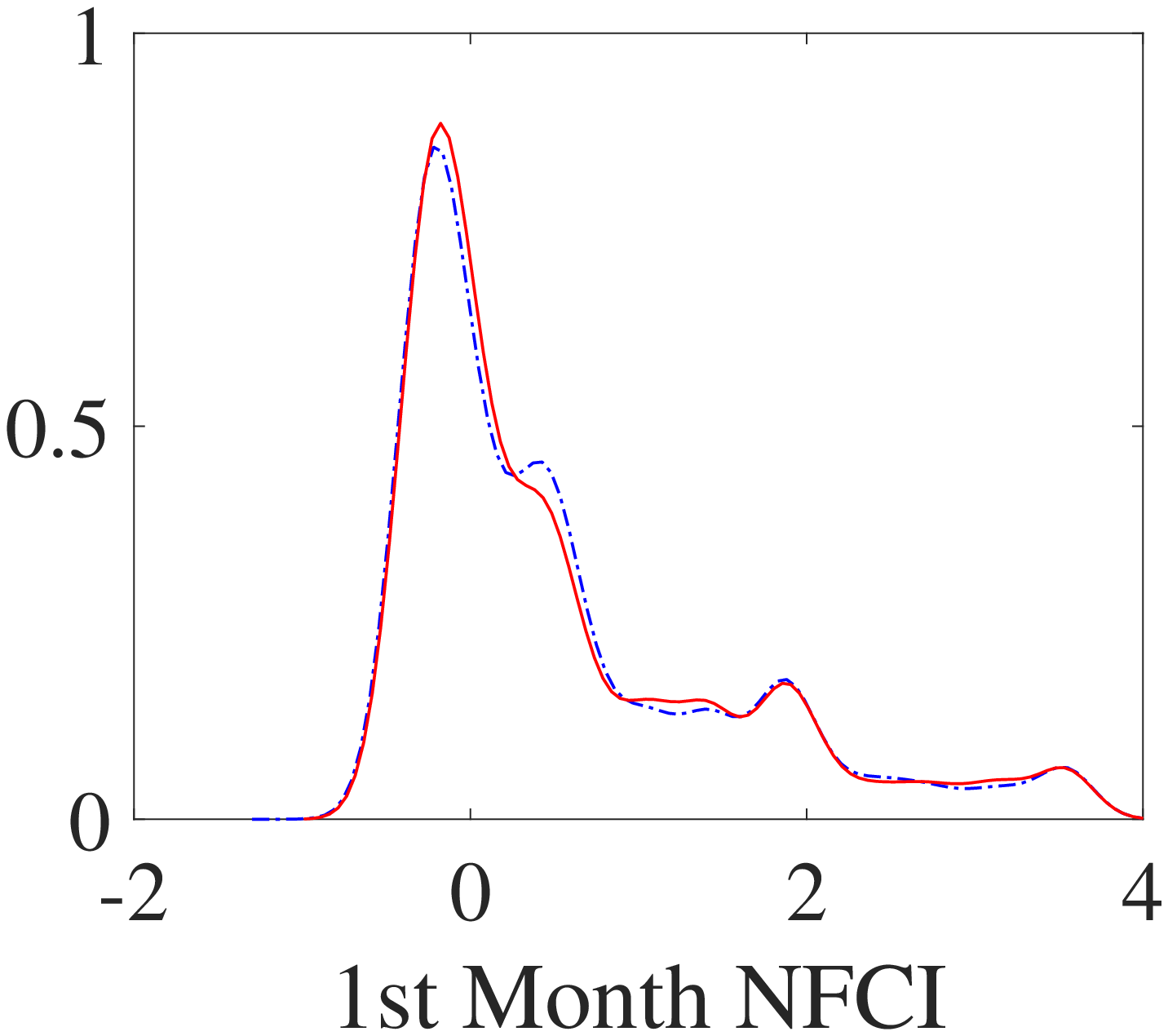}     		
		\end{subfigure}
		\begin{subfigure}[t]{0.24\textwidth}
			\centering	
			\includegraphics[width=0.95\textwidth]{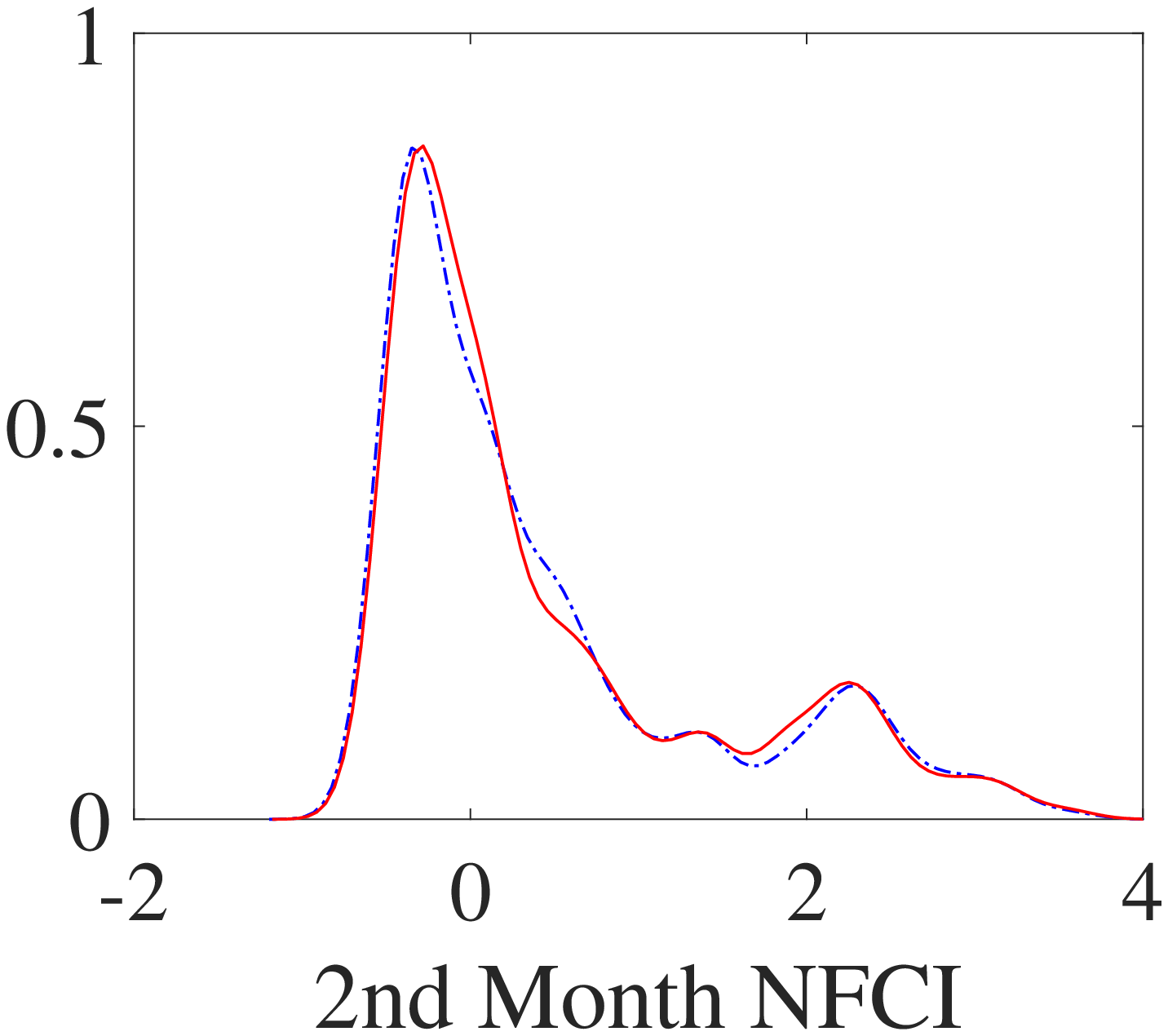}
		\end{subfigure}
		\begin{subfigure}[t]{0.24\textwidth}
			\centering	
			\includegraphics[width=0.95\textwidth]{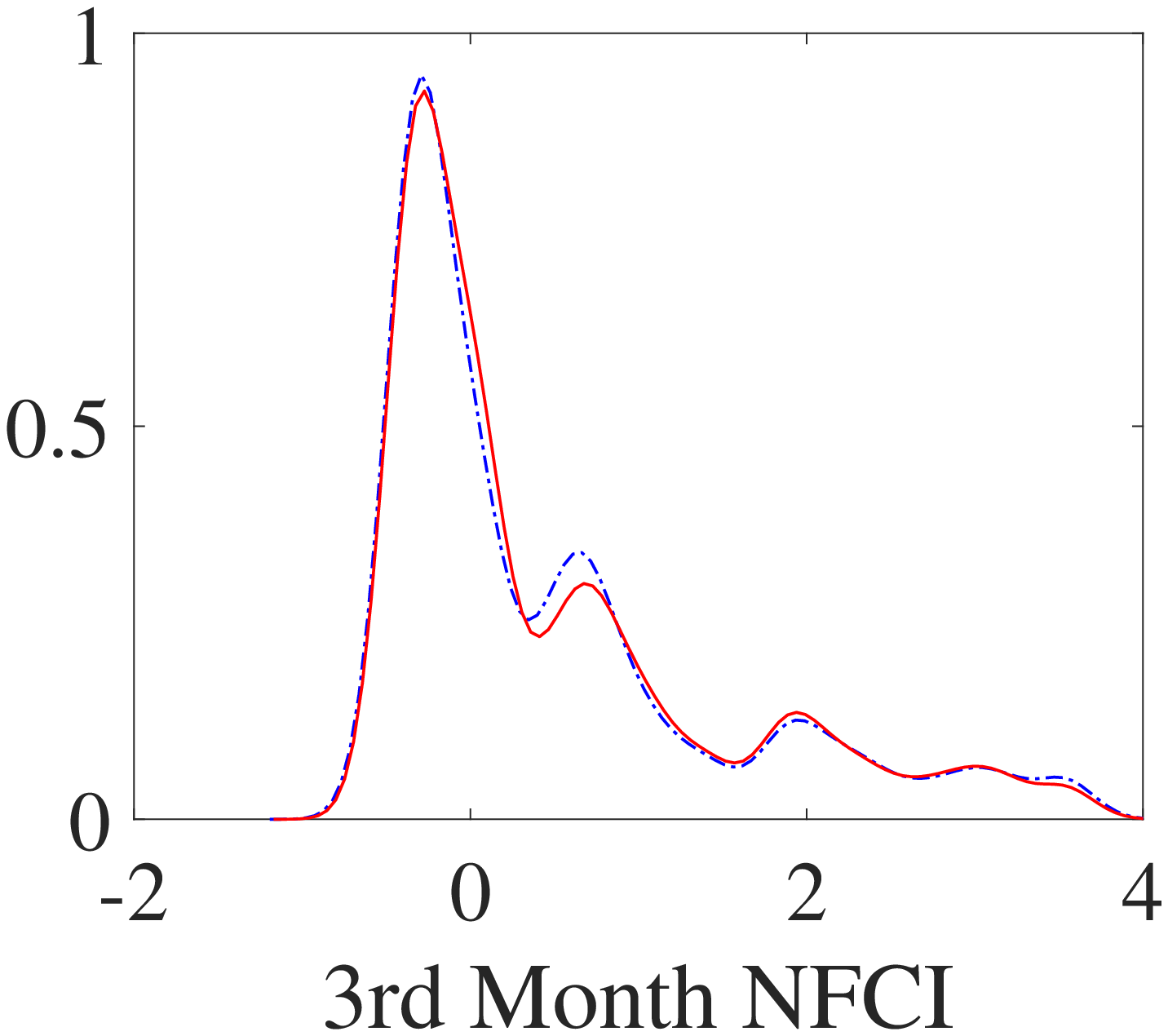}
		\end{subfigure}
		\begin{subfigure}[t]{0.24\textwidth}
			\centering	
			\includegraphics[width=0.95\textwidth]{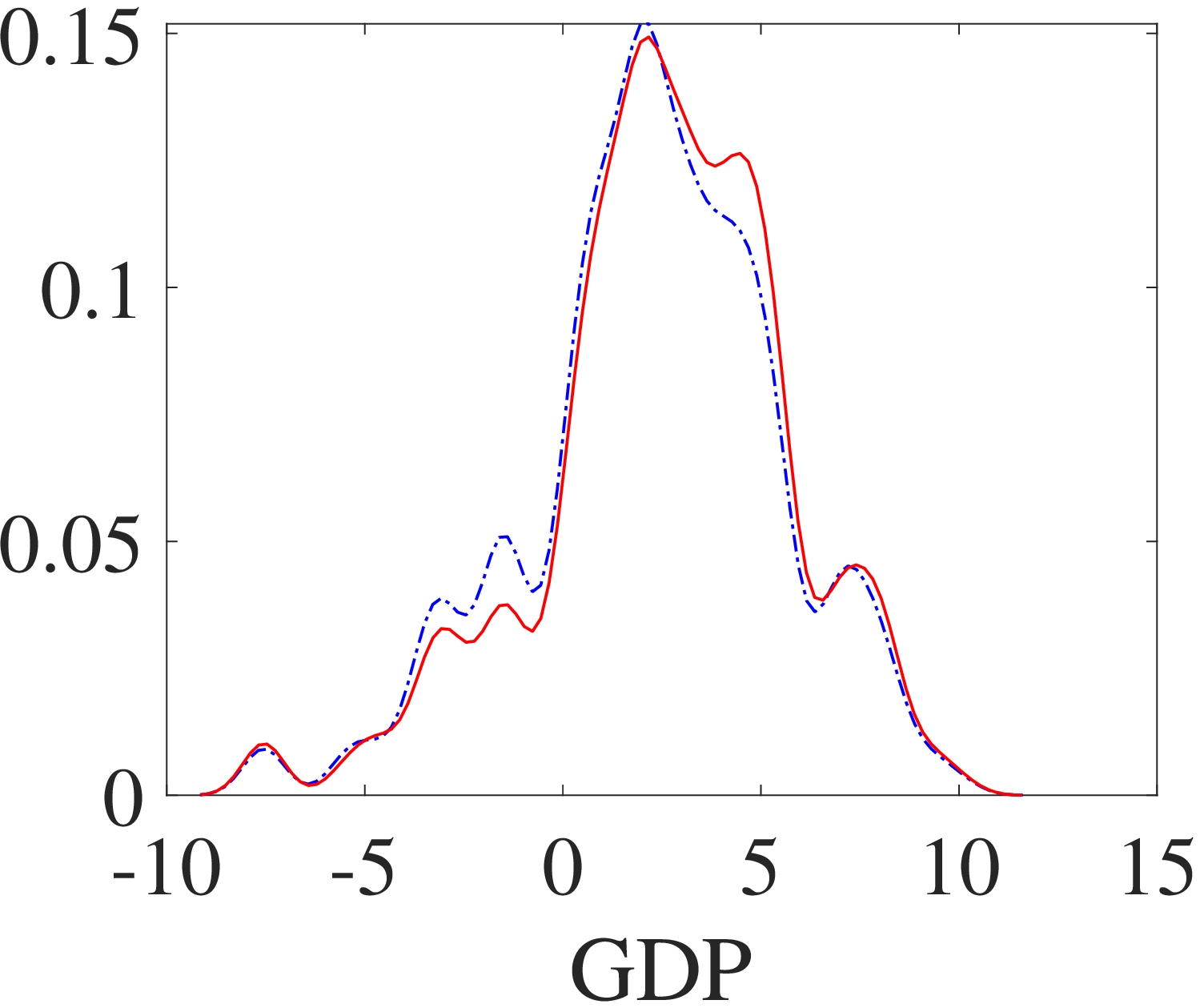}
		\end{subfigure}
	\vspace{0.2cm}
	\end{minipage}
			\begin{minipage}[t]{\textwidth}	
			\centering $h=3$\\
			\begin{subfigure}[t]{0.24\textwidth}
				\centering  		
				\includegraphics[width=0.95\textwidth]{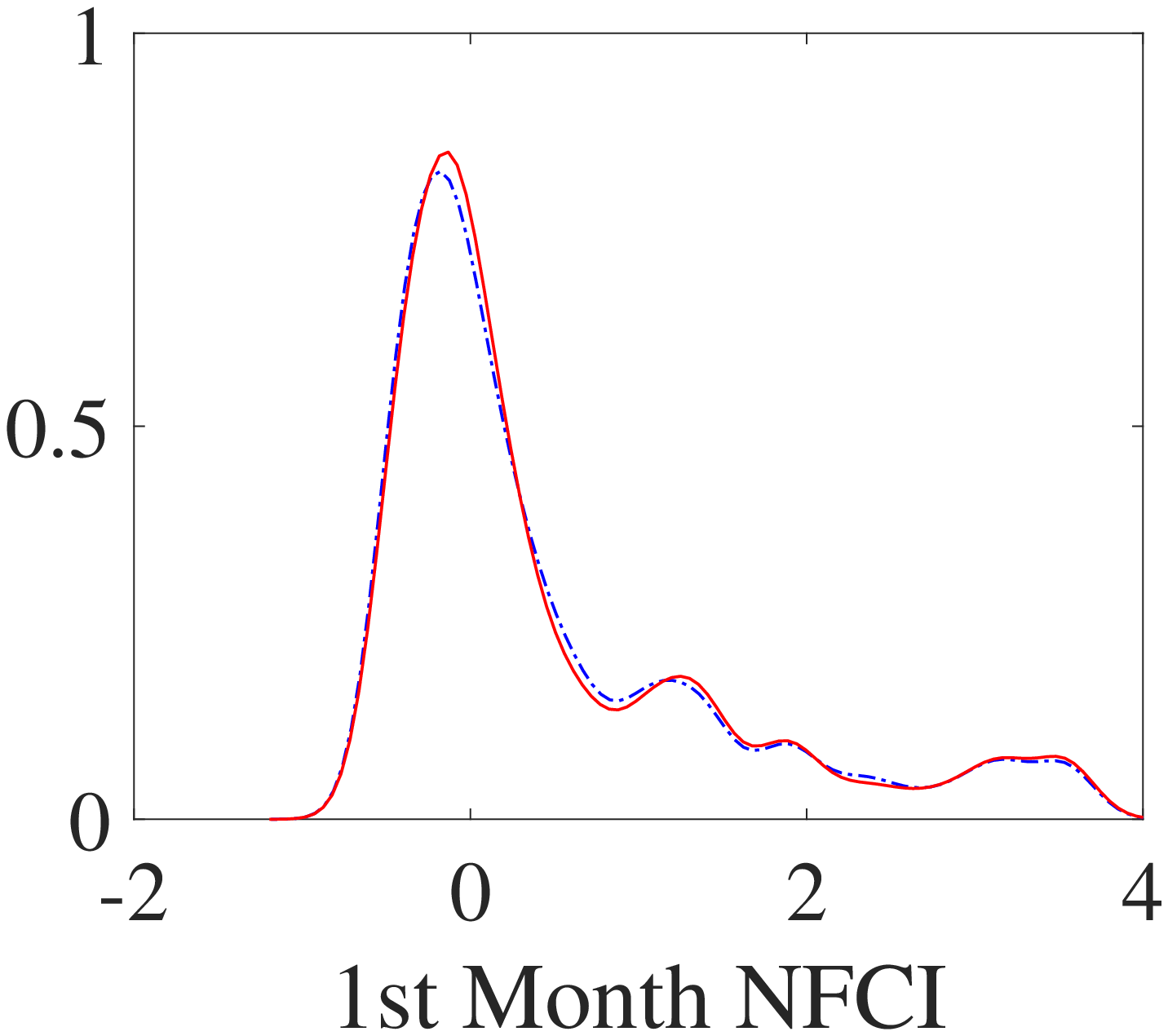}     		
			\end{subfigure}
			\begin{subfigure}[t]{0.24\textwidth}
				\centering	
				\includegraphics[width=0.95\textwidth]{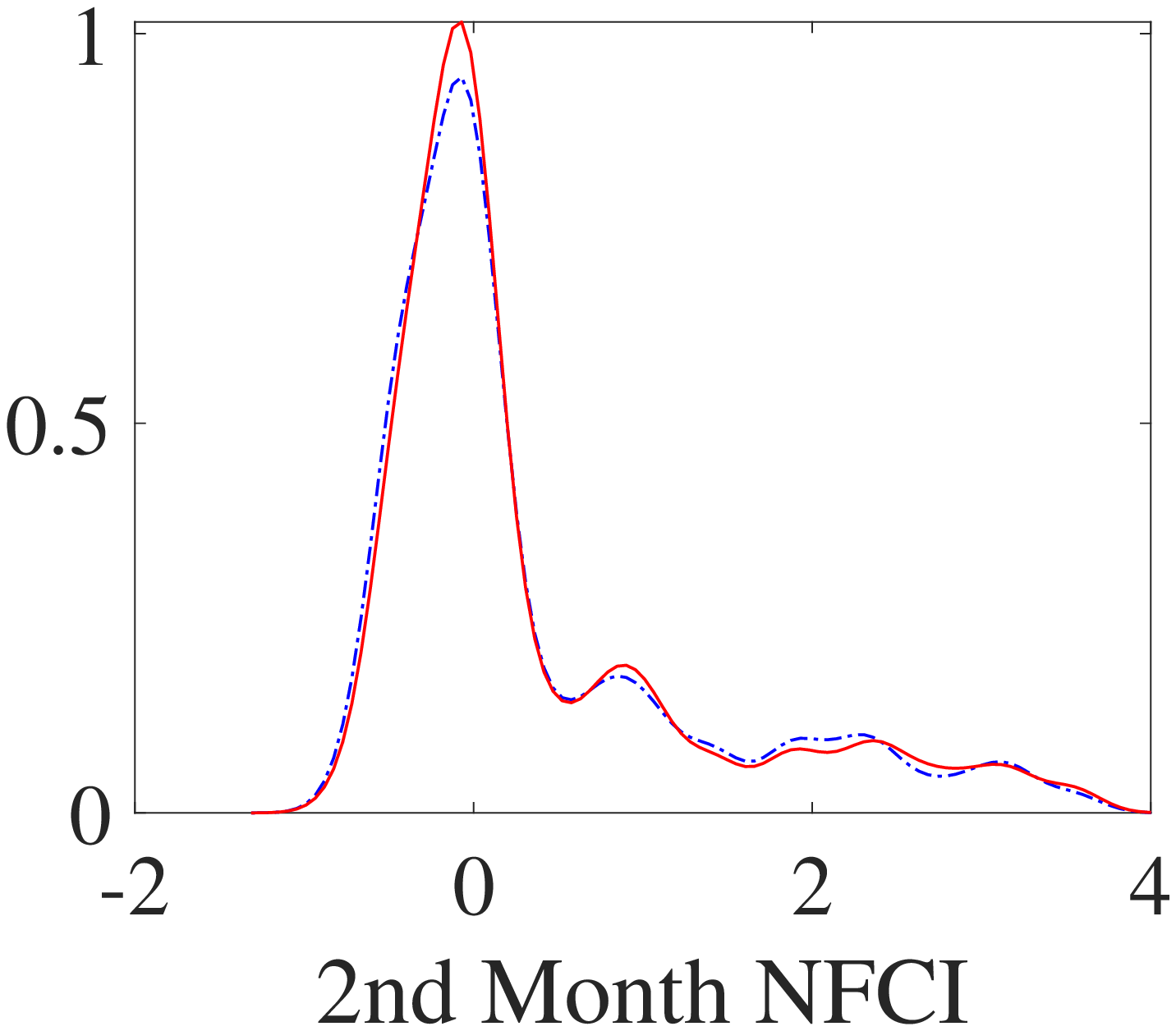}
			\end{subfigure}
			\begin{subfigure}[t]{0.24\textwidth}
				\centering	
				\includegraphics[width=0.95\textwidth]{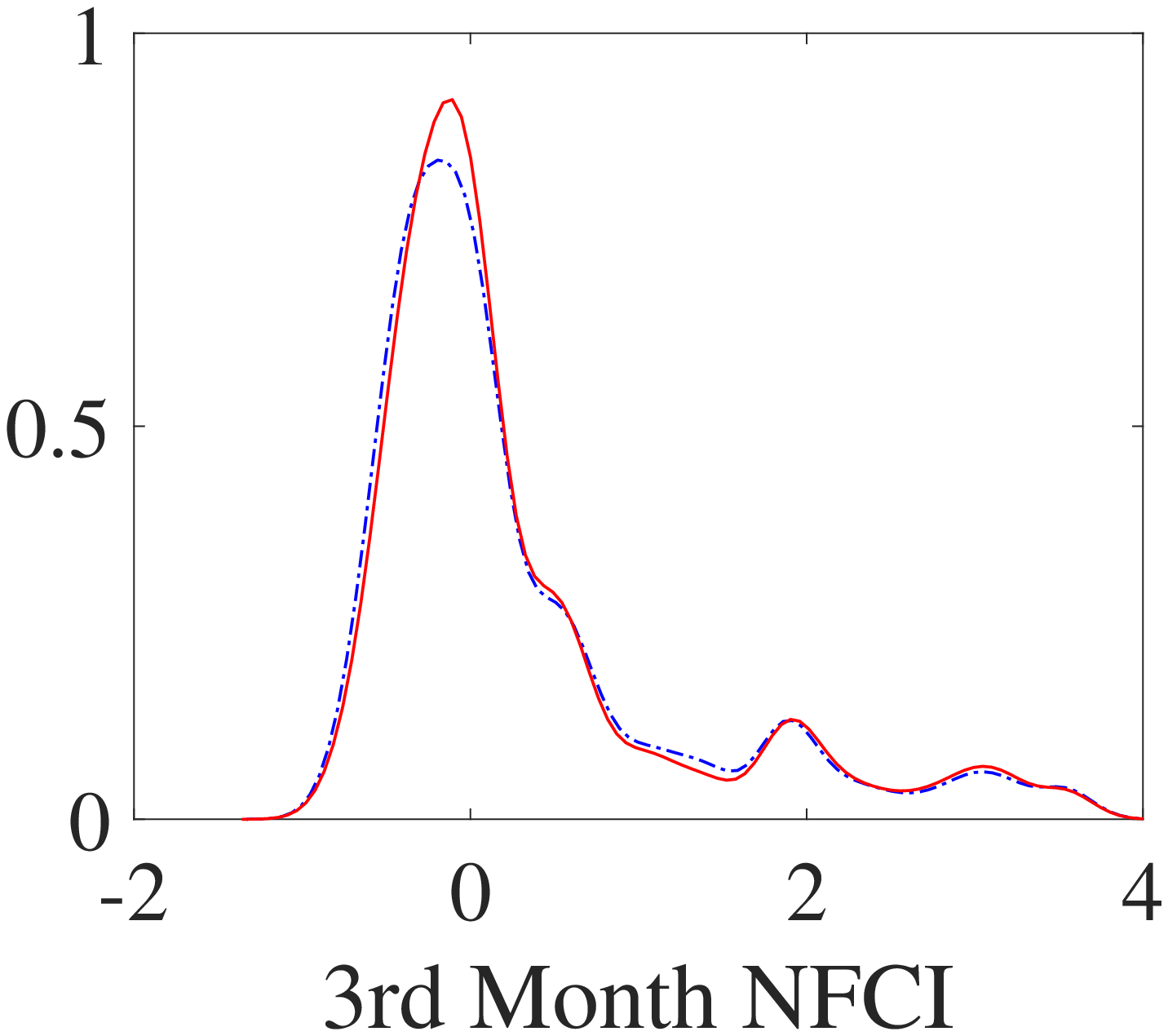}
			\end{subfigure}
			\begin{subfigure}[t]{0.24\textwidth}
				\centering	
				\includegraphics[width=0.95\textwidth]{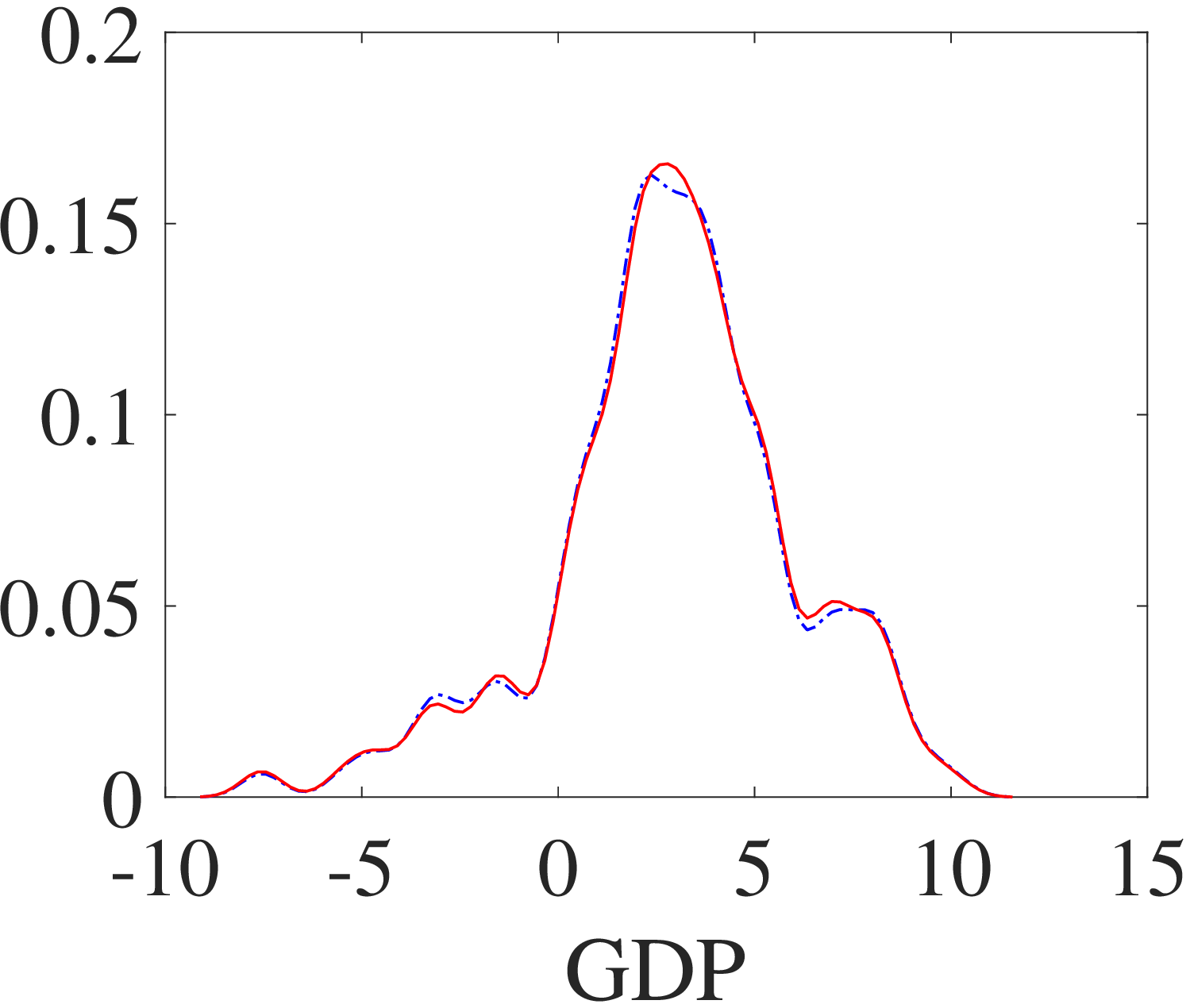}
			\end{subfigure}
		\vspace{0.2cm}
		\end{minipage}
			\begin{minipage}[t]{\textwidth}	
			\centering $h=4$\\
			\begin{subfigure}[t]{0.24\textwidth}
				\centering 		
				\includegraphics[width=0.95\textwidth]{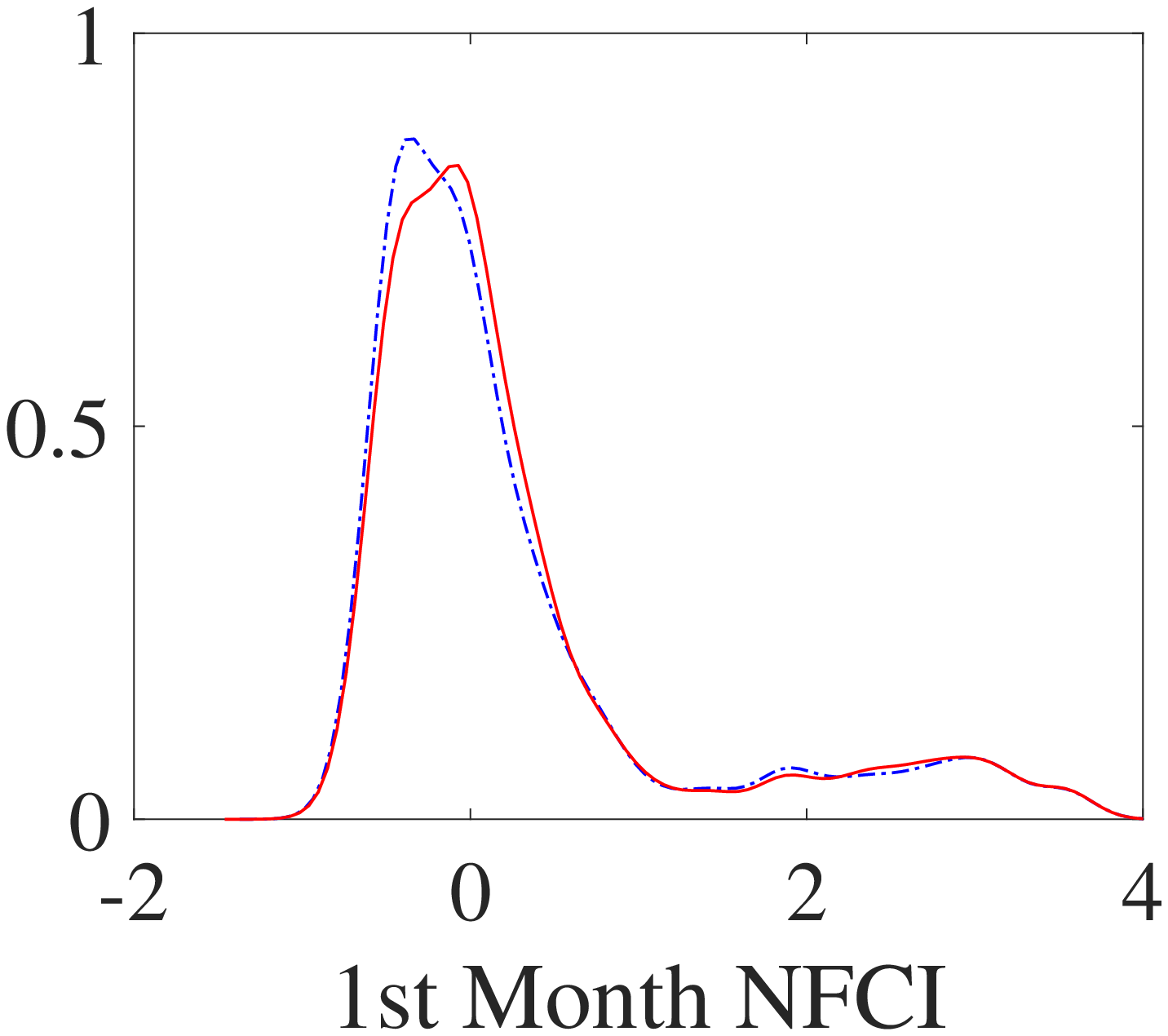}     		
			\end{subfigure}
			\begin{subfigure}[t]{0.24\textwidth}
				\centering	
				\includegraphics[width=0.95\textwidth]{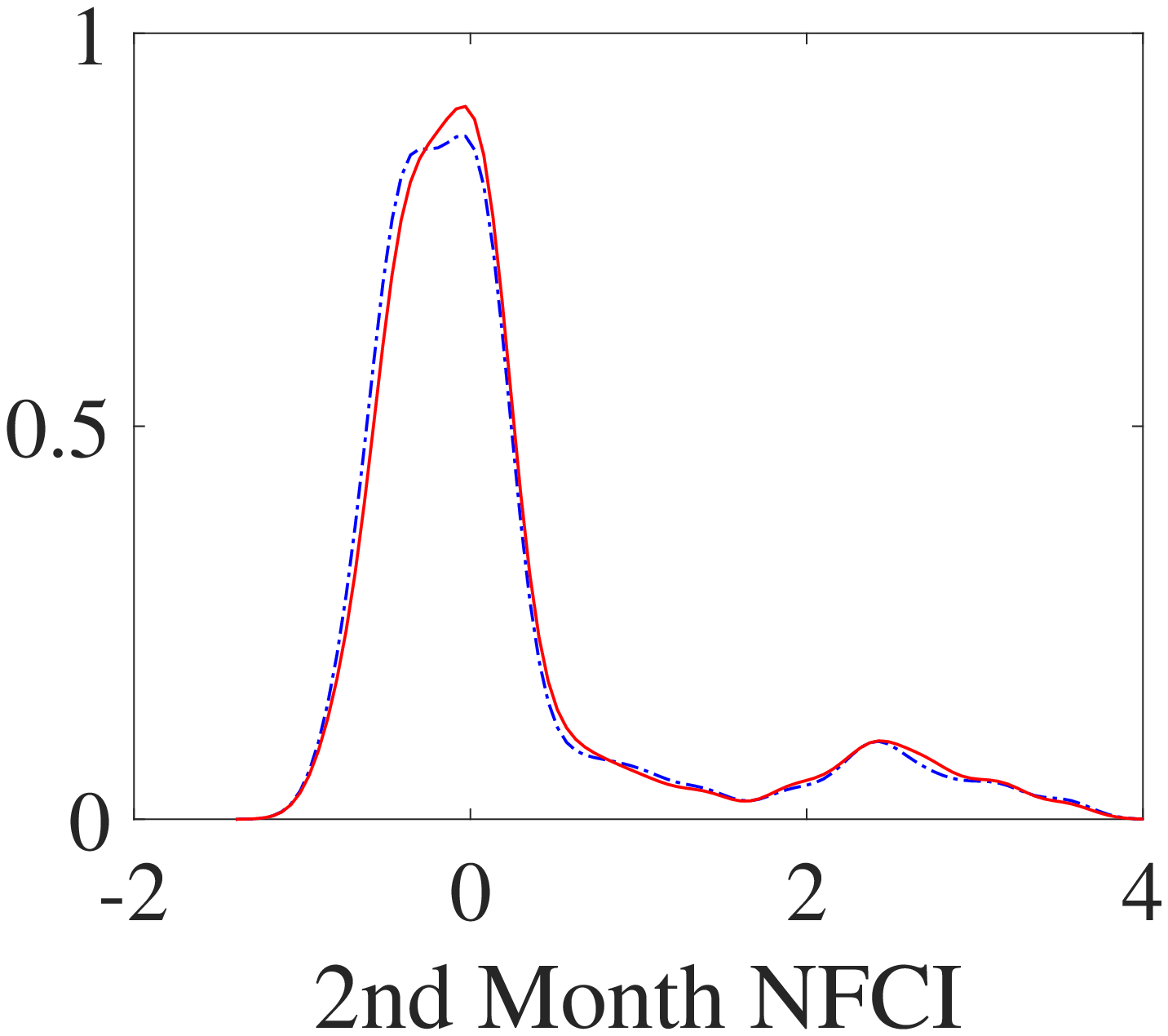}
			\end{subfigure}
			\begin{subfigure}[t]{0.24\textwidth}
				\centering	
				\includegraphics[width=0.95\textwidth]{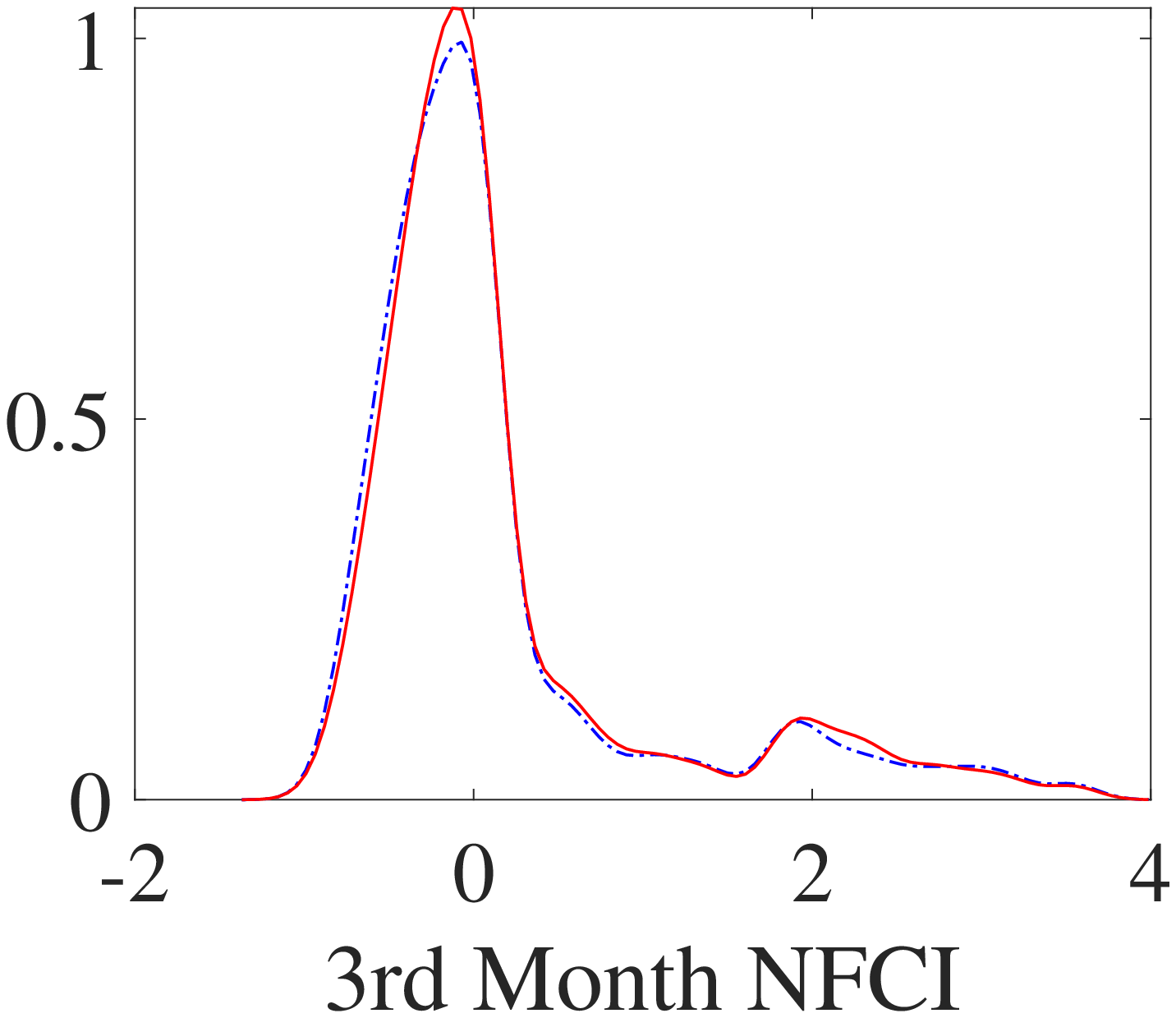}
			\end{subfigure}
			\begin{subfigure}[t]{0.24\textwidth}
				\centering	
				\includegraphics[width=0.95\textwidth]{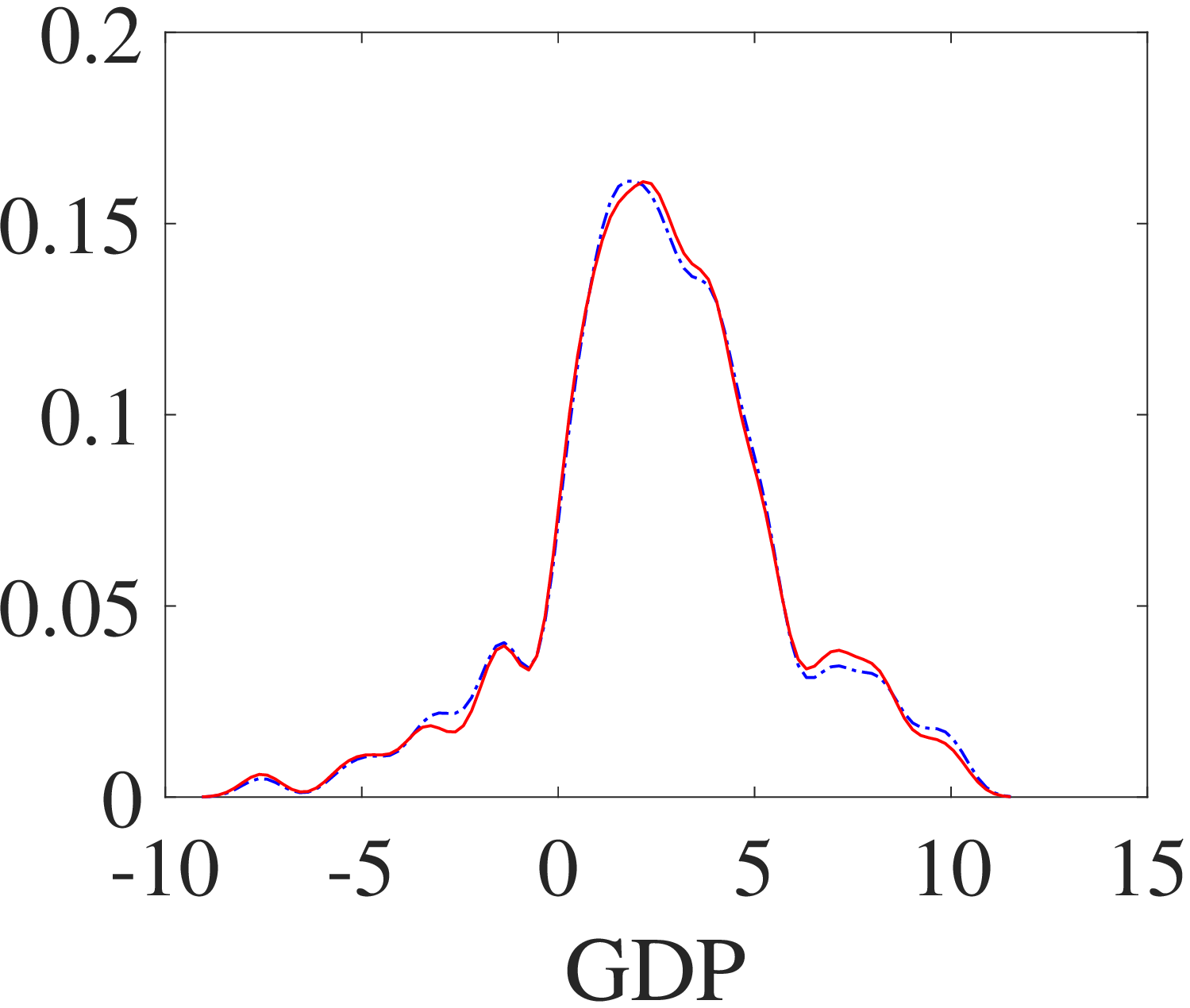}
			\end{subfigure}
		\end{minipage}
	\vspace{0.5cm}
	\begin{minipage}{.95\linewidth} 
		\linespread{1}\footnotesize
		\textit{Notes}: Refer to Figure \ref{fig: M-NFCI-response}.
	\end{minipage}
\end{figure}

To explore the impulse effect of GDP, we keep the conditional distribution of the three monthly financial conditions $Y_{jt}, j=1,2,3$ in 2008:Q4 as it is, and the same gamma counterfactual distribution is considered for real GDP growth $Y_{4t}$. The impulse response results for the entire distribution are presented in Figure \ref{fig: M-GDP-response}. Under this model, more evidence shows that the counterfactual distributions of real GDP growth at one and two quarters ahead exhibit a fatter right tail and somewhat thinner left tail. Still, the final conclusion derived from these results is consistent with those in the Bivariate model. 

	\begin{table}[H]
		\centering
		\footnotesize
		\caption{QIR to GDP Impulse\label{tab: DIR-MG-Quantile}}
		\begin{tabular}{llllllllllll}
			\hline \hline
			Quantiles & & \multicolumn{2}{c}{0.05} & \multicolumn{2}{c}{0.25} &  \multicolumn{2}{c}{0.5} &  \multicolumn{2}{c}{0.75} & \multicolumn{2}{c}{0.95}\\
			& $h$ & Base & Diff & Base & Diff & Base & Diff & Base & Diff & Base & Diff   \\
			\hline
			NFCI1 
			& 0 & 0.62  & 0.00  & 0.82  & 0.00 & 1.03  & 0.00 & 1.08 & 0.00  & 2.31 & 0.00 \\
			& 1 & -0.16 & -0.02 & 0.25  & 0.00 & 0.52  & 0.00 & 1.47 & -0.16 & 2.96 & 0.00 \\
			& 2 & -0.36 & 0.01  & -0.19 & 0.01 & 0.18  & 0.00 & 1.03 & 0.00  & 2.71 & 0.00 \\
			& 3 & -0.45 & 0.00  & -0.21 & 0.05 & 0.00  & 0.06 & 0.82 & 0.26  & 3.18 & 0.00 \\
			& 4 & -0.51 & 0.02  & -0.41 & 0.11 & -0.06 & 0.04 & 0.37 & 0.05  & 2.71 & 0.25\\
			\hline
			NFCI2 
			& 0 & 0.20  & 0.00 & 0.45  & 0.00 & 0.82  & 0.00 & 1.89 & 0.00  & 2.46 & 0.00 \\
			& 1 & -0.30 & 0.00 & 0.07  & 0.00 & 0.45  & 0.00 & 1.08 & -0.05 & 2.71 & 0.00 \\
			& 2 & -0.46 & 0.04 & -0.36 & 0.09 & 0.07  & 0.00 & 0.72 & 0.26  & 2.46 & 0.00 \\
			& 3 & -0.50 & 0.03 & -0.22 & 0.04 & -0.02 & 0.01 & 0.72 & 0.00  & 2.46 & 0.00 \\
			& 4 & -0.66 & 0.07 & -0.41 & 0.07 & -0.08 & 0.06 & 0.18 & 0.00  & 2.46 & 0.00\\
			\hline
			NFCI3 
			& 0 & -0.02 & 0.00 & 0.42  & 0.00 & 0.55  & 0.00 & 1.89 & 0.00  & 3.18 & 0.00 \\
			& 1 & -0.41 & 0.05 & -0.02 & 0.01 & 0.55  & 0.00 & 1.31 & -0.23 & 2.71 & 0.00 \\
			& 2 & -0.38 & 0.00 & -0.32 & 0.02 & 0.00  & 0.00 & 0.82 & 0.21  & 2.96 & 0.00 \\
			& 3 & -0.56 & 0.03 & -0.32 & 0.09 & -0.02 & 0.01 & 0.52 & -0.10 & 2.46 & 0.00 \\
			& 4 & -0.67 & 0.05 & -0.36 & 0.08 & -0.06 & 0.00 & 0.08 & 0.09  & 2.31 & 0.00\\
			\hline
			GDP 
			& 0 & -3.20 & 4.65 & 0.50  & 1.95 & 1.90 & 1.45 & 3.10 & 1.40 & 6.48 & 0.00 \\
			& 1 & -3.60 & 0.40 & -0.55 & 0.58 & 1.15 & 0.25 & 2.80 & 0.40 & 6.88 & 1.15 \\
			& 2 & -3.20 & 0.00 & 0.50  & 0.73 & 2.30 & 0.90 & 4.50 & 0.25 & 7.60 & 0.42 \\
			& 3 & -3.20 & 0.00 & 1.40  & 0.10 & 3.10 & 0.00 & 4.75 & 0.00 & 8.15 & 0.00 \\
			& 4 & -2.70 & 0.00 & 1.15  & 0.00 & 2.45 & 0.00 & 4.10 & 0.00 & 8.15 & 0.00\\
			\hline         
		\end{tabular}
		\begin{minipage}{.85\linewidth} 
			\linespread{1}\footnotesize
			\textit{Notes}: Refer to Table \ref{tab: DIR-MN-Quantile}.
		\end{minipage}
	\end{table}   
	
	\begin{table}[H]
		\centering
		\footnotesize
		\caption{MIR to GDP Impulse\label{tab: DIR-MG-Moment}}
		\begin{tabular}{llllllllll}
			\hline \hline
			Quantiles& & \multicolumn{2}{c}{Mean} & \multicolumn{2}{c}{Std} &  \multicolumn{2}{c}{Skewess} &  \multicolumn{2}{c}{Kurtosis} \\
			& $h$ & Base & Diff & Base & Diff & Base & Diff & Base & Diff   \\
			\hline
			NFCI1 
			& 0 & 1.17 & 0.00  & 0.48 & 0.00  & 1.62 & -0.02 & 5.60 & -0.08 \\
			& 1 & 0.94 & -0.02 & 1.01 & 0.00  & 1.05 & 0.06  & 2.86 & 0.03  \\
			& 2 & 0.53 & 0.01  & 0.97 & -0.01 & 1.40 & 0.02  & 4.24 & 0.10  \\
			& 3 & 0.47 & 0.00  & 1.03 & -0.03 & 1.52 & 0.03  & 4.44 & 0.19  \\
			& 4 & 0.26 & 0.06  & 0.97 & 0.02  & 1.85 & -0.05 & 5.56 & -0.32\\
			\hline
			NFCI2 
			& 0 & 1.09 & 0.00  & 0.78 & 0.01  & 0.76 & 0.01  & 2.42 & 0.01  \\
			& 1 & 0.75 & -0.01 & 0.97 & -0.02 & 1.08 & 0.07  & 2.95 & 0.21  \\
			& 2 & 0.43 & 0.07  & 0.99 & 0.02  & 1.28 & -0.15 & 3.51 & -0.39 \\
			& 3 & 0.35 & 0.01  & 0.97 & -0.02 & 1.57 & 0.10  & 4.60 & 0.35  \\
			& 4 & 0.17 & 0.05  & 0.93 & 0.01  & 1.99 & -0.07 & 6.23 & -0.40\\
			\hline
			NFCI3 
			& 0 & 1.05 & 0.01  & 0.98 & 0.02  & 0.87 & 0.01  & 2.68 & -0.01 \\
			& 1 & 0.76 & -0.01 & 0.97 & -0.03 & 0.97 & 0.07  & 3.21 & 0.26  \\
			& 2 & 0.46 & 0.06  & 1.02 & 0.03  & 1.44 & -0.13 & 4.20 & -0.48 \\
			& 3 & 0.29 & 0.01  & 0.94 & -0.01 & 1.69 & 0.17  & 5.22 & 0.54  \\
			& 4 & 0.15 & 0.04  & 0.88 & -0.02 & 1.97 & -0.11 & 6.32 & -0.55\\
			\hline
			GDP 
			& 0 & 1.77 & 3.93 & 2.68 & -1.04 & 0.06  & 0.91  & 4.33 & -1.06 \\
			& 1 & 1.22 & 0.45 & 2.81 & 0.34  & 0.33  & -0.05 & 3.83 & -0.75 \\
			& 2 & 2.27 & 0.59 & 3.17 & -0.11 & -0.43 & -0.17 & 3.48 & 0.46  \\
			& 3 & 2.95 & 0.10 & 3.08 & -0.06 & -0.49 & -0.03 & 3.92 & 0.18  \\
			& 4 & 2.61 & 0.01 & 2.99 & -0.02 & -0.16 & 0.00  & 3.76 & 0.09 \\
			\hline         
		\end{tabular}
		\begin{minipage}{.72\linewidth} \footnotesize
			\textit{Notes}: Refer to Table \ref{tab: DIR-MN-Moment}.
		\end{minipage}
	\end{table}

	\subsection{Confidence Bands based on Moving Block Bootstrap}\label{sec: appendix-B5}
	Moving block bootstrap is a nonparametric bootstrap procedure that can be applied to dependent time series observations. It consists in drawing blocks of fixed length randomly with replacement from the blocks of consecutive data, which can thus accounts for conditional heteroskedasticity.
	In this application, setting the length for each block is 8, we use 500 bootstrap replications to construct $95\%$ pointwise confidence intervals of the contempreneous correlation coefficients and DIRFs.
	\subsubsection{Distribution and Density IRFs}	
	\begin{figure}[H]
		\captionsetup[subfigure]{aboveskip=-2pt,belowskip=-5pt}
		\centering
		\caption{Confidence Intervals of DIRF for NFCI Impulse (Bivariate Model)}  		
		\includegraphics[width=0.9\textwidth]{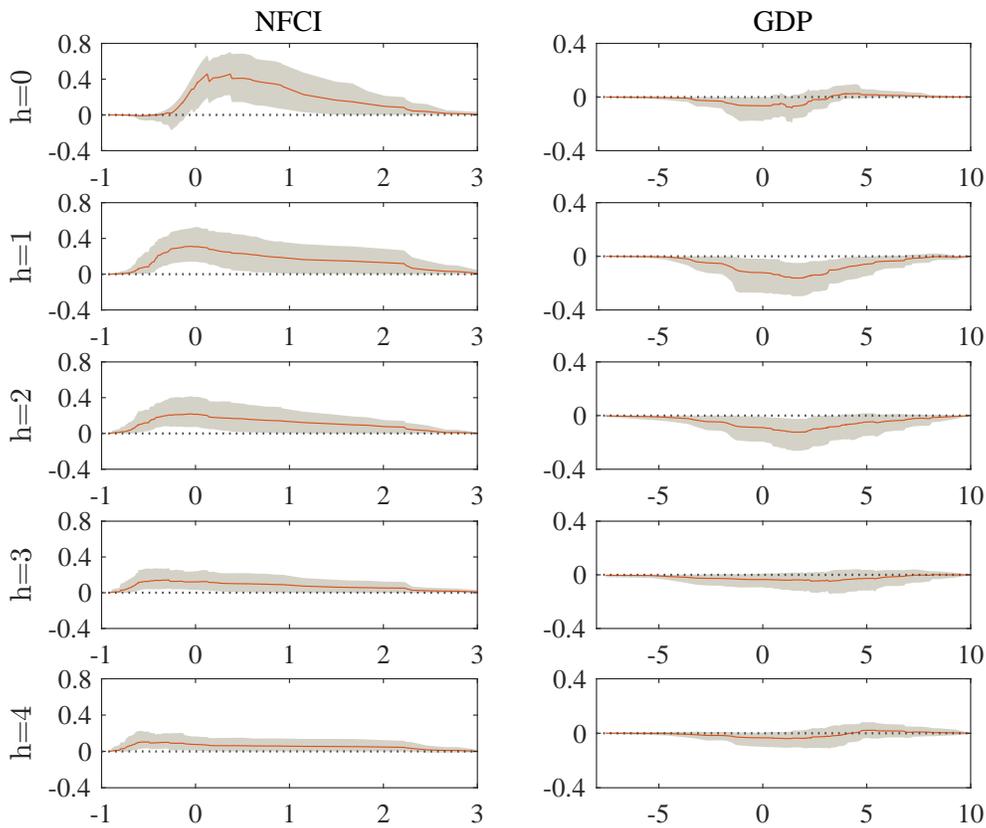}     	
   \end{figure}	

	\begin{figure}[H]
		\captionsetup[subfigure]{aboveskip=-2pt,belowskip=-5pt}
		\centering
		\caption{Confidence Intervals of Density IRF for NFCI Impulse (Bivariate Model)} 
		\includegraphics[width=0.9\textwidth]{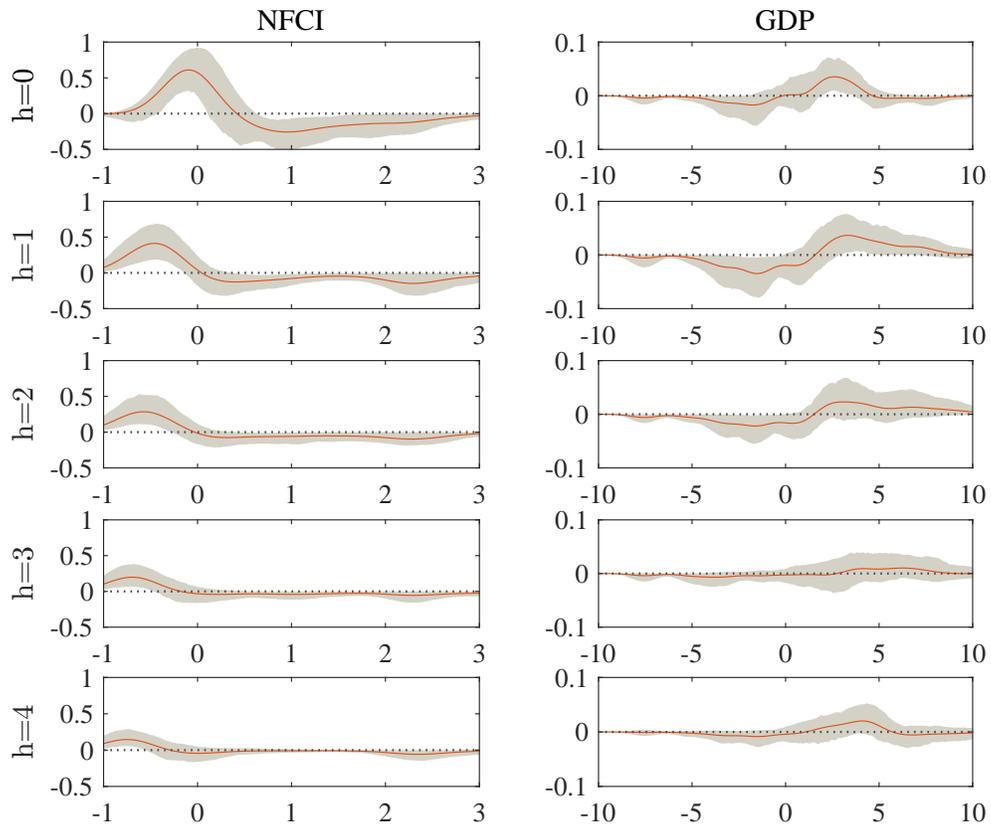}
	\end{figure}

	\begin{figure}[H]
		\captionsetup[subfigure]{aboveskip=-2pt,belowskip=-8pt}
		\centering
		\caption{Confidence Intervals of DIRFs for NFCI Impulse (Mixed-Frequency Model)} 
		\begin{subfigure}[t]{0.9\textwidth}
			\centering
			\caption{Distribution IRFs}     		
			\includegraphics[width=0.9\textwidth]{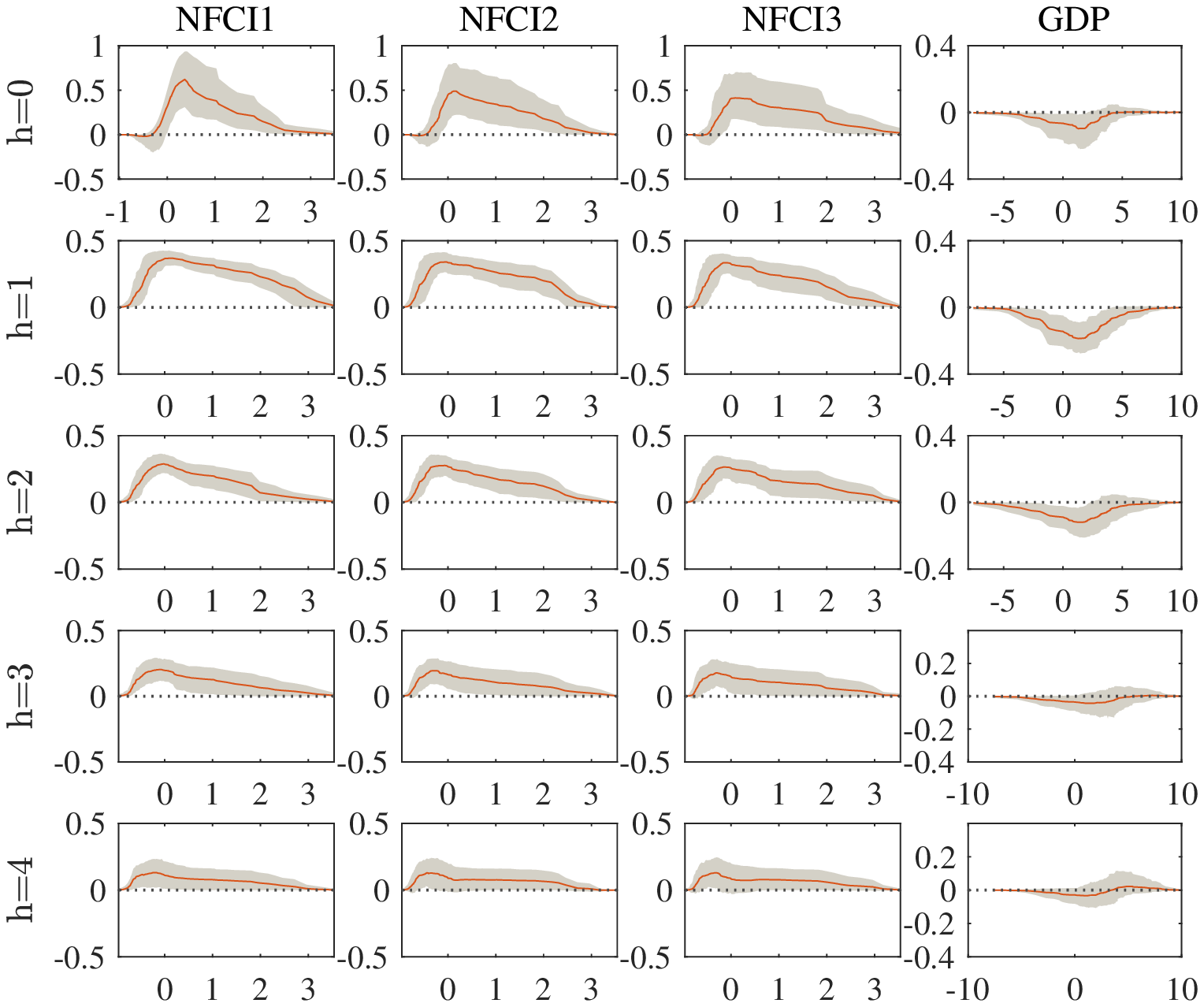}     		
		\end{subfigure}
		\begin{subfigure}[t]{0.9\textwidth}
			\centering
			\caption{Density IRFs}		
			\includegraphics[width=0.9\textwidth]{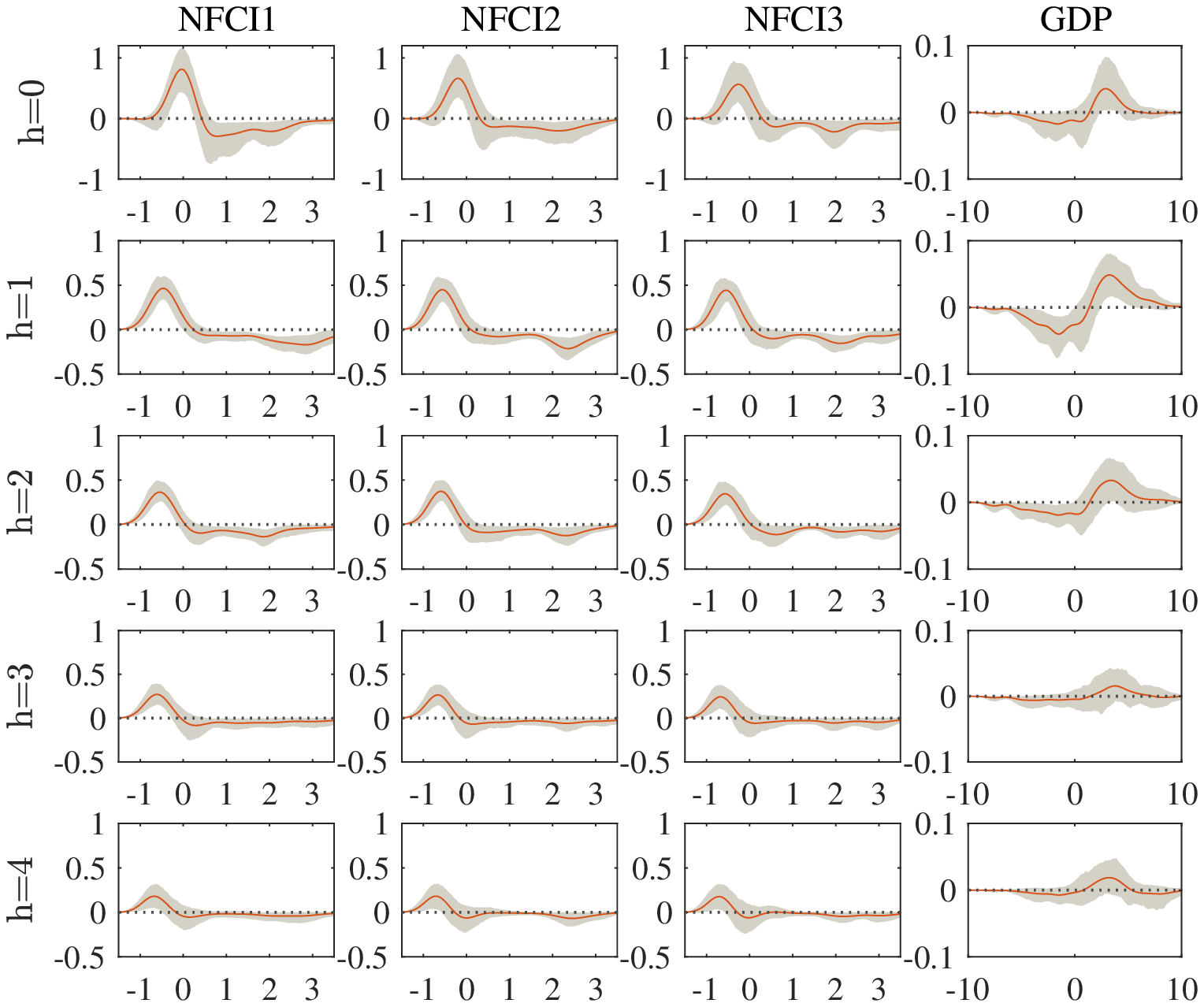}
		\end{subfigure}
	\end{figure}
	
	\begin{figure}[H]
		\captionsetup[subfigure]{aboveskip=-2pt,belowskip=-8pt}
		\centering
		\caption{Confidence Intervals of DIRFs for GDP Impulse (Bivariate Model)} 
		\begin{subfigure}[t]{0.9\textwidth}
			\centering
			\caption{Distribution IRFs}     		
			\includegraphics[width=0.9\textwidth]{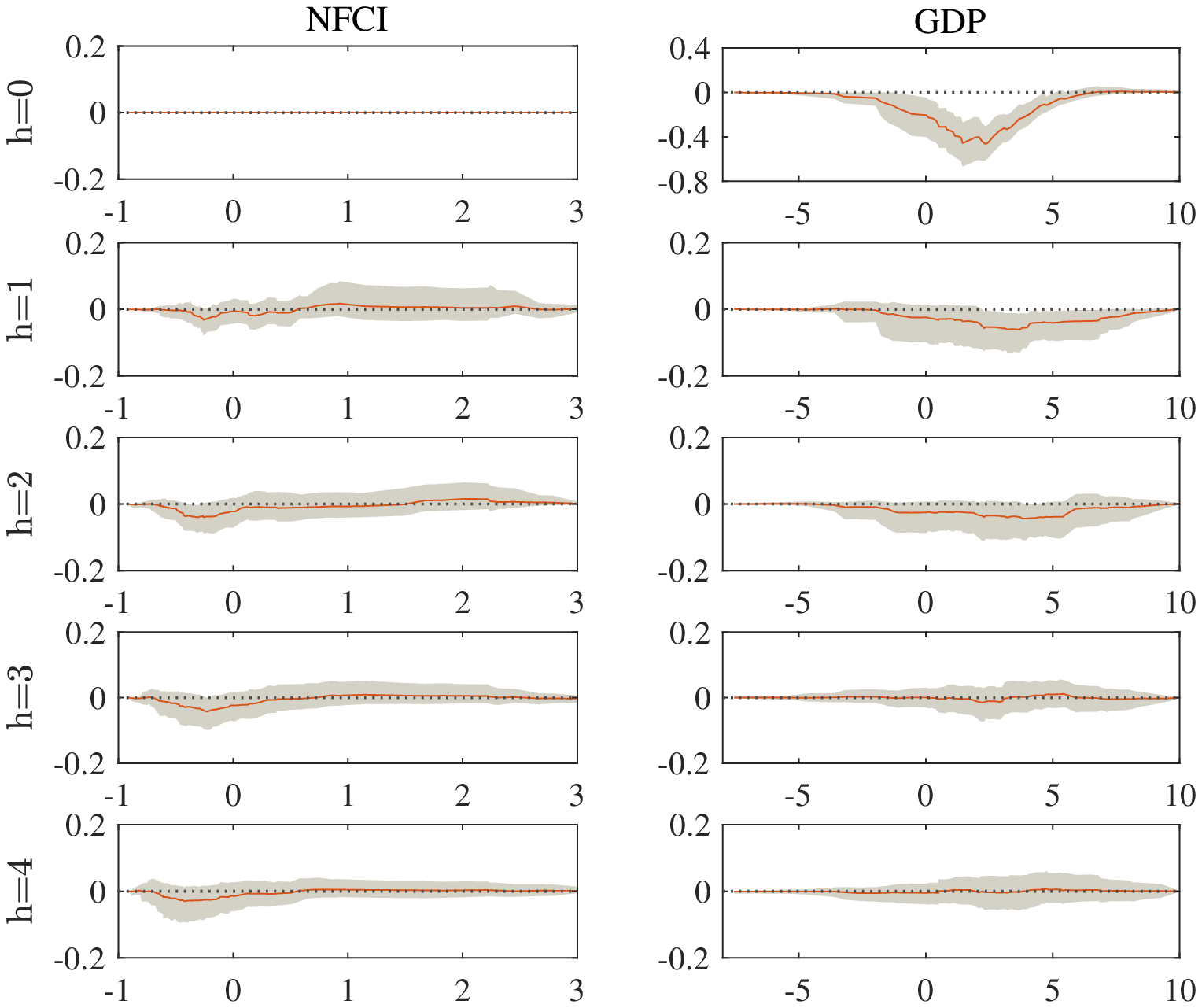}     		
		\end{subfigure}
		\begin{subfigure}[t]{0.9\textwidth}
			\centering
			\caption{Density IRFs}		
			\includegraphics[width=0.9\textwidth]{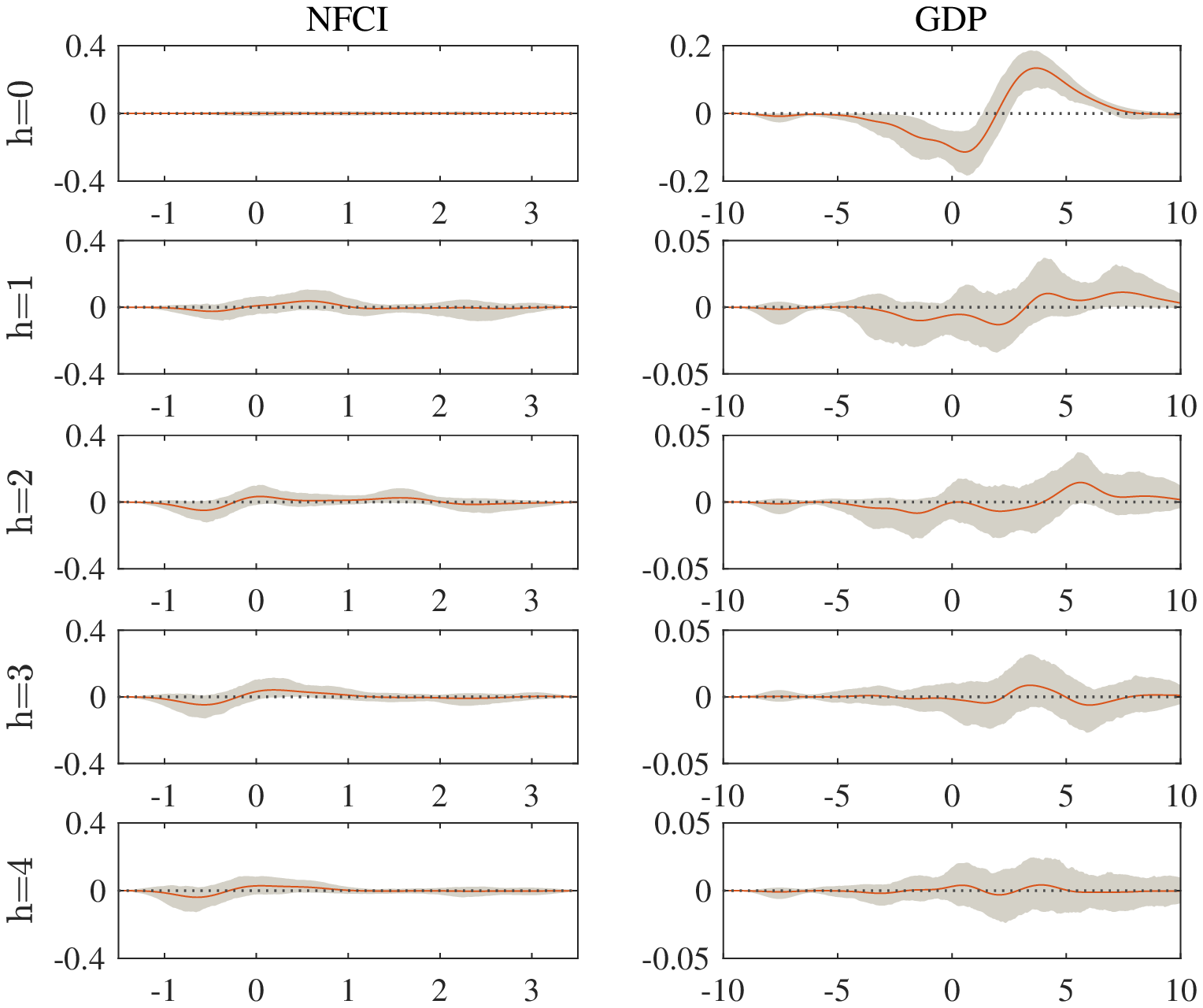}
		\end{subfigure}
	\end{figure}
	
	\begin{figure}[H]
		\captionsetup[subfigure]{aboveskip=-2pt,belowskip=-8pt}
		\centering
		\caption{Confidence Intervals of DIRFs for GDP Impulse (Mixed-Frequency Model)} 
		\begin{subfigure}[t]{0.9\textwidth}
			\centering
			\caption{Distribution IRFs}     		
			\includegraphics[width=0.9\textwidth]{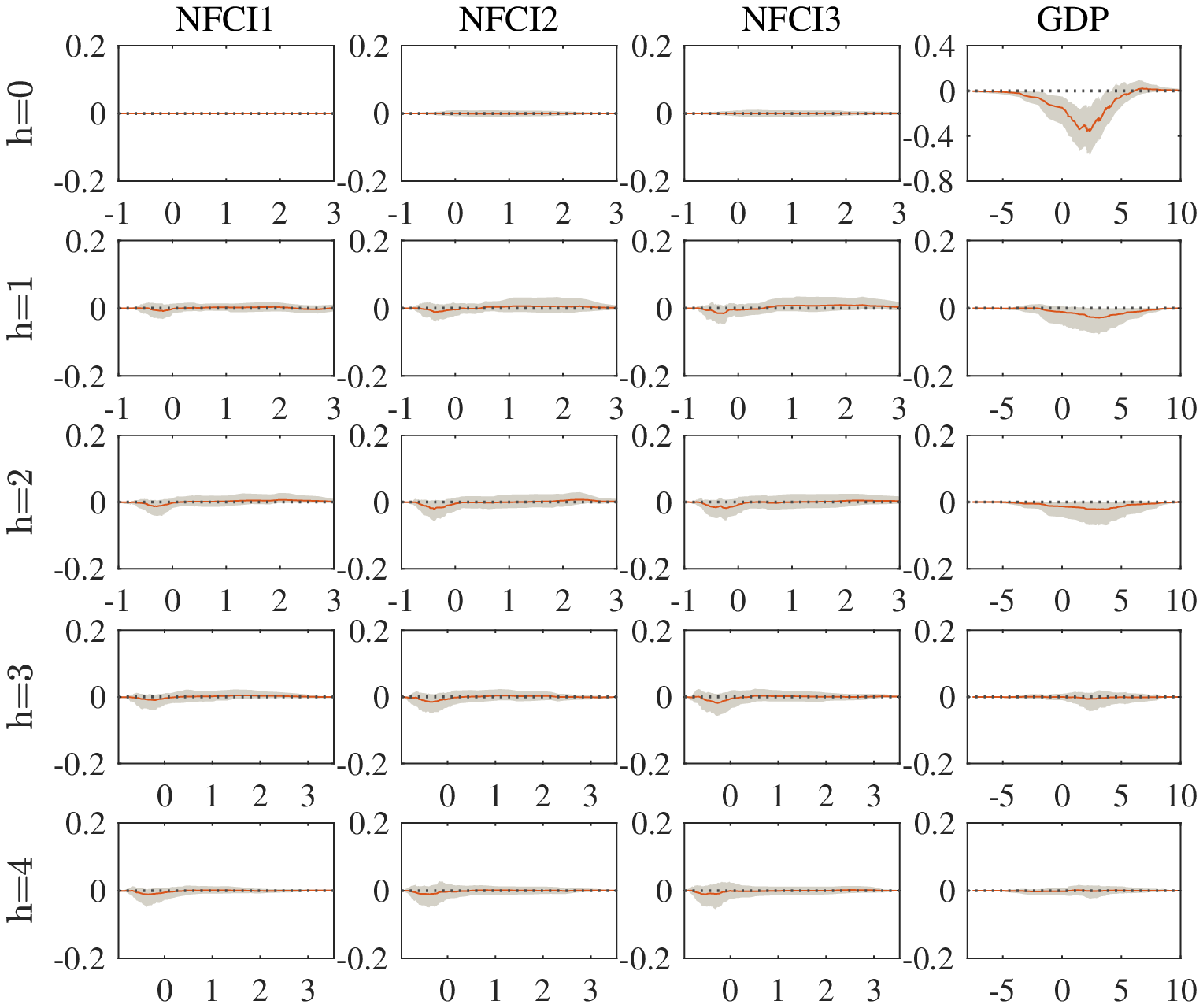}     		
		\end{subfigure}
		\begin{subfigure}[t]{0.9\textwidth}
			\centering
			\caption{Density IRFs}		
			\includegraphics[width=0.9\textwidth]{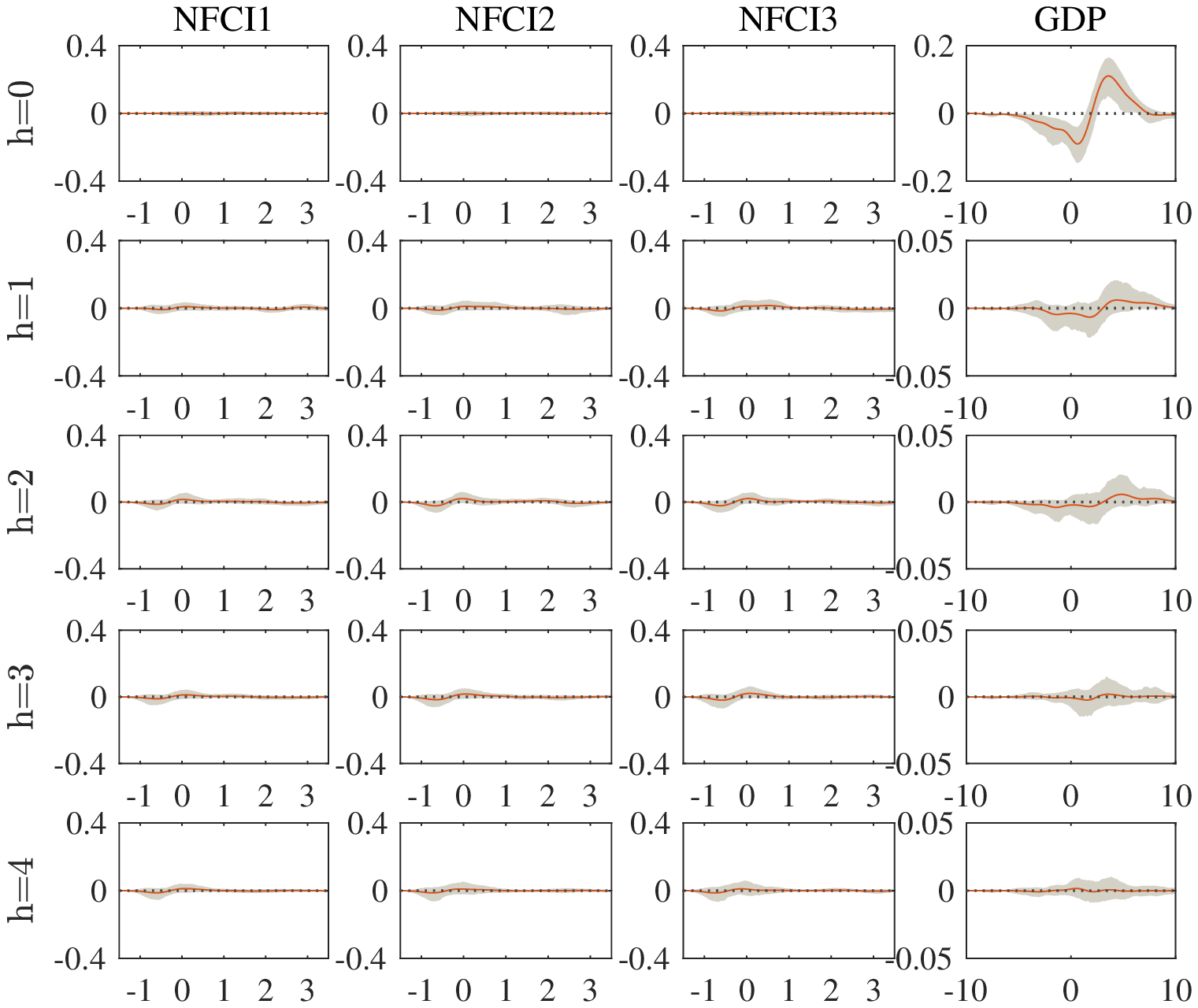}
		\end{subfigure}
	\end{figure}
 
 \subsubsection{Contempreneous Correlation Coefficient}
      \begin{figure}[H]
      	\captionsetup[subfigure]{aboveskip=-3pt,belowskip=0pt}
      	\centering
      	\caption{In-sample Contempreneous Correlation between NFCI and real GDP Growth\label{fig: correlation_boots}}
      	\includegraphics[width=0.8\textwidth]{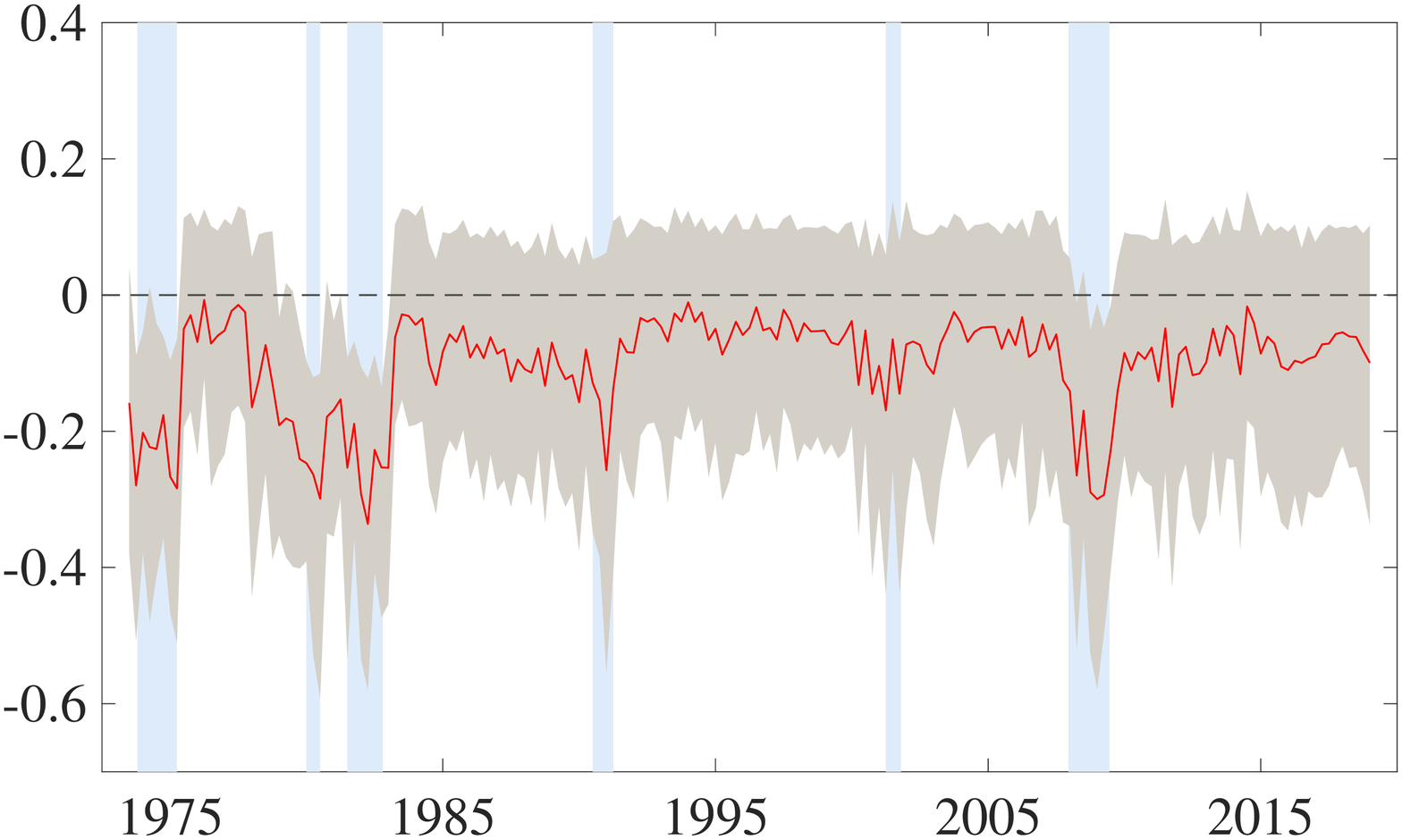}    
      	\begin{minipage}{.7\linewidth} 
      		\linespread{1}\footnotesize
      		\textit{Notes}: The figure plots the correlation coeficients based on the one-step-ahead forecasting distributions. The vertical shaded areas indicate U.S. recessions, and horizontal shaded area represents the 95\% interval.
      	\end{minipage} 
      \end{figure}



\end{document}